\documentclass[fleqn,10pt]{article}
\usepackage{latexsym, graphicx, epsfig, amsmath, notation, amssymb,amsfonts}
\usepackage{natbib,amsthm,version}
\usepackage{amsbsy,bm,multirow}
\usepackage[titletoc,page]{appendix}
\usepackage[mathscr]{eucal}
\usepackage{enumitem,color}
\usepackage{subfigure}

\newtheorem{theorem}{Theorem}[section]

\newtheorem{remark}{Remark}

\theoremstyle{definition}

\title{
Energy-Stable Boundary Conditions
Based on a Quadratic Form:
Applications to Outflow/Open-Boundary Problems in Incompressible Flows
} 
\author{
  Naxian Ni, Zhiguo Yang, Suchuan Dong\thanks{Author of correspondence;
    Email: sdong@purdue.edu} \\
  Center for Computational \& Applied Mathematics \\
  Department of Mathematics \\
  Purdue University \\
  West Lafayette, IN 47907, USA
 } 

\date{} %(\today)}
\begin{document}
\maketitle

%% double space
%\baselineskip 2em %2.2em

\graphicspath{{Figures/Cyl_2D/Yang/}, {Figures/Cyl_2D/}}

%%%%%%%%%%%%%%%%%%%%%%%%%%%%%%%%%%%%%%%%%%%%%%%%%%
%% Abstract

\begin{abstract}

We present a set of new energy-stable open boundary conditions
for tackling the backflow instability in simulations of
outflow/open boundary problems for incompressible flows.
These boundary conditions are developed through two steps:
(i) devise a general form of boundary conditions that ensure
the energy stability
by re-formulating the boundary contribution into a quadratic
form in terms of a symmetric matrix and computing an associated
eigen problem; and
(ii) require that, upon imposing the boundary conditions from the previous step, 
the scale of boundary dissipation should match a physical scale. 
These open boundary conditions can be re-cast into
the form of a traction-type condition, and therefore they can be
implemented numerically using the splitting-type algorithm from a 
previous work. The current boundary conditions can effectively
overcome the backflow instability typically encountered at moderate
and high Reynolds numbers.
These boundary conditions 
in general give rise to a non-zero traction on the entire 
open boundary, unlike previous related methods which only take effect
in the backflow regions of the boundary.
Extensive numerical experiments in two and three dimensions
are presented to test the effectiveness and 
performance of the presented methods, and simulation results are compared
with the available experimental data to demonstrate their accuracy.

\end{abstract}

%%%%%%%%%%%%%%%%%%%%%%%%%%%%%%%%%%%%%%%%%%%%%%%%%%

\vspace{0.05cm}

%%%
\section{Introduction}
\label{sec:intro}

Outflow/open-boundary problems are an important and challenging class of  
problems for incompressible flow simulations.
Several types of flows that are of
practical engineering/biological significance and
fundamental physical interest belong to this class,
such as wakes, jets,
shear layers,  cardiovascular and
respiratory flows.
The predominant challenge in the numerical simulations
of such problems lies in the treatment of the outflow
or open boundary~\cite{Gresho1991,SaniG1994}. If the Reynolds number is low,
a number of types of open/outflow boundary conditions (OBC)
can work well and lead to reasonable simulation results. 
But when the Reynolds number increases beyond some moderate
value, typically close to $2000$ (which can be as low as
several hundred depending on
the flow geometry), the so-called
backflow instability (see e.g.~\cite{Dong2015obc})
will become a severe issue, and
many open boundary conditions that work well for low Reynolds numbers
cease to work.
Backflow instability refers to the numerical instability
caused by the un-controlled energy influx into the domain
through the open/outflow boundary, often associated
with strong vortices or backflows on such boundaries.
A telltale symptom of this instability
is that an otherwise stable computation blows up instantly
when a vortex reaches the open/outflow
boundary~\cite{DongK2005,DongKER2006,VargheseFF2007,GhaisasSF2015}. 
It is observed that usual measures such as increasing
the mesh resolution or reducing the time step size
do not help with this instability~\cite{DongKC2014}.

% methods that can work and are effective
% common property: enforce certain energy stability
%   provide some control over energy influx

To tackle the backflow instability, the energy influx into
the domain through the open boundary, if any, 
must be controlled in some fashion. 
Employing a large enough computational domain
such that vortices can be sufficiently dissipated before
reaching the outflow/open boundary, or artificially
increasing the viscosity in a region near/at
the outflow boundary (so-called sponge) such
that vortices can be smoothed out or sufficiently weakened, 
are some measures in actual simulations.
These measures may not be desirable in terms of
e.g.~the increased computational cost due to
the larger domain or the negative influence on the accuracy
due to the artificially modified viscosity
in regions of the flow, and additionally they may not always be 
effective with the increase of Reynolds number.  

How to algorithmically control the energy influx through
the open boundary by devising effective open
boundary conditions seems a more attractive approach.
A number of researchers have contributed to
this area, and there appears to be a surging interest in recent years. 
In the early works (see e.g.~\cite{BruneauF1994,BruneauF1996})
the traction on the open boundary is modified
to include a term $\frac{1}{2}(\mathbf{n}\cdot\mathbf{u})^{-}\mathbf{u}$,
where $\mathbf{u}$ and $\mathbf{n}$ are respectively
the velocity on and the directional unit vector of
the boundary, and $(\mathbf{n}\cdot\mathbf{u})^{-}$
is equal to $\mathbf{n}\cdot\mathbf{u}$ if 
$\mathbf{n}\cdot\mathbf{u}<0$ and zero otherwise.
This open boundary condition has also appeared or is studied in
some later works; 
see e.g.~\cite{LanzendorferS2011,FeistauerN2013,Fouchet2014,BraackM2014}
among others. A variant of this form,
with a term $(\mathbf{n}\cdot\mathbf{u})^{-}\mathbf{u}$
in the traction (without the $1/2$ factor), has been investigated in
a number of 
works (see e.g.~\cite{BazilevsGHMZ2009,Moghadametal2011,PorporaZVP2012,GravemeierCYIW2012,IsmailGCW2014}).
It is noted that in \cite{Moghadametal2011} a form 
$\beta (\mathbf{n}\cdot\mathbf{u})^{-}\mathbf{u}$, with a constant $0<\beta<1$,
has also been considered.
%
% DongKC(2014)
In \cite{DongKC2014} an open boundary condition with a
modified traction term $\left[\frac{1}{2}|\mathbf{u}|^2\mathbf{n}\right]\Theta_0(\mathbf{n},\mathbf{u})$
is suggested, where $\Theta_0(\mathbf{n},\mathbf{u})$ is a smoothed step function
essentially taking the unit value if $\mathbf{n}\cdot\mathbf{u}<0$
and vanishing otherwise. So this additional traction term only takes effect in
regions of backflow on the open boundary, and has no effect in normal outflow regions
or if no backflow is present.
Note that this is very different from the total
pressure ($p+\frac{1}{2}|\mathbf{u}|^2$, where $p$ denotes the normalized static pressure) 
as discussed in e.g.~\cite{HeywoodRT1996},
which can lead to a similar term $\frac{1}{2}|\mathbf{u}|^2\mathbf{n}$ on
the boundary. Unlike that of \cite{DongKC2014}, 
this term in the total pressure 
precludes the energy from exiting the domain even in normal outflow situations, 
which results in poor and unphysical simulation results~\cite{HeywoodRT1996}.
In contrast, the boundary condition of
\cite{DongKC2014} has been shown to
ensure the energy stability on the open boundary
%, even in the presence of
%strong vortices or backflows,
and that it produces accurate simulation results
for outflow problems. 
%
% Dong and Shen (2015)
In \cite{DongS2015} a general form of open boundary conditions that
ensure the energy stability on the open boundary for
incompressible flows has been proposed. This form contains 
those of \cite{BruneauF1994,BazilevsGHMZ2009,GravemeierCYIW2012,IsmailGCW2014,DongKC2014}
as particular cases. More importantly, the general form 
suggests other forms of energy-stable open boundary conditions
involving terms such as
$
\frac{1}{4}\left[|\mathbf{u}|^2\mathbf{n} + (\mathbf{n}\cdot\mathbf{u})\mathbf{u} \right]
\Theta_0(\mathbf{n},\mathbf{u}),
$
$
\frac{1}{2}\left[|\mathbf{u}|^2\mathbf{n} + (\mathbf{n}\cdot\mathbf{u})\mathbf{u} \right]
\Theta_0(\mathbf{n},\mathbf{u}),
$
and
$
\left[|\mathbf{u}|^2\mathbf{n} \right]
\Theta_0(\mathbf{n},\mathbf{u}).
$
Several of these forms  have been studied in detail
in \cite{DongS2015}.
It is observed that 
the term $(\mathbf{n}\cdot\mathbf{u})\mathbf{u}$
(with or without the $1/2$ factor)
in the OBC tends to cause the vortices to move laterally
as they cross the open boundary, while the term
$|\mathbf{u}|^2\mathbf{n}$ tends to have the effect of
squeezing the vortices along the direction
normal to the open boundary.
%On the other hand, when both of these two terms are present, they appear
%able to somewhat counteract the negative effects on the vortices
%caused by each individual form,
%and allow the vortices to discharge from the domain in a more natural 
%way~\cite{DongS2015}. The OBC with the form
%$
%\frac{1}{2}\left[|\mathbf{u}|^2\mathbf{n} + (\mathbf{n}\cdot\mathbf{u})\mathbf{u} \right]
%\Theta_0(\mathbf{n},\mathbf{u})
%$
%is recommended in \cite{DongS2015} as the most favorable
%among the several forms investigated therein.
%
% Dong (2015)
In \cite{Dong2015obc} a convective-like energy-stable open boundary condition
is proposed, which contains an inertia term (velocity time-derivative)
and represents a Newton's second-law type relation on the open boundary.
Under certain situations it can be reduced to a form that is reminiscent of
the usual convective boundary condition, hence the name ``convective-like''
condition. This OBC not only ensures the energy stability
but also provides a control over the velocity on the outflow/open
boundary. It is observed in~\cite{Dong2015obc} that
%while ensuring stability in the presence of backflows on the open boundary, 
the inertia term in this OBC allows the vortices to 
discharge from the domain in a more smooth and natural fashion,
when compared with the previous energy-stable OBCs without the inertia 
term (see e.g.~those of \cite{DongS2015}).
A generalization of this condition to other forms of
convective-like energy-stable OBCs has also been provided in~\cite{Dong2015obc}.
%
% tangential velocity penalization
Besides the above methods, other open boundary conditions
that can work with the backflow instability also exist. We refer
the reader to e.g.~those of~\cite{BertoglioC2014,BertoglioC2016} which are given
based on a weak formulation of the Navier-Stokes equations, and also
to~\cite{Bertoglioetal2018} for a recent study of several methods
in the context of physiological flows.
%
% two-phase and multiphase OBCs
We also refer to~\cite{Dong2014obc,DongW2016,YangD2018}
for methods dealing with two-phase and multiphase outflows and
open boundaries and related issues.

% Nordstrom's strategy
% introduce what we are going to do in this paper

The principle 
for addressing the backflow instability issue lies in the management and control
of the boundary contribution of the open boundary 
to the energy balance of the system. 
A key strategy for achieving this 
is to devise boundary conditions such that the boundary contribution 
to the energy balance is dissipative (i.e.~negative semi-definite)
on the open/outflow boundary. This strategy has been employed
in the developments of \cite{DongKC2014,DongS2015,Dong2015obc}
and several other afore-mentioned methods.
Recently, a more systematic roadmap to formulating boundary conditions
to ensure the definiteness of the boundary contribution (in
the context of compressible Navier-Stokes equations) is proposed
in \cite{Nordstrom2017}. This roadmap involves three main steps:
(i) reformulate the boundary contribution into a quadratic form in terms of
a symmetric matrix, (ii) rotate the variables to diagonalize
the matrix, and (iii) formulate the boundary condition
in the form of the eigen-variables corresponding to the 
negative eigenvalues expressed in terms of the eigen-variables
corresponding to the positive eigenvalues.
This procedure is very recently applied in \cite{NordstromC2018}
to the incompressible
Navier-Stokes equations in two dimensions  to 
investigate the boundary conditions on solid walls and far fields that
can bound the energy of the system.
 
% what are you going to do in this paper?
% how is it different from Nordstrom and Cognata (2018)?

Inspired by the works of~\cite{Nordstrom2017,NordstromC2018},
we develop in this paper a set of new open/outflow boundary conditions 
for tackling the backflow instability for incompressible
flows in two and three dimensions
based on the procedure of \cite{Nordstrom2017}.
By formulating the boundary integral term in the energy balance equation
into a quadratic form involving a symmetric matrix,
we have derived a general form of boundary conditions
that ensure the energy dissipation on the open boundary.
It should be pointed out that, due to differences in the formulation
of the quadratic form and the symmetric matrix involved therein,  
the energy-stable boundary conditions obtained here are different from those of
\cite{NordstromC2018}, even though the procedure used for deriving 
the boundary conditions is similar.
%in the current work and \cite{NordstromC2018}.

%
% formulation into traction form
% importance of dissipation control

We find that the  energy-stable boundary conditions as devised above based on
the quadratic form can be re-formulated equivalently into 
a traction-type condition similar to those of \cite{DongKC2014,DongS2015},
albeit involving a different traction term.
More importantly, we observe that the boundary conditions 
as obtained above in general give rise to poor or even unphysical
results in numerical simulations of outflow problems,
even though the computations are indeed stable,
unless the algorithmic parameters take certain values for the given flow problem
under study.
We further observe that the values for the 
algorithmic parameters that can lead to ``good'' simulation results,
unfortunately, are flow-problem dependent. 

% dissipation on boundary is important

An investigation of this issue reveals that the resultant dissipation on
the open boundary after imposing these boundary conditions is crucial to and strongly
influences the accuracy of simulation results. 
By requiring that the scale of the boundary dissipation on the open boundary
should match a reasonable physical scale,
%e.g.~that of the traction-free condition
%under normal outflow situations,
we attain a set of open boundary
conditions in two and three dimensions that can 
effectively overcome the backflow instability and also
provide accurate simulation results.
This set of new open boundary conditions is different from 
and not equivalent to the family of conditions developed in~\cite{DongS2015,DongKC2014}.
For one thing, the new boundary conditions 
are active (i.e.~leading to generally non-zero traction) on the entire open boundary,
in both backflow regions and normal outflow regions.
In contrast, the previous methods only take effect in the backflow
regions of the open boundary, and give rise to a zero traction
in normal outflow regions due to the terms like $\Theta_0(\mathbf{n},\mathbf{u})$ 
or $(\mathbf{n}\cdot\mathbf{u})^{-}$.

Therefore, the current  energy-stable open
boundary conditions with physical accuracy are developed through two steps:
(i) devise energy-stable boundary conditions based on a quadratic
form in terms of a symmetric matrix, using the procedure of~\cite{Nordstrom2017};
(ii) require that the boundary dissipation with these
conditions should match a physical scale.
The boundary conditions resulting from the 
first step only can lead to poor or even unphysical
simulation results, even though the computations are stable.

% what are the novelties?

The contribution of this paper lies in the set of
energy-stable and physically-accurate open boundary conditions
developed herein
for incompressible flows. 
These open boundary conditions can be implemented numerically 
with the commonly-used splitting-type
schemes for the incompressible Navier-Stokes equations.
This is because the current conditions are
formulated in a traction form,
similar to those of \cite{DongKC2014,DongS2015}. 
This allows us to employ any of the algorithms developed in
the previous works (see~\cite{DongKC2014,DongS2015,Dong2015obc})
for simulations with the new open boundary conditions.
For the numerical experiments reported in the current work, the algorithm 
from \cite{Dong2015obc} has been employed.
%It is worth pointing out that,
%the new open boundary conditions can be readily modified to 
%form a new set of ``convective-like'' energy-stable open boundary conditions,
%as is discussed in the remarks of Section \ref{sec:obc}.

The current implementation of these open boundary conditions is
based on the $C^0$-continuous spectral element
method~\cite{SherwinK1995,KarniadakisS2005,ZhengD2011},
similar to previous works~\cite{DongKC2014,DongS2015,Dong2015obc}.
It should be pointed out that these boundary conditions 
are given on the continuum level, irrespective of 
the numerical methods employed for their implementation. They can also be
used with other popular techniques such as finite difference, finite element, or 
finite volume methods.

The rest of this paper is organized as follows.
In Section \ref{sec:method} we first derive the general forms of
energy-stable boundary conditions,  referred to as OBC-A,
based on the method of quadratic forms
in two and three dimensions. Then we impose the requirement that 
the scale of boundary dissipation  upon imposing
these conditions should match a physical scale,
%that of the traction-free condition,
and thus acquire another set of
open boundary conditions.  Two boundary conditions among this
set, referred to as OBC-B and OBC-C, are studied in more detail.
The numerical implementation of these boundary conditions is also discussed.
In Section \ref{sec:tests} we present extensive numerical simulations
using two canonical flows, the flow past a circular cylinder in two and three dimensions
and a jet impinging on a wall in two dimensions, 
to test the accuracy and performance of the three  open
boundary conditions OBC-A, OBC-B and OBC-C.
Section \ref{sec:summary} concludes the presentation with discussions and
some closing
remarks. In Appendix A we provide a proof of Theorem \ref{thm:thm_1}
used in Section \ref{sec:method} for the derivation of energy-stable
boundary conditions. 
Appendix B provides a summary of the numerical algorithm from \cite{Dong2015obc},
and provides some details on the numerical implementation of the
open boundary conditions developed in the main text of the paper.

%%%  schemes
\section{Energy-Stable Boundary Conditions for Incompressible Navier-Stokes
  Equations}
\label{sec:method}

\subsection{Navier-Stokes Equations and Energy Balance Relation}

Consider a flow domain $\Omega$ in two or three dimensions,
whose boundary is denoted by $\partial\Omega$,
and an incompressible flow on this domain. Let $L$ denote
a length scale, $U_0$ denote a velocity scale, and
$\nu_f$ denote the kinematic viscosity of the fluid. The flow
is described by the normalized incompressible Navier-Stokes
equations,
\begin{subequations}
  \begin{equation}
    \frac{\partial\mathbf{u}}{\partial t}
    + \mathbf{u}\cdot\nabla\mathbf{u}
    + \nabla p -\nu\nabla^2\mathbf{u}
    = \mathbf{f},
    \label{equ:nse}
  \end{equation}
  \begin{equation}
    \nabla\cdot\mathbf{u} = 0, \label{equ:continuity}
  \end{equation}
\end{subequations}
where $\mathbf{u}(\mathbf{x},t)$ is the velocity,
$p(\mathbf{x},t)$ is the pressure, $\mathbf{f}(\mathbf{x},t)$
is an external body force, and
$\mathbf{x}$ and $t$ are the spatial coordinate and time,
respectively. $\nu$ is the non-dimensional viscosity, given by
\begin{equation}
\nu = \frac{1}{Re} = \frac{\nu_f}{U_0L},
\end{equation}
where $Re$ is the Reynolds number.

The equations \eqref{equ:nse}--\eqref{equ:continuity}
are to be supplemented by appropriate boundary conditions
on $\partial\Omega$, which is the focus of this work
in subsequent sections, together with the following initial condition
for the velocity
\begin{equation}
\mathbf{u}(\mathbf{x},0) = \mathbf{u}_{in}(\mathbf{x}),
\label{equ:ic}
\end{equation}
where $\mathbf{u}_{in}$ is the initial velocity distribution
satisfying equation \eqref{equ:continuity} and
compatible with the boundary condition on $\partial\Omega$.

% energy balance equation

Taking the $L^2$ inner product between \eqref{equ:nse} and
$\mathbf{u}$ and using the integration by part,
the divergence theorem and equation \eqref{equ:continuity},
we arrive at the following
energy-balance equation
\begin{equation}
  \frac{\partial}{\partial t}\int_{\Omega} \frac{1}{2}|\mathbf{u}|^2
  = \int_{\Omega} \mathbf{f}\cdot\mathbf{u}
  - \nu\int_{\Omega} \left\|\nabla\mathbf{u} \right\|^2
  + \int_{\partial\Omega} \underbrace{\left[
  \mathbf{n}\cdot\mathbf{T}\cdot\mathbf{u}
  -\frac{1}{2}(\mathbf{u}\cdot\mathbf{u})(\mathbf{n}\cdot\mathbf{u})
  \right]}_{\text{boundary term (BT)}},
  \label{equ:energy_balance}
\end{equation}
where $\mathbf{n}$ is the outward-pointing unit vector
normal to $\partial\Omega$,
$\mathbf{T} = -p\mathbf{I} + \nu\nabla\mathbf{u}$
and $\mathbf{I}$ is the identity tensor.
$\mathbf{T}$ can be roughly considered as the fluid stress
tensor.
If the external body force is absent ($\mathbf{f}=0$),
the volume integral term on the right hand side (RHS)
of the above equation is always dissipative and will not
cause the system energy to increase over time.
The surface integral term, on the other hand,
is indefinite. Its contribution to the system energy will
depend on the boundary conditions imposed on
the domain boundary.

\subsection{Energy-Stable Boundary Conditions Based on a Quadratic Form}
\label{sec:esbc}

We are interested in the boundary conditions on $\partial\Omega$
such that the boundary integral term on RHS of the energy
balance equation \eqref{equ:energy_balance} will always
be non-positive. As such, the contribution of the surface integral
will not cause the system energy to increase over time.
Such boundary conditions are referred to as energy-stable boundary
conditions.

Inspired by the strategy of \cite{Nordstrom2017}
%(see also \cite{NordstromC2018})
to enforce the definiteness of the boundary contribution,
we will first reformulate the the boundary integral
term in \eqref{equ:energy_balance}
into a quadratic
form in terms of a symmetric matrix. Then by looking into
the eigenvalues and the associated eigenvectors of this matrix,
we formulate the boundary condition
in the form of a relation between those eigenvariables 
corresponding to the eigenvalues of different signs.
By imposing a proper condition on the coefficients involved in
this relation, the boundary condition can guarantee
the negative semi-definiteness of the quadratic form.

The following property about a particular form of symmetric
matrices will be extensively used subsequently:
\begin{theorem}
  \label{thm:thm_1}

  Let $\mathbf{G}$ denote an $m\times m$ ($m\geqslant 1$) real symmetric matrix,
$\mathbf{I}_m$ denote the $m\times m$ identity matrix,
and 
$
\mathbf{A} = \begin{bmatrix}
\mathbf{0} & -\mathbf{I}_m \\
-\mathbf{I}_m & \mathbf{G}
\end{bmatrix}.
$
Then
\begin{enumerate}[label=\emph{(\alph*)}]

\item
  The eigenvalues of $\mathbf{A}$ are real and non-zero.
  
\item
  $\lambda$ is an eigenvalue of $\mathbf{A}$ if
and only if $\left(\lambda-\frac{1}{\lambda}\right)$ is an eigenvalue
of $\mathbf{G}$.

\item
$\begin{bmatrix}\mathbf{Z} \\ -\lambda\mathbf{Z}  \end{bmatrix}$
is an eigenvector of $\mathbf{A}$ corresponding to
the eigenvalue $\lambda$ if and only if $\mathbf{Z}$ is
an eigenvector of $\mathbf{G}$ corresponding to
the eigenvalue $\left(\lambda - \frac{1}{\lambda}  \right)$.

\end{enumerate}

\end{theorem}
\noindent A proof of this property is provided in the Appendix A.
This theorem suggests that the eigenvalues and the eigenvectors of
the $2m\times 2m$ matrix $\mathbf{A}$ can be constructed based on
those of the $m\times m$ matrix $\mathbf{G}$.
Let $\xi$ denote an eigenvalue (real) of the symmetric matrix
$\mathbf{G}$. Then the corresponding eigenvalues of matrix $\mathbf{A}$
are given by
$
\lambda = \frac{\xi}{2} \pm \sqrt{\left(\frac{\xi}{2} \right)^2 +1}.
$
Therefore, half of the eigenvalues of $\mathbf{A}$ are positive
and  half are negative.

\subsubsection{Two Dimensions (2D)}
\label{sec:2d}

We first consider two dimensions in space.
Let $\mathbf{n}$ and $\bm{\tau}$ denote the unit vectors
normal (pointing outward) and tangential to the boundary $\partial\Omega$,
respectively,
and $\mathbf{n}\times\bm{\tau} = \mathbf{e}_z$, where $\mathbf{e}_z$
denotes the unit vector along the third (i.e.~$z$) direction normal 
to the two-dimensional plane.
Define the normal and tangent components of the fluid stress and the velocity
on the boundary by
\begin{equation}
  T_{nn} = \mathbf{n}\cdot\mathbf{T}\cdot\mathbf{n}, \quad
  T_{n\tau} = \mathbf{n}\cdot\mathbf{T}\cdot\bm{\tau}, \quad
  u_{n} = \mathbf{n}\cdot\mathbf{u}, \quad
  u_{\tau} = \bm{\tau}\cdot\mathbf{u},
  \quad \text{on} \ \partial\Omega.
  \label{equ:def_Tnn}
\end{equation}
Note that 
$\mathbf{u} = u_n\mathbf{n} + u_{\tau}\bm{\tau}$, and
$\mathbf{n}\cdot\mathbf{T} = T_{nn}\mathbf{n}+T_{n\tau}\bm{\tau}$ on $\partial\Omega$.
The boundary term in equation \eqref{equ:energy_balance}
can then be written as a
quadratic form with a symmetric matrix $\mathbf{A}$ as follows,
\begin{equation}
  \text{BT}  = -\frac{1}{2}
  \begin{bmatrix} T_{nn} \\ T_{n\tau} \\ u_n \\ u_{\tau} \end{bmatrix}^{T}
  \underbrace{
  \begin{bmatrix}
    0 & 0 & -1 & 0 \\
    0 & 0 & 0 & -1 \\
    -1 & 0 & u_n & \alpha u_{\tau} \\
    0 & -1 & \alpha u_{\tau} & \beta u_{n}
  \end{bmatrix}
  }_{\text{matrix}\ \mathbf{A}}
  \underbrace{
    \begin{bmatrix} T_{nn} \\ T_{n\tau} \\ u_n \\ u_{\tau} \end{bmatrix}
  }_{\mathbf{X}}
  =-\frac{1}{2}\mathbf{X}^T\mathbf{AX},
  \label{equ:bt_quad_form}
\end{equation}
where the superscript in $(\cdot)^T$ denotes transpose,
$\alpha$ a chosen constant satisfying $0\leqslant \alpha\leqslant \frac{1}{2}$,
and $\beta=1-2\alpha$.
This matrix has the form as given by Theorem \ref{thm:thm_1}
with $m=2$, and in this case
\begin{equation}
\mathbf{G} = \begin{bmatrix}
  u_n & \alpha u_{\tau} \\
  \alpha u_{\tau} & \beta u_n
\end{bmatrix}.
\label{equ:def_G_2d}
\end{equation}
%are two constants satisfying
%\begin{equation}
%  0\leqslant \alpha \leqslant \frac{1}{2}, \quad
%  \beta = 1-2\alpha.
%  \label{equ:def_alpha}
%\end{equation}
%
In what follows, we distinguish two cases:
(i) $0<\alpha\leqslant \frac{1}{2}$, and
(ii) $\alpha=0$,
and treat them individually.

\paragraph{\underline{Case $0<\alpha\leqslant \frac{1}{2}$.}}

% eigenvalues and eigenvectors of A

The matrix $\mathbf{A}$ defined in \eqref{equ:bt_quad_form} has
four distinct eigenvalues, 
\begin{equation}
  \left\{
  \begin{array}{ll}
    \lambda_1 = \frac{\xi_1}{2} - \sqrt{\left(\frac{\xi_1}{2}\right)^2 + 1}, &
    \lambda_2 = \frac{\xi_2}{2} - \sqrt{\left(\frac{\xi_2}{2}\right)^2 + 1}, \\
    \lambda_3 = \frac{\xi_1}{2} + \sqrt{\left(\frac{\xi_1}{2}\right)^2 + 1}, &
    \lambda_4 = \frac{\xi_2}{2} + \sqrt{\left(\frac{\xi_2}{2}\right)^2 + 1},
  \end{array}
  \right.
  \label{equ:def_eigenvalue}
\end{equation}
where $\xi_1$ and $\xi_2$ are the eigenvalues of
the matrix $\mathbf{G}$ defined in \eqref{equ:def_G_2d},
\begin{equation}
  \left\{
  \begin{split}
    &
  \xi_1 = 
    (1-\alpha) u_n + \alpha\sqrt{u_n^2 + u_{\tau}^2}
  =
    (1-\alpha) u_n + \alpha|\mathbf{u}| \\
  &
  \xi_2 = 
    (1-\alpha) u_n - \alpha\sqrt{u_n^2 + u_{\tau}^2}
  = 
    (1-\alpha) u_n - \alpha|\mathbf{u}|,
  \end{split}
  \right.
  \label{equ:def_xi}
\end{equation}
and $|\mathbf{u}|$ denotes the magnitude of the velocity.
Note that $\lambda_1, \lambda_2<0$, and
$\lambda_3, \lambda_4>0$.
The following relations about these eigenvalues will be
useful for subsequent discussions,
\begin{equation}
  \left\{
  \begin{split}
    &
  1-\lambda_1^2 = -\xi_1\lambda_1, \qquad
  1-\lambda_3^2 = -\xi_1\lambda_3;\\
  &
  1-\lambda_2^2 = -\xi_2\lambda_2, \qquad
  1-\lambda_4^2 = -\xi_2\lambda_4.
  \end{split}
  \right.
  \label{equ:eigen_relation}
\end{equation}
If $|\mathbf{u}|=0$, the contribution of the
boundary term in \eqref{equ:energy_balance} vanishes.
So we assume that $|\mathbf{u}|\neq 0$ in the following derivation
of the boundary conditions.
%If $\alpha=0$, $\lambda_1$ and $\lambda_2$,
%and $\lambda_3$ and $\lambda_4$ will reduce to two
%double eigenvalues.

The eigenvectors of $\mathbf{G}$ corresponding to the eigenvalues
$\xi_1$ and $\xi_2$ have two representations,  given by
\begin{equation}
  \left\{
  \begin{split}
    &
    \begin{bmatrix} 1 \\ \eta \end{bmatrix}, \ \
    \begin{bmatrix} -\eta \\ 1 \end{bmatrix}, \ \
    \text{if} \ u_n\geqslant 0; \\
    &
    \begin{bmatrix} \eta \\ 1 \end{bmatrix}, \ \
    \begin{bmatrix} 1 \\ -\eta \end{bmatrix}, \ \
    \text{if} \ u_n < 0,
  \end{split}
  \right.
\end{equation}
where
\begin{equation}
\eta = \frac{u_{\tau}}{|\mathbf{u}|+|u_n|}.
\label{equ:def_eta}
\end{equation}
If $u_{\tau}\neq 0$, both representations of the eigenvectors are equivalent.
But when $u_{\tau}=0$ only one of these two representations
is suitable, depending on the sign of $u_n$ as given above.
Based on Theorem \ref{thm:thm_1},
the four eigenvectors of the matrix $\mathbf{A}$ are given by
\begin{equation}
  \begin{bmatrix} 1 \\ \eta
    \\ -\lambda_1\\ -\eta\lambda_1 \end{bmatrix},
  \qquad
  \begin{bmatrix} -\eta \\ 1\\
    \eta\lambda_2  \\ -\lambda_2 \end{bmatrix},
  \qquad
  \begin{bmatrix} 1 \\ \eta
    \\ -\lambda_3\\ -\eta\lambda_3  \end{bmatrix},
  \qquad
  \begin{bmatrix} -\eta  \\ 1\\
    \eta\lambda_4  \\ -\lambda_4 \end{bmatrix},
  \quad \text{if} \ u_n\geqslant 0,
  \label{equ:eigen_vector_1}
\end{equation}
%where $\eta_1 = \frac{u_{\tau}}{|\mathbf{u}|+u_n}$.
%
and  by
\begin{equation}
  \begin{bmatrix} \eta \\ 1\\
    -\eta\lambda_1 \\ -\lambda_1 \end{bmatrix},
  \qquad
  \begin{bmatrix} 1 \\ -\eta
    \\ -\lambda_2\\ \eta\lambda_2 \end{bmatrix},
  \qquad
  \begin{bmatrix} \eta \\ 1\\
    -\eta\lambda_3 \\ -\lambda_3 \end{bmatrix},
  \qquad
  \begin{bmatrix} 1 \\ -\eta
    \\ -\lambda_4\\ \eta\lambda_4 \end{bmatrix},
  \quad \text{if} \ u_n>0.
  \label{equ:eigen_vector_2}
\end{equation}
%When solving for these eigenvectors the relations
%\eqref{equ:eigen_relation} have been used.

We use the four eigenvectors of $\mathbf{A}$
to form an orthogonal matrix $\hat{\mathbf{P}}$.
For $u_n\geqslant 0$,
\begin{equation}
  \hat{\mathbf{P}} =
  \underbrace{
  \begin{bmatrix}
    1 & -\eta & 1 & -\eta \\
    \eta & 1 & \eta & 1 \\
    -\lambda_1 & \eta\lambda_2 & -\lambda_3 & \eta\lambda_4 \\
    -\eta\lambda_1 & -\lambda_2 & -\eta\lambda_3 & -\lambda_4
  \end{bmatrix}
  }_{\mathbf{P}}
  \underbrace{
  \frac{1}{\sqrt{1+\eta^2}}
  \text{diag}\left(\frac{1}{\sqrt{1+\lambda_1^2}},\cdots, \frac{1}{\sqrt{1+\lambda_4^2}}\right)
  }_{\mathbf{N}}
  = \mathbf{P}\mathbf{N}
  \label{equ:def_P_mat_1}
\end{equation}
and for $u_n<0$,
\begin{equation}
  \hat{\mathbf{P}} =
  \underbrace{
  \begin{bmatrix}
    \eta & 1 & \eta & 1 \\
    1 & -\eta & 1 & -\eta \\
    -\eta\lambda_1 & -\lambda_2 & -\eta\lambda_3 & -\lambda_4 \\
    -\lambda_1 & \eta\lambda_2 & -\lambda_3 & \eta\lambda_4
  \end{bmatrix}
  }_{\mathbf{P}}
  \underbrace{
  \frac{1}{\sqrt{1+\eta^2}}
  \text{diag}\left(\frac{1}{\sqrt{1+\lambda_1^2}},\cdots, \frac{1}{\sqrt{1+\lambda_4^2}}\right)
  }_{\mathbf{N}}
  =\mathbf{PN}.
  \label{equ:def_P_mat_2}
\end{equation}
Then the matrix $\mathbf{A}$ in \eqref{equ:bt_quad_form}
can be written as
\begin{equation}
  \mathbf{A} = \mathbf{PN}\begin{bmatrix}
    \lambda_1 & & & \\
    & \lambda_2 & & \\
    & & \lambda_3 & \\
    & & & \lambda_4
    \end{bmatrix}
  \mathbf{N}^T\mathbf{P}^T
  = \mathbf{P}\bm{\Lambda}\mathbf{P}^T
  \label{equ:A_rep}
\end{equation}
where
\begin{equation}
  \bm{\Lambda} = \begin{bmatrix} \bm{\Lambda}^{-} & 0 \\
    0 & \bm{\Lambda}^{+} \end{bmatrix}, \quad
  \bm{\Lambda}^{+}=
  \frac{1}{2(1+\eta^2)}\text{diag}\left(
    \frac{1}{\sqrt{(\xi_1/2)^2+1}}, \frac{1}{\sqrt{(\xi_2/2)^2+1}}
    \right),
  \quad
  \bm{\Lambda}^{-}=-\bm{\Lambda}^{+},
  \label{equ:def_Lambda}
\end{equation}
and we have used the relations
$\lambda_1\lambda_3 = -1$ and
$\lambda_2\lambda_4 = -1$.

The quadratic form in \eqref{equ:bt_quad_form} is transformed
into
\begin{equation}
  \begin{split}
  BT &= -\frac{1}{2}\mathbf{X}^T\mathbf{P}\bm{\Lambda}\mathbf{P}^T\mathbf{X}
  = -\frac{1}{2}\mathbf{W}^T\bm{\Lambda}\mathbf{W}
  = -\frac{1}{2}\begin{bmatrix}(\mathbf{W}^{-})^T & (\mathbf{W}^{+})^T \end{bmatrix}
  \begin{bmatrix}
    \bm{\Lambda}^{-} & 0 \\
    0 & \bm{\Lambda}^{+}
  \end{bmatrix}
  \begin{bmatrix} \mathbf{W}^{-} \\ \mathbf{W}^{+} \end{bmatrix} \\
  &
  = -\frac{1}{2} (\mathbf{W}^{-})^T\bm{\Lambda}^{-}\mathbf{W}^{-}
  -\frac{1}{2} (\mathbf{W}^{+})^T\bm{\Lambda}^{+}\mathbf{W}^{+} 
  \end{split}
\label{equ:quad_form_1}
\end{equation}
where
\begin{equation}
  \mathbf{W} = \mathbf{P}^T\mathbf{X}
  = \begin{bmatrix} \mathbf{W}^{-} \\ \mathbf{W}^{+}  \end{bmatrix}.
\end{equation}
%
% explicit form for W^{-} and W^{+}
Define the matrix formed by the eigenvectors of $\mathbf{G}$ as
\begin{equation}
  \mathbf{S} = \left\{
  \begin{split}
&
  \begin{bmatrix}
    1 & -\eta \\
    \eta & 1
  \end{bmatrix},
  & \text{if} \ u_n\geqslant 0, \\
&
  \begin{bmatrix}
    \eta & 1 \\
    1 & -\eta
  \end{bmatrix},
  & \text{if} \ u_n< 0.
  \end{split}
  \right.
%  \quad \text{and} \ \mathbf{B} = \mathbf{S}^T.
\label{equ:def_B}
\end{equation}
Then $\mathbf{W}^{-}$ and $\mathbf{W}^{+}$ are
specifically given by
\begin{equation}
\left\{
\begin{split}
&
\mathbf{W}^{-} = \mathbf{S}^T\begin{bmatrix} T_{nn} \\ T_{n\tau}  \end{bmatrix}
-\begin{bmatrix}\lambda_1 & \\ & \lambda_2  \end{bmatrix}\mathbf{S}^T
\begin{bmatrix}u_n \\ u_{\tau}  \end{bmatrix}, \\
&
\mathbf{W}^{+} = \mathbf{S}^T\begin{bmatrix} T_{nn} \\ T_{n\tau}  \end{bmatrix}
-\begin{bmatrix}\lambda_3 & \\ & \lambda_4  \end{bmatrix}\mathbf{S}^T
\begin{bmatrix}u_n \\ u_{\tau}  \end{bmatrix}.
\end{split}
\right.
\label{equ:def_W_pm}
\end{equation}
$\mathbf{W}^{-}$ and $\mathbf{W}^{+}$ are the eigenvariables
corresponding to the negative and the positive eigenvalues of
matrix $\mathbf{A}$, respectively.

Following the strategy of \cite{Nordstrom2017},
we consider boundary conditions of the form
\begin{equation}
\mathbf{W}^{-} = \mathbf{RW}^{+}, \quad
\text{with} \ \mathbf{R} = \begin{bmatrix}
a_{11} & a_{12} \\
a_{21} & a_{22}
\end{bmatrix}
\label{equ:def_bc}
\end{equation}
where $\mathbf{R}$ is a chosen constant matrix satisfying the
conditions to be specified below.
Substitute these boundary conditions into \eqref{equ:quad_form_1},
and the quadratic form becomes
\begin{equation}
BT = -\frac{1}{2}(\mathbf{W}^{+})^T 
\underbrace{\left(
\mathbf{R}^T\bm{\Lambda}^{-}\mathbf{R} + \bm{\Lambda}^{+}
\right)
}_{\mathbf{Q}}
 \mathbf{W}^{+}
=-\frac{1}{2}(\mathbf{W}^{+})^T \mathbf{QW}^{+}
\label{equ:quad_form_2}
\end{equation}
We require that the matrix $\mathbf{R}$ be chosen such that
the matrix $\mathbf{Q}$ as defined above 
is symmetric positive semi-definite (semi-SPD).
As such, the surface integral term in \eqref{equ:energy_balance}
will always be non-positive, and the energy stability
of the system is guaranteed.
Therefore, 
equation \eqref{equ:def_bc} represents a class of energy-stable
boundary conditions.

Let us look into the semi-SPD requirement on $\mathbf{Q}$
in more detail. Let
\begin{equation}
z_1 = \frac{1}{\sqrt{(\xi_1/2)^2+1}}, \quad
z_2 = \frac{1}{\sqrt{(\xi_2/2)^2+1}}, \quad
\mathbf{M} = \text{diag}(z_1, z_2).
%\begin{bmatrix} z_1 & \\  & z_2  \end{bmatrix}.
\label{equ:def_M}
\end{equation}
In light of \eqref{equ:def_Lambda} and \eqref{equ:quad_form_2}, we have
$ 
\mathbf{Q} = 
\frac{1}{2(1+\eta^2)}\left(-\mathbf{R}^T\mathbf{MR} + \mathbf{M} \right).
$ 
Therefore we only need to find constant matrix $\mathbf{R}$ such that 
the matrix $\mathbf{Q}_1=\mathbf{M}-\mathbf{R}^T\mathbf{MR}$
be symmetric positive semi-definite for all $u_n\in (-\infty,\infty)$,
$u_{\tau}\in (-\infty,\infty)$, and $|\mathbf{u}|>0$.
Requiring that the eigenvalues of $\mathbf{Q}_1$ be non-negative
is equivalent to the following conditions:
\begin{subequations}
\begin{equation}
z_1(a_{11}^2 + a_{12}^2 - 1) + z_2(a_{21}^2 + a_{22}^2 - 1) \leqslant 0,
\label{equ:cond_1}
\end{equation}
\begin{equation}
-z_1^2a_{12}^2 - z_2^2a_{21}^2 + z_1z_2\left[
  (a_{11}a_{22} - a_{12}a_{21})^2 + 1 - (a_{11}^2 + a_{22}^2)
\right] \geqslant 0,
\label{equ:cond_2}
\end{equation}
\end{subequations}
for all $u_n, u_{\tau} \in(-\infty,\infty)$ and $|\mathbf{u}|>0$.
Noting that $z_1\in(0,1]$, $z_2\in(0,1]$, and $\frac{z_2}{z_1}\in (0,\infty)$,
we conclude that
\begin{equation}
a_{12} = 0, \quad a_{21} = 0, \quad
a^2_{11} \leqslant 1, \quad a_{22}^2 \leqslant 1.
\label{equ:cond_R_1}
\end{equation}
This is one set of conditions the matrix $\mathbf{R}$ must satisfy.

Substituting the expressions of \eqref{equ:def_W_pm} into
the boundary conditions \eqref{equ:def_bc} leads to
\begin{equation}
(\mathbf{I}_2-\mathbf{R})\mathbf{S}^T\begin{bmatrix}T_{nn}\\ T_{n\tau}  \end{bmatrix}
= \left(
-\mathbf{R}\begin{bmatrix}\lambda_3 & \\ & \lambda_4  \end{bmatrix}
+ \begin{bmatrix}\lambda_1 & \\ & \lambda_2  \end{bmatrix}
\right)
\mathbf{S}^T
\begin{bmatrix} u_n \\ u_{\tau}  \end{bmatrix},
\label{equ:bc_trans_1}
\end{equation}
where $\mathbf{I}_2$ is the identity matrix of dimension two.
We impose the requirement that 
$(\mathbf{I}_2-\mathbf{R})$ be non-singular, i.e.
\begin{equation}
a_{11} \neq 1, \quad 
a_{22} \neq 1.
\label{equ:cond_R_2}
\end{equation}
This is another set of conditions for $\mathbf{R}$.
Equation \eqref{equ:bc_trans_1} is then transformed into
\begin{equation}
\begin{split}
\begin{bmatrix}T_{nn}\\ T_{n\tau}  \end{bmatrix}
&= \mathbf{S}^{-T}(\mathbf{I}_2-\mathbf{R})^{-1}
\left(
-\mathbf{R}\begin{bmatrix}\lambda_3 & \\ & \lambda_4  \end{bmatrix}
+ \begin{bmatrix}\lambda_1 & \\ & \lambda_2  \end{bmatrix}
\right)
\mathbf{S}^T\begin{bmatrix}u_n \\ u_{\tau}  \end{bmatrix} 
= \mathbf{S}^{-T}\begin{bmatrix}
K_1 & \\
 & K_2
\end{bmatrix}
\mathbf{S}^T\begin{bmatrix}u_n \\ u_{\tau}  \end{bmatrix}.
\end{split}
\label{equ:bc_trans_2}
\end{equation}
where
\begin{equation}
K_1 = \frac{\lambda_1-a_{11}\lambda_3}{1-a_{11}}, \quad
K_2 = \frac{\lambda_2-a_{22}\lambda_4}{1-a_{22}}.
\label{equ:def_K12}
\end{equation}

Substituting the expressions \eqref{equ:def_B} 
for $\mathbf{S}$ into \eqref{equ:bc_trans_2},
we have the boundary conditions in the following form.
For $u_n\geqslant 0$,
\begin{subequations}
\begin{equation}
T_{nn} = \frac{K_1+K_2\eta^2}{1+\eta^2} u_n
+ \frac{\eta(K_1-K_2)}{1+\eta^2} u_{\tau}
= f_1(u_n,u_{\tau}),
\label{equ:bc_p_nn}
\end{equation}
\begin{equation}
T_{n\tau} = \frac{\eta(K_1-K_2)}{1+\eta^2} u_n
+ \frac{K_1\eta^2 + K_2}{1+\eta^2} u_{\tau}
= f_2(u_n,u_{\tau}).
\label{equ:bc_p_nt}
\end{equation}
\end{subequations}
For $u_n<0$,
\begin{subequations}
\begin{equation}
  T_{nn} = \frac{K_1\eta^2+K_2}{1+\eta^2}u_n
+ \frac{\eta(K_1-K_2)}{1+\eta^2} u_{\tau}
= f_1(u_n,u_{\tau}),
\label{equ:bc_n_nn}
\end{equation}
\begin{equation}
T_{n\tau} = \frac{\eta(K_1-K_2)}{1+\eta^2} u_n
+ \frac{K_1 + K_2\eta^2}{1+\eta^2} u_{\tau}
= f_2(u_n,u_{\tau}).
\label{equ:bc_n_nt}
\end{equation}
\end{subequations}
In the above equations $K_1$ and $K_2$ are
given by \eqref{equ:def_K12} and $\eta$ is
given by \eqref{equ:def_eta}. 
The eigenvalues $\lambda_i$ ($1\leqslant i\leqslant 4$)
are given by \eqref{equ:def_eigenvalue}.
The parameters $a_{11}$ and $a_{22}$ are
chosen constants satisfying the following conditions,
in light of 
equations \eqref{equ:cond_R_1} and \eqref{equ:cond_R_2},
\begin{equation}
-1\leqslant a_{11} < 1, \quad
-1\leqslant a_{22} < 1.
\label{equ:cond_R}
\end{equation}
These boundary conditions ensure the energy stability
of the system.

% dissipation functions

Let us next look into
the boundary term \eqref{equ:quad_form_2}
associated with these boundary conditions. 
The matrix $\mathbf{Q}$ is reduced to
$
\mathbf{Q} = \frac{1}{2(1+\eta^2)}\text{diag}\left(
z_1(1-a_{11}^2), z_2(1-a_{22}^2)
\right)
$
in light of equations \eqref{equ:cond_R_1} and \eqref{equ:cond_R_2}.
Let
$
\mathbf{W}^{+} = \begin{bmatrix} W_1 \\ W_2 \end{bmatrix}.
$
Then we have, in light of equations
\eqref{equ:def_W_pm} and \eqref{equ:bc_trans_2},
\begin{equation}
\begin{bmatrix} W_1\\ W_2  \end{bmatrix}
= \mathbf{W}^{+}
= \begin{bmatrix} -\frac{2}{z_1(1-a_{11})} & \\ & -\frac{2}{z_2(1-a_{22})}  \end{bmatrix}
\mathbf{S}^T\begin{bmatrix} u_n \\ u_{\tau}  \end{bmatrix}.
\label{equ:def_W12}
\end{equation}
Substituting the above expression and the expression \eqref{equ:def_B}
into \eqref{equ:quad_form_2}, we have
\begin{equation}
BT = \left\{
\begin{split}
&
-\frac{1}{2}\left[
\frac{1+a_{11}}{1-a_{11}}\frac{2}{z_1(1+\eta^2)}(u_n+\eta u_{\tau})^2
+ \frac{1+a_{22}}{1-a_{22}}\frac{2}{z_2(1+\eta^2)}(-\eta u_n + u_{\tau})^2 
\right], \quad \text{if} \ u_n\geqslant 0, \\
&
-\frac{1}{2}\left[
\frac{1+a_{11}}{1-a_{11}}\frac{2}{z_1(1+\eta^2)}(\eta u_n+ u_{\tau})^2
+ \frac{1+a_{22}}{1-a_{22}}\frac{2}{z_2(1+\eta^2)}(u_n - \eta u_{\tau})^2
\right], \quad \text{if} \ u_n< 0.
\end{split}
\right.
\label{equ:bt_form_1}
\end{equation}
According to the above expression, for the boundary conditions
given by \eqref{equ:bc_p_nn}--\eqref{equ:bc_n_nt},
the amount of dissipation on the boundary
is controlled by the parameters $a_{11}$ and $a_{22}$.
The larger $a_{11}$ and $a_{22}$ are,
the more dissipative these boundary conditions are.
When $a_{11} = a_{22}=-1$, the boundary dissipation vanishes completely.
In other words, no energy can be
convected through the boundary
(into or out of the domain) where these boundary
conditions are imposed.
When $a_{11}\rightarrow 1$ or $a_{22}\rightarrow 1$,
the dissipation on the boundary will become infinitely large.

% comment on absorbing the case |u| = 0.
\begin{remark}
\label{remark_1}

When deriving the boundary conditions \eqref{equ:bc_p_nn}--\eqref{equ:bc_n_nt},
we have assumed that locally $|\mathbf{u}|\neq 0$ on the boundary.
The boundary conditions \eqref{equ:bc_p_nn}--\eqref{equ:bc_n_nt},
however, can also accommodate the case when $|\mathbf{u}|=0$ locally
on the boundary, if we modify the definition of $\eta$ in \eqref{equ:def_eta}
as follows to make it well defined for $|\mathbf{u}|=0$,
\begin{equation}
\eta = \frac{u_{\tau}}{|\mathbf{u}|+|u_n| + \epsilon},
%  \eta = \left\{
%\begin{split}
%&
%0, \quad \text{if} \ |\mathbf{u}| = 0, \\
%&
%\frac{u_{\tau}}{|\mathbf{u}|+|u_n|}, \quad \text{if} \ |\mathbf{u}| \neq 0.
%\end{split}
%\right.
\label{equ:def_eta_mod}
\end{equation}
where $\epsilon$ is a small positive number on
the order of magnitude of the machine zero or smaller
(e.g.~$\epsilon=10^{-18}$).
With this modified definition for $\eta$, 
when $\mathbf{u}=0$ locally at any point on the boundary, 
the boundary conditions are reduced to
$T_{nn} = T_{n\tau} = 0$. 

\end{remark}

% vector form of BCs

The boundary conditions \eqref{equ:bc_p_nn}--\eqref{equ:bc_n_nt} can be
written into a vector form,
\begin{equation}
\left\{
\begin{split}
&
\mathbf{n}\cdot\mathbf{T} = \mathbf{E}(\mathbf{u},\partial\Omega), \ \text{or} \\
&
-p\mathbf{n} + \nu\mathbf{n}\cdot\nabla\mathbf{u}
- \mathbf{E}(\mathbf{u},\partial\Omega) = 0,
\end{split}
\right.
\label{equ:bc_form}
\end{equation}
where in two dimensions
\begin{equation}
  \mathbf{E}(\mathbf{u},\partial\Omega)=
  f_1(u_n,u_{\tau})\mathbf{n} + f_2(u_n,u_{\tau})\bm{\tau},
%  \left\{
%\begin{split}
%&
%f_1(u_n,u_{\tau})\mathbf{n} + g_1(u_n,u_{\tau})\bm{\tau}, \quad 
%\text{if} \ u_n\geqslant 0, \\
%&
%f_2(u_n,u_{\tau})\mathbf{n} + g_2(u_n,u_{\tau})\bm{\tau}, \quad
%\text{if} \ u_n < 0,
%\end{split}
%\right.
\label{equ:bc_vec_1}
\end{equation}
and $f_1$, $f_2$ are given in
\eqref{equ:bc_p_nn}--\eqref{equ:bc_n_nt}.

% what else to discuss here?

% alpha = 0

\paragraph{\underline{Case $\alpha=0$.}}

The matrix $\mathbf{A}$ has two double eigenvalues,
\begin{equation}
\left\{
\begin{split}
&
\lambda_1 = \lambda_2 = \frac{u_n}{2} - \sqrt{\left(\frac{u_n}{2} \right)^2+1}, \\
&
\lambda_3 = \lambda_4 = \frac{u_n}{2} + \sqrt{\left(\frac{u_n}{2} \right)^2+1}.
\end{split}
\right.
\label{equ:def_eigenvalue_0}
\end{equation}
Note that these eigenvalues satisfy the relation
$ %\begin{equation}
u_n - \lambda_i = -\frac{1}{\lambda_i}\ (1\leqslant i\leqslant 4).
$ %\end{equation}
The corresponding eigenvectors are
\begin{equation}
\begin{bmatrix}
1 \\ 0 \\ -\lambda_1 \\ 0
\end{bmatrix}, \quad
\begin{bmatrix}
0 \\ 1 \\ 0 \\ -\lambda_1
\end{bmatrix}, \quad
\begin{bmatrix}
  1 \\ 0 \\ -\lambda_3 \\0
\end{bmatrix}, \quad
\begin{bmatrix}
0 \\ 1 \\ 0 \\ -\lambda_3
\end{bmatrix}.
\label{equ:def_eigenvector_0}
\end{equation}
Matrix $\mathbf{A}$ can then be expressed as
\begin{equation}
\mathbf{A} = \mathbf{P}\bm{\Lambda} \mathbf{P}^T
= \mathbf{P} \begin{bmatrix} \bm{\Lambda}^{-} & \\ & \bm{\Lambda}^{+}  \end{bmatrix}
\mathbf{P}^T,
\end{equation}
where
%($\mathbf{I}$ denoting the $2\times 2$ identity matrix)
\begin{equation}
\mathbf{P} = \begin{bmatrix}
1 & 0 & 1 & 0 \\
0 & 1 & 0 & 1 \\
-\lambda_1 & 0 & -\lambda_3 & 0 \\
0 & -\lambda_1 & 0 & -\lambda_3
\end{bmatrix}
= \begin{bmatrix}
\mathbf{I}_2 & \mathbf{I}_2 \\
-\lambda_1\mathbf{I}_2 & -\lambda_3\mathbf{I}_2
\end{bmatrix},
\quad
\bm{\Lambda}^{-} = -\bm{\Lambda}^{+} = -\frac{1}{2\sqrt{(u_n/2)^2+1}}\mathbf{I}_2.
\label{equ:def_P_mat}
\end{equation}

Accordingly, the boundary term in \eqref{equ:bt_quad_form}
is transformed into
\begin{equation}
\begin{split}
BT &= -\frac{1}{2}\mathbf{X}^T\mathbf{P}\bm{\Lambda}\mathbf{P}^T\mathbf{X}
= -\frac{1}{2}\mathbf{W}^T\bm{\Lambda}\mathbf{W}
= -\frac{1}{2}\begin{bmatrix} \mathbf{W}^{-} \\ \mathbf{W}^{+}  \end{bmatrix}^T
\begin{bmatrix} \bm{\Lambda}^{-} & 0\\ & \bm{\Lambda}^{+}  \end{bmatrix}
\begin{bmatrix} \mathbf{W}^{-} \\ \mathbf{W}^{+}  \end{bmatrix} \\
&
= -\frac{1}{2}\left[(\mathbf{W}^{-})^T\bm{\Lambda}^{-}\mathbf{W}^{-}
+ (\mathbf{W}^{+})^T\bm{\Lambda}^{+}\mathbf{W}^{+}\right],
\end{split}
\label{equ:bt_trans_1}
\end{equation}
where 
$
\mathbf{W} = \mathbf{P}^T\mathbf{X} 
= \begin{bmatrix} \mathbf{W}^{-} \\ \mathbf{W}^{+}  \end{bmatrix}.
$
$\mathbf{W}^{-}$ and $\mathbf{W}^{+}$ are vectors of dimension two and are
given specifically by
\begin{equation}
\left\{
\begin{split}
&
\mathbf{W}^{-} = \begin{bmatrix}T_{nn} \\ T_{n\tau}  \end{bmatrix}
 - \lambda_1\begin{bmatrix} u_n \\ u_{\tau}  \end{bmatrix}, \\
&
\mathbf{W}^{+} = \begin{bmatrix}T_{nn} \\ T_{n\tau}  \end{bmatrix}
 - \lambda_3\begin{bmatrix} u_n \\ u_{\tau}  \end{bmatrix}.
\end{split}
\right.
\label{equ:def_W_pm_0}
\end{equation}
We again introduce boundary conditions in the form of
equation \eqref{equ:def_bc}, where the $2\times 2$ constant matrix %$\mathbf{R}$
$
\mathbf{R} = \begin{bmatrix} a_{11} & a_{12} \\ a_{21} & a_{22}  \end{bmatrix}
$ 
is to be determined.
Therefore, the boundary term in \eqref{equ:bt_trans_1}
can be transformed into the same form as
equation \eqref{equ:quad_form_2},
in which the matrix 
\begin{equation}
\mathbf{Q} = \mathbf{R}^T\bm{\Lambda}^{-}\mathbf{R}+\bm{\Lambda}^{+}
= \frac{1}{2\sqrt{(u_n/2)^2+1}}(\mathbf{I}_2 - \mathbf{R}^T\mathbf{R})
\label{equ:def_Q}
\end{equation}
is required to be symmetric positive semi-definite.
Requiring that the eigenvalues of $\mathbf{Q}$ be
non-negative leads to the following conditions:
\begin{subequations}
\begin{equation}
a_{11}^2 + a_{12}^2 + a_{21}^2 + a_{22}^2 \leqslant 2, \ \text{and}
\label{equ:cond_R_1_0}
\end{equation}
\begin{equation}
a_{11}^2 + a_{12}^2 + a_{21}^2 + a_{22}^2 \leqslant 1 + (a_{11}a_{22}-a_{12}a_{21})^2.
\label{equ:cond_R_2_0}
\end{equation}
\end{subequations}
A sufficient condition to guarantee both \eqref{equ:cond_R_1_0}
and \eqref{equ:cond_R_2_0} is
\begin{equation}
a_{11}^2 + a_{12}^2 + a_{21}^2 + a_{22}^2 \leqslant 1.
\label{equ:suff_cond_R}
\end{equation}
This indicates that when $a_{ij}$ ($i,j=1,2$) are 
chosen to be sufficiently small
the matrix $\mathbf{R}$ will guarantee the positive semi-definiteness
of the matrix $\mathbf{Q}$ and the non-positivity 
of the surface integral term in \eqref{equ:energy_balance}.

In light of equation \eqref{equ:def_W_pm_0},
the boundary condition in the form of equation
\eqref{equ:def_bc} is transformed into
\begin{equation}
(\mathbf{I}-\mathbf{R})\begin{bmatrix}T_{nn} \\ T_{n\tau}  \end{bmatrix}
= (\lambda_1\mathbf{I}-\lambda_3\mathbf{R})
\begin{bmatrix} u_n\\ u_{\tau}  \end{bmatrix}.
\label{equ:bc_trans_1_0}
\end{equation}
We impose the requirement that $(\mathbf{I}-\mathbf{R})$
be non-singular, i.e.
\begin{equation}
\mathscr{K} = \text{det}(\mathbf{I}-\mathbf{R})
= (1-a_{11})(1-a_{22}) - a_{12}a_{21} \neq 0.
\label{equ:cond_R_3_0}
\end{equation}
This is another condition the matrix $\mathbf{R}$ must satisfy.
Consequently, the boundary condition \eqref{equ:bc_trans_1_0}
becomes
\begin{equation}
\begin{bmatrix}T_{nn} \\ T_{n\tau}  \end{bmatrix}
= (\mathbf{I}-\mathbf{R})^{-1}(\lambda_1\mathbf{I}-\lambda_3\mathbf{R})
\begin{bmatrix}u_n\\ u_{\tau}  \end{bmatrix}.
\label{equ:bc_trans_2_0}
\end{equation}
In component forms they are
\begin{subequations}
\begin{equation}
T_{nn} = \frac{1}{\mathscr{K}}\left[ 
(1-a_{22})(\lambda_1-\lambda_3a_{11})-\lambda_3a_{12}a_{21}
\right] u_n
+ \frac{1}{\mathscr{K}}(\lambda_1-\lambda_3)a_{12}u_{\tau}
= f_1(u_n,u_{\tau}),
\label{equ:bc_1_0}
\end{equation}
\begin{equation}
T_{n\tau} = \frac{1}{\mathscr{K}}(\lambda_1-\lambda_3)a_{21}u_n
+ \frac{1}{\mathscr{K}}\left[
(1-a_{11})(\lambda_1-\lambda_3a_{22}) - \lambda_3a_{12}a_{21}
\right] u_{\tau}
= f_2(u_n,u_{\tau}),
\label{equ:bc_2_0}
\end{equation}
\end{subequations}
where $\mathscr{K}$ is given by \eqref{equ:cond_R_3_0},
$\lambda_1$ and $\lambda_3$ are given by \eqref{equ:def_eigenvalue_0},
and the chosen constants $a_{ij}$ ($i,j=1,2$) satisfy
the conditions \eqref{equ:cond_R_1_0}, \eqref{equ:cond_R_2_0}
and \eqref{equ:cond_R_3_0}.
These are the energy-stable boundary conditions
for the case $\alpha=0$.

% what about the dissipation?

Let us now consider a simplified case:
$\mathbf{R}$ is assumed to be a diagonal matrix.
The conditions \eqref{equ:cond_R_1_0}, \eqref{equ:cond_R_2_0}
and \eqref{equ:cond_R_3_0} are then reduced to
\begin{equation}
a_{12} = a_{21} = 0, \quad
-1\leqslant a_{11} < 1, \quad
-1\leqslant a_{22} < 1.
\label{equ:cond_R_simple}
\end{equation}
These are the same as those conditions for $\mathbf{R}$
in the case $0<\alpha\leqslant \frac{1}{2}$;
see equation \eqref{equ:cond_R}. 
The boundary conditions \eqref{equ:bc_1_0}--\eqref{equ:bc_2_0}
are reduced to 
\begin{equation}
\left\{
\begin{split}
&
T_{nn} = \frac{\lambda_1-\lambda_3a_{11}}{1-a_{11}}u_n, \\
&
T_{n\tau} = \frac{\lambda_1-\lambda_3a_{22}}{1-a_{22}}u_{\tau}.
\end{split}
\right.
\label{equ:bc_0_simple}
\end{equation}
With the above condition, the boundary term becomes
\begin{equation}
BT = -\frac{1}{2}\sqrt{u_n^2+4}\left(
\frac{1+a_{11}}{1-a_{11}}u_n^2
+ \frac{1+a_{22}}{1-a_{22}}u_{\tau}^2
\right).
\label{equ:bt_0_simple}
\end{equation}
For this simplified case, 
the amount of boundary dissipation  is controlled
by the constants $a_{11}$ and $a_{22}$.
It is more dissipative with
increasing  $a_{11}$ and $a_{22}$.
%as in the case with $0<\alpha\leqslant \frac{1}{2}$.

The boundary conditions \eqref{equ:bc_1_0}--\eqref{equ:bc_2_0}
can be cast into the same vectorial form as given by \eqref{equ:bc_form},
in which $\mathbf{E}(\mathbf{u},\partial\Omega)$ is given by
\begin{equation}
\mathbf{E}(\mathbf{u},\partial\Omega) = f_1(u_n,u_{\tau})\mathbf{n}
+ f_2(u_n,u_{\tau})\bm{\tau},
\end{equation}
where $f_1(u_n,u_{\tau})$ and $f_2(u_n,u_{\tau})$ are defined
by \eqref{equ:bc_1_0} and \eqref{equ:bc_2_0}.

% what else to discuss here?

\subsubsection{Three Dimensions (3D)}
\label{sec:3d}

We next consider three dimensions in space.
Let $\mathbf{n}$ denote the outward-pointing
unit vector  normal to the boundary
$\partial\Omega$, and
$\bm\tau$ and $\mathbf{s}$ denote the unit vectors along
the two independent directions tangent
to $\partial\Omega$, such that $(\mathbf{n},\bm\tau,\mathbf{s})$
are mutually orthogonal and form a right-handed system.
Define the three components along the
$(\mathbf{n},\bm\tau,\mathbf{s})$ directions for the
stress vector $\mathbf{n}\cdot\mathbf{T}$ and
the velocity $\mathbf{u}$,
\begin{equation}
  T_{nn} = \mathbf{n}\cdot\mathbf{T}\cdot\mathbf{n}, \quad
  T_{n\tau} = \mathbf{n}\cdot\mathbf{T}\cdot\bm{\tau}, \quad
  T_{ns} = \mathbf{n}\cdot\mathbf{T}\cdot\mathbf{s}, \quad
  u_{n} = \mathbf{n}\cdot\mathbf{u}, \quad
  u_{\tau} = \bm{\tau}\cdot\mathbf{u}, \quad
  u_{s} = \mathbf{s}\cdot\mathbf{u}.
  \label{equ:def_Tnn_3d}
\end{equation}

In three dimensions
the boundary term in the energy balance equation can be written as
\begin{equation}
  \text{BT}  = -\frac{1}{2}
  \begin{bmatrix} T_{nn} \\ T_{n\tau} \\ T_{ns} \\ u_n \\ u_{\tau} \\ u_s \end{bmatrix}^{T}
  \underbrace{
  \begin{bmatrix}
    0 & 0 & 0 & -1 & 0 & 0 \\
    0 & 0 & 0 & 0 & -1 & 0 \\
    0 & 0 & 0 & 0 & 0 & -1 \\
    -1 & 0 & 0 & u_n & \alpha_1 u_{\tau} & \alpha_2 u_{s} \\
    0 & -1 & 0 & \alpha_1 u_{\tau} & \beta_1 u_n & 0 \\
    0 &  0 & -1 & \alpha_2 u_{s} & 0 & \beta_2 u_{n}
  \end{bmatrix}
  }_{\text{matrix}\ \mathbf{A}}
  \underbrace{
    \begin{bmatrix} T_{nn} \\ T_{n\tau} \\ T_{ns} \\ u_n \\ u_{\tau} \\ u_s \end{bmatrix}
  }_{\mathbf{X}}
  =-\frac{1}{2}\mathbf{X}^T\mathbf{AX},
  \label{equ:bt_quad_form_3d}
\end{equation}
where $\alpha_1$ and $\alpha_2$ are chosen constants
satisfying $0\leqslant \alpha_1\leqslant \frac{1}{2}$
and $0\leqslant \alpha_2\leqslant\frac{1}{2}$,
and $\beta_1 = 1-2\alpha_1$ and $\beta_2 = 1-2\alpha_2$.
%Because the matrix $\mathbf{A}$ is symmetric and 
%$\text{det}(\mathbf{A})\neq 0$, all the eigenvalues of
%$\mathbf{A}$ are non-zero real numbers.
The matrix $\mathbf{A}$ has the form as discussed in Theorem \ref{thm:thm_1}, 
and in this case
$
\mathbf{G}=\begin{bmatrix}
u_n & \alpha_1 u_{\tau} & \alpha_2 u_s \\
\alpha_1 u_{\tau} & \beta_1 u_n & 0 \\
\alpha_2 u_s & 0 & \beta_2 u_n
\end{bmatrix}.
$

Let $\xi_1$, $\xi_2$ and $\xi_3$ denote the three (real) eigenvalues
of $\mathbf{G}$, and $\mathbf{Z}_1$, $\mathbf{Z}_2$ and $\mathbf{Z}_3$
denote the corresponding orthonormal eigenvectors, and
$\mathbf{S}=\begin{bmatrix} \mathbf{Z}_1 \ \mathbf{Z}_2\ \mathbf{Z}_3 \end{bmatrix}$
denote the orthogonal matrix formed by these eigenvectors.
According to Theorem \ref{thm:thm_1}, the eigenvalues of
$\mathbf{A}$ are
\begin{equation}
\left\{
\begin{array}{lll}
\lambda_1 = \frac{\xi_1}{2} - \sqrt{\left(\frac{\xi_1}{2}\right)^2+1}, & 
\lambda_2 = \frac{\xi_2}{2} - \sqrt{\left(\frac{\xi_2}{2}\right)^2+1}, & 
\lambda_3 = \frac{\xi_3}{2} - \sqrt{\left(\frac{\xi_3}{2}\right)^2+1}, \\
\lambda_4 = \frac{\xi_1}{2} + \sqrt{\left(\frac{\xi_1}{2}\right)^2+1}, &
\lambda_5 = \frac{\xi_2}{2} + \sqrt{\left(\frac{\xi_2}{2}\right)^2+1}, &
\lambda_6 = \frac{\xi_3}{2} + \sqrt{\left(\frac{\xi_3}{2}\right)^2+1}.
\end{array}
\right.
\label{equ:def_lambda_3d}
\end{equation}
The corresponding eigenvectors are given by
\begin{equation}
  \begin{bmatrix} \mathbf{Z}_1 \\ -\lambda_1\mathbf{Z}_1  \end{bmatrix}, \quad
  \begin{bmatrix} \mathbf{Z}_2 \\ -\lambda_2\mathbf{Z}_2  \end{bmatrix}, \quad
  \begin{bmatrix} \mathbf{Z}_3 \\ -\lambda_3\mathbf{Z}_3  \end{bmatrix}, \quad
  \begin{bmatrix} \mathbf{Z}_1 \\ -\lambda_4\mathbf{Z}_1  \end{bmatrix}, \quad
  \begin{bmatrix} \mathbf{Z}_2 \\ -\lambda_5\mathbf{Z}_2  \end{bmatrix}, \quad
  \begin{bmatrix} \mathbf{Z}_3 \\ -\lambda_6\mathbf{Z}_3  \end{bmatrix}.
\end{equation}
Let $\mathbf{C}_1=\text{diag}(\lambda_1,\lambda_2,\lambda_3)$,
$\mathbf{C}_2=\text{diag}(\lambda_4,\lambda_5,\lambda_6)$, and
$
\mathbf{P} =\begin{bmatrix}
\mathbf{S} & \mathbf{S} \\
-\mathbf{SC}_1 & -\mathbf{SC}_2
\end{bmatrix}.
$
Then matrix $\mathbf{A}$ can be represented as
\begin{equation}
  \mathbf{A} = \mathbf{P}\begin{bmatrix}
    \frac{\lambda_1}{1+\lambda_1^2} & & \\
    & \ddots & \\
    & & \frac{\lambda_6}{1+\lambda_6^2}
  \end{bmatrix}
  \mathbf{P}^T
  = \mathbf{P}\bm{\Lambda} \mathbf{P}^T
  = \mathbf{P}\begin{bmatrix} \bm{\Lambda}^{-} & \mathbf{0} \\
    \mathbf{0} & \bm{\Lambda}^{+} \end{bmatrix}\mathbf{P}^T,
  \label{equ:A_3d_trans}
\end{equation}
where
$
\bm{\Lambda}^{-} = -\bm{\Lambda}^{+}
= -\text{diag}\left(
\frac{1}{2\sqrt{(\xi_1/2)^2+1}},
\frac{1}{2\sqrt{(\xi_2/2)^2+1}},
\frac{1}{2\sqrt{(\xi_3/2)^2+1}}
\right).
$

Let
\begin{equation}
\begin{bmatrix} \mathbf{W}^{-} \\ \mathbf{W}^{+}  \end{bmatrix}
= \mathbf{W} = \mathbf{P}^T \mathbf{X}
= \begin{bmatrix}
  \mathbf{S}^T & -\mathbf{C}_1\mathbf{S}^T \\
  \mathbf{S}^T & -\mathbf{C}_2\mathbf{S}^T
\end{bmatrix}
\begin{bmatrix}
  \mathbf{T}_n \\ \mathbf{U}
\end{bmatrix}
= \begin{bmatrix}
  \mathbf{S}^T\mathbf{T}_n - \mathbf{C}_1\mathbf{S}^T\mathbf{U} \\
  \mathbf{S}^T\mathbf{T}_n - \mathbf{C}_2\mathbf{S}^T\mathbf{U} 
\end{bmatrix},
\label{equ:def_W_3d}
\end{equation}
where
$
\mathbf{T}_n = \begin{bmatrix}T_{nn} \\ T_{n\tau} \\ T_{ns}  \end{bmatrix}
$
and
$
\mathbf{U} = \begin{bmatrix} u_n \\ u_{\tau} \\ u_s  \end{bmatrix}.
$
Analogous to the two-dimensional case, we consider
boundary conditions of the form
$
\mathbf{W}^{-} = \mathbf{R} \mathbf{W}^{+},
$
where $\mathbf{R}$ is a chosen $3\times 3$ constant matrix.
The boundary  term \eqref{equ:bt_quad_form_3d} is then 
transformed into
\begin{equation}
  \begin{split}
  BT &= -\frac{1}{2}\left[
    (\mathbf{W}^{-})^T\bm{\Lambda}^{-}\mathbf{W}^{-}
    + (\mathbf{W}^{+})^T\bm{\Lambda}^{+}\mathbf{W}^{+}
    \right]
  = -\frac{1}{2}(\mathbf{W}^{+})^T \underbrace{\left[
    \mathbf{R}^T\bm{\Lambda}^{-}\mathbf{R}
    + \bm{\Lambda}^{+}
    \right]}_{\mathbf{Q}} \mathbf{W}^{+} \\
  &= -\frac{1}{2}(\mathbf{W}^{+})^T\mathbf{Q}\mathbf{W}^{+}.
  \end{split}
  \label{equ:bt_trans_1_3d}
\end{equation}
We then require that $\mathbf{R}$ be chosen such that
the symmetric matrix
$\mathbf{Q} = \mathbf{R}^T\bm{\Lambda}^{-}\mathbf{R} + \bm{\Lambda}^{+}$
is positive semi-definite to
ensure the non-positivity of the boundary term.

Since the closed form for $\xi_i$ ($i=1,2,3$) is unknown,
it is difficult to determine
the general form for the matrix $\mathbf{R}$ that ensures
the positive semi-definiteness of $\mathbf{Q}$. 
In the current work we consider only the following special case
for three dimensions: We assume that  $\mathbf{R}$ is
a diagonal matrix,
i.e.~$\mathbf{R} = \text{diag}(a_{11}, a_{22}, a_{33})$.
With this assumption,
the requirement that $\mathbf{Q}$ be symmetric positive semi-definite
leads to the conditions
\begin{equation}
  a_{11}^2 \leqslant 1, \quad
  a_{22}^2 \leqslant 1, \quad
  a_{33}^2 \leqslant 1.
  \label{equ:cond_R_1_3d}
\end{equation}
Substitution of the expressions for $\mathbf{W}^{-}$ and
$\mathbf{W}^{+}$ in \eqref{equ:def_W_3d} into
the boundary condition leads to
\begin{equation}
  (\mathbf{I}_3-\mathbf{R})\mathbf{S}^T\mathbf{T}_n =
  (\mathbf{C}_1 - \mathbf{R}\mathbf{C}_2)\mathbf{S}^T\mathbf{U}.
  \label{equ:bc_3d_1}
\end{equation}
We further require that
$(\mathbf{I}_3-\mathbf{R})$ be non-singular.
%$
%\text{det}(\mathbf{I}-\mathbf{R}) \neq 0
%$.
With this and the condition \eqref{equ:cond_R_1_3d},
we arrive at the matrix $\mathbf{R}$ for this
special case,
\begin{equation}
  \mathbf{R} = \text{diag}(a_{11}, a_{22}, a_{33}), \
  \text{where} \ -1\leqslant a_{11}, a_{22}, a_{33} < 1.
  \label{equ:cond_R_3d}
\end{equation}

With this $\mathbf{R}$ matrix the boundary condition \eqref{equ:bc_3d_1}
becomes
\begin{equation}
  \begin{split}
   \mathbf{T}_n 
  = \mathbf{S}(\mathbf{I}_3-\mathbf{R})^{-1}
  (\mathbf{C}_1-\mathbf{RC}_2)\mathbf{S}^T\mathbf{U}
  = \mathbf{S}\begin{bmatrix}
    \frac{\lambda_1-\lambda_4a_{11}}{1-a_{11}} & & \\
    & \frac{\lambda_2-\lambda_5a_{22}}{1-a_{22}} & \\
    & & \frac{\lambda_3-\lambda_6a_{33}}{1-a_{33}} 
  \end{bmatrix} \mathbf{S}^T\mathbf{U}.
  \end{split}
  \label{equ:bc_3d_2}
\end{equation}
Equivalently, it can be written as
\begin{equation}
  \begin{bmatrix}T_{nn} \\ T_{n\tau} \\ T_{ns}  \end{bmatrix}
  = \mathbf{L}\begin{bmatrix}u_n \\ u_{\tau} \\ u_s  \end{bmatrix}
  =\begin{bmatrix} L_{11}u_n + L_{12}u_{\tau} + L_{13}u_s \\
  L_{21}u_n+L_{22}u_{\tau}+L_{23}u_s \\
  L_{31}u_n + L_{32}u_{\tau}+L_{33}u_s\end{bmatrix}
  = \begin{bmatrix} g_1(u_n,u_{\tau},u_s) \\ g_2(u_n,u_{\tau},u_s)
    \\ g_3(u_n,u_{\tau},u_s)  \end{bmatrix}
  \label{equ:bc_3d}
\end{equation}
where
\begin{equation}
  \begin{bmatrix} L_{11} & L_{22} & L_{33} \\
  L_{21} & L_{22} & L_{23} \\ L_{31} & L_{32} & L_{33}\end{bmatrix}
  = \mathbf{L}
  = \mathbf{S}\begin{bmatrix}
    \frac{\lambda_1-\lambda_4a_{11}}{1-a_{11}} & & \\
    & \frac{\lambda_2-\lambda_5a_{22}}{1-a_{22}} & \\
    & & \frac{\lambda_3-\lambda_6a_{33}}{1-a_{33}} 
  \end{bmatrix} \mathbf{S}^T.
  \label{equ:def_L_3d}
\end{equation}
In vector form, this boundary condition has the same form
as given by \eqref{equ:bc_form}, but here
in three dimensions $\mathbf{E}(\mathbf{u},\partial\Omega)$
is given by
\begin{equation}
  \mathbf{E}(\mathbf{u},\partial\Omega)
  = g_1(u_n,u_{\tau},u_s)\mathbf{n} + g_2(u_n,u_{\tau},u_s)\bm{\tau}
  + g_3(u_n,u_{\tau},u_s)\mathbf{s},
  \label{equ:def_E_3d}
\end{equation}
where $g_1$, $g_2$ and $g_3$ are defined by \eqref{equ:bc_3d}.
The boundary term
\eqref{equ:bt_trans_1_3d} is accordingly transformed into
\begin{equation}
  BT = -\frac{1}{2}(\mathbf{S}^T\mathbf{U})^T
  \begin{bmatrix}
    \frac{2(1+a_{11})\sqrt{\left(\xi_1/2 \right)^2+1}}{1-a_{11}} & & \\
    & \frac{2(1+a_{22})\sqrt{\left(\xi_2/2 \right)^2+1}}{1-a_{22}} & \\
    & & \frac{2(1+a_{33})\sqrt{\left(\xi_3/2 \right)^2+1}}{1-a_{33}}
  \end{bmatrix}
  (\mathbf{S}^T\mathbf{U}).
  \label{equ:bt_3d}
\end{equation}
This expression indicates that the dissipativeness of
the boundary condition \eqref{equ:bc_3d} is controlled by
the coefficients $a_{ii}$ ($i=1,2,3$). The larger
the values for $a_{ii}$, the more dissipative the boundary condition
is. When $a_{11}=a_{22}=a_{33}=-1$, the boundary dissipation vanishes.
When $a_{ii}\rightarrow 1$ ($i=1,2,3$), the boundary dissipation
approaches infinity.

In summary,
the procedure for computing  $\mathbf{E}(\mathbf{u},\partial\Omega)$
in the 3D boundary condition is as follows.
Given domain boundary $\partial\Omega$ (with normal and tangent vectors
$\mathbf{n}$, $\bm{\tau}$, $\mathbf{s}$), the velocity $\mathbf{u}$
on $\partial\Omega$, the chosen constants $a_{ii}$ ($i=1,2,3$),
 $\alpha_1$ and $\alpha_2$,
we take the following steps:
\begin{itemize}

\item
  Compute $u_n$, $u_{\tau}$, $u_s$ based on equation \eqref{equ:def_Tnn_3d}; 

\item
  Form matrix $\mathbf{G}$.
  Compute the eigenvalues $\xi_i$ ($i=1,2,3$) and the eigenvectors of $\mathbf{G}$.
  Use the eigenvectors to form the orthogonal matrix $\mathbf{S}$.

\item
  Compute $\lambda_i$ ($i=1,\cdots,6$) by equation \eqref{equ:def_lambda_3d}.
  Compute matrix $\mathbf{L}$ by equation \eqref{equ:def_L_3d}.

\item
  Compute $g_1$, $g_2$, $g_3$ according to equation \eqref{equ:bc_3d},
  \begin{equation}
    \left\{
    \begin{split}
      &
      g_1(u_n,u_{\tau},u_s) = L_{11}u_n + L_{12}u_{\tau} + L_{13}u_s, \\
      &
      g_2(u_n,u_{\tau},u_s) = L_{21}u_n+L_{22}u_{\tau}+L_{23}u_s, \\
      &
      g_3(u_n,u_{\tau},u_s) = L_{31}u_n + L_{32}u_{\tau}+L_{33}u_s.
    \end{split}
    \right.
    \label{equ:def_fgh_3d}
  \end{equation}

\item
  Form $\mathbf{E}(\mathbf{u},\partial\Omega)$ based on equation \eqref{equ:def_E_3d}.
  
\end{itemize}

% what else to discuss here?

\subsection{Open/Outflow Boundary Conditions for Incompressible Flows}
\label{sec:obc}

% 2 boundary types
% comment on dissipativeness
% specify open boundary conditions, 2 types:
%    (i) without Theta function
%    (ii) with Theta function
% compare with OBCs of Dong (2015) etc
% remarks on convective-type, traction type
% discuss numerical algorithm; provide algorithm in Appendix
%
% 2 OBCs:
% (i) a_ii varied, such that dissipation is 0.5*u_n*u^2
% (ii) a_ii are fixed, small negative values

The class of boundary conditions obtained in
the previous section ensures that the boundary contribution
in the energy balance equation
will not cause the system energy
to increase over time.
We next apply these boundary conditions to specifically deal with
outflow or open boundaries. 

We assume that two types of boundaries (which are non-overlapping)
are present in the domain:
$\partial\Omega = \partial\Omega_d\cup\partial\Omega_o$.
$\partial\Omega_d$ is the Dirichlet type boundary,
on which the velocity is given,
\begin{equation}
  \mathbf{u} = \mathbf{w}(\mathbf{x},t), \quad
  \text{on} \ \partial\Omega_d,
  \label{equ:dbc}
\end{equation}
where $\mathbf{w}$ is the boundary velocity.
$\partial\Omega_o$ is the open or outflow boundary,
on which neither of the flow variables (velocity, pressure)
is known. 

On the outflow/open boundary $\partial\Omega_o$ we impose
the family of boundary conditions from Section \ref{sec:esbc},
\begin{equation}
  -p\mathbf{n} + \nu\mathbf{n}\cdot\nabla\mathbf{u}
  - \mathbf{E}(\mathbf{u},\partial\Omega_o) = 0,
  \quad \text{on} \ \partial\Omega_o,
  \label{equ:obc}
\end{equation}
where
\begin{equation}
  \mathbf{E}(\mathbf{u},\partial\Omega_o) = \left\{
  \begin{array}{ll}
    f_1(u_n,u_{\tau})\mathbf{n} + f_2(u_n,u_{\tau})\bm{\tau}, & \text{in 2D}, \\
    g_1(u_n,u_{\tau},u_s)\mathbf{n} + g_2(u_n,u_{\tau},u_s)\bm{\tau}
    + g_3(u_n,u_{\tau},u_s)\mathbf{s}, & \text{in 3D}.
  \end{array}
  \right.
  \label{equ:def_obc_E}
\end{equation}
In the above expressions $f_1$ and $f_2$ are given by
\eqref{equ:bc_p_nn}--\eqref{equ:bc_n_nt} or
\eqref{equ:bc_1_0}--\eqref{equ:bc_2_0}, and
$g_i$ ($i=1,2,3$) are given by \eqref{equ:def_fgh_3d}.
In two dimensions, $(\mathbf{n},\bm\tau)$ are the local unit 
vectors normal 
and tangent to $\partial\Omega_o$, and $(u_n,u_{\tau})$
are the local velocity components in these directions.
In three dimensions,
$(\mathbf{n},\bm\tau, \mathbf{s})$ are the local unit vectors
normal  to $\partial\Omega_o$
and along the two tangent directions of $\partial\Omega_o$,
and $(u_n,u_{\tau},u_s)$ are the local velocity components
in these directions.

The dissipation on the boundary, upon imposing
the condition \eqref{equ:def_obc_E},
is determined by the coefficients $\frac{1+a_{ii}}{1-a_{ii}}$
($i=1,2,3$). When $a_{ii}\rightarrow -1$ the 
boundary dissipation will vanish, and
when $a_{ii}\rightarrow 1$ the boundary dissipation will become infinite.
Both of these cases are obviously unphysical, even though they
may be energy stable.
So we expect
that the physical accuracy of the simulation results will be poor for these
cases. Indeed, numerical experiments seem to suggest that
the best results appear to correspond to values of $a_{ii}$ somewhat negative
and not too far from zero.

If no backflow occurs on the open boundary $\partial\Omega_o$, an
often-used boundary condition is the traction free condition,
i.e.~$
\mathbf{n}\cdot\mathbf{T}=-p\mathbf{n} + \nu\mathbf{n}\cdot\nabla\mathbf{u}=0,
$
which produces reasonable simulation results
but is unstable if backflow occurs at moderate and high Reynolds numbers.
With the traction-free condition,
the boundary term in \eqref{equ:energy_balance} becomes
$ %\begin{equation}
  BT = -\frac{1}{2}(\mathbf{n}\cdot\mathbf{u})|\mathbf{u}|^2,
%  \label{equ:dissipation_tract_free}
$ %\end{equation}
which physically means that the kinetic energy is convected out of
the domain by the normal velocity (when $u_n>0$).
This suggests that a reasonable scale for the magnitude of dissipation 
on the open boundary $\partial\Omega_o$ is comparable to
\begin{equation}
  |BT| = \frac{1}{2} |u_n| |\mathbf{u}|^2.
  \label{equ:dissipation_scale}
\end{equation}

% Note:
%  a_ii conditions are derived based on the velocity at locally
%  a single point. At different points on \partial\Omega_o, the a_ii
%  in principle does not have to be the same.
%  that is, a_ii can be a function, as long as it satisfies the conditions from
%  previous subsection.
%  so a_ii can vary from point to point on the boundary, and it can vary in time,
%  as long as it satisfies the conditions from previous subsection
%  on each point

For further development
it is important to realize another point. 
The conditions for the $\mathbf{R}$ matrix that ensure
the energy dissipation on the boundary derived in the previous section are
based on the local point-wise values
of the flow fields on each individual point of the boundary. 
Energy stability is guaranteed 
as long as the coefficient $a_{ii}$ satisfies
the conditions given in Section \ref{sec:esbc}
for an $\alpha$ coefficient (or the coefficients
$\alpha_1$ and $\alpha_2$ in 3D)
on each individual point of the boundary $\partial\Omega_o$.
On different points of the boundary and over time, however,
$a_{ii}$ and $\alpha$ do not have to assume the same value.
In other words, the coefficients can be e.g.~prescribed
field distributions $a_{ii}(\mathbf{x},t)$
and $\alpha(\mathbf{x},t)$ (or $\alpha_1(\mathbf{x},t)$
and $\alpha_2(\mathbf{x},t)$ in 3D),
as long as they satisfy
the conditions from Section \ref{sec:esbc} on each point of $\partial\Omega_o$
at all time.

In light of the above observations,
with the boundary conditions \eqref{equ:obc} and \eqref{equ:def_obc_E}
we will consider two configurations for $a_{ii}$ ($i=1,2,3$ for 3D and
$i=1,2$ for 2D) and $\alpha$:
\begin{itemize}

\item
  $a_{ii}$ and $\alpha$ (or in 3D, $\alpha_1$ and $\alpha_2$)
  are  uniform constants on the entire boundary
  $\partial\Omega_o$ and over time,
  which satisfy the conditions from Section \ref{sec:esbc}.

\item
  $a_{ii}$ and $\alpha$ (or in 3D, $\alpha_1$ and $\alpha_2$)
  may be field distributions, which (i) satisfy
  the conditions from Section \ref{sec:esbc} for energy stability
  and further (ii) are such that the magnitude of boundary
  dissipation on $\partial\Omega_o$ satisfies equation
  \eqref{equ:dissipation_scale}. 

\end{itemize}
These two configurations lead to two different sets of open boundary conditions. 
%We refer to the boundary condition corresponding to the first configuration as
%OBC-A, and that corresponding to the second configuration OBC-B.

Let us look into the second set of boundary conditions in more detail.
There are many means to choose $a_{ii}$ and $\alpha$
to satisfy \eqref{equ:dissipation_scale}.
We specifically consider two ways below.
In the first, we note that in the 3D equation \eqref{equ:bt_3d}
$\mathbf{S}$ is an orthogonal matrix and
$(\mathbf{S}^T\mathbf{U})^T(\mathbf{S}^T\mathbf{U})=\mathbf{U}^T\mathbf{U}
= |\mathbf{u}|^2.
$
Therefore the boundary term \eqref{equ:bt_3d} reduces
to \eqref{equ:dissipation_scale} if
\begin{equation}
  \frac{1+a_{ii}}{1-a_{ii}}2\sqrt{\left(\frac{\xi_i}{2}\right)^2+1} = |u_n|,
  \quad \text{or} \ \
  a_{ii} = -\frac{\sqrt{\xi_i^2+4} - |u_n|}{\sqrt{\xi_i^2+4} + |u_n|},
  \label{equ:opt_R}
\end{equation}
for $i=1,2,3$, where $\xi_i$ ($i=1,2,3$) are the
eigenvalues of the matrix $\mathbf{G}$ and depends on $\alpha_1$
and $\alpha_2$.
Similarly, for two dimensions the boundary terms in
equations \eqref{equ:bt_form_1} and \eqref{equ:bt_0_simple}
will reduce to the form given
by equation \eqref{equ:dissipation_scale} if $a_{ii}$ 
are given by the same expression as in \eqref{equ:opt_R}
for $i = 1,2$, noting that $\xi_i$ ($i=1,2$)  are now
the eigenvalues of the matrix $\mathbf{G}$ in two dimensions
and depend on $\alpha$.
%and that $\xi_1=\xi_2 = u_n$ for the case $\alpha=0$.
%
% more simplified form for optimal R
%
Substitution of the $a_{ii}$ expression \eqref{equ:opt_R} into
equations \eqref{equ:bc_p_nn}--\eqref{equ:bc_n_nt}
and also equations \eqref{equ:def_fgh_3d} results in
greatly simplified expressions for $f_i$ ($i=1,2$) and $g_i$ ($i=1,2,3$)
in equation \eqref{equ:def_obc_E},
as given below,
\begin{equation}
  \text{(2D)} \left\{
  \begin{split}
    &
    f_1(u_n,u_{\tau}) = \frac{1}{2}\left[
      (u_n - |u_n|)u_n + \alpha u_{\tau}^2
      \right], \\
    &
    f_2(u_n,u_{\tau}) = \frac{1}{2}u_{\tau}\left[
      (1-\alpha)u_n - |u_n|
      \right];
  \end{split}
  \right.
  \label{equ:opt_R_2d}
\end{equation}
\begin{equation}
  \text{(3D)} \left\{
  \begin{split}
    &
    g_1(u_n,u_{\tau},u_s) = \frac{1}{2}\left[
      (u_n-|u_n|)u_n + \alpha_1u_{\tau}^2 + \alpha_2u_s^2
      \right], \\
    &
    g_2(u_n,u_{\tau},u_s) = \frac{1}{2}u_{\tau}\left[
      (1-\alpha_1)u_n - |u_n|
      \right], \\
    &
    g_3(u_n,u_{\tau},u_s) = \frac{1}{2}u_s\left[
      (1-\alpha_2)u_n - |u_n|
      \right].
  \end{split}
  \right.
  \label{equ:opt_R_3d}
\end{equation}

We fix the $\alpha$ coefficient in 2D
(or $\alpha_1$, $\alpha_2$ in 3D) as follows.
In 2D we will
determine $\alpha$ to try to make $a_{ii}$ as large as possible.
This requirement is based on the following observation:
Numerical simulations with $a_{ii}$ and $\alpha$ as uniform constants
suggest that larger $a_{ii}$ values ($-1\leqslant a_{ii}<1$)
appear to be able to reduce the lateral meandering or distortion of
the vortex street
as the vortices cross the open boundary.
In light of the expression \eqref{equ:opt_R}, we will find $\alpha$
such that the
$
\min(|\xi_1|, |\xi_2|)
$
is minimized and that $\alpha$ will be well defined for all $|\mathbf{u}|> 0$.
Substituting the $\xi_1$
and $\xi_2$ expressions in \eqref{equ:def_xi} into this condition, we get
\begin{equation}
  \alpha = \frac{|u_n|}{|\mathbf{u}| + |u_n| + \epsilon},
  \qquad \text{(for 2D)}
  \label{equ:def_alpha_2d}
\end{equation}
where $\epsilon$ is a small positive number on the order of machine zero or smaller
(e.g.~$\epsilon\sim 10^{-18}$) to make the above expression well-defined
even if $|\mathbf{u}|=0$. Note that this expression satisfies
$0\leqslant \alpha\leqslant \frac{1}{2}$.
For three dimensions, the closed form for
$\xi_i$ ($i=1,2,3$) as a function of
$\alpha_1$ and $\alpha_2$ is unknown, and the coefficients
$\alpha_1$ and $\alpha_2$ cannot be determined as such.
Inspired by the 2D result of \eqref{equ:def_alpha_2d},
we will employ this same expression for
$\alpha_1$ and $\alpha_2$ in 3D in this work, that is,
\begin{equation}
  \alpha_1 = \alpha_2 = \frac{|u_n|}{|\mathbf{u}| + |u_n| + \epsilon},
  \qquad \text{(for 3D)}.
  \label{equ:def_alpha_3d}
\end{equation}
The boundary condition \eqref{equ:obc}, with $E(\mathbf{n},\partial\Omega_o)$
given by \eqref{equ:def_obc_E} and $f_i$ and $g_i$
given by \eqref{equ:opt_R_2d} and \eqref{equ:opt_R_3d},
in which $\alpha$ is given by \eqref{equ:def_alpha_2d}
and $\alpha_1$ and $\alpha_2$ are given by \eqref{equ:def_alpha_3d},
is energy stable and the magnitude of boundary dissipation on $\partial\Omega_o$
satisfies equation \eqref{equ:dissipation_scale}.

As a second way to satisfy equation \eqref{equ:dissipation_scale},
we assume that all $a_{ii}$ ($i=1,2,3$) are identical.
In 3D, let 
$
\begin{bmatrix}V_1 \\ V_2 \\ V_3  \end{bmatrix}
=\mathbf{S}^T\mathbf{U} = \mathbf{S}^T\begin{bmatrix} u_n \\ u_{\tau} \\ u_s \end{bmatrix}
$
and $Y_i = 2\sqrt{(\xi_i/2)^2 + 1}$ ($i=1,2,3$). 
Substitution of the boundary term  \eqref{equ:bt_3d}
into equation \eqref{equ:dissipation_scale} results in
\begin{equation}
a_{11} = a_{22} = a_{33} = \frac{J_t-1}{J_t+1}, 
\quad \text{where} \
J_t = \frac{|u_n||\mathbf{u}|^2}{\left(Y_1V_1^2 + Y_2V_2^2 + Y_3V_3^2\right) + \epsilon}.
\label{equ:obc_C_3d}
\end{equation}
In the above equation $\epsilon$ is a small positive number 
on the order of machine zero or smaller
to make the $J_t$ expression well defined when $|\mathbf{u}|=0$.
In 2D, let
$
\begin{bmatrix} V_1 \\ V_2 \end{bmatrix}
=\frac{1}{\sqrt{1+\eta^2}}\mathbf{S}^T\begin{bmatrix} u_n \\ u_{\tau}  \end{bmatrix}
$
and $Y_i = 2\sqrt{(\xi_i/2)^2 + 1}$ ($i=1,2$),
where the matrix $\mathbf{S}$ is given by \eqref{equ:def_B}.
Then the coefficients $a_{ii}$ are given by
\begin{equation}
a_{11} = a_{22} = \frac{J_t-1}{J_t+1}, 
\quad \text{where} \
J_t = \frac{|u_n||\mathbf{u}|^2}{\left(Y_1V_1^2 + Y_2V_2^2\right) + \epsilon}.
\label{equ:obc_C_2d}
\end{equation}
In 2D
we again determine $\alpha$ to try to
make $a_{ii}$ (or equivalently $J_t$) as large as
possible. Note that
$|\mathbf{u}|^2=V_1^2+V_2^2$ in 2D, 
and (omitting the $\epsilon$)
$
J_t = \frac{|u_n|}{Y_1\frac{V_1^2}{V_1^2+V_2^2}
  +Y_2\frac{V_2^2}{V_1^2+V_2^2} }
\leqslant \frac{|u_n|}{\min(Y_1,Y_2) }.
$
We minimize $\min(Y_1,Y_2)$, or equivalently $\min(|\xi_1|,|\xi_2|)$,
and obtain again the expression \eqref{equ:def_alpha_2d} for $\alpha$.
For 3D we will again employ the expressions in \eqref{equ:def_alpha_3d}
for $\alpha_1$ and $\alpha_2$.
%
%Note that $a_{ii}$ given by \eqref{equ:obc_C_3d} or \eqref{equ:obc_C_2d}
%are field distributions on $\partial\Omega_o$. 
%Substitution of these $a_{ii}$ expressions into \eqref{equ:obc}--\eqref{equ:def_obc_E}
%gives rise to an energy-stable boundary condition whose boundary
%dissipation satisfies the condition \eqref{equ:dissipation_scale}.
The boundary condition \eqref{equ:obc} with
$\mathbf{E}(\mathbf{u},\partial\Omega_o)$ given by
\eqref{equ:def_obc_E}, in which
$a_{ii}$ is given by \eqref{equ:obc_C_2d} or \eqref{equ:obc_C_3d}
and $\alpha$ (or $\alpha_1$ and $\alpha_2$) are given by
\eqref{equ:def_alpha_2d} or \eqref{equ:def_alpha_3d},
is another energy-stable open boundary condition
whose boundary dissipation satisfies equation \eqref{equ:dissipation_scale}.

%It should be noted that
%the $a_{ii}$ given by \eqref{equ:opt_R} is only a sufficient condition
%for meeting the requirement that the boundary dissipation satisfy
%equation \eqref{equ:dissipation_scale}.
%It is not the only way to satisfy this requirement.

% comment on another way to achieve a dissipation scale by
%   multiplying E expression by steps function

\begin{remark}

  One can multiply the $\mathbf{E}(\mathbf{n},\partial\Omega_o)$ expression
  in \eqref{equ:def_obc_E} by
  the smoothed step function introduced in~\cite{DongKC2014,DongS2015}
  to approximately
  enforce the requirement that the magnitude of boundary dissipation
  should be comparable to that given by
  \eqref{equ:dissipation_scale}.
   The modified $\mathbf{E}(\mathbf{n},\partial\Omega_o)$ is given by
\begin{equation}
  \mathbf{E}(\mathbf{u},\partial\Omega_o) =
  \left\{
  \begin{array}{ll}
    \left[f_1(u_n,u_{\tau})\mathbf{n} + f_2(u_n,u_{\tau})\bm{\tau}\right]\Theta(\mathbf{n},\mathbf{u}),
    & \text{in 2D}, \\
    \left[g_1(u_n,u_{\tau},u_s)\mathbf{n} + g_2(u_n,u_{\tau},u_s)\bm{\tau}
      + g_3(u_n,u_{\tau},u_s)\mathbf{s}\right]\Theta(\mathbf{n},\mathbf{u}),
      & \text{in 3D},
  \end{array}
  \right.
  \label{equ:def_obc_E2}
\end{equation}
where $\Theta(\mathbf{n},\mathbf{u})$ is a smoothed step function
given by (see \cite{DongS2015})
\begin{equation}
  \Theta(\mathbf{n},\mathbf{u}) = \frac{1}{2}\left(
  1 - \tanh\frac{\mathbf{n}\cdot\mathbf{u}}{U_0\delta}
  \right), \quad \text{and} \
  \lim_{\delta\rightarrow 0} \Theta(\mathbf{n},\mathbf{u})
= \Theta_0(\mathbf{n},\mathbf{u})
= \left\{
\begin{array}{ll}
1, & \text{if} \ \mathbf{n}\cdot\mathbf{u} < 0, \\
0, & \text{otherwise}.
\end{array}
\right.
\end{equation}
In the above expression, $U_0$ is the velocity scale, and
$\delta>0$ is a small positive constant that
controls the sharpness of the smoothed step function.
As $\delta\rightarrow 0$, $\Theta(\mathbf{n},\mathbf{u})$
approaches the unit step function $\Theta_0(\mathbf{n},\mathbf{u})$,
taking unit value if $u_n<0$ and vanishing otherwise.
This modified $\mathbf{E}(\mathbf{u},\partial\Omega_o)$
enforces the following requirement:
(i) if locally there is no backflow on $\partial\Omega_o$ (i.e.~$u_n\geqslant 0$),
then the boundary condition \eqref{equ:obc} should reduce
to the traction-free condition;
(ii) if backflow occurs locally on $\partial\Omega_o$ (i.e.~$u_n<0$),
then the boundary condition \eqref{equ:obc} shall reduce to
the form with $\mathbf{E}(\mathbf{u},\partial\Omega_o)$ given by
equation \eqref{equ:def_obc_E}.

\end{remark}

%%%%%%%%%%%%%%%%%%%%%%%%%%%%%%%
\begin{comment}

% comment on convective type esobc by adding du/dt term to OBC

\begin{remark}
\label{remark:mark_3}

  The boundary condition \eqref{equ:obc} can be considered as
  a traction-type energy-stable
  open boundary condition, because it imposes a traction force on
  $\partial\Omega_o$,
  $
  \mathbf{n}\cdot\mathbf{T} = -p\mathbf{n} + \nu\mathbf{n}\cdot\nabla\mathbf{u}
  = \mathbf{E}(\mathbf{u},\partial\Omega_o).
  $
  One can also devise a corresponding convective-type
  energy-stable open boundary condition,
  by using the idea of \cite{Dong2015obc}, as follows,
  \begin{equation}
    \nu D_0\frac{\partial\mathbf{u}}{\partial t}
    -p\mathbf{n} + \nu\mathbf{n}\cdot\nabla\mathbf{u}
    - \mathbf{E}(\mathbf{u},\partial\Omega_o) = 0,
    \label{equ:convective_obc}
  \end{equation}
  where $D_0\geqslant 0$ is a constant and $\frac{1}{D_0}$ (for $D_0>0$) represents
  a convection velocity scale at the open boundary $\partial\Omega_o$,
  and $\mathbf{E}(\mathbf{u},\partial\Omega_o)$ is given by
  \eqref{equ:def_obc_E}, or \eqref{equ:opt_R_2d} and \eqref{equ:opt_R_3d},
  or \eqref{equ:def_obc_E2}.
  The boundary condition \eqref{equ:convective_obc} can be shown to
  ensure that the contribution of the boundary integral term
  in the energy balance equation is always non-positive, and
  therefore this boundary condition ensures energy stability.
  
\end{remark}

\end{comment}
%%%%%%%%%%%%%%%%%%%%%%%%%%%%%%%%%%%%%

In the current work, we will concentrate on
three open boundary conditions (referred to as OBC-A, OBC-B and OBC-C, respectively)
corresponding to the two configurations for $a_{ii}$ as discussed above.
More specifically,
we will concentrate on the open boundary condition \eqref{equ:obc} with
$\mathbf{E}(\mathbf{u},\partial\Omega_o)$ given by \eqref{equ:def_obc_E},
in which $f_i$ ($i=1,2$) and $g_i$ ($i=1,2,3$) take three different forms:
\begin{enumerate}[label=(OBC-\Alph*)]

\item
  $f_i$ ($i=1,2$) are given by \eqref{equ:bc_p_nn}--\eqref{equ:bc_n_nt}
  or \eqref{equ:bc_1_0}--\eqref{equ:bc_2_0}
  and $g_i$ ($i=1,2,3$) are given by \eqref{equ:def_fgh_3d},
  in which $a_{ii}$ and $\alpha$ (or $\alpha_1$ and $\alpha_2$)
  are uniform  constants on the entire $\partial\Omega_o$.

\item
  $f_i$ ($i=1,2$) are given by equation \eqref{equ:opt_R_2d},
  in which $\alpha$ is given by \eqref{equ:def_alpha_2d}, and
  $g_i$ ($i=1,2,3$) are given by equation \eqref{equ:opt_R_3d},
  in which $\alpha_1$ and $\alpha_2$ are given by \eqref{equ:def_alpha_3d}.

\item
  $f_i$ ($i=1,2$) are given by \eqref{equ:bc_p_nn}--\eqref{equ:bc_n_nt}
  and $g_i$ ($i=1,2,3$) are given by \eqref{equ:def_fgh_3d},
  in which $a_{ii}$ are given by \eqref{equ:obc_C_2d} for 2D and
  \eqref{equ:obc_C_3d} for 3D, and $\alpha$ is given by \eqref{equ:def_alpha_2d}
  and $\alpha_1$ and $\alpha_2$ are given by \eqref{equ:def_alpha_3d}.

\end{enumerate}

With OBC-A the dissipation on the
open boundary depends on
the constants $a_{ii}$ and the flow field at $\partial\Omega_o$.
It is anticipated that the algorithmic parameters $a_{ii}$ will
influence the accuracy of simulation results, and that certain 
$a_{ii}$ values may lead to poor physical results
(e.g.~$a_{ii}$ close to $1$ and $-1$).
%Moreover, the values of $a_{ii}$ that lead to reasonable physical 
%results for one problem may give rise to poor physical results
%for a different problem. In other words, the best values for
%$a_{ii}$ with OBC-A may be problem-dependent. 
It is also likely the case that the best values for $a_{ii}$
with OBC-A will be flow-problem dependent.
These points will indeed be observed and confirmed from the numerical experiments
in Section \ref{sec:tests}.
With OBC-B and OBC-C, on the other hand, the dissipation on
$\partial\Omega_o$ matches the scale given by
equation \eqref{equ:dissipation_scale}. We
anticipate that OBC-B and OBC-C will lead to more accurate
simulation results. This will indeed be demonstrated by
the numerical simulations in Section \ref{sec:tests}.

% numerical algorithm

Let us finally discuss how to implement the class of open boundary
conditions represented by \eqref{equ:obc} in numerical simulations.
The equations \eqref{equ:nse}--\eqref{equ:continuity}, supplemented
by the boundary conditions \eqref{equ:dbc}--\eqref{equ:obc}
and the initial condition \eqref{equ:ic}, constitute
the system to be solved for in numerical simulations.
This system of equations and the boundary conditions
 are similar in form to those considered in our previous
works~\cite{DongS2015,Dong2015obc}.
Therefore one can employ the algorithms developed in~\cite{DongS2015} 
or \cite{Dong2015obc}
to numerically solve the current system of equations.
In this work, we employ the scheme from \cite{Dong2015obc}
(presented in Section 2.4 of \cite{Dong2015obc}) to simulate
the system consisting of \eqref{equ:nse}--\eqref{equ:continuity},
\eqref{equ:dbc}--\eqref{equ:obc} and \eqref{equ:ic}.
This is a velocity-correction type splitting scheme, 
in which the computations for the pressure and the 
velocity are de-coupled.
For the sake of completeness, we have provided a summary of
this algorithm in Appendix B, in which some details
on the implementation of these boundary conditions are given.
In the 2D implementation, the spatial discretization is
performed using a high-order spectral 
element method~\cite{SherwinK1995,KarniadakisS2005,ZhengD2011}.
In the 3D implementation, we restrict our attention to 
flow domains with 
at least one homogeneous direction (designated as $z$ direction),
while in the other two directions the domain can be arbitrarily
complex. Therefore for spatial discretizations
we employ a Fourier spectral expansion
of the field variables along the homogeneous $z$ direction,
and a spectral element expansion within the non-homogeneous $x$-$y$ planes.
%The spatial discretization 
%is based on a combined spectral element method within
%the $x$-$y$ plane and a Fourier spectral method along
%the $z$ direction;
We refer to e.g.~\cite{DongK2005,DongKER2006,Dong2007}
for more detailed discussions of the hybrid discretization of
the Navier-Stokes equations with
Fourier spectral and spectral-element methods.
All three boundary conditions, OBC-A, OBC-B and OBC-C, 
have been implemented in 2D,
and in 3D only OBC-B and OBC-C are implemented.

% what else to discuss here?

%%% Test problems: 
\section{Representative Numerical Examples}
\label{sec:tests}

In this section we present 
numerical simulations for several representative
flow problems in two and three dimensions to test the performance
of the energy-stable open boundary conditions developed in
Section \ref{sec:method}.
All these problems involve open/outflow boundaries,
and the  open boundary conditions
are critical to the stability
of computation at moderate and high Reynolds numbers for
these flows.

\subsection{Convergence Rates}

We first use a manufactured analytic solution
to the incompressible Navier-Stokes equations on a domain with
open boundaries to test the spatial and temporal
convergence rates of our method together with the energy-stable
boundary conditions from Section \ref{sec:method}.

\paragraph{Two Dimensions (2D)}

\begin{figure}
  \centerline{
    \includegraphics[height=2in]{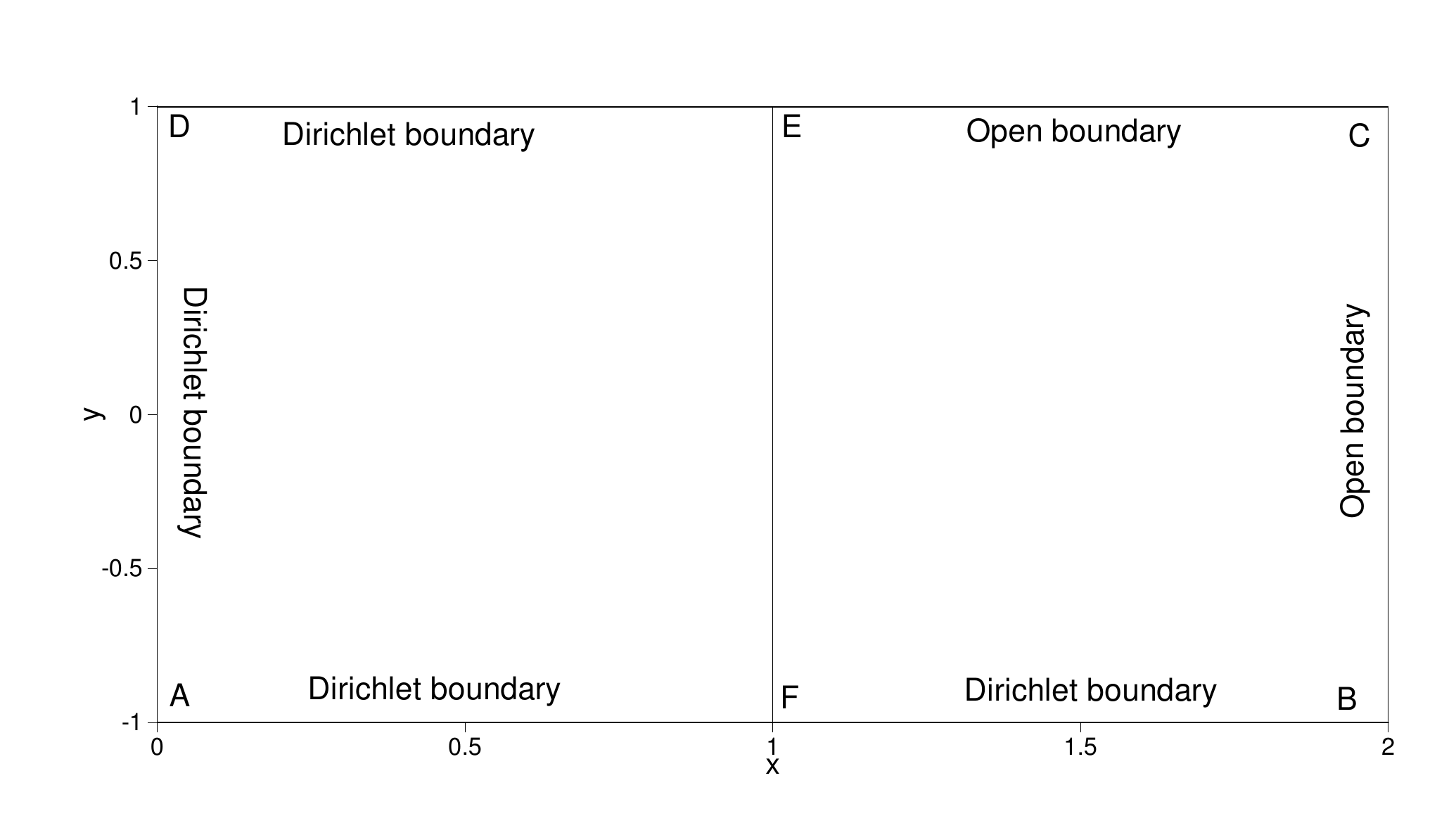}(a)
  }
  \centerline{
    \includegraphics[width=3in]{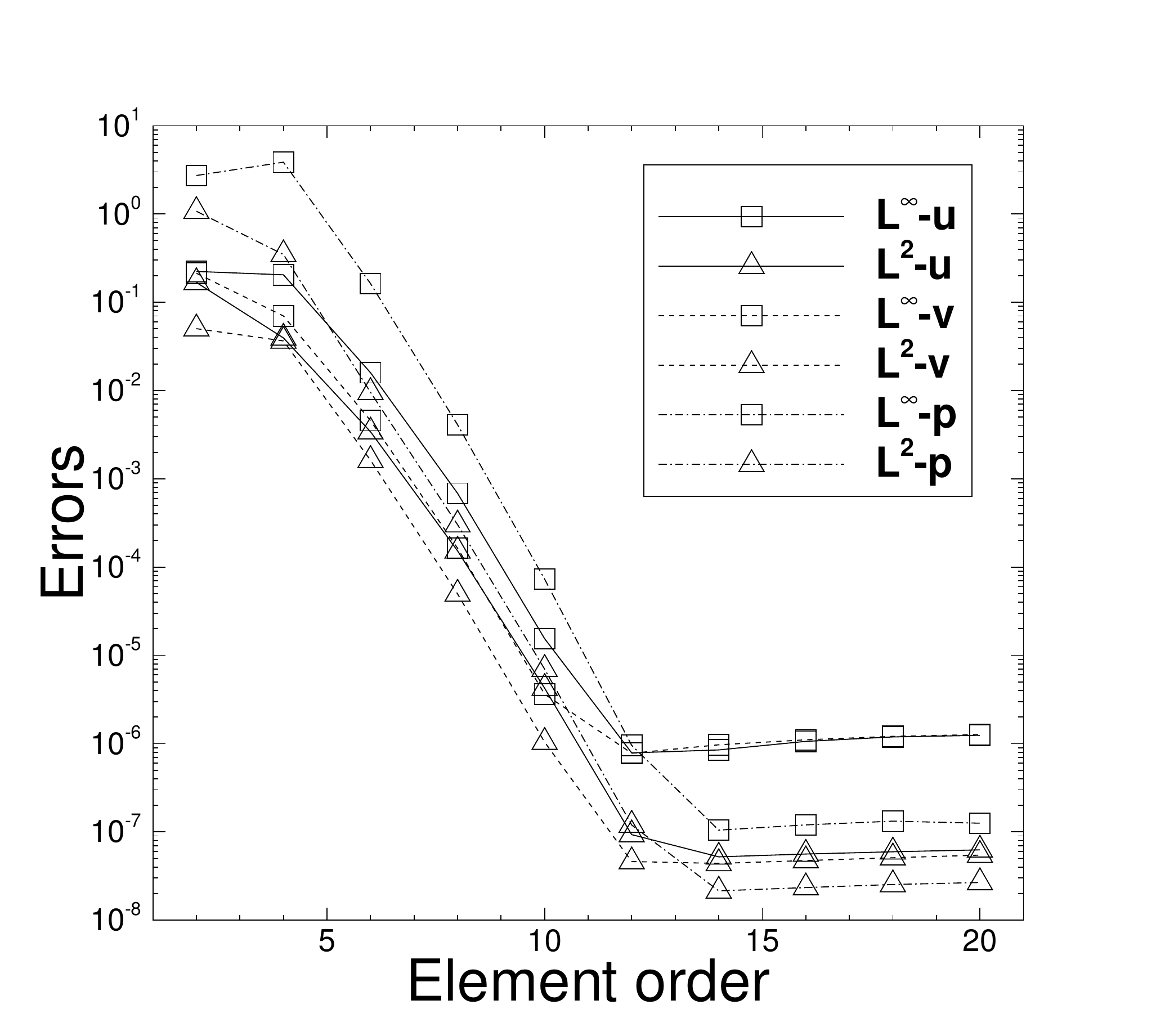}(b)
    \includegraphics[width=3in]{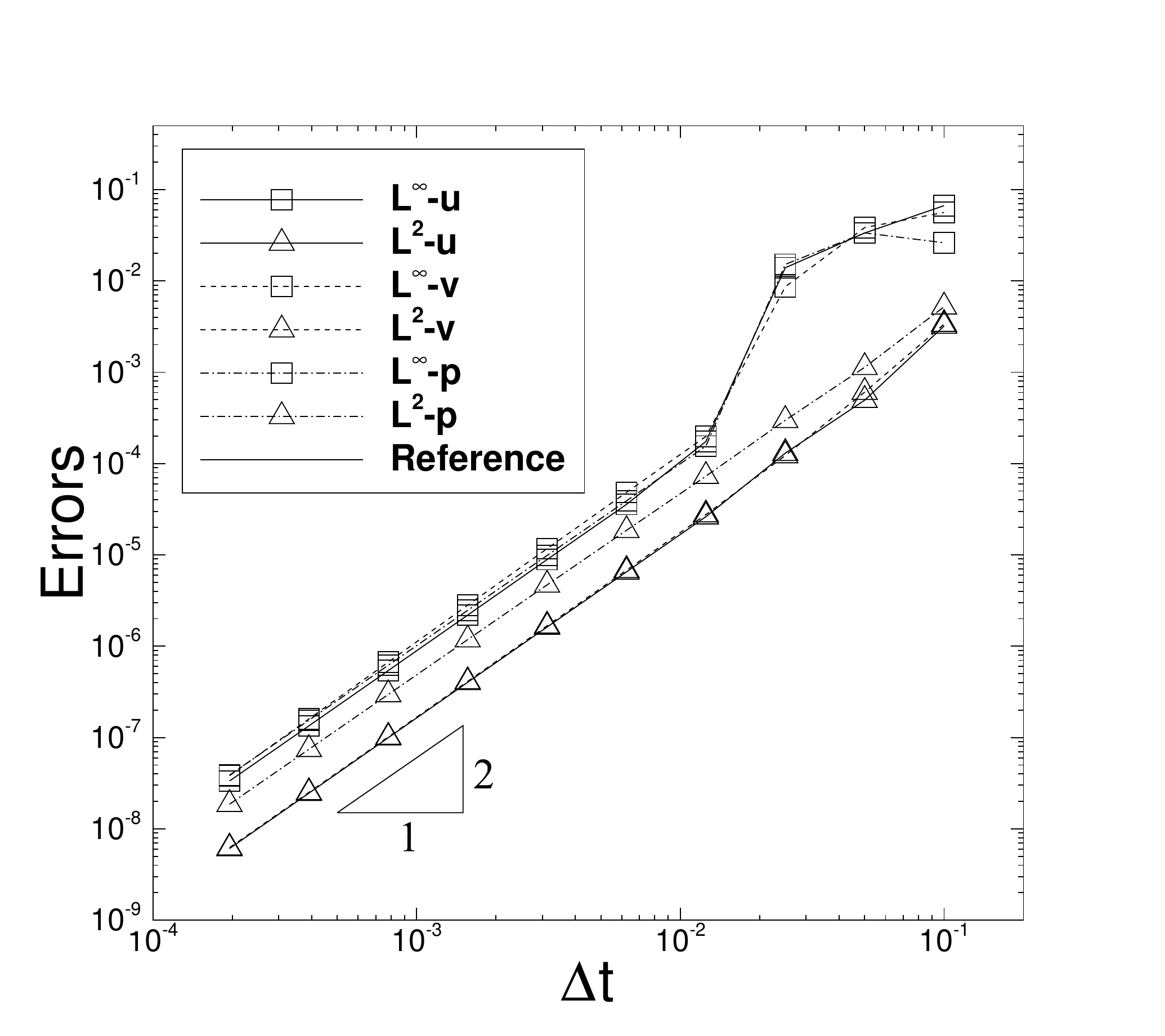}(c)
  }
  \caption{
    Convergence tests (2D): (a) Mesh and configuration.
    $L^{\infty}$ and $L^2$ errors of the flow variables
    as a function
    of the element order with fixed $t_f=0.1$ and $\Delta t=0.001$ (b),
    and of the time step size $\Delta t$ with fixed element order $16$ and
    $t_f=0.5$ (c). OBC-C has been used on
    the open boundaries.
  }
  \label{fig:conv_2d}
\end{figure}

We consider the 2D rectangular domain $\overline{ABCD}$
shown in Figure \ref{fig:conv_2d}(a),
$0\leqslant x\leqslant 2$ and $-1\leqslant y\leqslant 1$,
and the following analytic solution to the incompressible
Navier-Stokes equations,
\begin{equation}
  \left\{
  \begin{split}
    &
    u = 2\cos(\pi y)\sin(\pi x)\sin t \\
    &
    v = -2\sin(\pi y)\cos(\pi x)\sin t \\
    &
    p = 2\sin(\pi y)\sin(\pi x)\cos t
  \end{split}
  \right.
  \label{equ:anal_soln_2d}
\end{equation}
where $\mathbf{u} = (u,v)$.
The external force $\mathbf{f}$ in \eqref{equ:nse} is chosen such that
the expressions in \eqref{equ:anal_soln_2d} satisfy \eqref{equ:nse}.

The domain is discretized using two uniform quadrilateral
spectral elements ($\overline{AFED}$ and $\overline{FBCE}$)
as shown in Figure \ref{fig:conv_2d}(a).
On the boundaries $\overline{AB}$, $\overline{AD}$ and $\overline{DE}$
Dirichlet boundary condition \eqref{equ:dbc} is imposed,
in which the boundary velocity $\mathbf{w}(\mathbf{x},t)$
is set according to the analytic expressions from \eqref{equ:anal_soln_2d}.
On the boundaries $\overline{BC}$ and $\overline{CE}$
the open boundary condition \eqref{equ:obc_mod} (in Appendix B) is imposed,
in which $\mathbf{f}_b$ is chosen such that
the analytic expressions from \eqref{equ:anal_soln_2d}
satisfy the equation \eqref{equ:obc_mod} on $\partial\Omega_o$.
The initial condition is given by \eqref{equ:ic}, in which
the initial velocity $\mathbf{u}_{in}$ is obtained
according to the analytic solution in \eqref{equ:anal_soln_2d}
by setting $t=0$.

% algorithm, simulation parameters

The Navier-Stokes equations together with the open boundary
conditions presented in Section \ref{sec:obc} are
solved using the numerical algorithm given in the Appendix B.
The velocity and pressure fields are computed in
time from $t=0$ to $t=t_f$ ($t_f$ to be specified later).
Then the flow variables at $t=t_f$ from numerical simulations
are compared with the analytic solutions in \eqref{equ:anal_soln_2d},
and the numerical errors in various norms are computed.
The element order and the time step size $\Delta t$
are varied systematically to study their effects on the
numerical errors in spatial and temporal convergence
tests, respectively. The non-dimensional viscosity
in the Navier-Stokes equation \eqref{equ:nse}
is fixed at $\nu = 0.01$ in the following tests.

Figure \ref{fig:conv_2d}(b) demonstrates the  results for
the 2D spatial convergence tests. Here we have employed
a fixed $t_f=0.1$ and $\Delta t=0.001$ (i.e.~$100$ time steps)
in the test. The element order is varied systematically
between $2$ and $20$, and for each element order we have performed
simulations and computed the errors of the numerical
solutions at $t=t_f$ against the analytic solution.
Figure \ref{fig:conv_2d}(b) shows the numerical errors of
the velocity and pressure in
$L^{\infty}$ and $L^2$ norms as a function of the element order
from this set of tests, in which OBC-C is employed on the open boundaries.
It can be observed that
the numerical errors decrease exponentially
with increasing element order as the order is below $12$.
For element orders $12$ and beyond, a saturation of
the numerical errors can be observed at a level
$10^{-7} \sim 10^{-6}$ due to the temporal truncation errors. 

Figure \ref{fig:conv_2d}(c) demonstrates the results
for the 2D temporal convergence tests.
Here we have employed a fixed $t_f=0.5$ and element
order $16$. The time step size $\Delta t$ is varied
systematically between $\Delta t=0.1$
and $\Delta t=1.953125e-4$.
Figure \ref{fig:conv_2d}(c) shows the $L^{\infty}$ and
$L^2$ errors of the flow variables at $t=t_f$ as a function of
$\Delta t$ in this group of tests.
The open boundary condition is again OBC-C
in these tests.
The rate of convergence with respect to $\Delta t$
is observed to be second order when $\Delta t$ is sufficiently small.

\paragraph{Three Dimensions (3D)}

\begin{figure}
  \centerline{
    \includegraphics[width=2.5in]{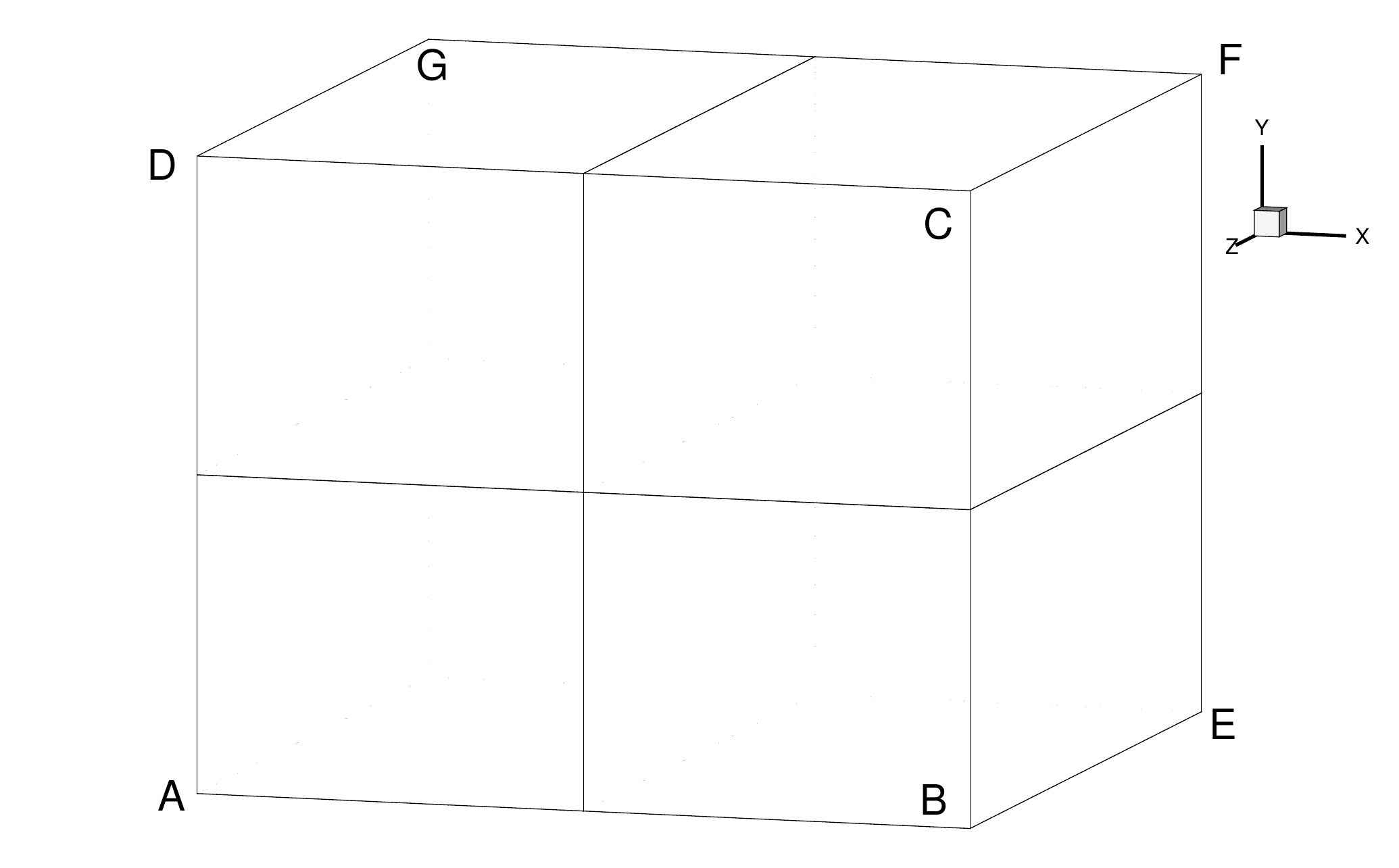}(a)
  }
  \centerline{
    \includegraphics[width=3in]{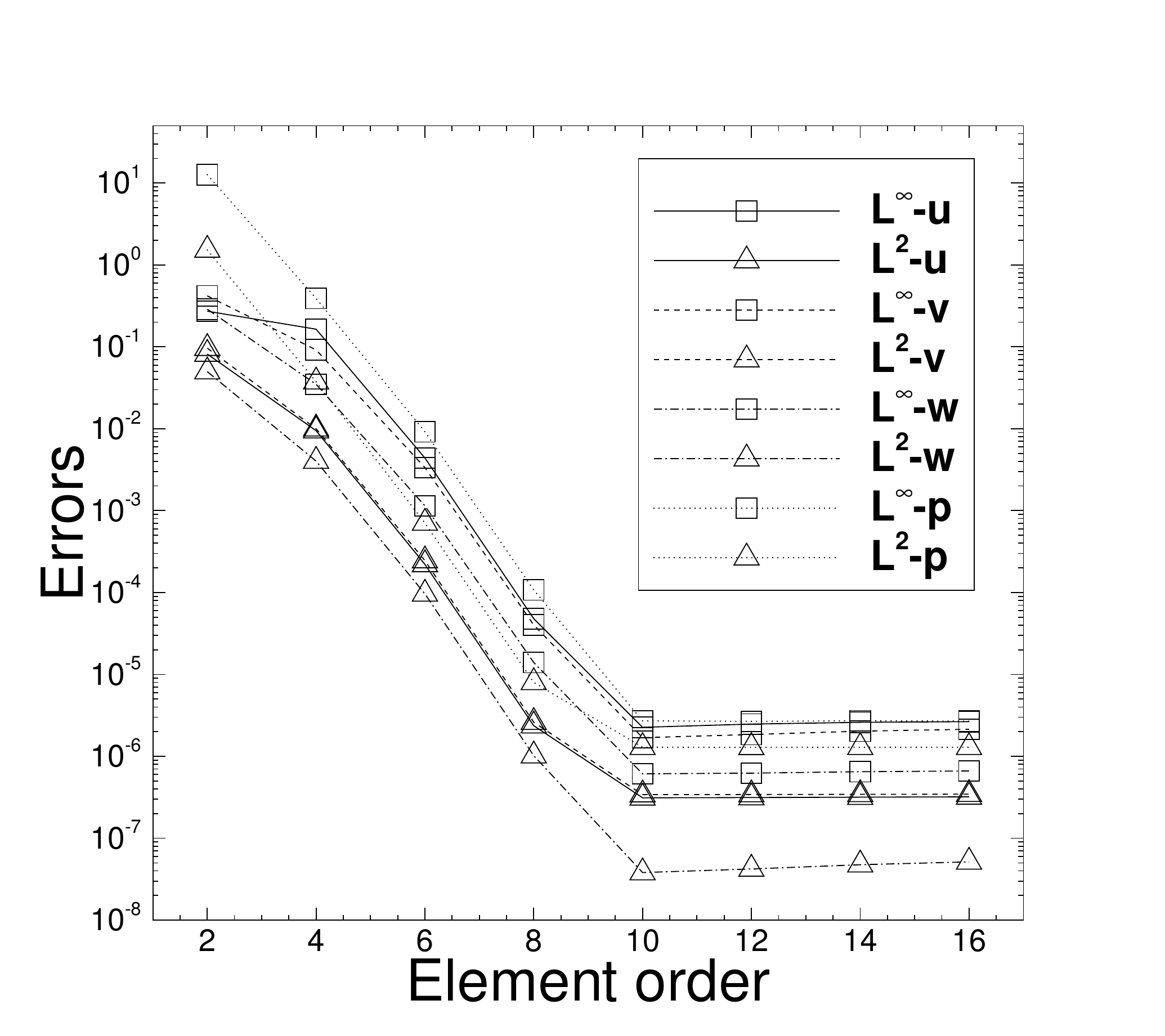}(b)
    \includegraphics[width=3in]{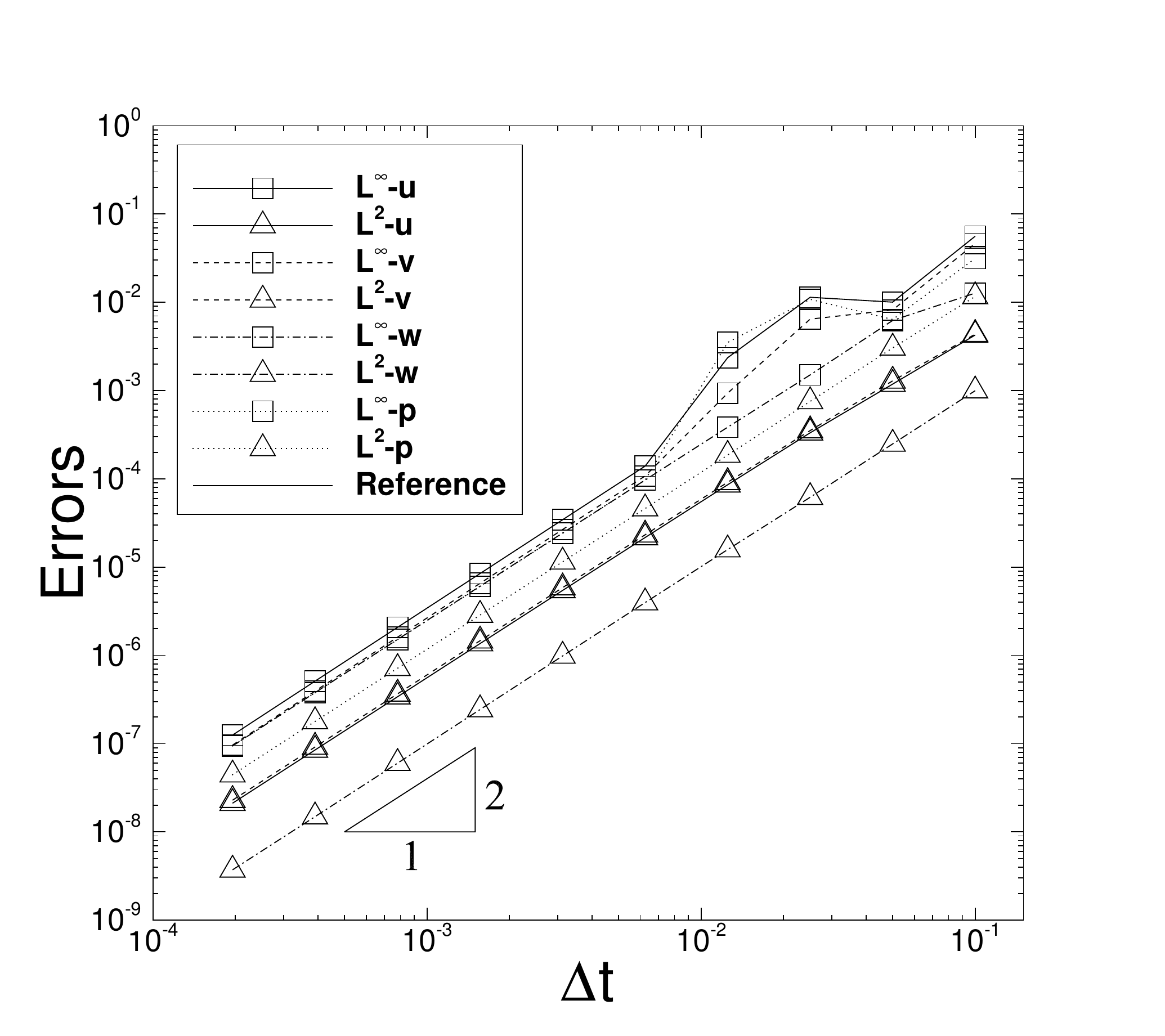}(c)
  }
  \caption{
    Convergence tests (3D): (a) mesh of four hexahedral elements.
    $L^{\infty}$ and $L^2$ errors of the flow variables
    as a function
    of the element order with fixed $t_f=0.1$ and $\Delta t=0.001$ (b),
    and of the time step size $\Delta t$ with a fixed element order $12$ and
    $t_f=0.2$ (c).
    The faces $\overline{DCFG}$ and $\overline{BEFC}$ are open boundaries, on which the
    OBC-B is employed for the boundary condition.
    %$Nz=4$, optimal R matrix, $\alpha_1=\alpha_2=1/2$ in tests.
  }
  \label{fig:conv_3d}
\end{figure}

We consider the 3D domain as sketched in
Figure \ref{fig:conv_3d}(a),
$0.25\leqslant x\leqslant 1.25$,
$-1\leqslant y\leqslant 1$, and $0\leqslant z\leqslant 2$,
and the following analytic solutions to the
Navier-Stokes equations on this domain,
\begin{equation}
  \left\{
  \begin{split}
    &
    u = 2\cos(2\pi x)\cos(\pi y)\cos(\pi z)\sin t \\
    &
    v = 2\sin(2\pi x)\sin(\pi y)\cos(\pi z)\sin t \\
    &
    w = 2\sin(2\pi x)\cos(\pi y)\sin(\pi z)\sin t \\
    &
    p = 2\sin(2\pi x)\sin(\pi y)\sin(\pi z)\cos t
  \end{split}
  \right.
  \label{equ:conv_3d}
\end{equation}
where $\mathbf{u} = (u,v,w)$ in 3D.
The external force $\mathbf{f}$ is chosen such that
the analytic expressions in \eqref{equ:conv_3d}
satisfy the equation \eqref{equ:nse}.
We assume that the domain and all the flow variables are
periodic along the $z$ direction (on the faces $z=0$ and $z=2$).
The faces $\overline{BEFC}$ and $\overline{DCFG}$
are open boundaries, on which the open boundary condition
\eqref{equ:obc} will be imposed.

% discretization, BCs, algorithm, simulation parameters

As discussed in Section \ref{sec:obc}, we employ a hybrid
Fourier spectral method and spectral element method to
discretize the 3D domain. Fourier spectral expansions
are employed along the homogeneous $z$ direction,
and spectral element expansions are employed
in the $x$-$y$ planes. In the numerical tests that follow
four Fourier planes are employed along the $z$ direction, and
four quadrilateral elements are employed to discretize
each plane. The boundary conditions \eqref{equ:dbc} and
\eqref{equ:obc_mod} are imposed on the Dirichlet and open
boundaries, in which the boundary velocity $\mathbf{w}$ is
set in accordance with the analytic expressions \eqref{equ:conv_3d}
and the function $\mathbf{f}_b$ is chosen such that
the analytic solutions in \eqref{equ:conv_3d}
satisfy \eqref{equ:obc_mod} on the open boundaries.
The algorithm from the Appendix B is employed to
solve the incompressible Navier-Stokes equations, together with
the Dirichlet and open boundary conditions.
The flow fields are obtained from $t=0$ to $t=t_f$, and
the errors of the numerical solution at $t=t_f$ are
computed against the analytic solution given in \eqref{equ:conv_3d}.
The non-dimensional viscosity is $\nu=0.01$
in the following tests.

% test results

In the spatial convergence tests, we fix the integration time
at $t_f=0.1$ and the time step size at $\Delta t=0.001$,
and then vary the element order systematically between $2$
and $16$. Figure \ref{fig:conv_3d}(b) shows the numerical errors
of the velocity and pressure at $t=t_f$ as a function of
the element order, obtained using OBC-B 
for the open boundaries.
An exponential decrease in the numerical errors can be
observed when the element order is below $10$,
and a saturation in the numerical errors is observed
for element orders beyond $10$ due to the temporal
truncation error.

In the temporal convergence tests we employ a fixed
$t_f=0.2$ and an element order $12$, and then
vary the time step size systematically between
$\Delta t=0.1$ and $\Delta t=1.953125e-4$.
We have computed the errors of the numerical solution at $t=t_f$
corresponding to each $\Delta t$ with OBC-B as the open boundary condition, 
and in Figure \ref{fig:conv_3d}(c) these errors are plotted
as a function of $\Delta t$ (in logarithmic scales)
from these tests.
The results signify a second-order rate of convergence
in time.

To summarize,
the 2D and 3D results of this section demonstrate that,
with the open boundary conditions developed herein,
the method exhibits a spatial exponential convergence
rate and a temporal second-order accuracy for incompressible
flows on domains with open/outflow boundaries.

\subsection{Flow Past a Circular Cylinder}

We focus on a canonical wake flow, 
the flow past a circular cylinder, in two and three
dimensions in this section.
At moderate and high Reynolds numbers,
how to deal with the outflow boundary in this flow
is critical to the stability of simulations. 
We employ this canonical problem to test the open/outflow
boundary conditions developed in the current work.

\begin{figure}
  \centerline{
    \includegraphics[width=2.5in]{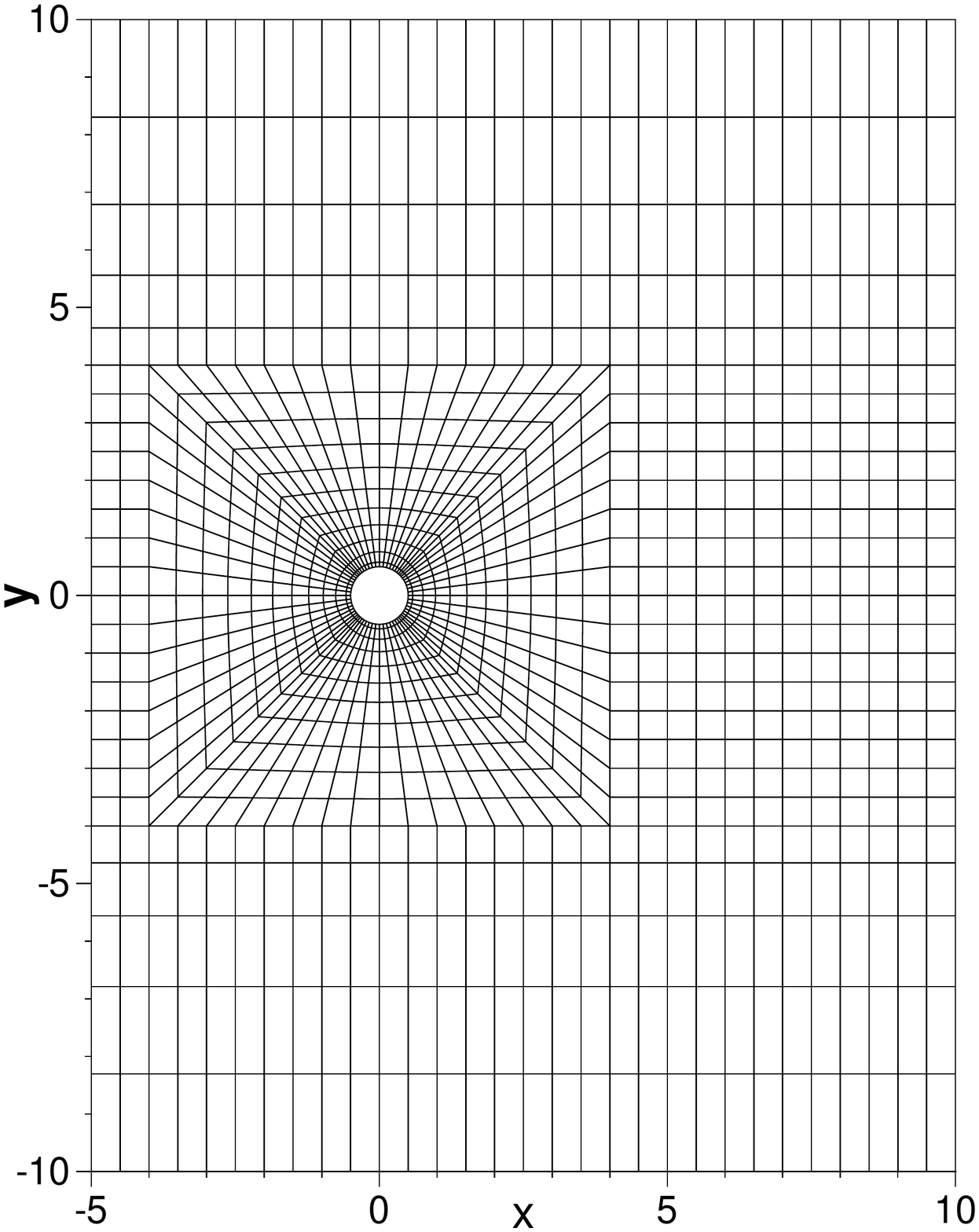}(a)
    \includegraphics[width=2.5in]{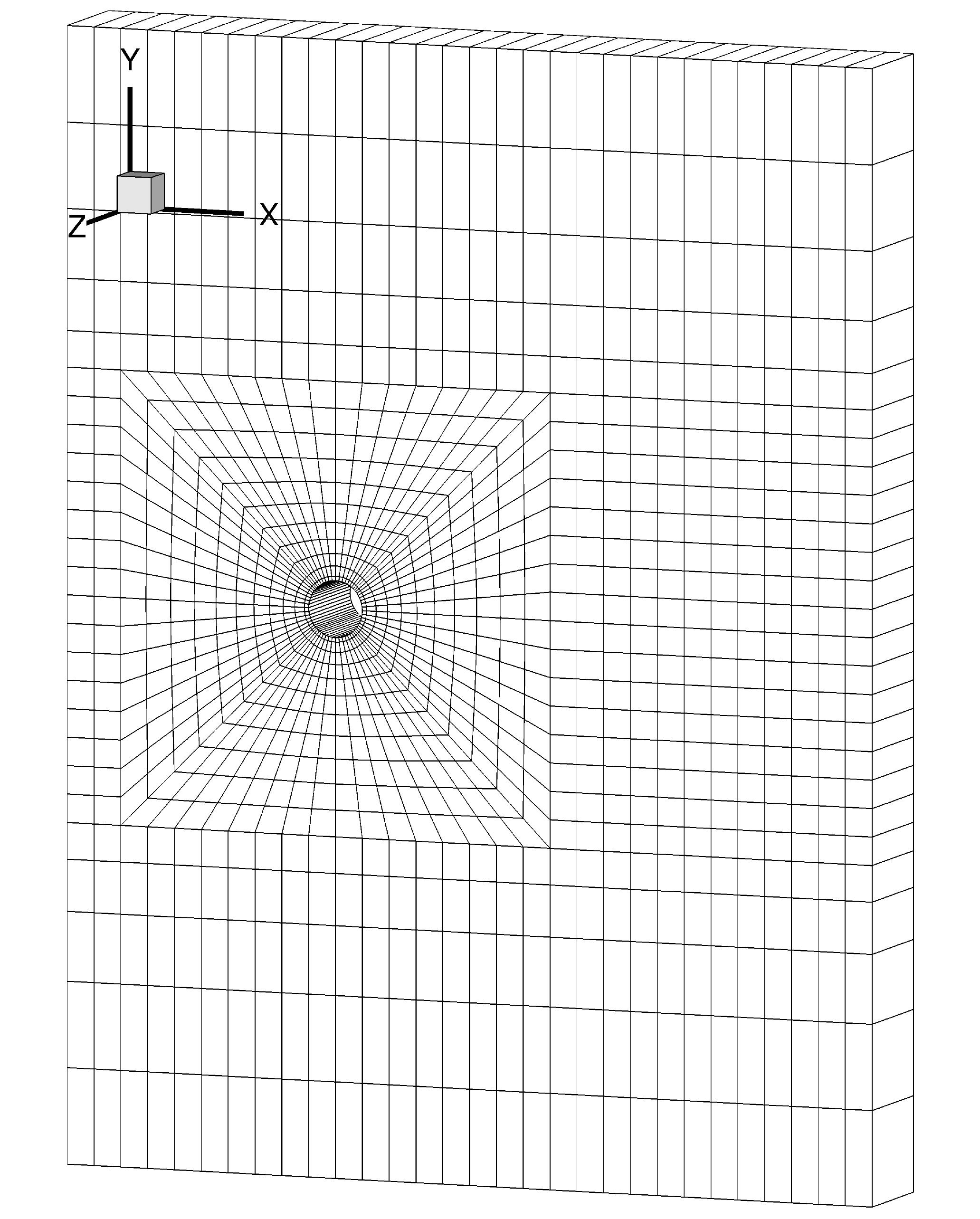}(b)
  }
  \caption{
    Flow past a circular cylinder: spectral element meshes in
    (a) 2D, and (b) 3D. The 2D mesh and each $x$-$y$ plane of
    the 3D mesh contains $1228$ quadrilateral
    elements.
  }
  \label{fig:cyl_mesh}
\end{figure}

\subsubsection{Two-Dimensional Simulations}
\label{sec:cyl_2d}

Let us first investigate the cylinder flow numerically in two dimensions.
Consider the domain in Figure \ref{fig:cyl_mesh}(a),
$-5\leqslant x/d\leqslant 10$ and $-10\leqslant y/d\leqslant 10$,
where $d$ is the cylinder diameter. The center of the
cylinder coincides with the origin of the coordinate
system. The top and bottom of the domain ($y=\pm 10d$)
are assumed to be periodic. A uniform flow (free-stream velocity $U_0$,
along the $x$ direction) enters the domain from the left side,
and the wake exits the domain through the right boundary at $x=10d$.
We assume that no external body force is present,
and thus $\mathbf{f}=0$ in equation \eqref{equ:nse}.
In the following simulations all the length
variables are normalized based on the cylinder
diameter $d$, and all the velocity variables are normalized by
the free stream velocity $U_0$. So the Reynolds number
is defined based on $U_0$ and $d$.
All the other variables are normalized accordingly
in a consistent way.
%and the pressure is normalized
%by $\rho_fU_0^2$ ($\rho_f$ denotes the fluid density).

We discretize the domain using the spectral element mesh shown in
Figure \ref{fig:cyl_mesh}(a), which contains $1228$
quadrilateral elements. The algorithm from
the Appendix B is used to numerically solve the
incompressible Navier-Stokes equations together with
the boundary conditions specified as follows.
On the cylinder surface no-slip condition is imposed,
i.e.~the Dirichlet boundary condition \eqref{equ:dbc}
with $\mathbf{w}=0$. At the inlet ($x/d=-5$) we impose
the Dirichlet condition \eqref{equ:dbc} where the boundary
velocity $\mathbf{w}$ is set based on the free-stream velocity.
Periodic conditions are imposed at the top and bottom
of the domain for all flow variables.
At the outflow boundary ($x/d=10$) the boundary condition \eqref{equ:obc}
from Section \ref{sec:obc} is imposed, where OBC-A, 
OBC-B and OBC-C are all employed and
the various algorithmic parameters have been tested.

\begin{figure}
  \centering
  \includegraphics[width=3in]{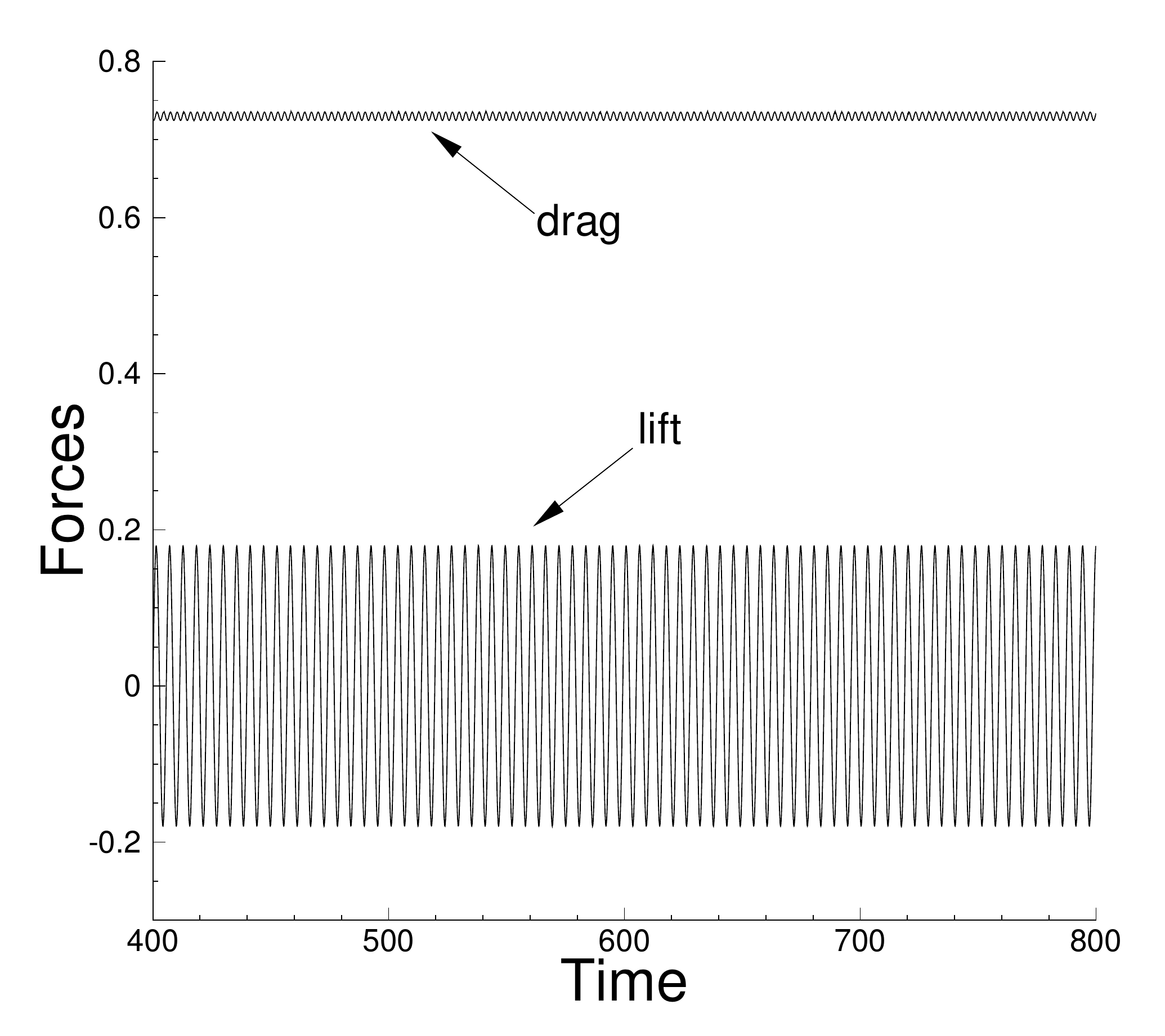}
  \caption{
    Time histories of drag ($x$ component of force, $f_x$)
    and lift ($y$ component of force, $f_y$) on the cylinder
    at $Re=100$, computed using an element order $8$ and
    OBC-B    as the outflow boundary condition in 2D.
  }
  \label{fig:force_re100}
\end{figure}

\begin{table}
  \centering
  \begin{tabular}{l| c lll l}
    \hline
    $Re$ & element order & mean-$f_x$ & rms-$f_x$  & rms-$f_y$ \\ \hline
    $30$ & $4$ & $0.968$ & $0$  & $0$ \\
    & $6$ & $0.968$ & $0$  & $0$ \\
    & $8$ & $0.968$ & $0$  & $0$\\
    & $10$ & $0.968$ & $0$ & $0$\\ \hline
    $100$ & $4$ & $0.729$ & $0.00374$  & $0.126$ \\
    & $6$ & $0.730$ & $0.00377$  & $0.127$ \\
    & $8$ & $0.730$ & $0.00377$ & $0.127$ \\
    & $10$ & $0.730$ & $0.00377$ & $0.127$ \\
    \hline
  \end{tabular}
  \caption{
    Flow past a cylinder (2D): forces on the cylinder
    computed using several element orders.
    OBC-B  is used for the outflow
    boundary condition.
  }
  \label{tab:cyl_conv}
\end{table}

% simulation parameters, element order, dt
% grid convergence tests

Figure \ref{fig:force_re100} illustrates a long-time simulation
of the 2D cylinder flow at Reynolds number $Re=100$ with a window
of time histories of the forces (drag $f_x$, and lift $f_y$) acting on
the cylinder obtained from current simulations. The results are
obtained using an element order $8$ and OBC-B 
as the outflow boundary condition.
Periodic vortex shedding into the wake induces
a fluctuating drag and lift force exerting on the cylinder.
Based on these histories we can compute the statistical
quantities such as the time-averaged mean
and root-mean-square (rms) forces.
Table \ref{tab:cyl_conv} lists the mean and rms
forces  on the cylinder
at Reynolds numbers $Re=30$ and $100$ obtained using
several element orders ranging from $4$ to $10$.
The mean lift is not shown in the table because they
are all zeros at $Re=30$ and all essentially zeros at $Re=100$.
Note that the flow is in a steady state at $Re=30$, and
so no averaging is performed with the forces at this Reynolds number.
When the element order is sufficiently large ($6$ or above),
the forces obtained from the simulations are essentially the same,
suggesting convergence of the simulation results with
respect to the grid resolution.
The majority of simulations in subsequent discussions are
performed using an element order $8$ and a time step
size $\Delta t=2.5e-4$, and at lower
Reynolds numbers (below $Re=60$) an element order $6$
and $\Delta t=1e-3$ have also been employed.
A range of Reynolds numbers (to be specified below)
has been simulated and studied for this problem.

% compare with experiments etc

\begin{figure}
  \centerline{
    \includegraphics[width=3in]{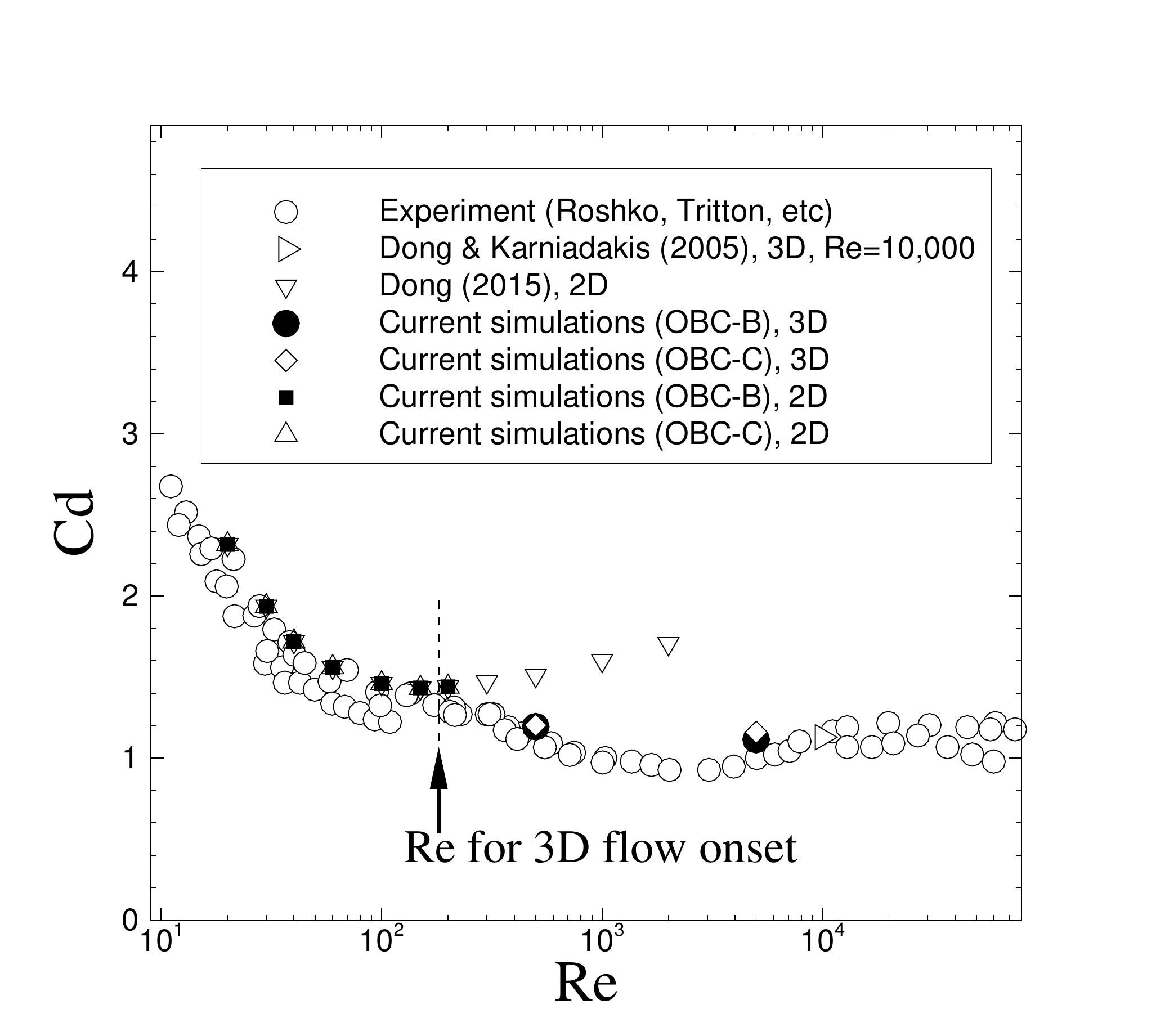}(a)
    \includegraphics[width=3in]{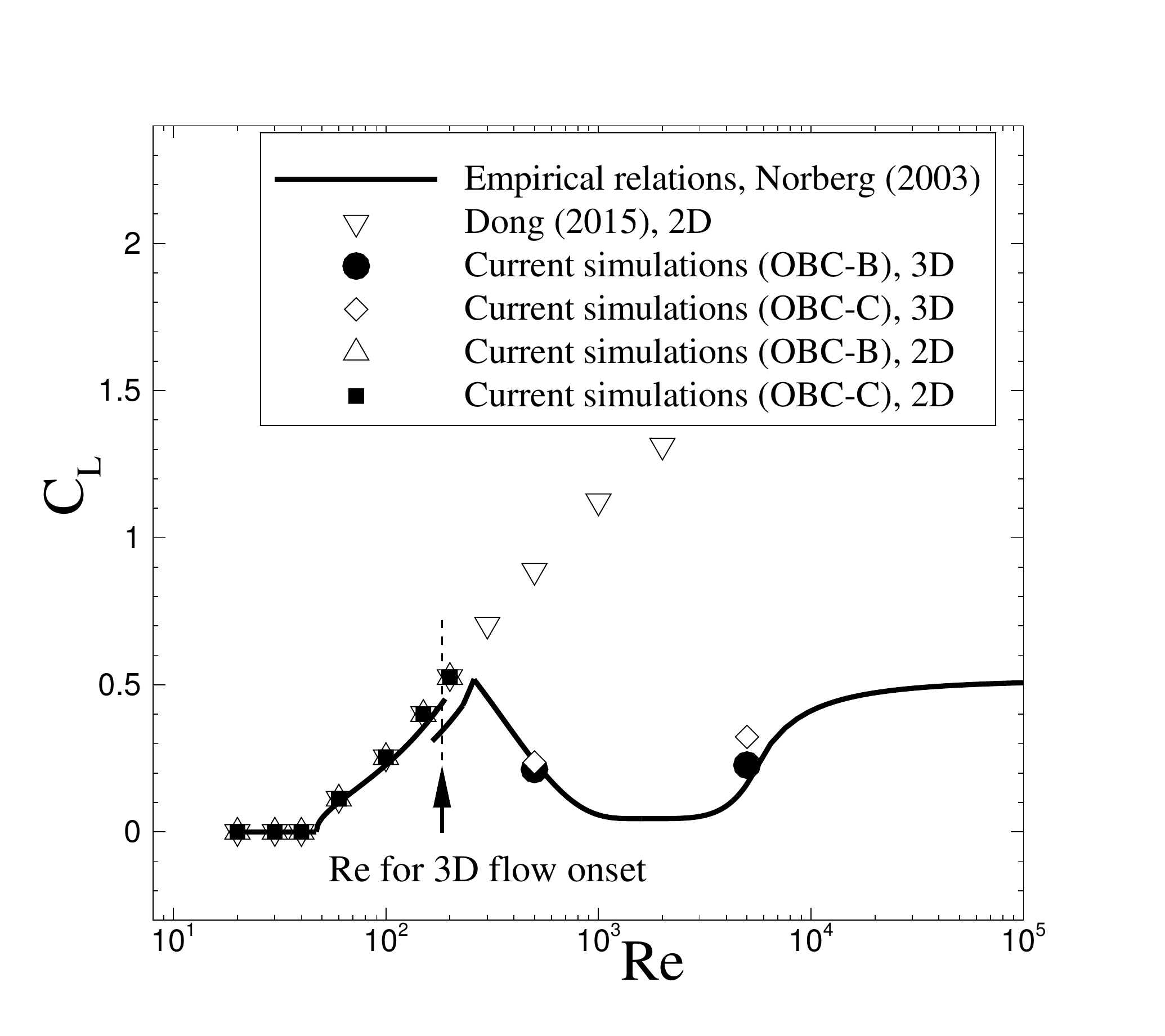}(b)
  }
  \caption{
    Flow past a cylinder:
    Comparison of the drag coefficient (a) and the rms lift coefficient (b)
    as a function of the Reynolds number between current simulations (2D/3D)
    and the experimental measurements.
    OBC-B and OBC-C are used as
    the outflow boundary condition with current simulations.
  }
  \label{fig:compare_exp}
\end{figure}

We first study this flow for a range of low Reynolds numbers
($Re=200$ and below). The physical flow is two-dimensional and
is either at a steady state
(for $Re\lesssim 45$) or unsteady with periodic vortex shedding
(for $Re\lesssim 185$) \cite{Williamson1996}.
We have conducted simulations at several Reynolds numbers in
this range, and computed the corresponding forces on the cylinder.
Figures \ref{fig:compare_exp}(a) and (b) are comparisons of
the drag coefficients ($C_d$) and rms lift coefficients ($C_L$)
obtained from current simulations with those from
the experimental measurements and simulations from the
literature~\cite{Wieselsberger1921,Finn1953,Tritton1959,Roshko1961,Norberg2003,DongKER2006,Dong2015obc}.
These coefficients are defined by
\begin{equation}
  C_d = \frac{\bar{f}_x}{\frac{1}{2}\rho_fU_0^2}, \quad
  C_L = \frac{f_y^{\prime}}{\frac{1}{2}\rho_fU_0^2}
\end{equation}
where $\bar{f}_x$ denotes the time-averaged (mean) drag, $f_y^{\prime}$
denotes the rms-lift on the cylinder, and
$\rho_f$ is the fluid density.
Note that the plots also include results from the
three-dimensional simulations, which will be discussed subsequently in
Section \ref{sec:cyl_3d}.
The current 2D results are obtained using the OBC-B and OBC-C
as the outflow boundary conditions.
We observe that in this range of the Reynolds numbers
the current 2D simulation results
are in good agreement with the experimental data, and they also
agree well with the results from~\cite{Dong2015obc}.
When the Reynolds number is beyond this range (above
$Re\approx 185 \sim 260$),
the physical flow will undergo a transition
and become three-dimensional~\cite{Williamson1996}.
So there will be a large discrepancy between the drag/lift coefficients
from 2D simulations and the experimentally observed
values~\cite{DongK2005,DongS2015,Dong2015obc}.

\begin{table}[htbp]
\centering 
\begin{tabular}{l | l | r | c c  c} 
\hline 
$Re$ & method & parameters  & mean-$f_x$ & rms-$f_x$ & rms-$f_y$  \\  
\hline
10 & OBC-A & $a_{11}=a_{22}=$ 0.95&  1.652              & 0 & 0\\
& & 0.9 &		1.652	& 0 & 0\\
& & 0.5 &	1.648		 & 0 &0\\
& & 0.2 & 1.644		 & 0&0\\
& & 0.1& 	1.641		 &0  &0\\
& & 0.0 &	1.639		 & 0 & 0\\
& & -0.1&	1.635	 & 0 & 0\\
& & -0.2 & 1.630	& 0 & 0\\
& & -0.5 &1.593		& 0 & 0\\
& & -0.9 & 1.445	& 0 & 0\\
& & -0.95& 1.427	 & 0 & 0\\ \cline{2-6}
& OBC-B & & 1.631 & 0 & 0 \\ \cline{2-6} 
& OBC-C & & 1.631 & 0 & 0 \\ \cline{2-6}
& Traction-free OBC & & 1.631		& 0 & 0\\
\hline
20 & OBC-A & $a_{11}=a_{22}=$ 0.95& 1.173                & 0 & 0\\
& & 0.9 & 1.173 & 0 & 0\\
& & 0.5 & 1.171 & 0 &0\\
& & 0.2 & 1.168 & 0&0\\
& & 0.1& 1.166 &0  &0\\
& & 0.0 & 1.164 & 0 & 0\\
& & -0.1&1.162 & 0 & 0\\
& & -0.2 & 1.158 & 0 & 0\\
& & -0.5 &1.121 & 0 & 0\\
& & -0.9 & 0.967 & 0 & 0\\
& & -0.95& 0.952 & 0 & 0\\ \cline{2-6}
& OBC-B& & 1.159 & 0 &0 \\ \cline{2-6}
& OBC-C& & 1.159 & 0 &0 \\ \cline{2-6}
& Traction-free OBC & & 1.159 & 0 & 0\\
\hline
100 & OBC-A & $a_{11}=a_{22}=$ 0.95&  0.734   & 0.00412   & 0.128   \\
& & 0.9 &  0.734   &0.00412    & 0.127 \\
& & 0.5 & 0.732    &0.00392    &0.125 \\
& & 0.2 &0.731     &0.00378    &0.126 \\
& & 0.1 & 0.731    &0.00377    &0.126 \\
& & 0.0 & 0.731    &0.00377    &0.126 \\
& & -0.1 &0.730    &0.00378    &0.127 \\
& & -0.2 &0.730    &0.00382    &0.127  \\
& & -0.5 &0.723    &0.00425    &0.131 \\
& & -0.9 & (unstable) \\
& & -0.95 & (unstable) \\ \cline{2-6}
& OBC-B& & 0.730 & 0.00377 & 0.127 \\ \cline{2-6}
& OBC-C& & 0.730 & 0.00378 & 0.127 \\ \cline{2-6}
& Traction-free OBC &  &0.729   &0.00381   &0.127      \\
\hline
\end{tabular}
\caption{
  2D cylinder flow: comparison of the mean and rms forces on the cylinder
  obtained using OBC-A (with $\alpha=1/2$ and various parameters $a_{11}=a_{22}$), OBC-B, and OBC-C
  for several Reynolds numbers.
  Results from
  the traction-free open boundary condition are included as a reference.
}
\label{tab:obc-A} 
\end{table}

\begin{figure}[tb]
 \subfigure[at $x=1.0$]{ \includegraphics[scale=.42]{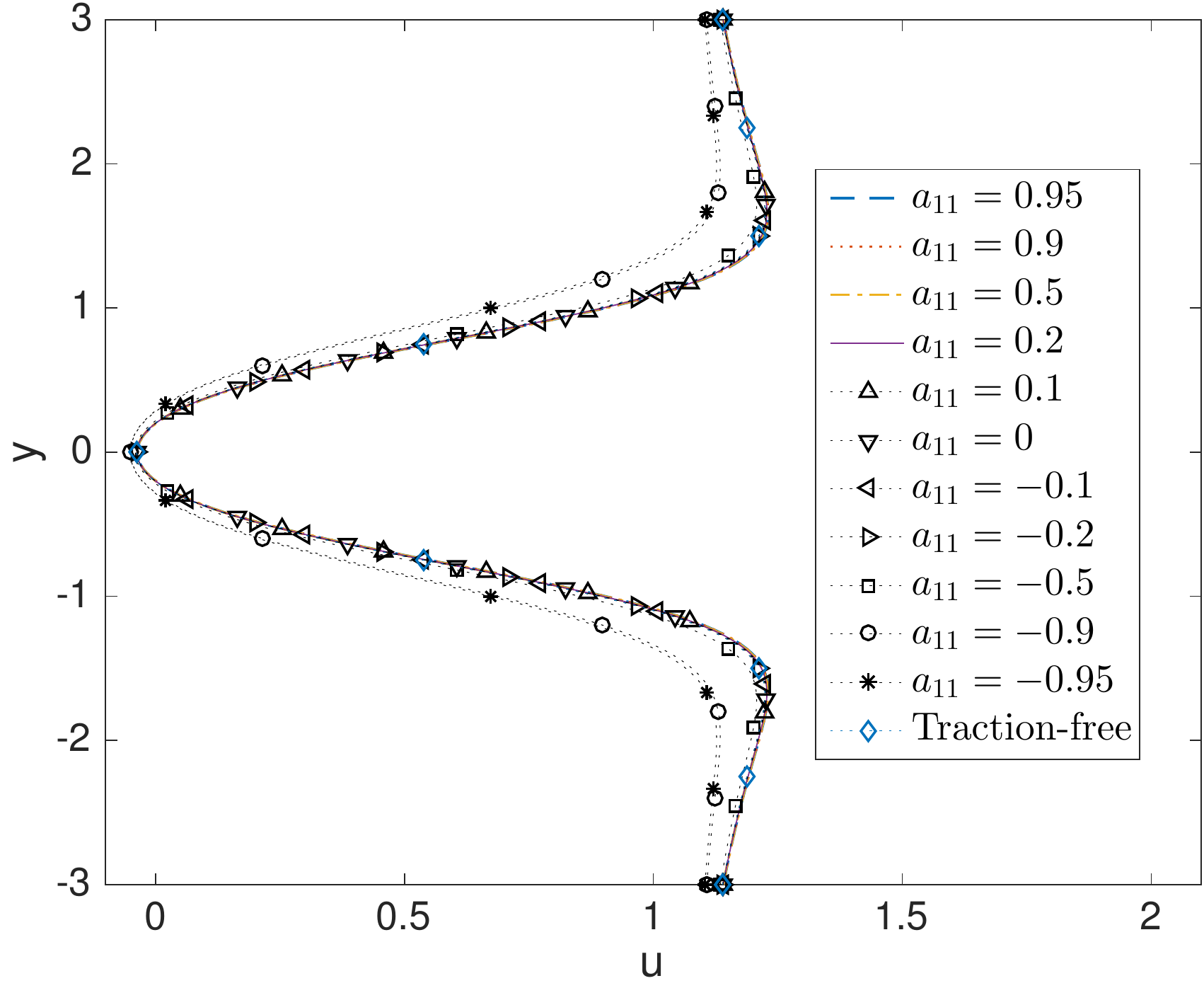}} \quad
 \subfigure[at $x=5.0$]{ \includegraphics[scale=.42]{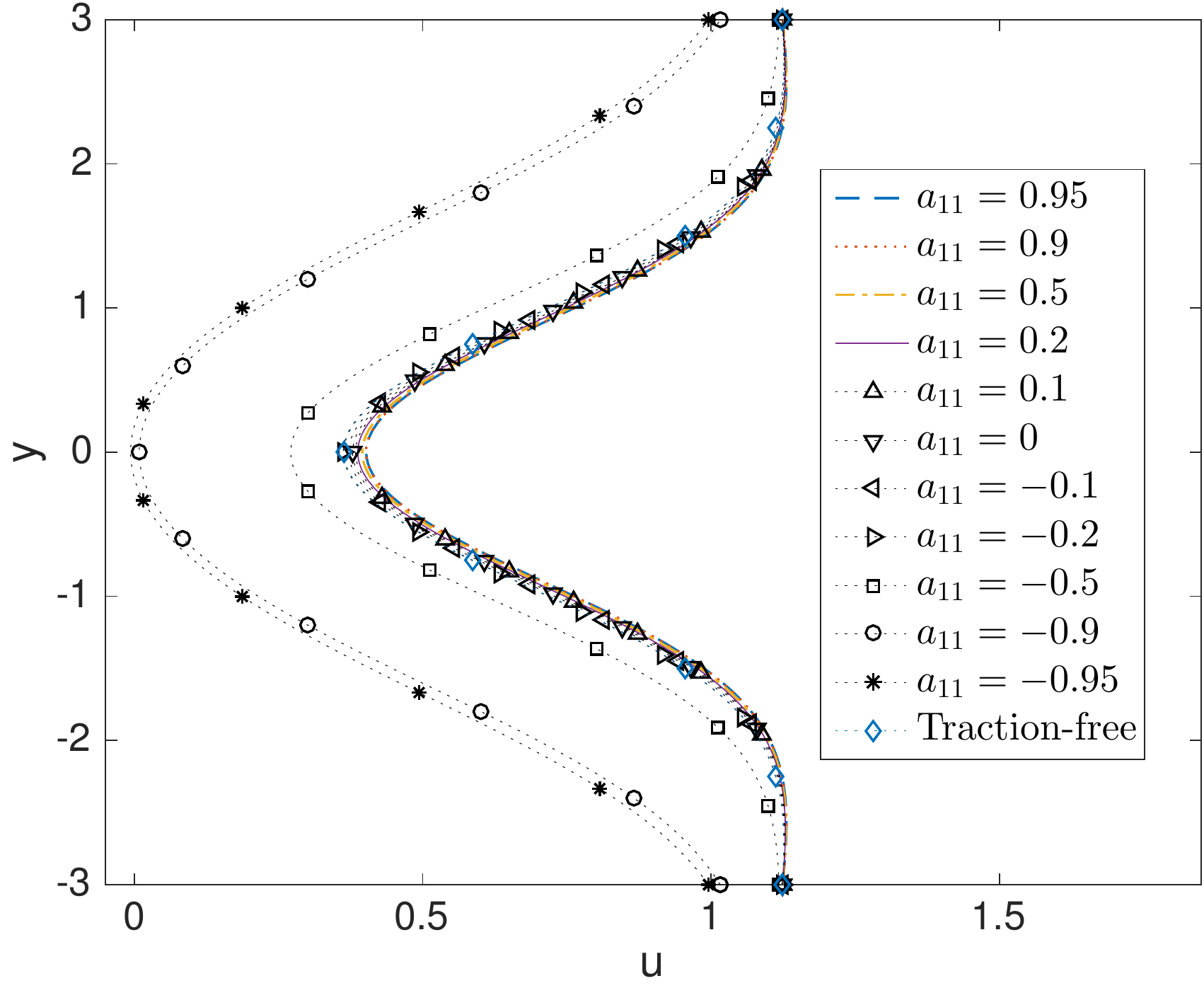}} \\
  \subfigure[at $x=10.0$ (outlet)]{ \includegraphics[scale=.42]{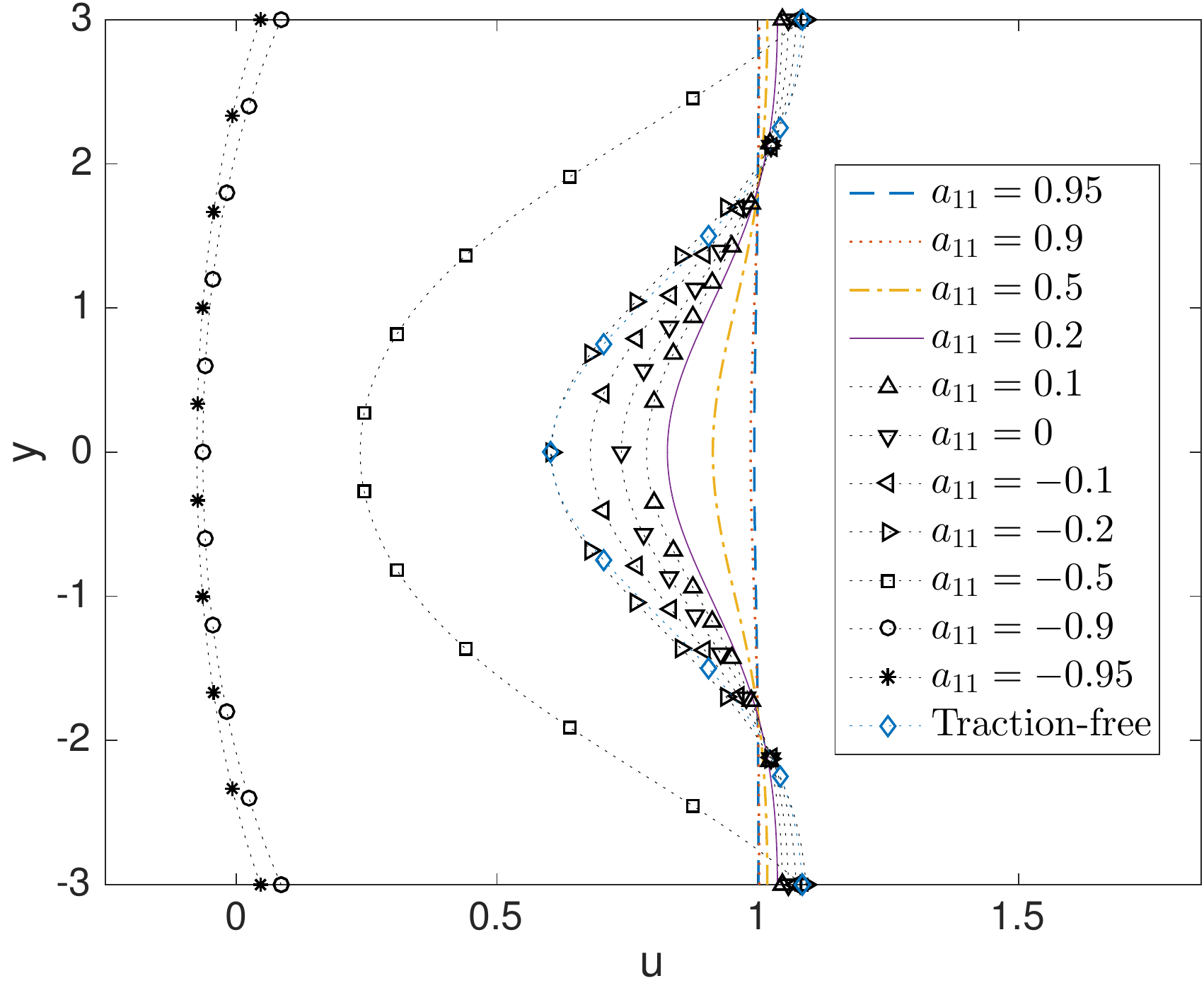}} \quad
 \subfigure[ along centerline $y=0$]{ \includegraphics[scale=.42]{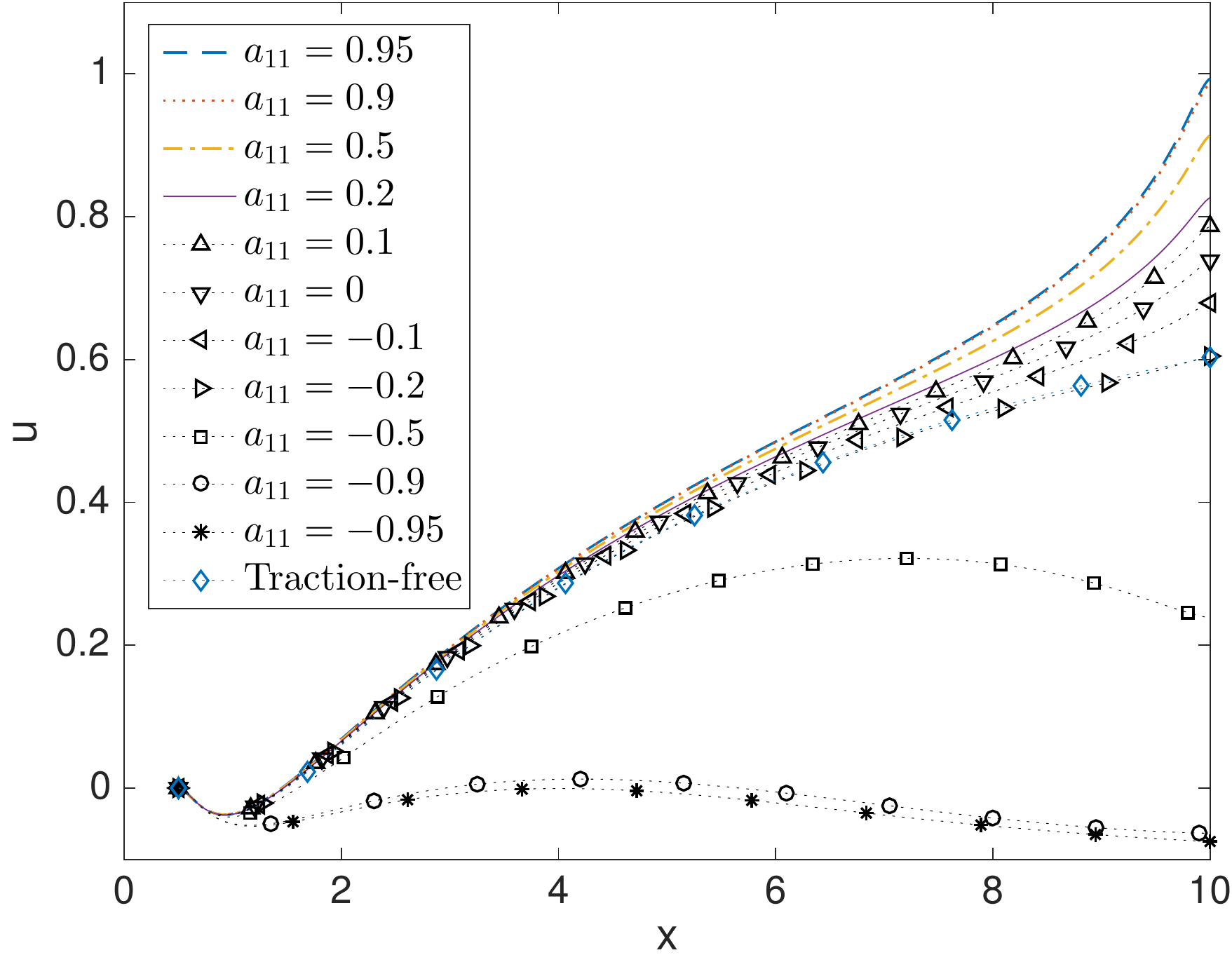}} 
 \caption{
   Cylinder flow ($Re=20$): comparison of streamwise velocity (i.e.~$x$ velocity
   component) profiles
   at several downstream locations in the wake and along the centerline
   computed using OBC-A with $\alpha=1/2$ and
   different $a_{11}$ (and $a_{22}$, with $a_{22}=a_{11}$) values.
   Results from traction-free condition
   are included for comparison.
 }
\label{fig:vel_profile_re20}
\end{figure}

At these low Reynolds numbers it is relatively easy to carry out
a  study of how
the algorithmic parameters affect the simulation results.
OBC-A, OBC-B and OBC-C have all been employed and tested
with the open boundary condition \eqref{equ:obc}
in current simulations.

Let us first concentrate on OBC-A.
In Table \ref{tab:obc-A} we list the (time-averaged) mean
and rms forces on the cylinder at three
Reynolds numbers ($Re=10$, $20$ and $100$),
obtained using OBC-A as the outflow boundary condition
under a range of values for $a_{11}$ (and $a_{22}$, with $a_{22}=a_{11}$)
and with a fixed $\alpha=\frac{1}{2}$.
Since the flow is steady at $Re=10$ and $20$,
no time-averaging is performed for these two Reynolds numbers.
As a reference for comparison,
we have also included the results computed
using the traction-free condition on the outflow boundary, namely,
\begin{equation}
  -p\mathbf{n} + \nu \mathbf{n}\cdot\nabla\mathbf{u} = 0,
  \quad \text{on} \ \partial\Omega_o.
  \label{equ:obc_tractfree}
\end{equation}

% what are the results?

A trend can be discerned from the data obtained using OBC-A.
The mean drag on the cylinder computed using OBC-A
tends to decrease with decreasing $a_{11}$ (and $a_{22}$) values.
When compared with the results based on the traction-free condition,
the best results with OBC-A seem to correspond to
a value around $a_{11}=a_{22}\approx -0.2$ for the cylinder flow.
In an interval around this best value, 
the computed forces seem to be not sensitive to $a_{11}$ ($a_{22}$)
and they are very close to the forces corresponding
to the traction-free condition. Even when $a_{11}=a_{22}=0.95$,
the discrepancy in the mean drag seems to be around $1\%$.
% a_11 --> -1
But as $a_{11}=a_{22} \rightarrow -1$, the discrepancy
in the mean drag seems to grow rapidly and becomes very substantial.
For example, with $a_{11}=a_{22}=-0.95$ the difference
in the mean drag values produced by OBC-A and the traction-free condition
is approximately $18\%$ at $Re=20$.
In addition, we observe that at $Re=100$, with $a_{11}=a_{22}=-0.9$ and smaller,
the computation with OBC-A is unstable. 
We recall that as $a_{11}=a_{22} \rightarrow 1$
the amount of dissipation on $\partial\Omega_o$ with OBC-A
becomes infinite,
and as $a_{11}=a_{22} \rightarrow -1$
the amount of dissipation approaches zero.
The above results with the computed forces suggest that, 
while the best $a_{11}$ ($a_{22}$) values seem to be
around $-0.2$, larger values appear not harmful,
but it can be detrimental to the accuracy
if $a_{11}$ ($a_{22}$) is too small.

% velocity profiles comparison for different R with OBC-A

The velocity distribution in the cylinder wake demonstrates
the effects of $a_{11}$ (and $a_{22}$) on the simulation results
even more clearly.
Figure \ref{fig:vel_profile_re20} is a comparison of the
steady-state streamwise velocity ($x$ velocity) profiles
along the vertical direction at downstream locations
$x/d=1.0$, $5.0$ and $10.0$ (plots (a), (b) and (c)), and
along the centerline (plot (d)).
The different curves correspond to OBC-A as the outflow boundary
condition with $\alpha=\frac{1}{2}$ and a set of
$a_{11}$ (and $a_{22}$, with $a_{22}=a_{11}$) values ranging from
$-0.95$ to $0.95$.
For the purpose of comparison, the velocity profiles
computed using the traction-free condition \eqref{equ:obc_tractfree}
%and the OBC-B with $\alpha=\frac{1}{2}$ 
are also included in these plots.
We have the following observations:
\begin{itemize}

\item
  The velocity profiles corresponding to OBC-A with
  $a_{11}=a_{22}=-0.5$ and below
  exhibit a large discrepancy when compared with
  the rest of the profiles in essentially the entire wake region.

\item
  The profiles corresponding to OBC-A with $a_{11}=a_{22}=-0.2$
  and above are quite close to those resulting from
  the traction-free boundary condition \eqref{equ:obc_tractfree}
  in the near wake ($x/d\lesssim 5$).
  Further downstream ($x/d\gtrsim 6$) the discrepancy in
  all the profiles (except the one with $a_{11}=a_{22}=-0.2$), 
  when compared with the traction-free condition,
  becomes very pronounced.

\item
  Among the set of $a_{11}$ ($a_{22}$) values tested for OBC-A, the best
  profile corresponds to $a_{11}=a_{22}=-0.2$, in terms of
  the comparison with results based on the traction-free condition.

%\item
%  The velocity profiles resulting from the OBC-B with $\alpha=\frac{1}{2}$
%  essentially overlap with those from the traction-free condition
%  in the entire wake region examined.

\end{itemize}

The effects of the $a_{11}$ and $a_{22}$ parameters in OBC-A on
the simulation results are investigated with a fixed $\alpha=\frac{1}{2}$
in the above. Studies of the $a_{11}$ ($a_{22}$) effect
with other $\alpha$ values are also performed, but not as systematically.
The above observed behaviors of OBC-A with respect to $a_{11}$ and $a_{22}$
appear to also apply to other $\alpha$ values.

% effect of R_21, R_12 with alpha=0

\begin{figure}
  \centerline{
    \includegraphics[width=2in]{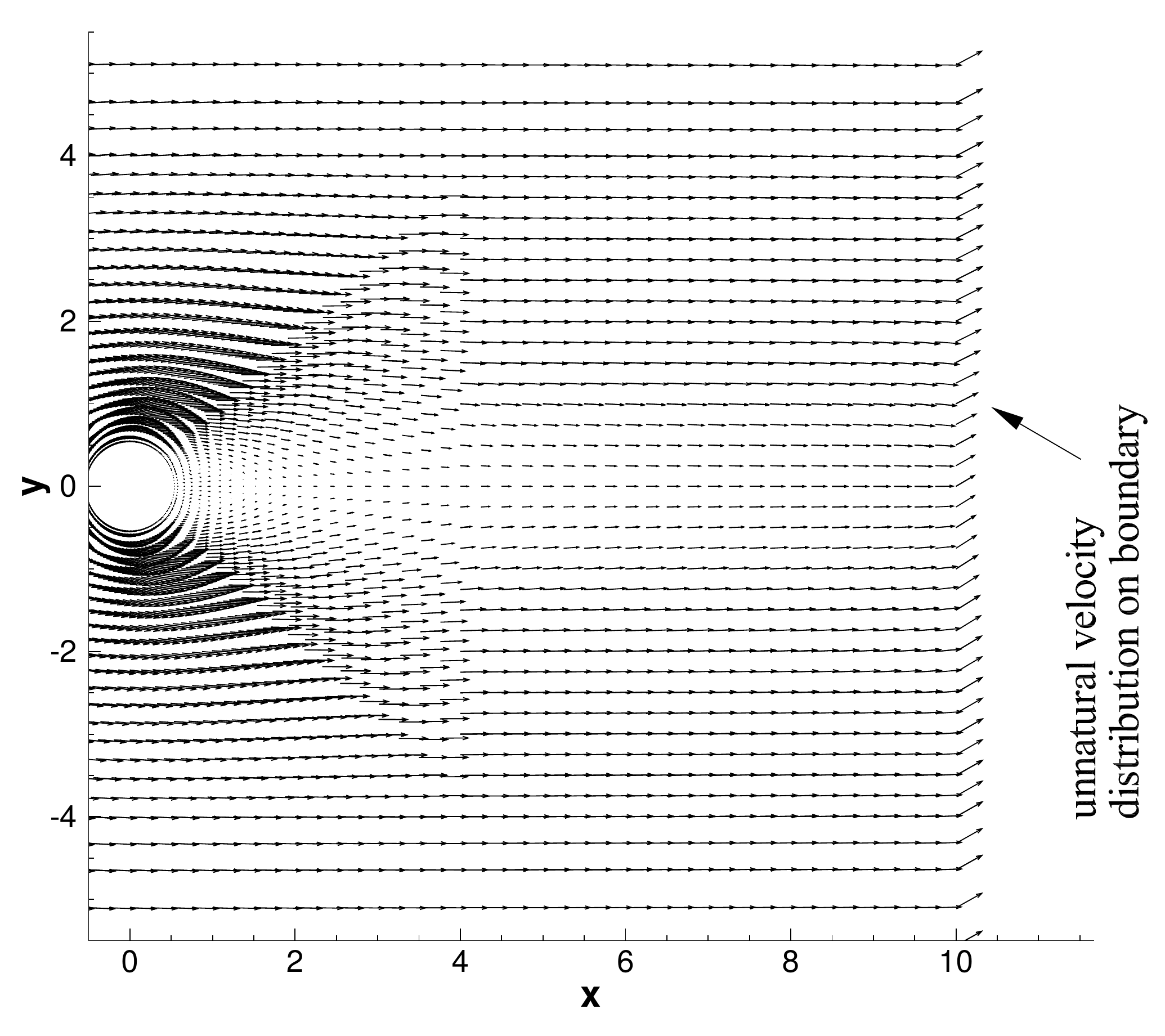}(a)
    \includegraphics[width=2in]{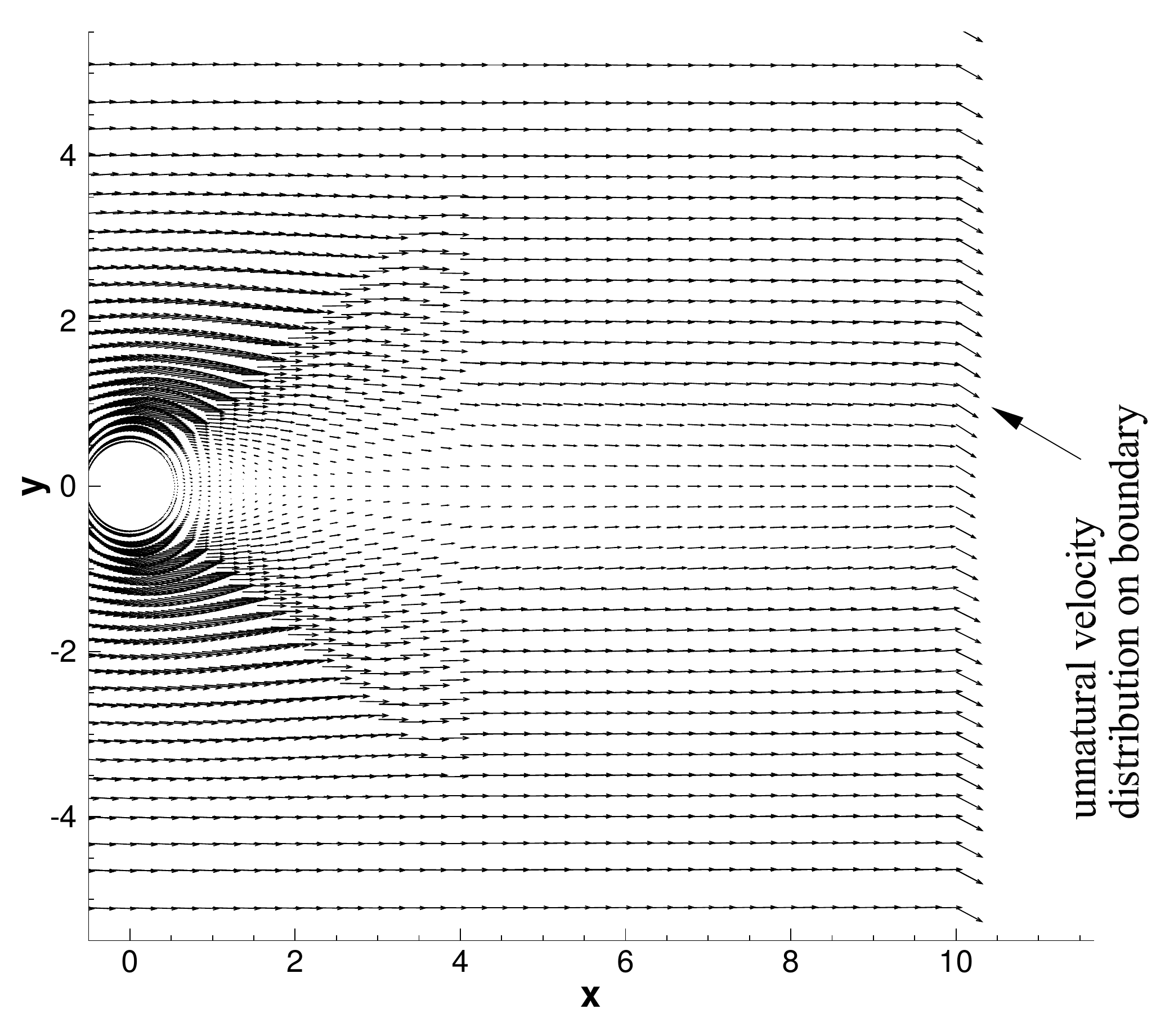}(b)
    \includegraphics[width=2in]{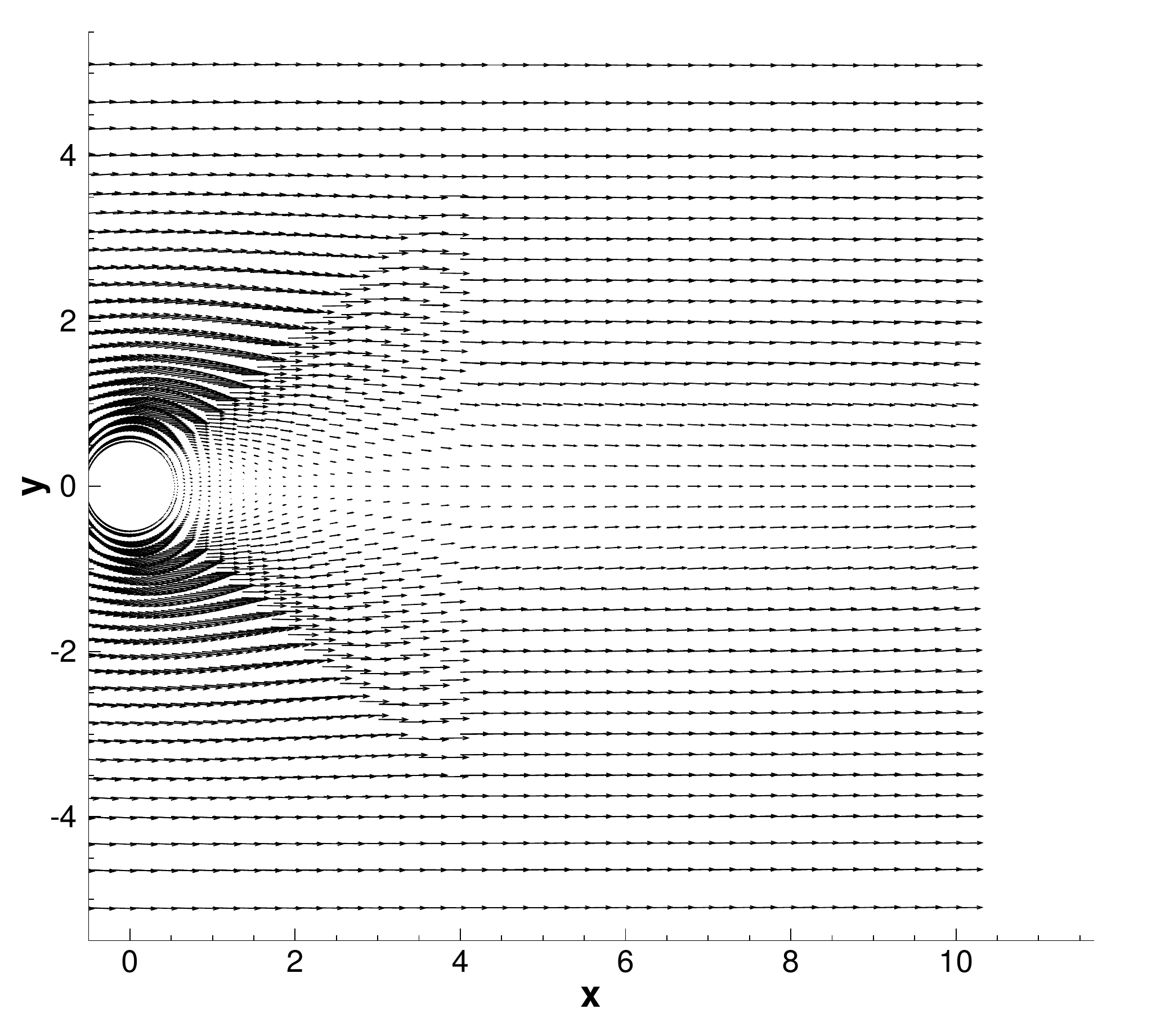}(c)
  }
  \caption{
    2D Cylinder flow ($Re=20$): velocity fields computed using
    OBC-A ($\alpha=0$, $a_{11}=a_{22}=a_{12}=0$) with different
    $a_{21}$ values:
    (a) $a_{21}=-0.4$, (b) $a_{21}=0.4$, and (c) $a_{21}=0$.
    Un-physical velocity distributions can be observed at
    the outflow boundary if $a_{21}\neq 0$ with OBC-A.
    Velocity vectors are plotted on every eighth quadrature points in each
    direction within each element.
  }
  \label{fig:a21_effect}
\end{figure}

With OBC-A, when $\alpha=0$, the $\mathbf{R}$ matrix may not be diagonal in 2D,
as long as its elements $a_{ij}$ ($i,j=1,2$) satisfy the conditions
\eqref{equ:cond_R_1_0}, \eqref{equ:cond_R_2_0}
and \eqref{equ:cond_R_3_0}.
We observe however that non-zero off-diagonal elements ($a_{21}$ and $a_{12}$),
especially when $a_{21}\neq 0$,
can result in
poor or unphysical simulation results with OBC-A.
This point is demonstrated by the velocity distributions (steady-state)
in Figure \ref{fig:a21_effect} for Reynolds number $Re=20$,
which are computed using OBC-A with $\alpha=0$ and several
$a_{21}$ values ($a_{21}=-0.4$, $0.0$ and $0.4$), while $a_{11}=a_{22}=a_{12}=0$
in the $\mathbf{R}$ matrix.
At this Reynolds number the velocity is expected to be approximately in
the horizontal direction at the outflow boundary.
To one's surprise, when $a_{21}\neq 0$, the computed velocity 
at the outflow boundary points to an oblique direction, even though
all the velocity vectors are approximately along the horizontal direction
inside the domain; see Figures \ref{fig:a21_effect}(a)-(b).
The angle of the velocity vectors on
the boundary depends on the sign and the magnitude of $a_{21}$.
If $a_{21}=0$, on the other hand, the computed velocity is
approximately along the horizontal direction as expected
(Figure \ref{fig:a21_effect}(c)).
The above unphysical results can be understood by considering
equation \eqref{equ:bc_2_0} for OBC-A,
which in this case is reduced to the following on the boundary,
\begin{equation}
  \nu \frac{\partial v}{\partial x}
  = \frac{\lambda_1-\lambda_3}{\mathscr{K}}a_{21}u
  + \frac{\lambda_1}{\mathscr{K}}v
\end{equation}
where $u$ and $v$ are the $x$ and $y$ components of
the velocity. This equation indicates that the horizontal velocity
$u$ will contribute to the vertical velocity $v$ at the outflow
boundary when $a_{21}\neq 0$.
Therefore, even if $v=0$ inside the domain, a non-zero $v$
will be generated on the outflow boundary due to
the boundary condition, leading to poor velocity distributions.
%
% a_12
By considering equation \eqref{equ:bc_1_0}, one can infer that
the parameter $a_{12}$ has an analogous effect.
It induces a contribution of the tangent velocity $u_{\tau}$
to the normal velocity $u_n$ on the open boundary.
In practical simulations, $a_{12}$ seems not as
detrimental to the results as $a_{21}$ does, which is probably
because of the pressure term involved in $T_{nn}$ in equation \eqref{equ:bc_1_0}.

%% OBC-B

\begin{figure}[tb]
  \centering
    \subfigure[at $x=1.0$]{ \includegraphics[width=2.8in]{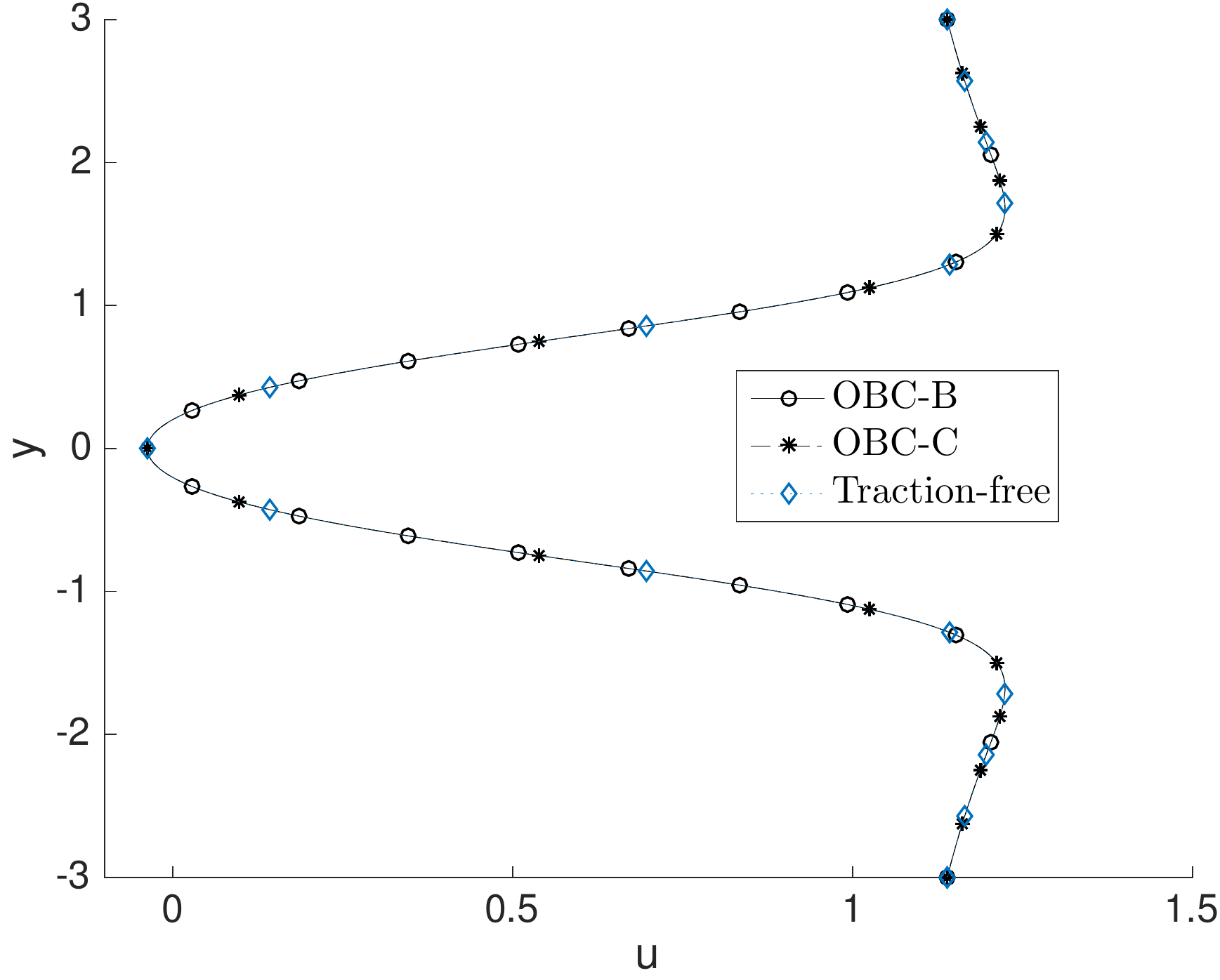}} 
    \subfigure[at $x=5.0$]{ \includegraphics[width=2.8in]{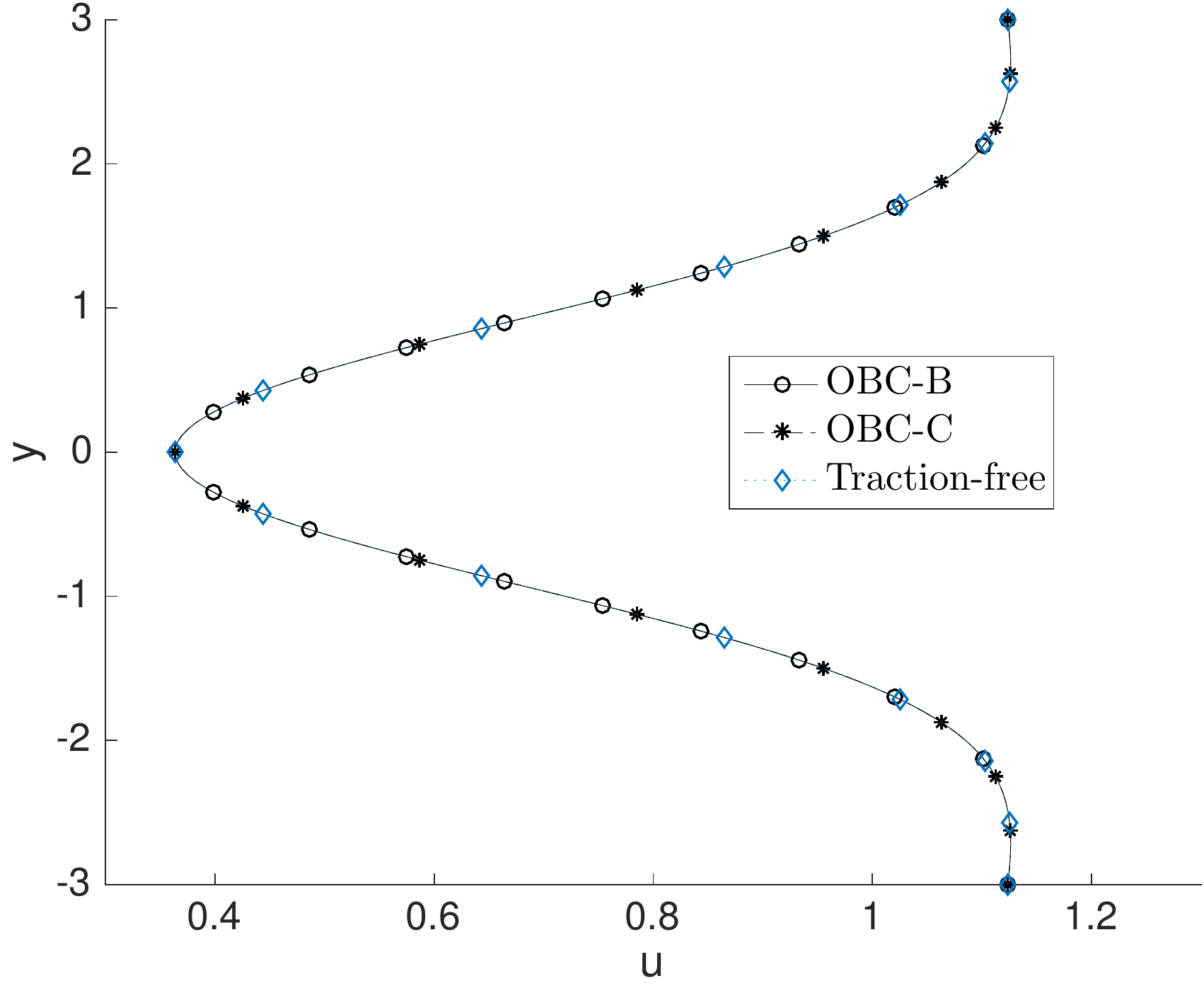}} \\
    \subfigure[at $x=10.0$ (outlet)]{ \includegraphics[width=2.8in]{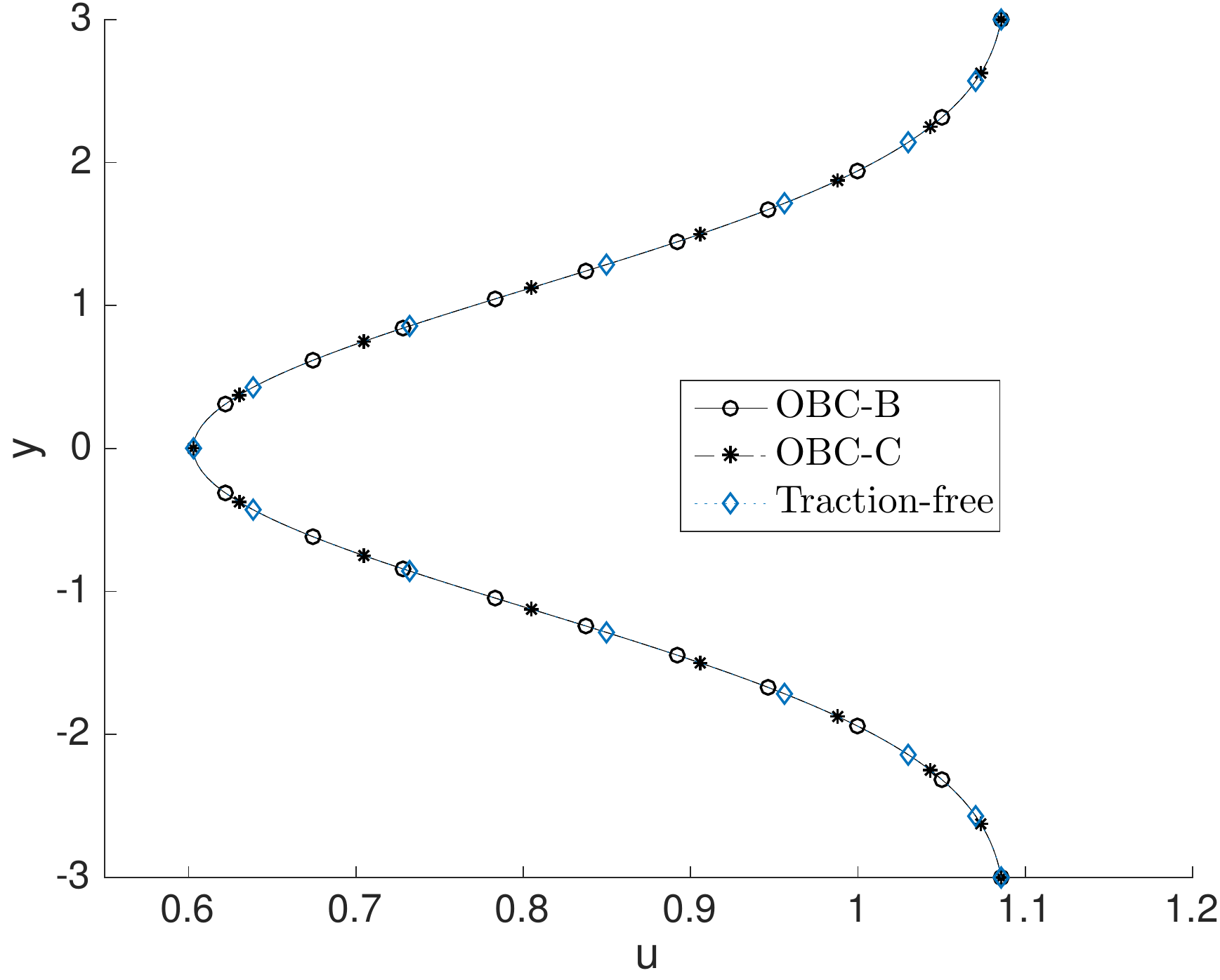}} 
    \subfigure[along centerline $y=0$]{ \includegraphics[width=2.8in]{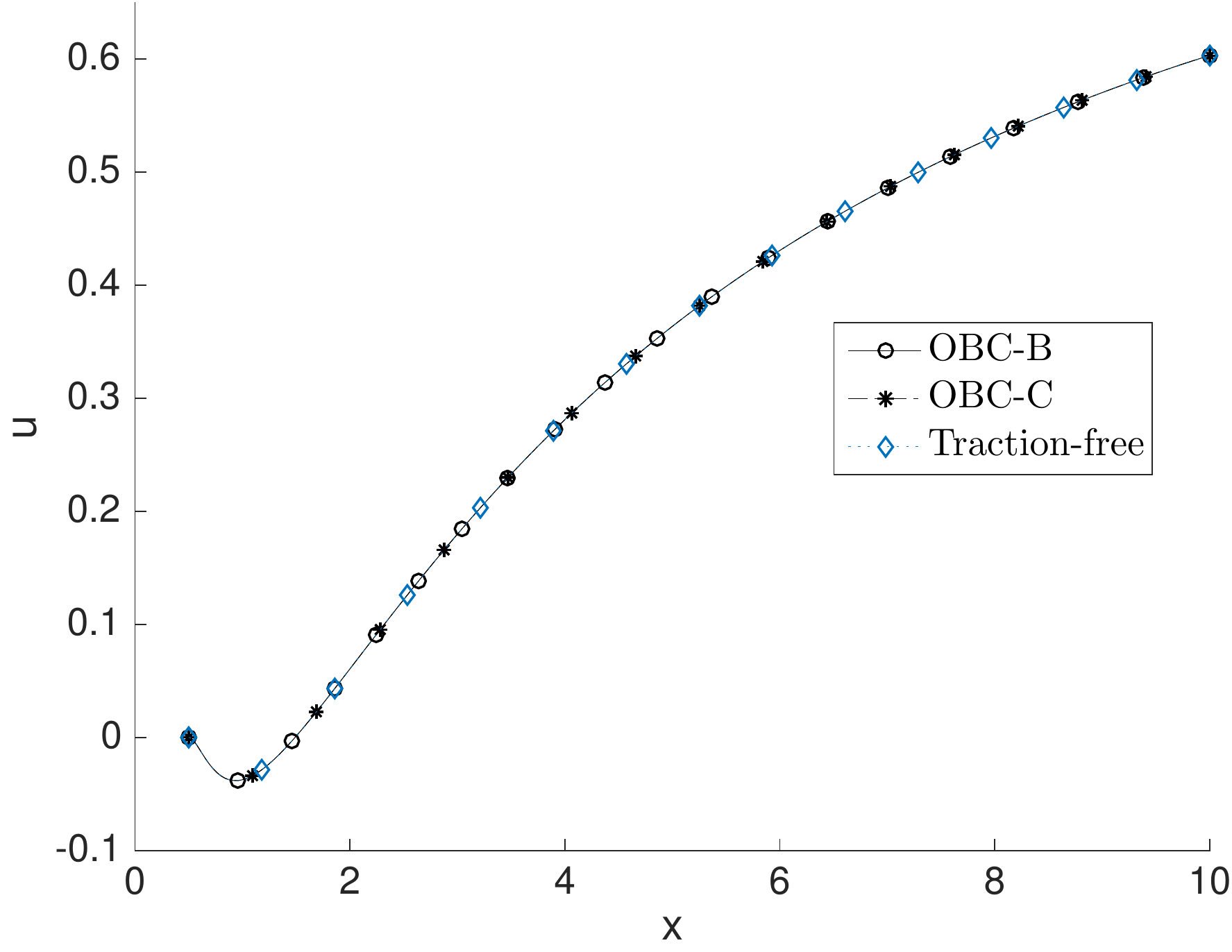}}
 \caption{
   2D Cylinder flow ($Re=20$): Streamwise velocity profiles at
   several downstream locations in the wake and along the centerline of domain,
   computed using OBC-B and OBC-C 
    as the outflow boundary
   condition. Velocity profiles obtained with the traction-free condition
   are included for comparison.
 }
\label{fig:vel_prof_re20_obcB}
\end{figure}

Let us next consider OBC-B and OBC-C. 
Table \ref{tab:obc-A} also lists the mean and rms forces on the
cylinder at the Reynolds numbers $Re=10$, $20$ and $100$ that are
computed using OBC-B and OBC-C
as the outflow boundary condition.
%For comparison, the results corresponding to the traction-free
%condition are also included in the table.
%With OBC-B and OBC-C, 
%in addition to the range $0\leqslant \alpha\leqslant \frac{1}{2}$,
%as considered in Section \ref{sec:method} when deriving
%the boundary conditions, we have included several
%values outside this range ($\alpha=1.0$, $-0.25$, $-0.5$ and $-1.0$)
%in the numerical tests. 
%Note that without the restriction $\alpha \leqslant \frac{1}{2}$,
%the conditions about the matrix $\mathbf{R}$
%as represented by \eqref{equ:cond_R_1}
%are only sufficient conditions (no longer necessary conditions)
%for the positive semi-definiteness of the matrix $\mathbf{Q}$.
%Once the matrix $\mathbf{R}$ satisfies \eqref{equ:cond_R_1},
%the parameter $\alpha$ can in principle take values outside
%the range $0\leqslant\alpha\leqslant\frac{1}{2}$.
It is observed that the computed forces based on OBC-B and OBC-C 
are identical to those based on the traction-free
condition for $Re=10$ and $20$.
For $Re=100$, the forces obtained using OBC-B and OBC-C 
are essentially the same as that from the traction-free condition,
with only a negligible difference.
%The data for $Re=100$ also seem to suggest that the result from OBC-B and OBC-C 
%may deteriorate,
%albeit only negligibly for this problem,
%when the magnitude of $\alpha$ becomes too large. 
%The best results appear to be produced with 
%$-\frac{1}{2}\leqslant\alpha\leqslant\frac{1}{2}$ for OBC-B and OBC-C.

% velocity profiles with OBC-B

Figure \ref{fig:vel_prof_re20_obcB} shows a comparison of
the streamwise velocity profiles along the vertical direction
at three downstream locations ($x/d=1.0$, $5.0$ and $10.0$)
and along the centerline ($y/d=0$)
among results computed using OBC-B, OBC-C, and the traction-free condition
at Reynolds number $Re=20$.
Note that $x/d=10$ is the outflow boundary in this problem.
%Figure \ref{fig:vel_prof_re20_obcC} is a comparison of the corresponding
%velocity profiles obtained using OBC-C at this Reynolds number.
%The profiles computed using
%the traction-free condition have been included in these plots
%for comparison. 
We observe that all the velocity profiles
computed using OBC-B and OBC-C and 
the traction-free condition exactly overlap with
one another. These results 
suggest that both OBC-B and OBC-C result in 
the same flow distributions as the traction-free condition.

% what else to discuss about OBC-B?

%%%%%%%%%%%%%%%%%%%%%%%%%%%%%%%%%%%%%%%%%%%%%%%%
% high Reynolds numbers

% force histories

Let us next consider the cylinder flow at higher
Reynolds numbers ($Re\geqslant 2000$). At these Reynolds numbers
the vortices shed from the cylinder can persist far downstream
into the wake, and thus may cross the outflow boundary
and exit the domain. 
This can cause severe difficulties and instabilities (backflow instability~\cite{DongKC2014}) 
to conventional methods.
Energy-stable boundary conditions are critical to overcoming
the backflow instability for
successful simulations at these Reynolds numbers.
We have conducted long-time simulations 
at several Reynolds numbers ranging from $Re=2000$
to $Re=10000$ using the methods developed herein
to test their performance.
Note that the traction-free boundary condition 
is unstable for simulations in this range of these Reynolds numbers.

% velocity snapshots at RE=5000

\begin{figure}
  \centerline{
    \includegraphics[width=2.1in]{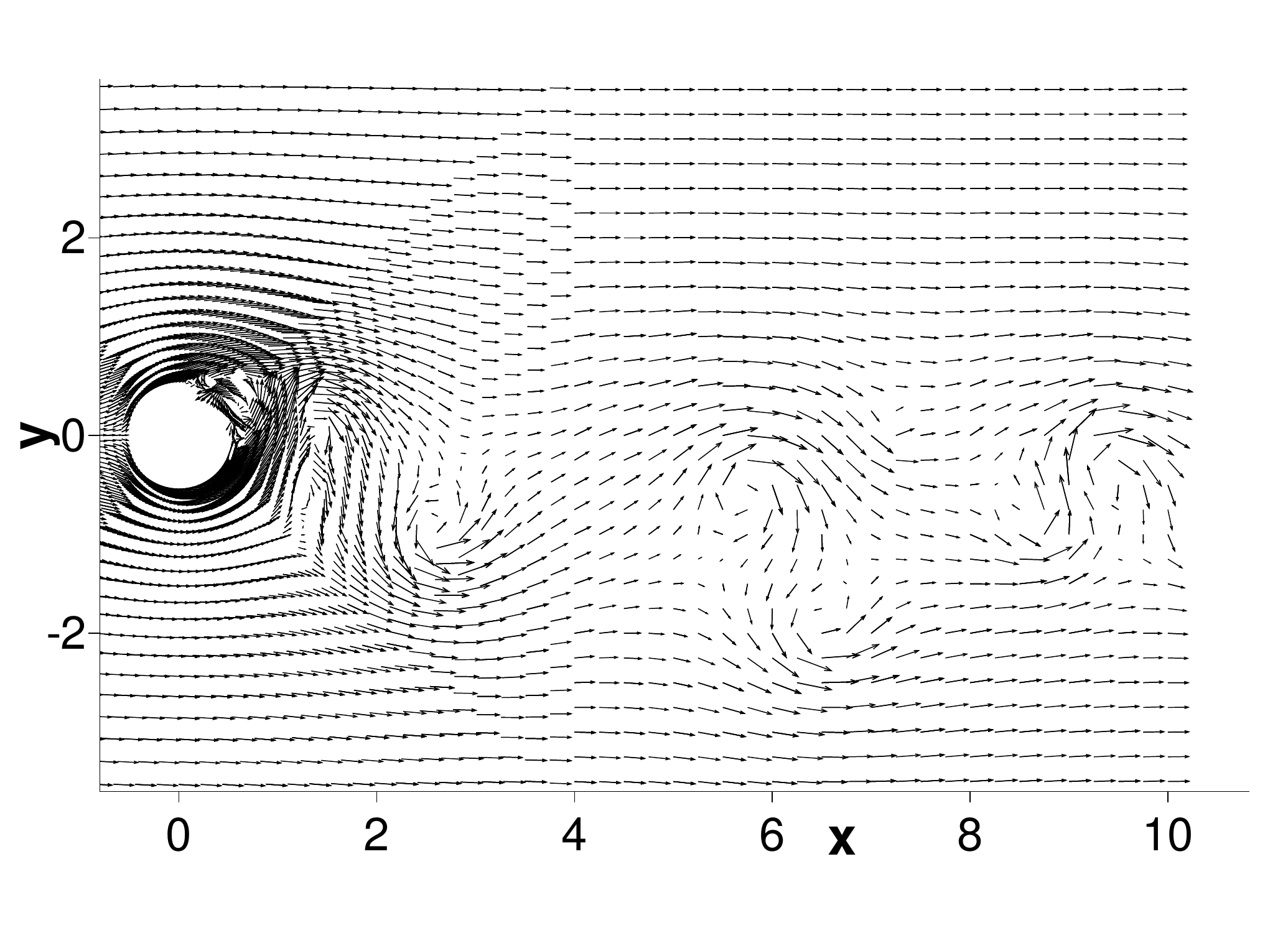}(a)
    \includegraphics[width=2.1in]{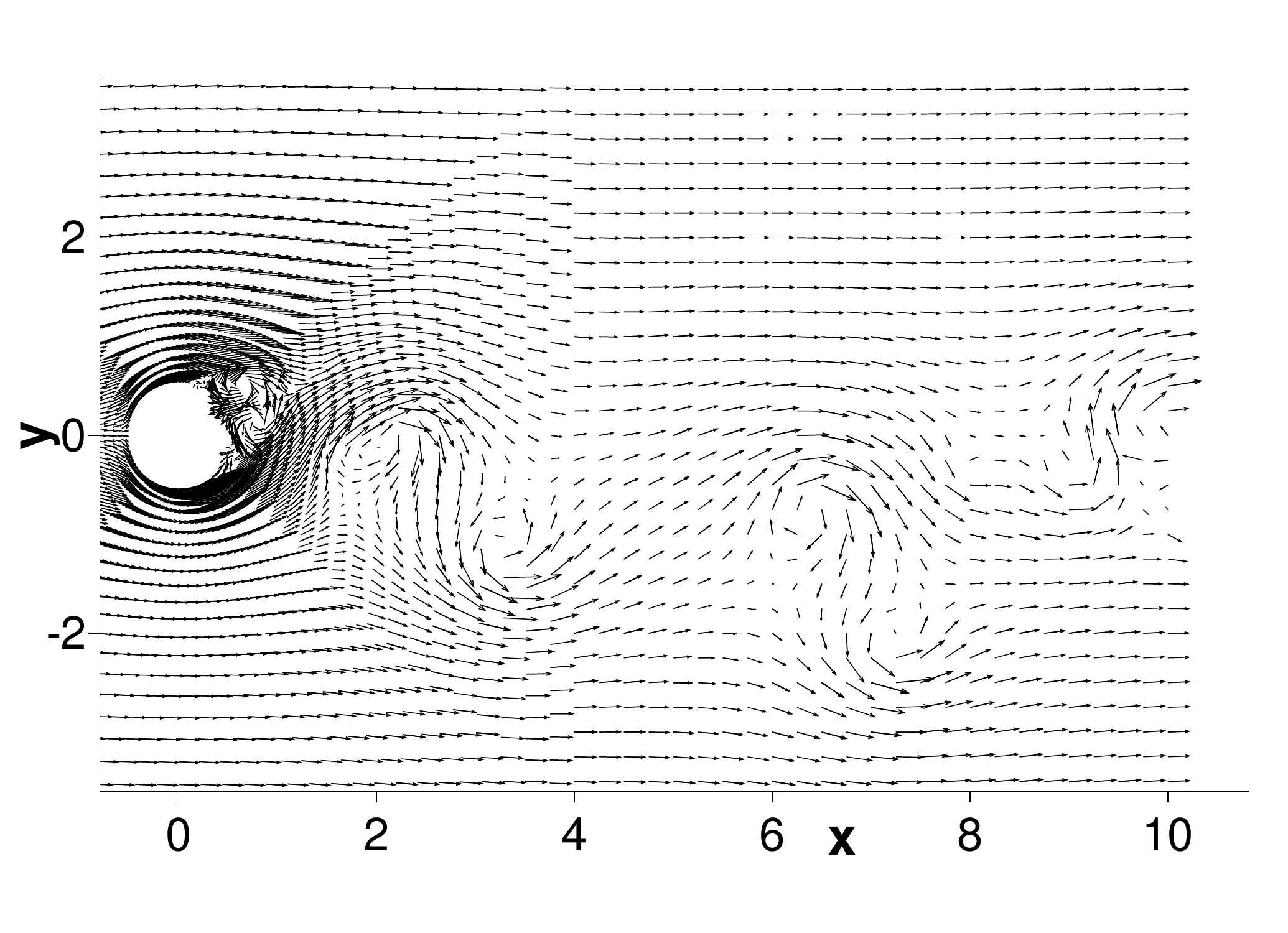}(b)
    \includegraphics[width=2.1in]{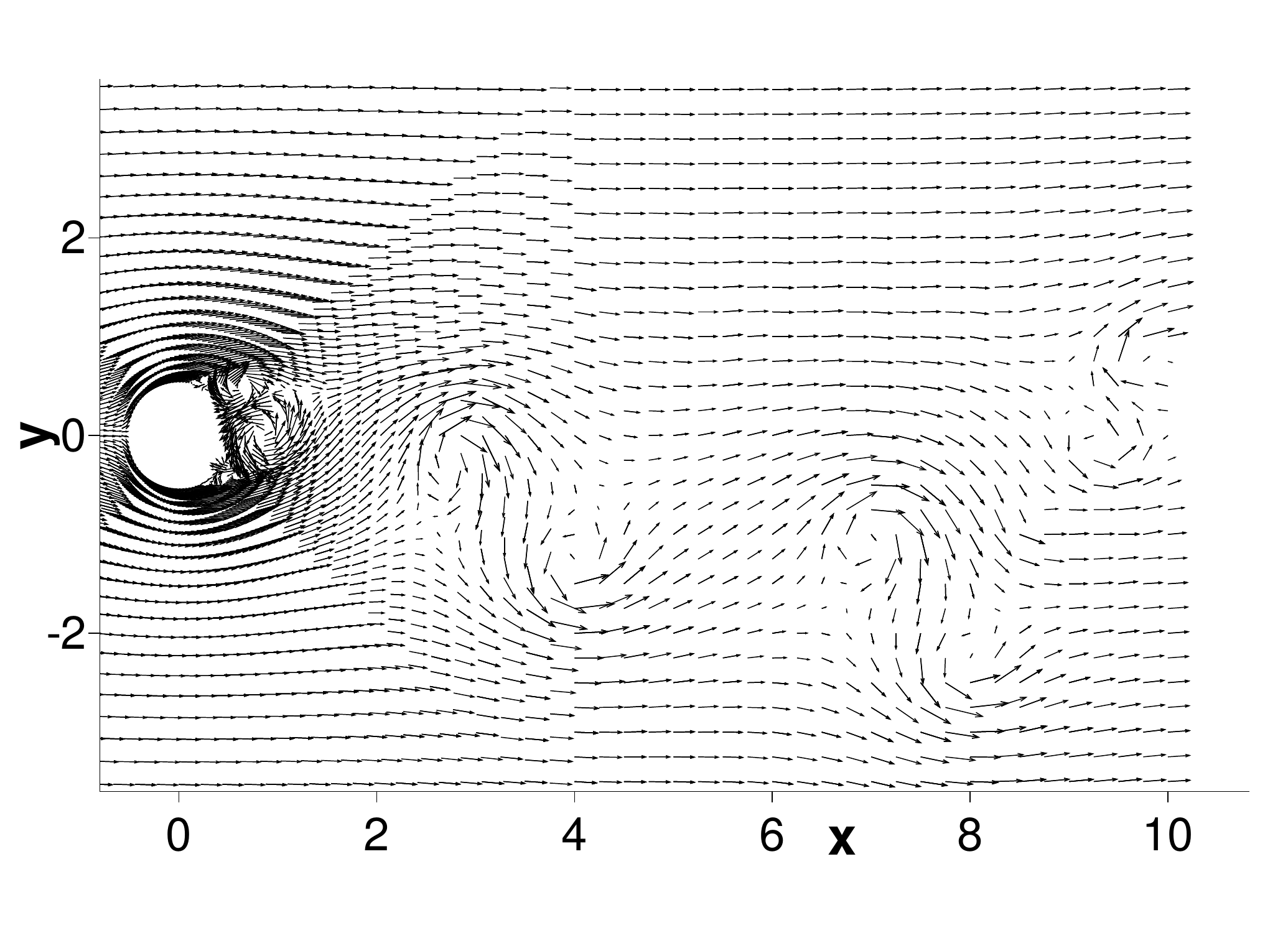}(c)
  }
  \centerline{
    \includegraphics[width=2.1in]{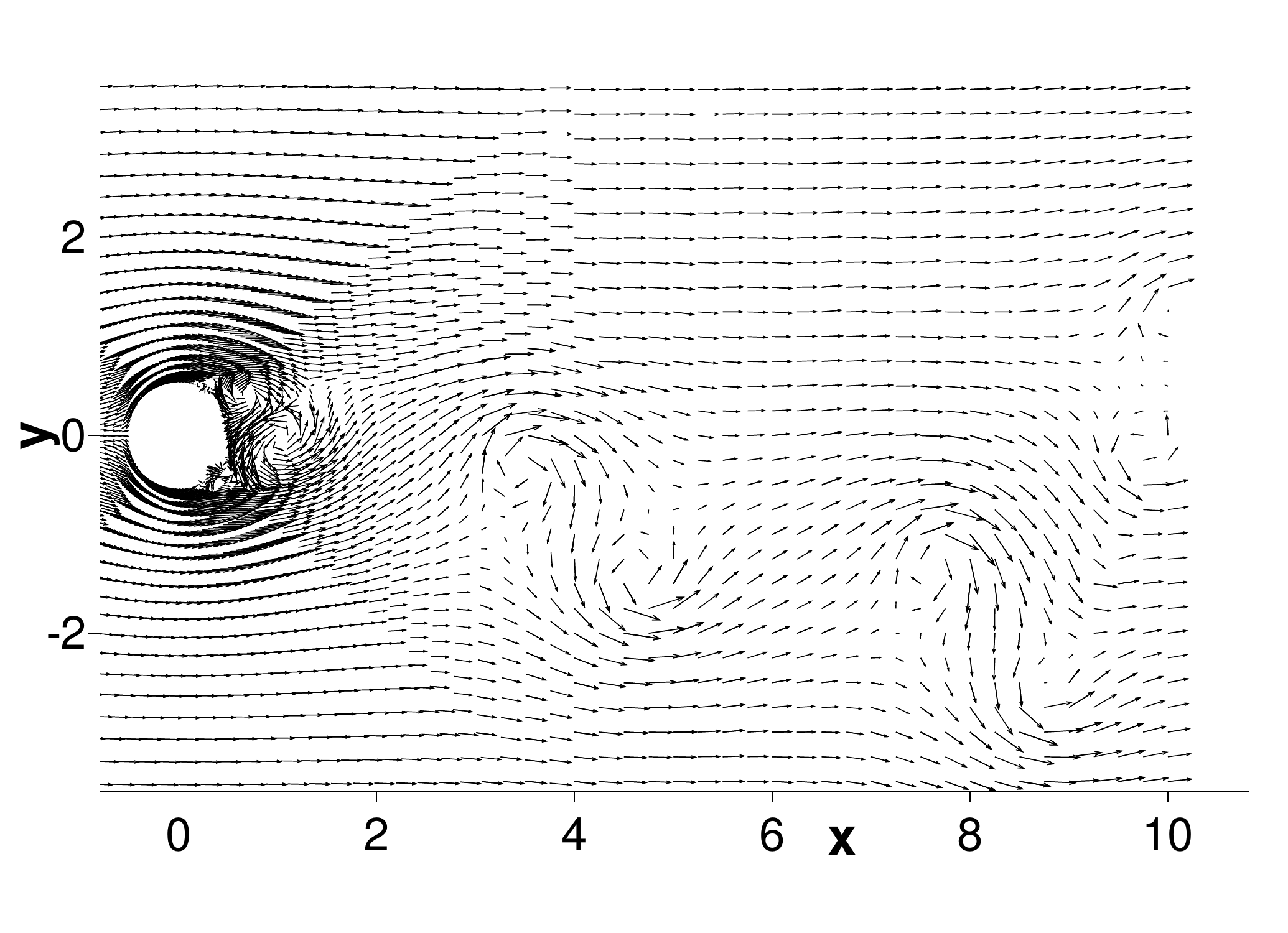}(d)
    \includegraphics[width=2.1in]{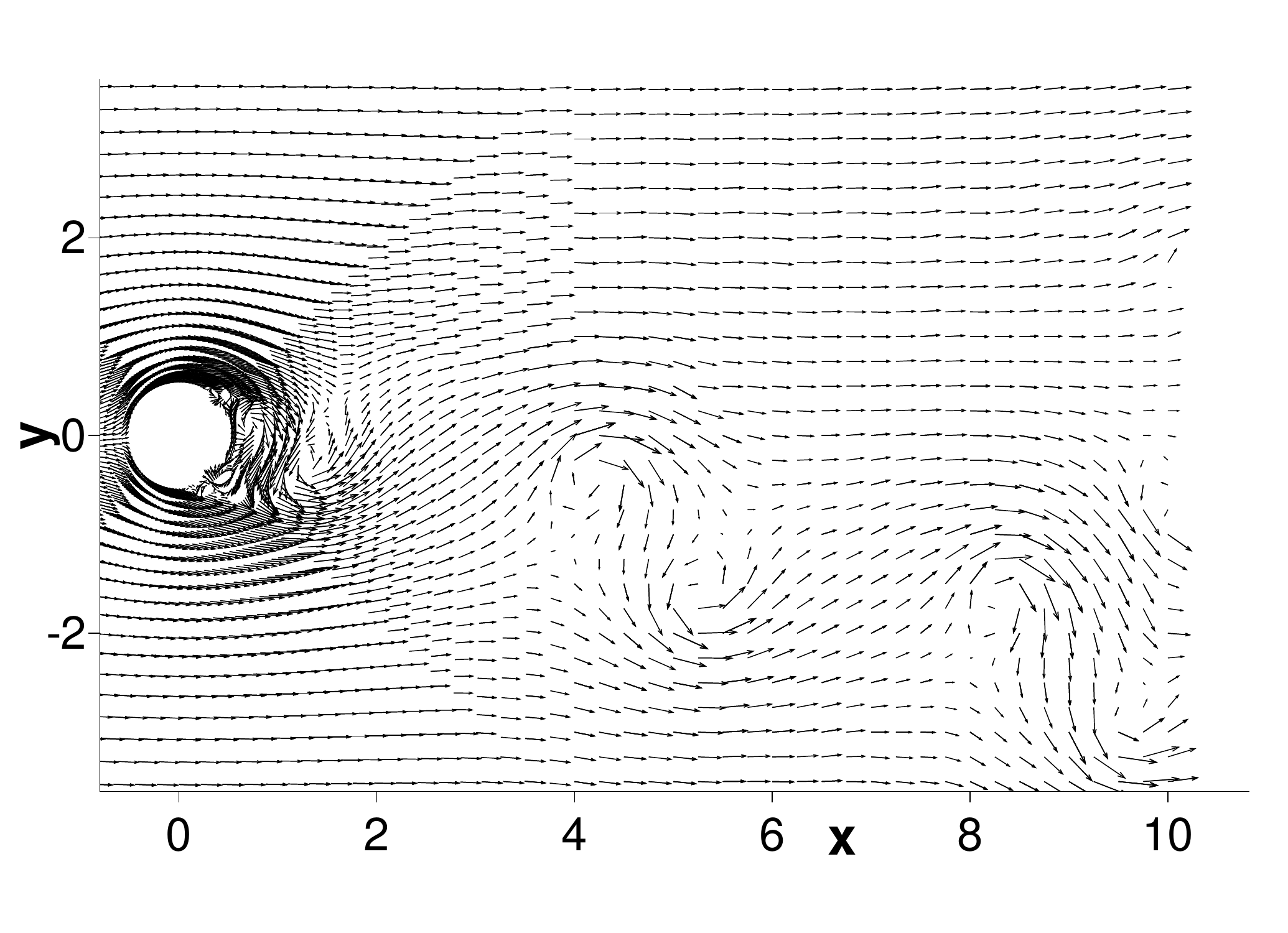}(e)
    \includegraphics[width=2.1in]{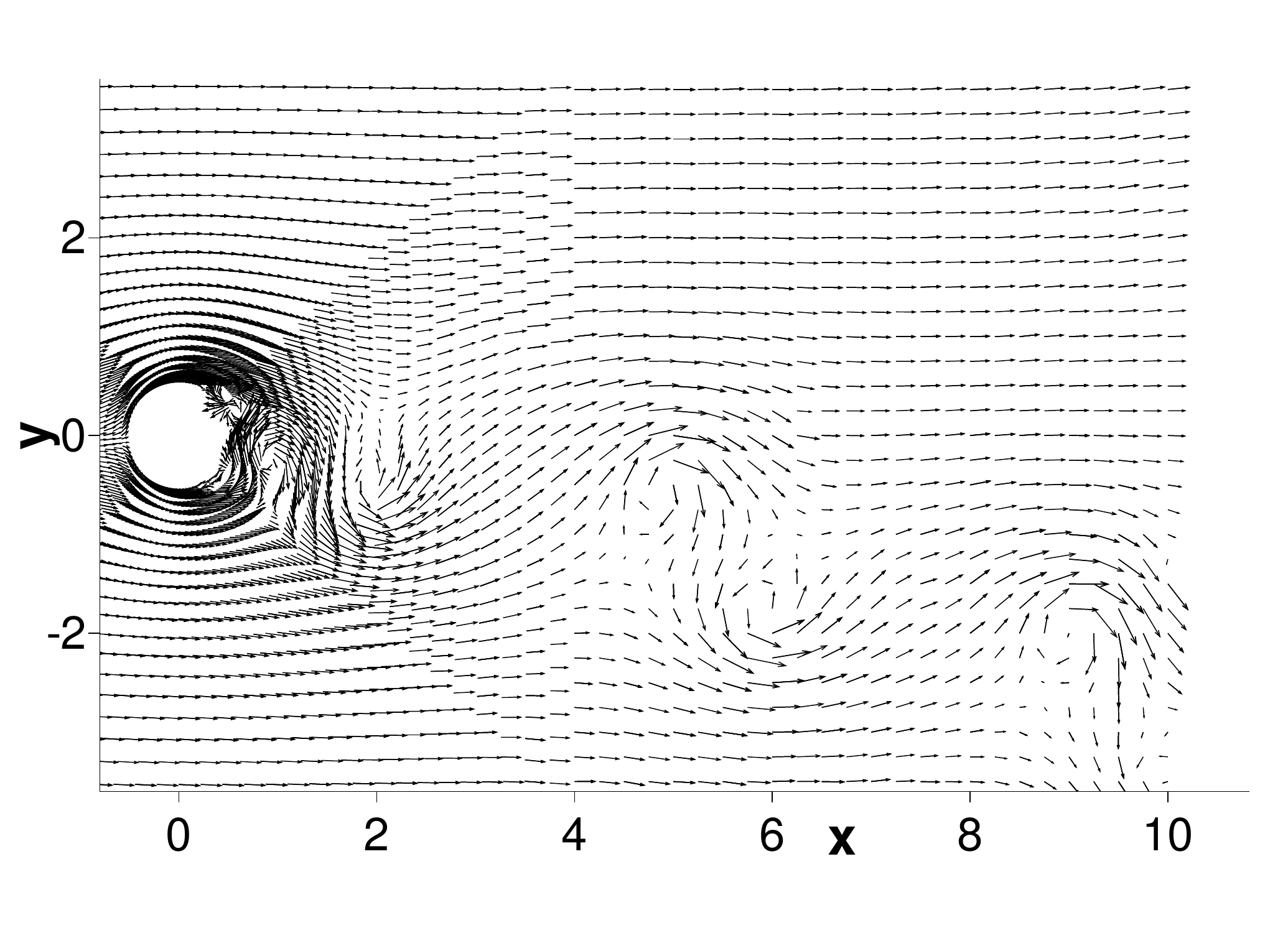}(f)
  }
  \centerline{
    \includegraphics[width=2.1in]{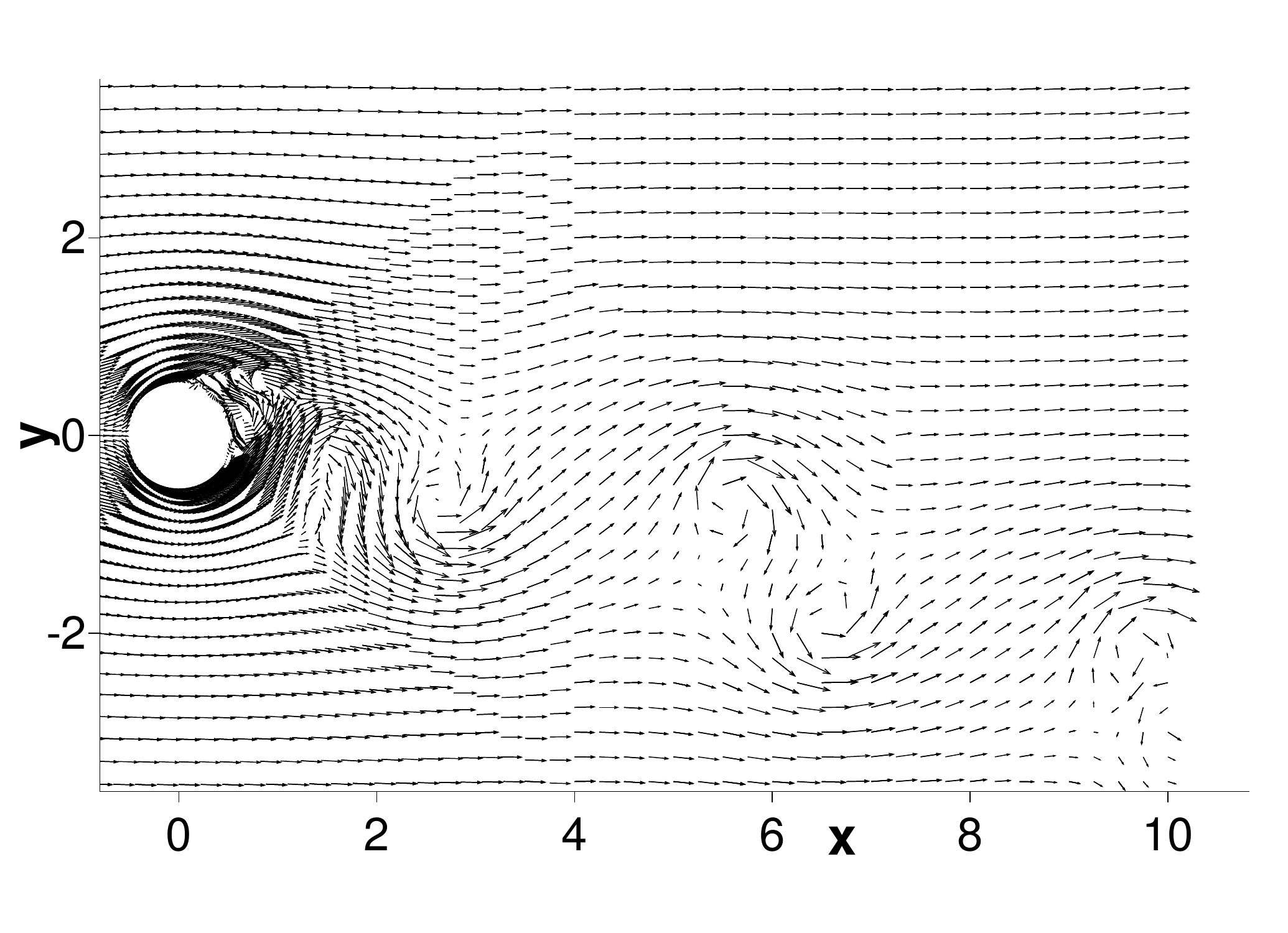}(g)
    \includegraphics[width=2.1in]{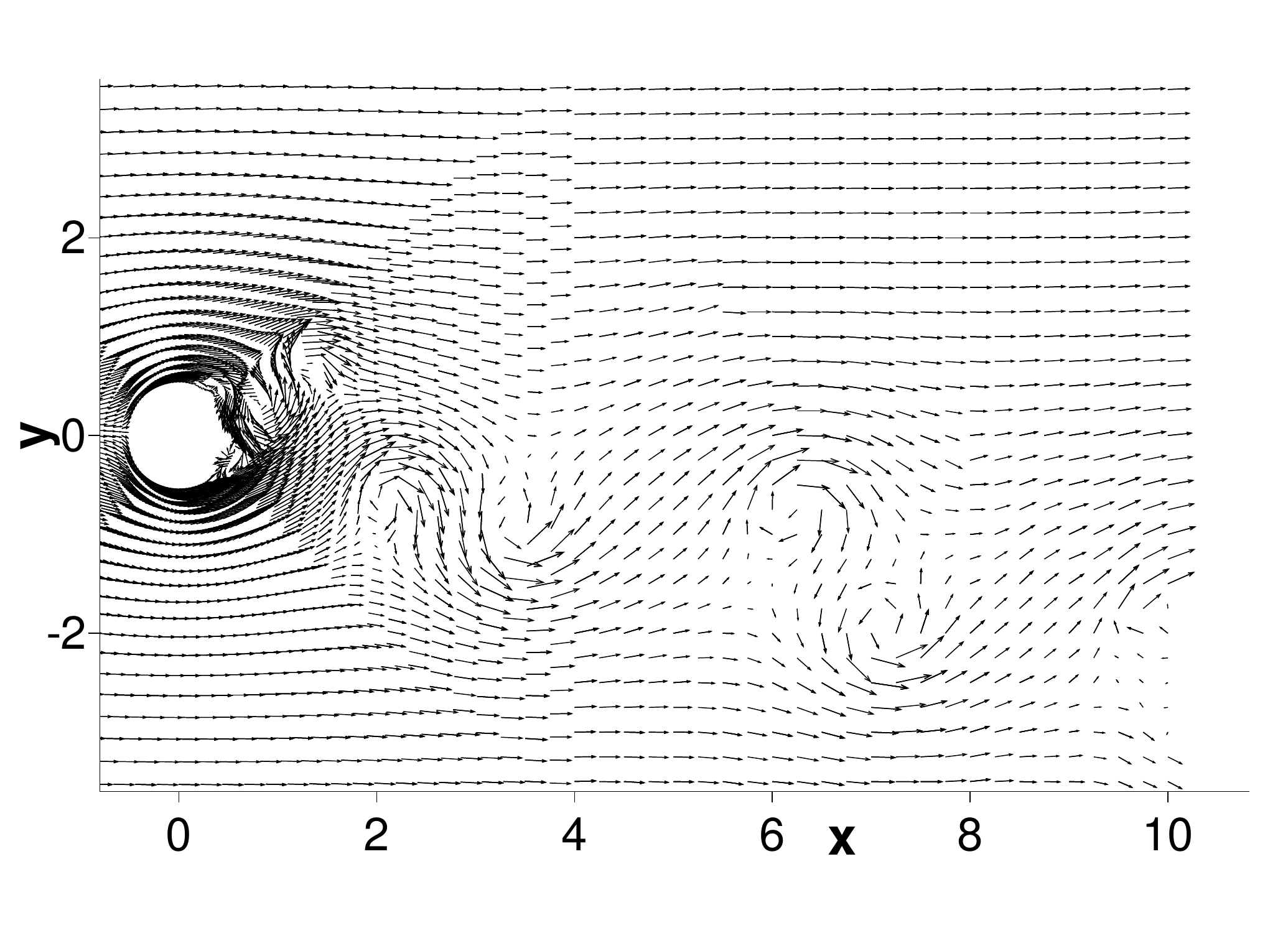}(h)
    \includegraphics[width=2.1in]{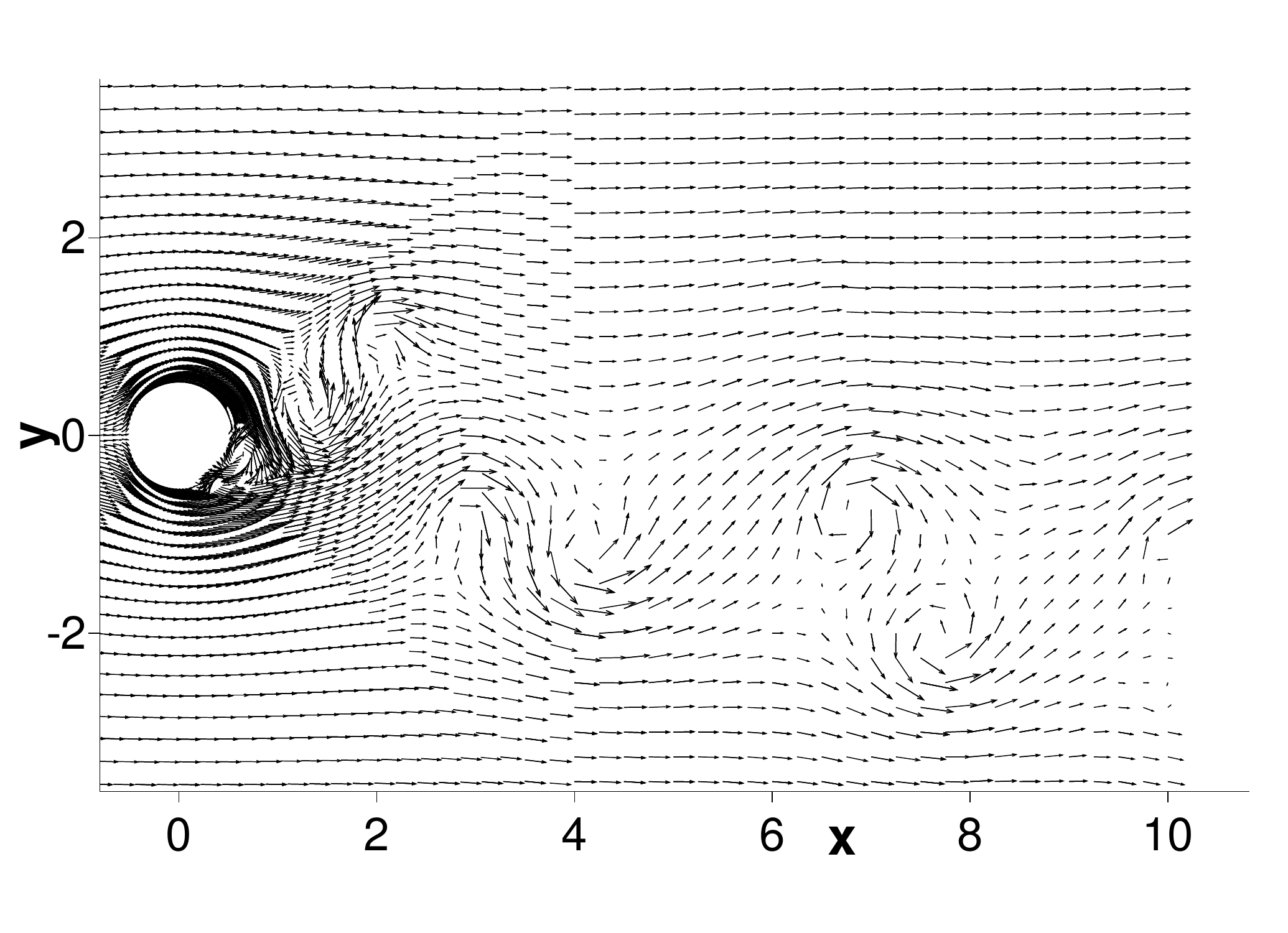}(i)
  }
  \centerline{
    \includegraphics[width=2.1in]{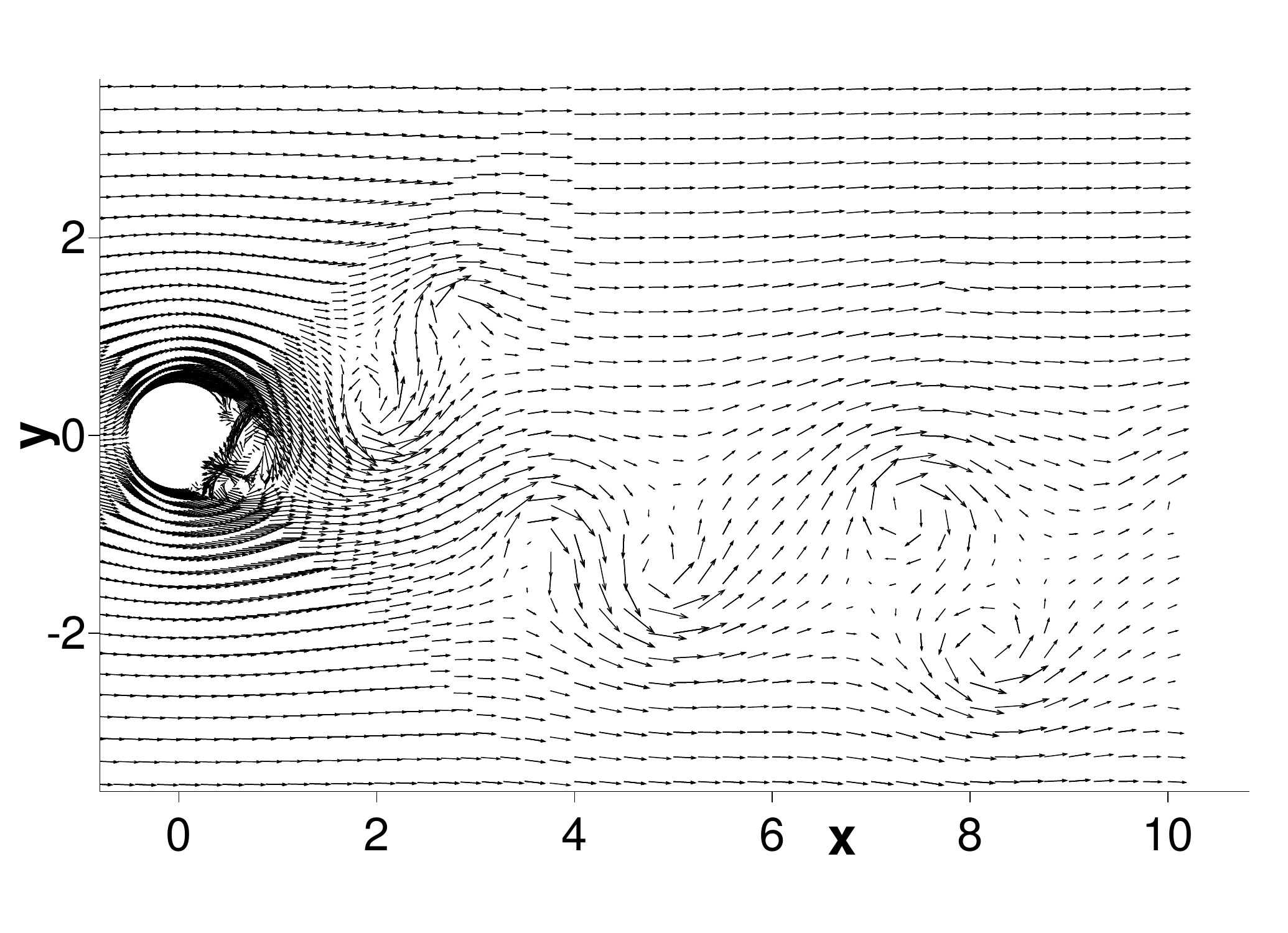}(j)
    \includegraphics[width=2.1in]{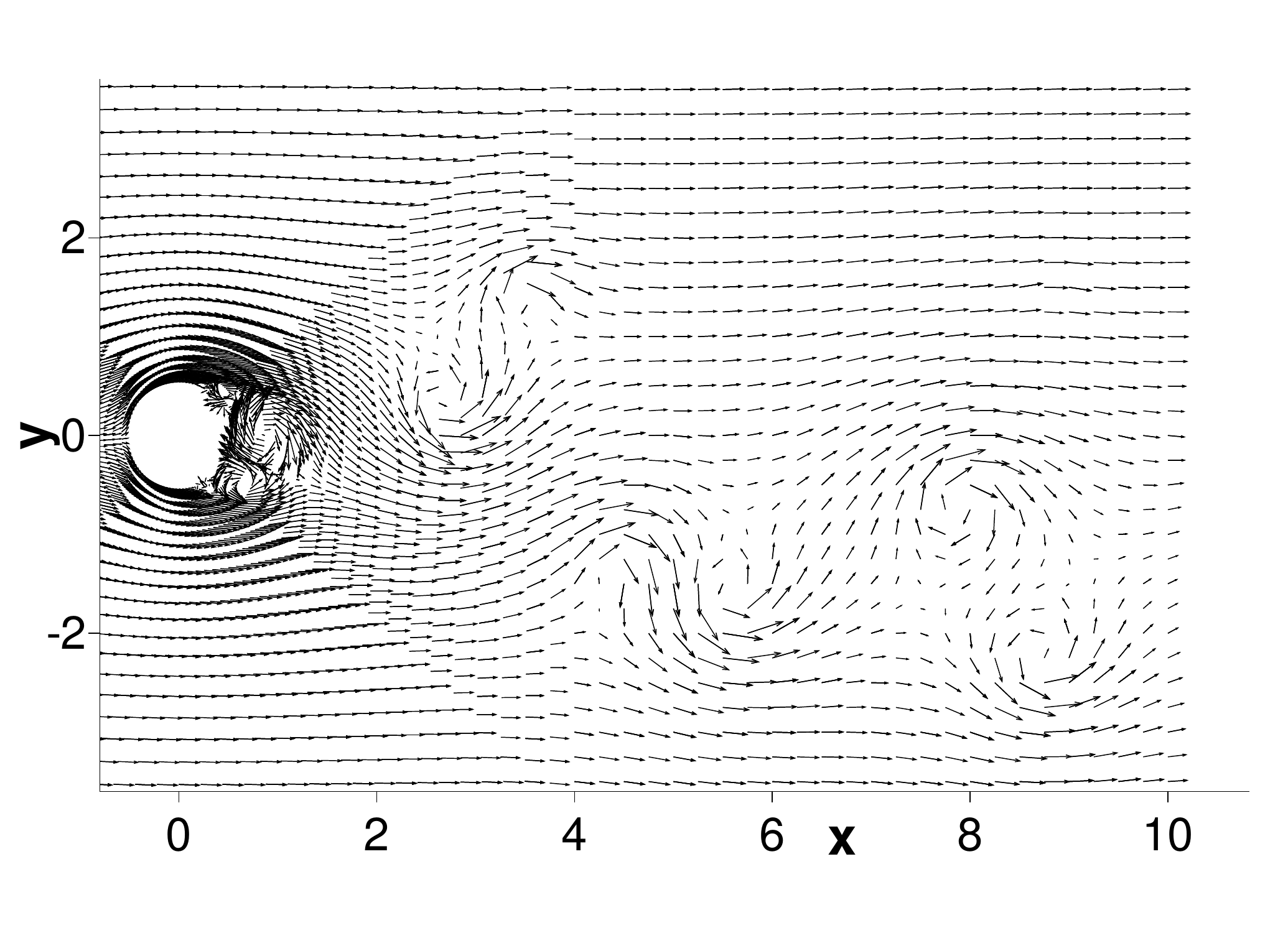}(k)
    \includegraphics[width=2.1in]{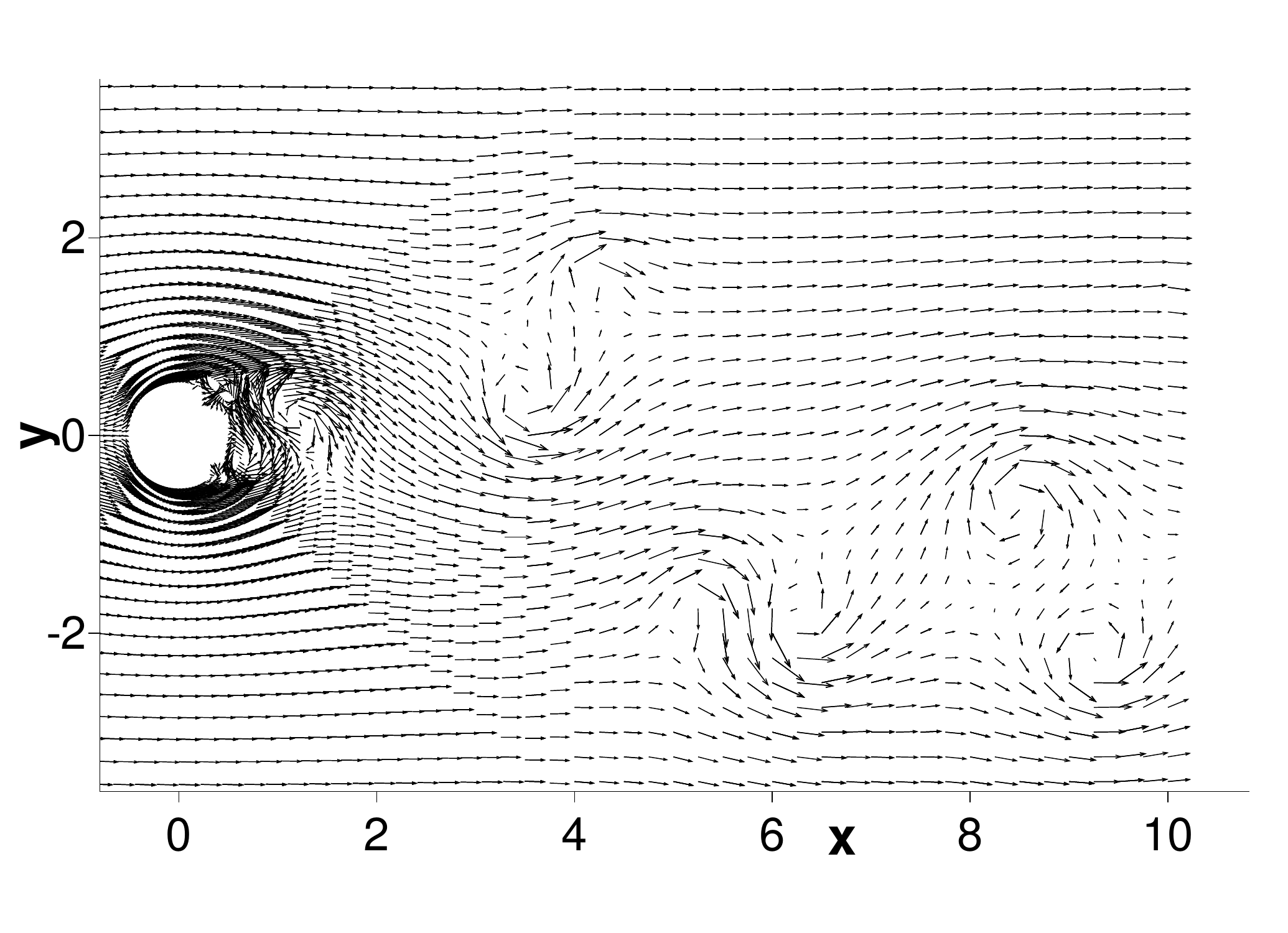}(l)
  }
  \caption{
    Temporal sequence of snapshots of velocity fields ($Re=5000$):
    (a) $t=443.2$,
    (b) $t=444.2$,
    (c) $t=445.2$,
    (d) $t=446.2$,
    (e) $t=447.2$,
    (f) $t=448.2$,
    (g) $t=449.2$,
    (h) $t=450.2$,
    (i) $t=451.2$,
    (j) $t=452.2$,
    (k) $t=453.2$,
    (l) $t=454.2$.
    Velocity vectors are plotted on every eighth
    quadrature points in each direction
    within each element.
    Results are obtained using OBC-B as the
    outflow boundary condition.
  }
  \label{fig:cyl_snap_re5k}
\end{figure}

Figure \ref{fig:cyl_snap_re5k} shows a temporal sequence of snapshots of
the instantaneous velocity fields at $Re=5000$, illustrating the dynamics
of the cylinder wake based on two-dimensional simulations.
These results are obtained using OBC-B 
as the outflow boundary condition, and the element order
is $8$ and $\Delta t=2.5e-4$ in the simulations.
One can observe pairs of vortices shed from the cylinder.
These vortices are convected downstream and persist
in the entire wake region. 
The vortices successively approach and pass through 
the outflow boundary, and discharge from the domain.
It is observed that our method is able to allow the vortices
to cross the outflow/open boundary and exit the domain
in a fairly natural way (see Figures \ref{fig:cyl_snap_re5k}(a)-(e)
and \ref{fig:cyl_snap_re5k}(f)-(i)).
But some distortion
to the vortices can also be observed
as they pass through the outflow boundary.

% force histories  

\begin{figure}
  \centerline{
    \includegraphics[width=3.2in]{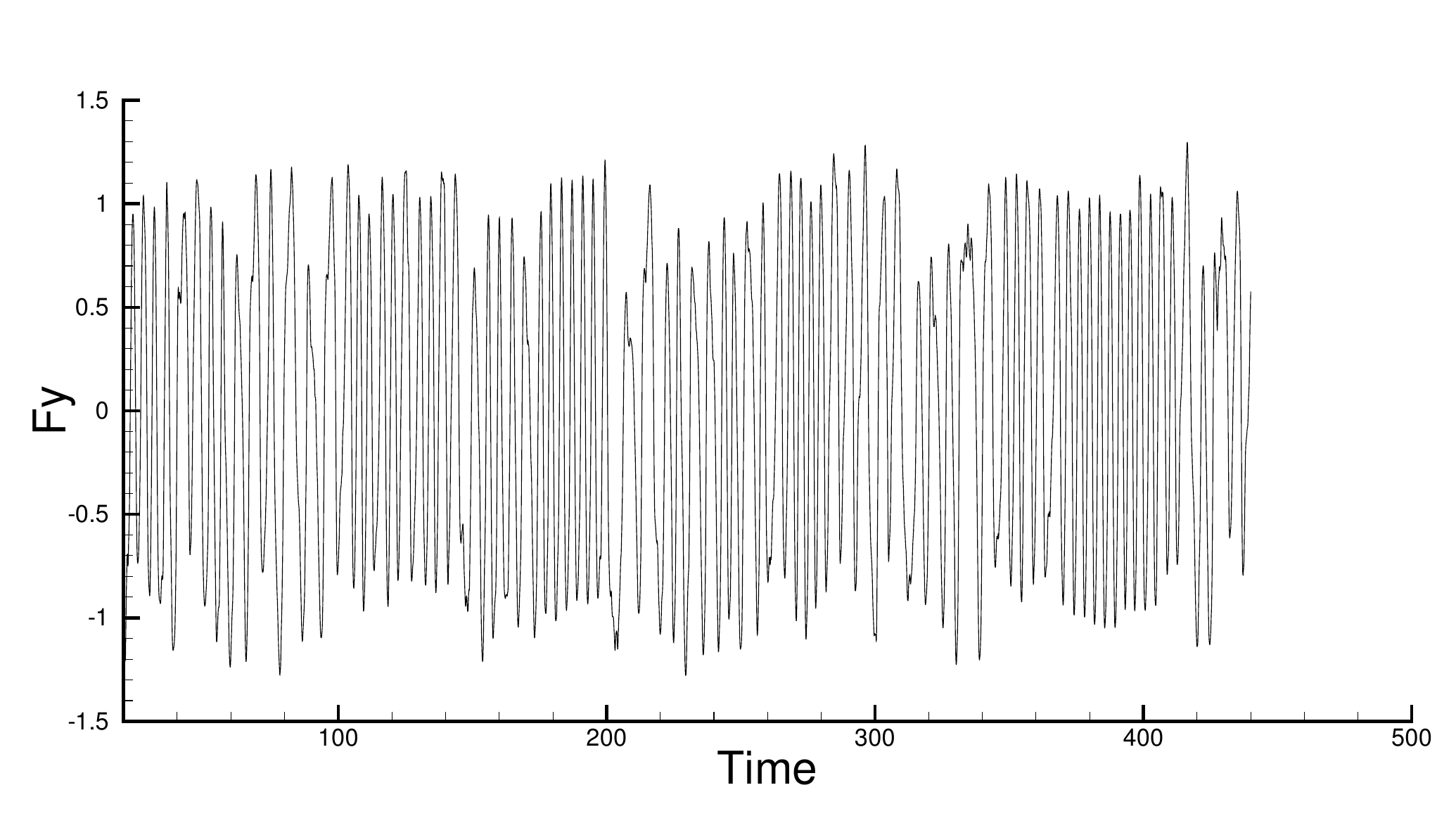}(a)
    \includegraphics[width=3.2in]{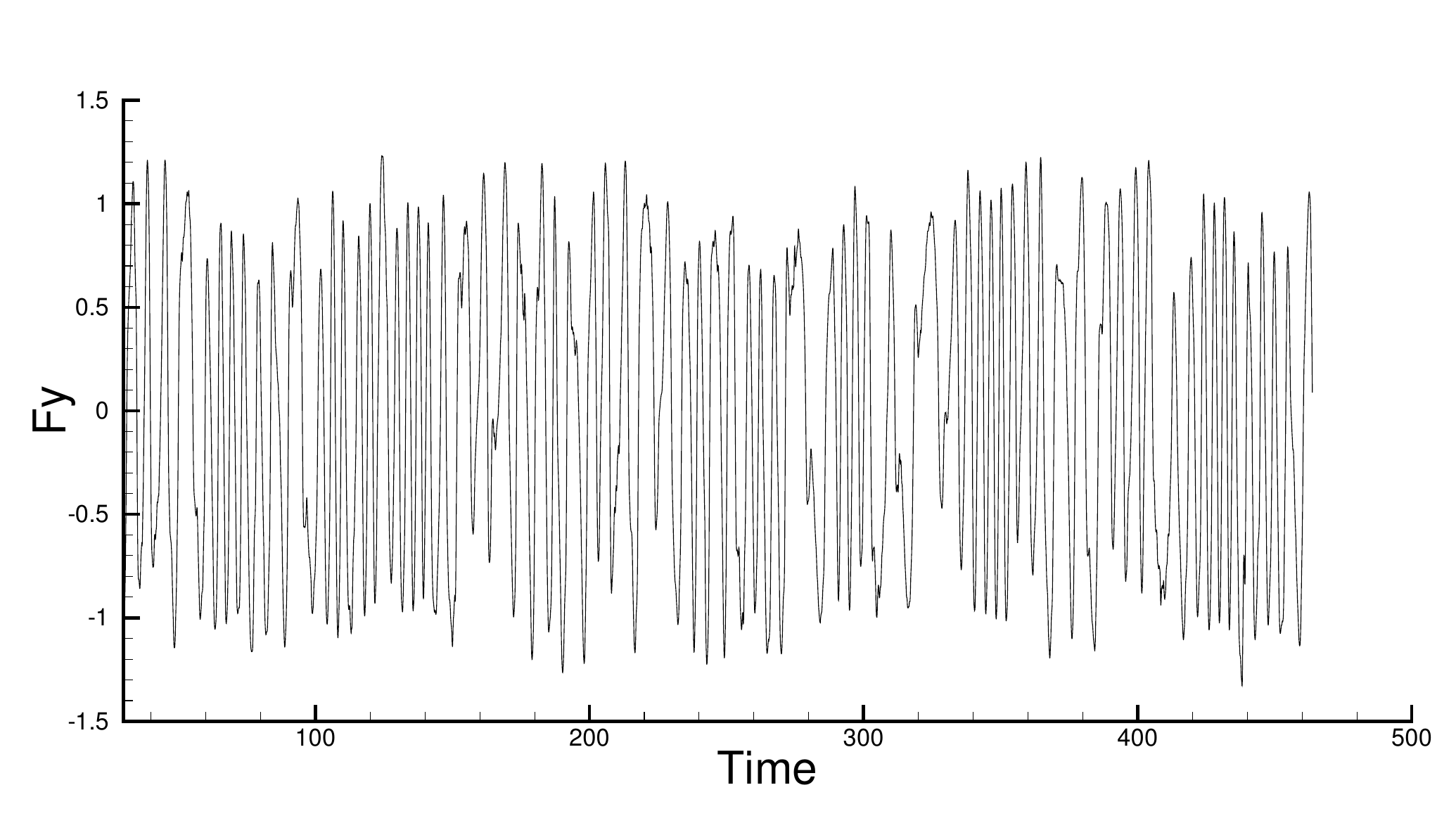}(b)
  }
  \centerline{
    \includegraphics[width=3.2in]{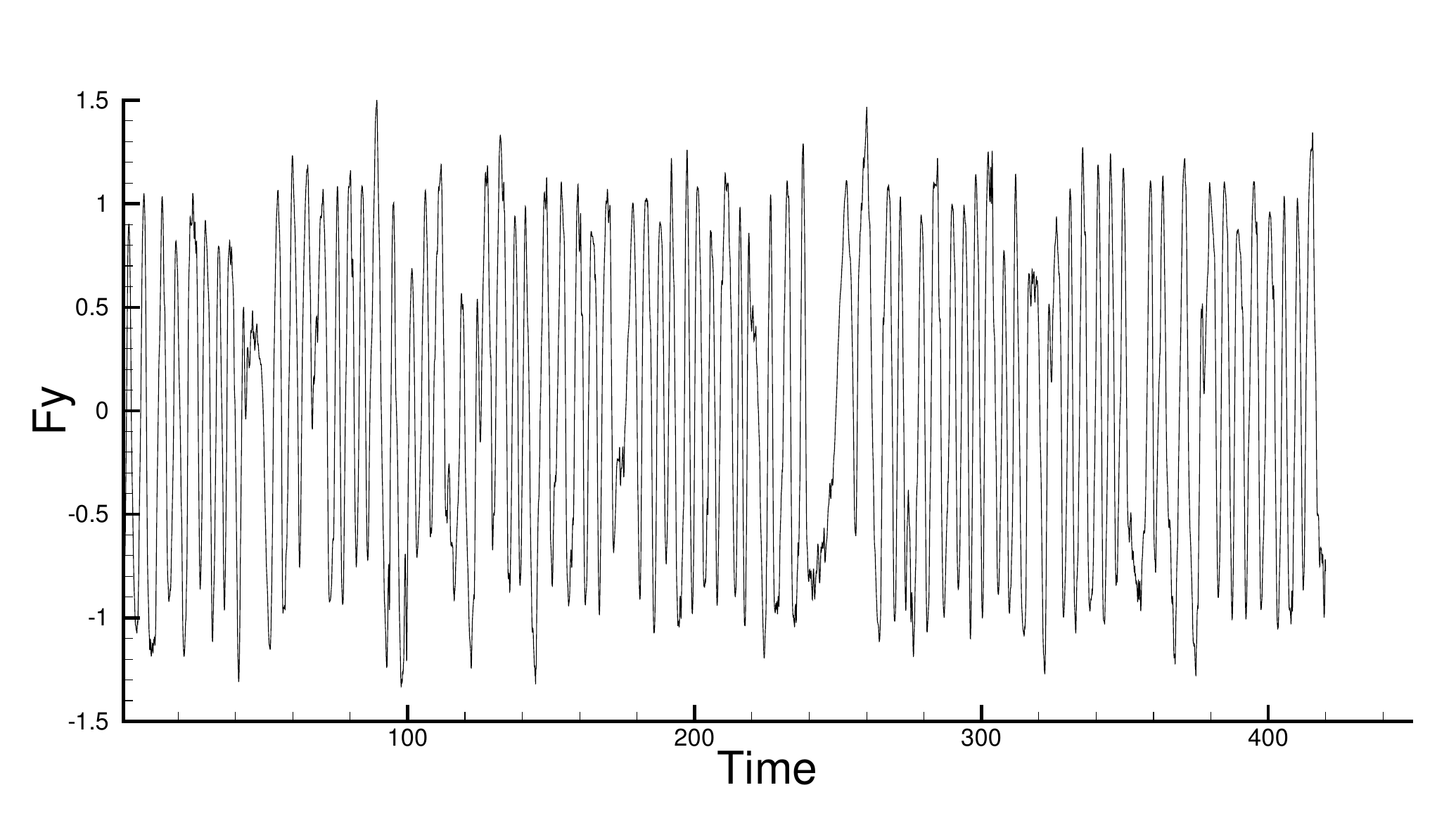}(c)
    \includegraphics[width=3.2in]{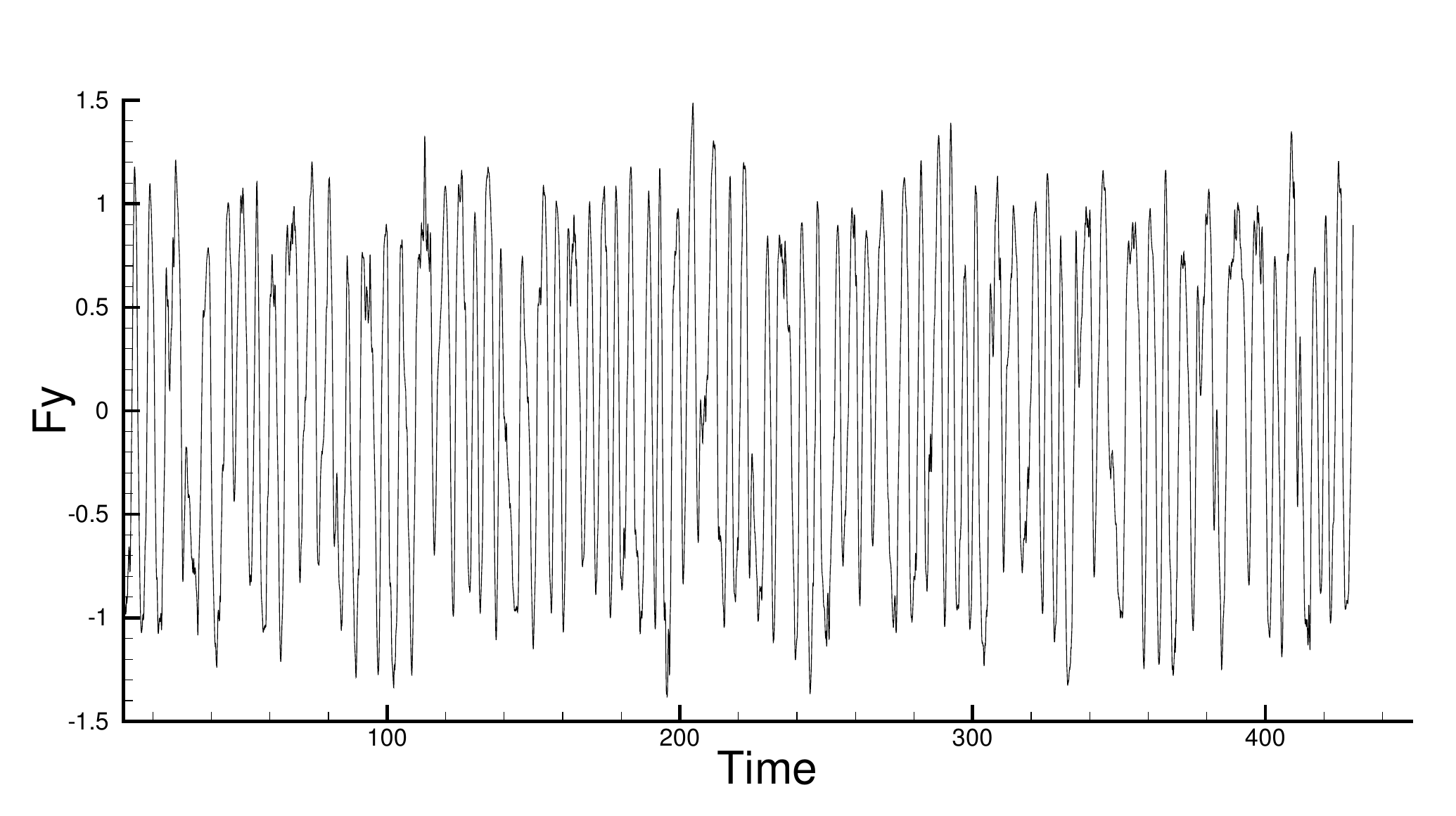}(d)
  }  
  \caption{
    Time histories of the lift on the cylinder at Reynolds numbers
     $Re=5000$ ((a) and (b)) and $Re=10000$ ((c) and (d)).
    Results in (a) and (c) are computed using OBC-B, and
    those in (b) and (d) are computed using OBC-C, as the
    outflow boundary condition.
  }
  \label{fig:cyl_fy_hist_high_re}
\end{figure}

Long-time simulations have been performed and our
methods are stable for these high Reynolds numbers.
The long-term stability of the method is demonstrated
by Figure \ref{fig:cyl_fy_hist_high_re},
which plots the time histories of the lift on
the cylinder at Reynolds numbers $Re=5000$ (Figure \ref{fig:cyl_fy_hist_high_re}(a)-(b))
and $Re=10000$ (Figure \ref{fig:cyl_fy_hist_high_re}(c)-(d)).
These simulations are conducted using OBC-B (Figure \ref{fig:cyl_fy_hist_high_re}(a) and (c)) 
and OBC-C (Figure \ref{fig:cyl_fy_hist_high_re}(b) and (d))
as the outflow boundary condition.
The long-term stability of the simulations and
the chaotic nature of flow are evident from the time signals.

% what else to discuss here?
% what other results to include here?

\subsubsection{Three-Dimensional Simulations}
\label{sec:cyl_3d}

% only with opt-R, alpha = 1/2
% (1) two data points on comparison with experiments
% (2) force histories for two Reynolds numbers
% (3) comparison of force histories between 2D and 3D at same Re
% (4) sequence of snapshots for 3D visualization of flow fields
%     for Re=5000 or Re=2000

We next look into the simulation of the cylinder flow
in three dimensions. Consider the 3D domain sketched in
Figure \ref{fig:cyl_mesh}(b),
$-5d\leqslant x\leqslant 10d$, $-10d\leqslant y\leqslant 10d$,
and $0\leqslant z\leqslant L_{z}$, where $d$ again denotes
the cylinder diameter and $L_z$ is the domain dimension along the $z$ direction. 
The cylinder axis is assumed to coincide with
the $z$ axis of the coordinate system.
The top and bottom of domain ($y=\pm 10d$) are assumed to be periodic.
We also assume that all the flow variables and the domain are homogeneous
along the $z$ direction and are periodic at $z=0$ and $z=L_z$, and therefore
a Fourier expansion of the field variables in $z$ can be
carried out. A uniform inflow with a free stream velocity $U_0$
enters the domain at $x=-5d$ along the $x$ direction,
and the wake discharges from the domain through the
boundary at $x=10d$. As in 2D simulations, all length variables are normalized
by the cylinder diameter $d$ and all velocity variables are 
normalized by the free stream velocity $U_0$. Therefore,
the Reynolds number is defined based on $U_0$ and $d$.
 
% discretization

We consider two Reynolds numbers $Re=500$ and $5000$
for 3D simulations in this paper.
We employ a domain dimension $L_z/d=1.0$ along the $z$ direction for $Re=500$
and a dimension $L_z/d=2.0$ for $Re=5000$.
The domain is discretized using $32$ uniform points (i.e.~$32$ Fourier planes)
along the $z$ direction, and each of the plane ($x$-$y$ plane) is discretized
using a mesh of $1228$ quadrilateral spectral elements with an element order $6$.
Figure \ref{fig:cyl_mesh}(b) is a sketch of the 3D domain and the spectral element mesh
within the $x$-$y$ planes. In the current work
the mesh used in each $x$-$y$ plane for
the 3D simulations is exactly the same as that of Figure \ref{fig:cyl_mesh}(a)
for the 2D simulations in Section \ref{sec:cyl_2d}.
We impose the no-slip condition (i.e.~zero velocity) on the cylinder surface,
and the Dirichlet condition \eqref{equ:dbc} on the left boundary ($x=-5d$), 
in which the boundary velocity
is set according to the free stream velocity.
On the top/bottom boundaries ($y=\pm 10d$) periodic boundary conditions
are imposed. Along the $z$ direction a periodic condition is enforced
because of the Fourier expansions of the field variables.
On the outflow boundary $x=10d$ the open boundary condition \eqref{equ:obc}
from Section \ref{sec:method} is imposed. Both OBC-B and OBC-C
are employed for 3D simulations. 
Long-time simulations are performed and the flow has reached
a statistically stationary state. So the initial conditions will have
no effect on the state of the flow.
%We have considered two Reynolds numbers ($Re=500$ and $5000$)
%for the 3D simulations, and
The normalized time step size is $\Delta t=2.5e-4$ in the simulations.

\begin{figure}
  \centering
  \includegraphics[width=4.5in]{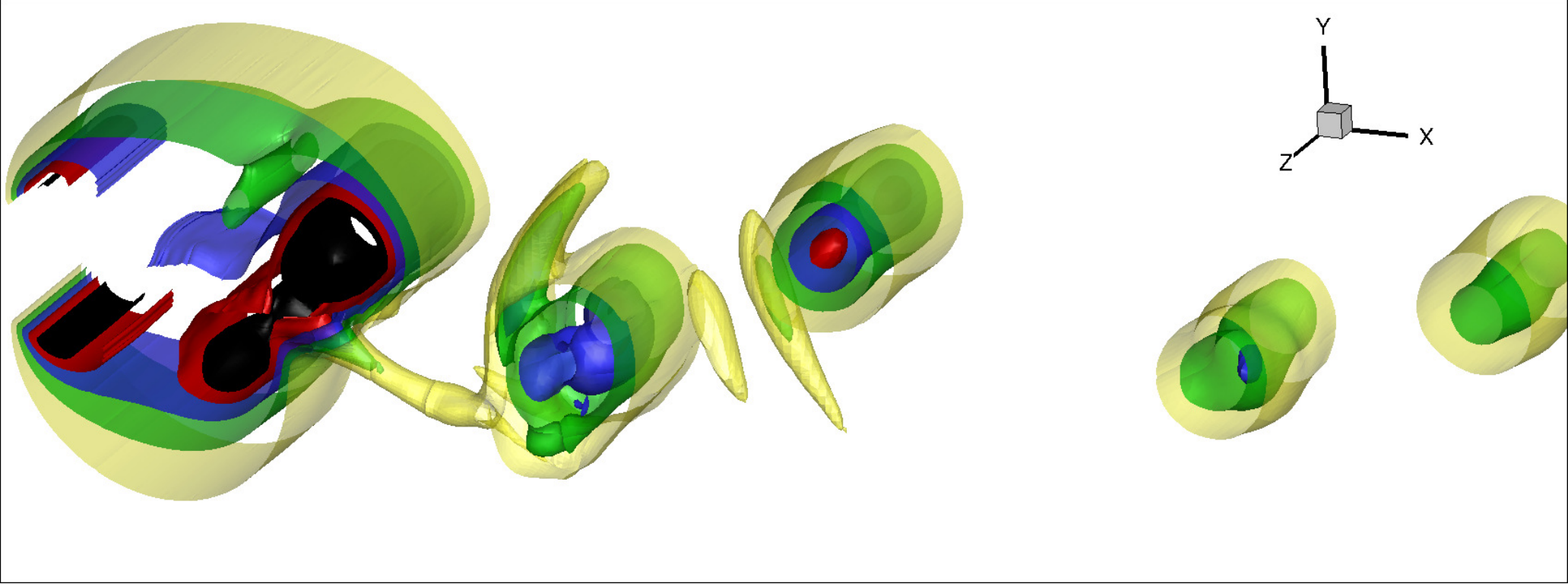}(a)
  \includegraphics[width=4.5in]{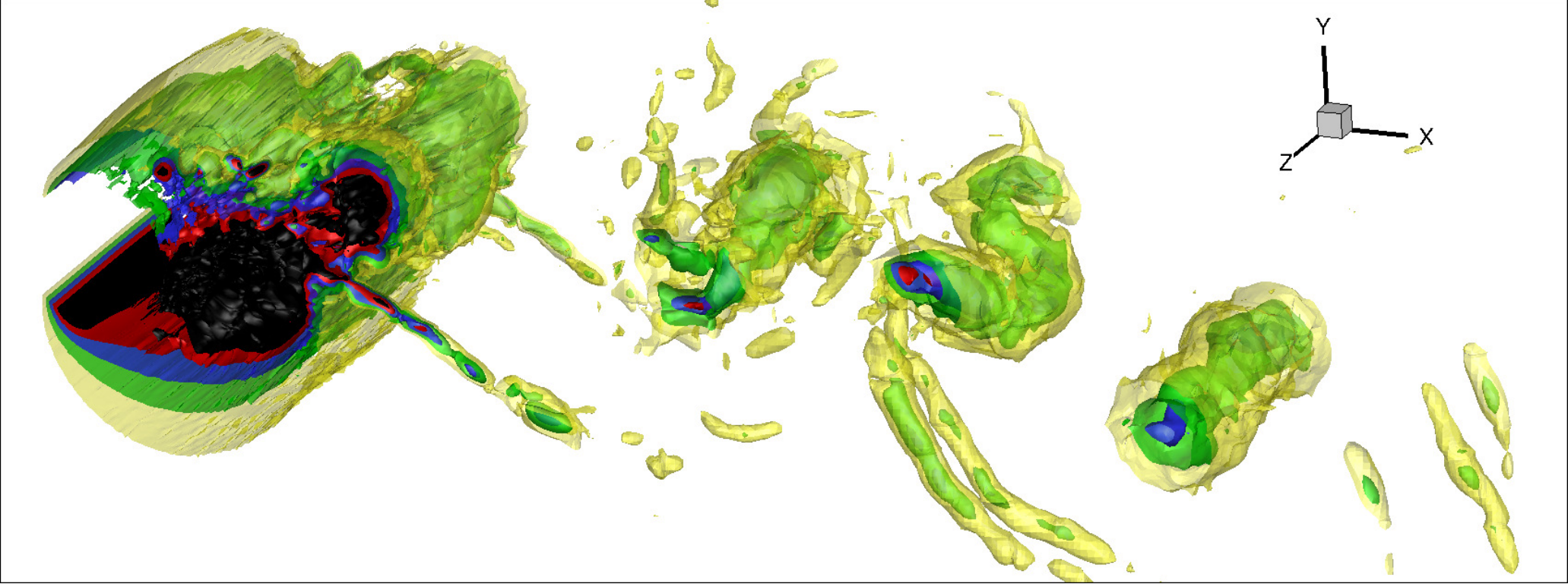}(b)
  \caption{ (color online)
    Visualization of vortices in 3D cylinder flow:
    Pressure isosurfaces (five uniform levels between $p=-0.6$ and $p=-0.2$)
    at Reynolds numbers (a) $Re=500$ and (b) $Re=5000$.
    Results are obtained with OBC-C as the outflow boundary condition.
  }
  \label{fig:p_contour_3d}
\end{figure}

Figure \ref{fig:p_contour_3d} shows a visualization of
the vortices in the cylinder wake by plotting the iso-surfaces
of the pressure fields at $Re=500$ (plot (a)) and
$Re=5000$ (plot (b)).
These results are obtained using OBC-C as the outflow boundary condition.
In addition to the spanwise vortices (``rollers'') in the wake, 
3D flow structures along the streamwise direction can be clearly observed.
With the larger Reynolds number, the flow
structures exhibit notably finer length scales, and the flow
field is much noisier.

\begin{figure}
  \centerline{
    \includegraphics[width=3in]{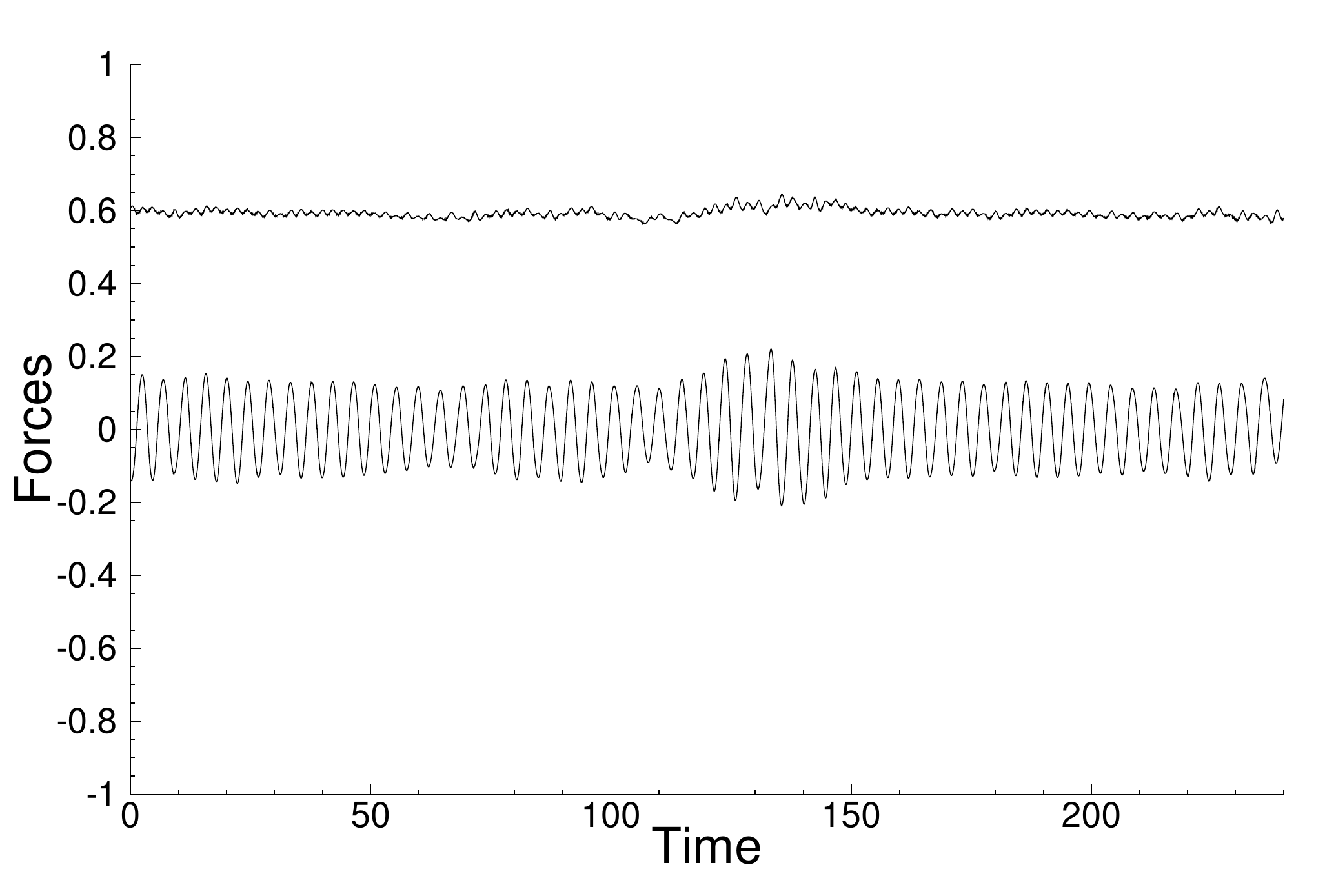}(a)
    \includegraphics[width=3in]{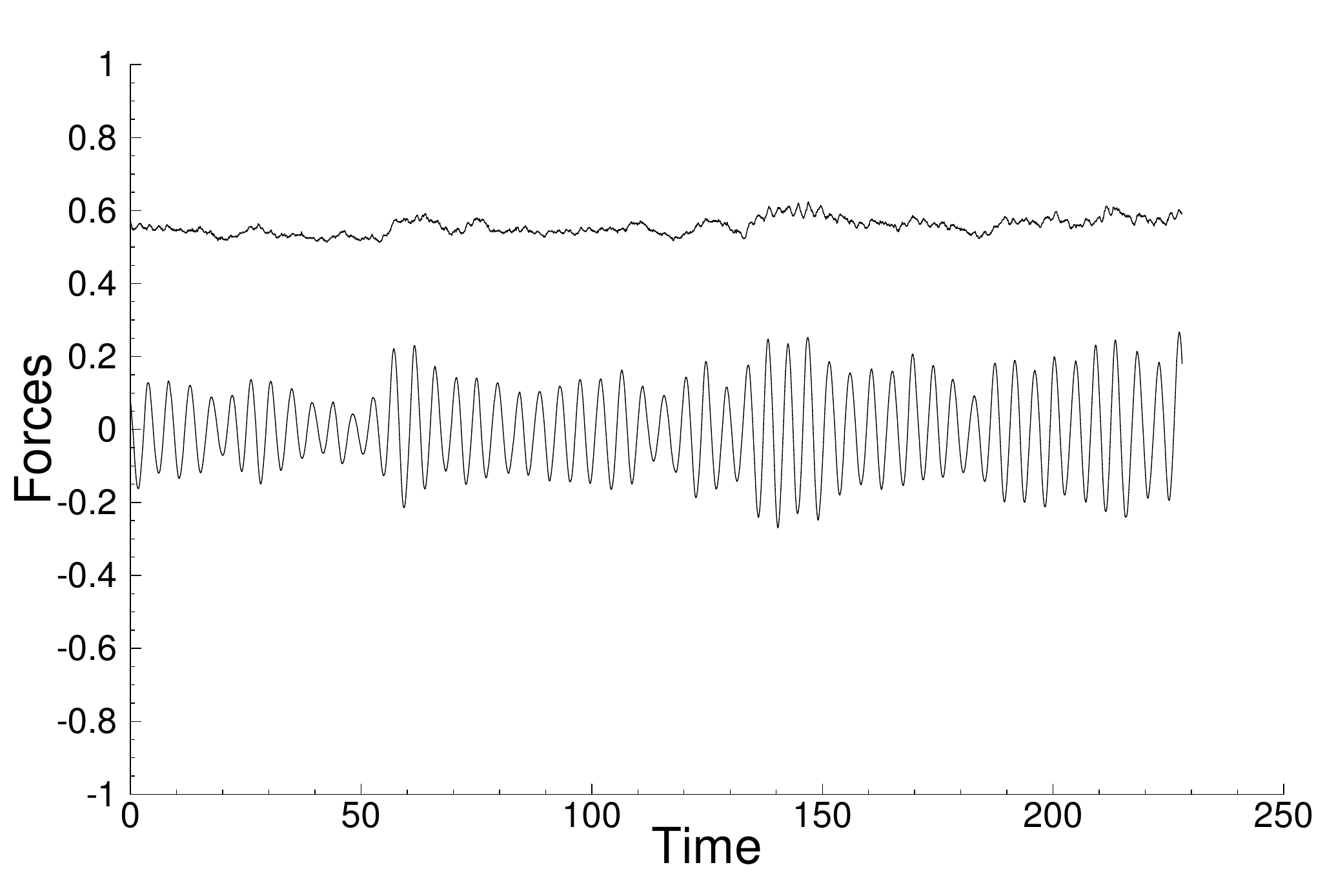}(b)
  }
  \caption{
    3D cylinder flow: time histories of drag and lift on
    the cylinder at Reynolds numbers (a) $Re=500$ and (b) $Re=5000$.
    Results correspond to OBC-B as the outflow boundary condition.
  }
  \label{fig:cyl3d_force_hist}
\end{figure}

Figure \ref{fig:cyl3d_force_hist} shows the time histories of
the drag and lift on the cylinder at the two Reynolds numbers
$Re=500$ (plot (a)) and $Re=5000$ (plot (b))
from the 3D simulations, which are obtained using OBC-B
as the outflow boundary condition.
The history signals show that the flow has reached a statistically stationary
state. They also demonstrate the long-term stability of
the methods developed herein.
The energy-stable boundary conditions are critical to
the stability of 3D simulations at moderate and high Reynolds numbers.
It is observed that  with the traction-free outflow boundary condition
the 3D simulation is unstable at the higher Reynolds number $Re=5000$.
One can also compare the lift history in Figure \ref{fig:cyl3d_force_hist}(b)
from 3D simulations
with that in Figure \ref{fig:cyl_fy_hist_high_re}(a)
from 2D simulations, both at Reynolds number $Re=5000$
and corresponding to OBC-B as the outflow boundary condition.
It can be observed that the 2D simulation leads to much larger lift
amplitudes (and correspondingly larger rms lift coefficient)
than the 3D simulation for the same Reynolds number,
which is well-known in the literature~\cite{DongK2005,DongKER2006}.

We have computed the drag coefficient and the rms lift coefficient
based on the force histories at $Re=500$ and $Re=5000$.
These data from 3D simulations are
included in Figure \ref{fig:compare_exp} for comparison with
the experimentally determined coefficient values.
It is observed that the current 3D simulation results are
in reasonably good agreement with the values from the experimental measurements.
In contrast, 2D simulations grossly over-predict both
the drag and the rms-lift coefficients
in the regime where the flow is physically three-dimensional.

\subsection{Jet Impinging on a Wall}
\label{sec:jet}

\begin{figure}
  \centerline{
    \includegraphics[width=3in]{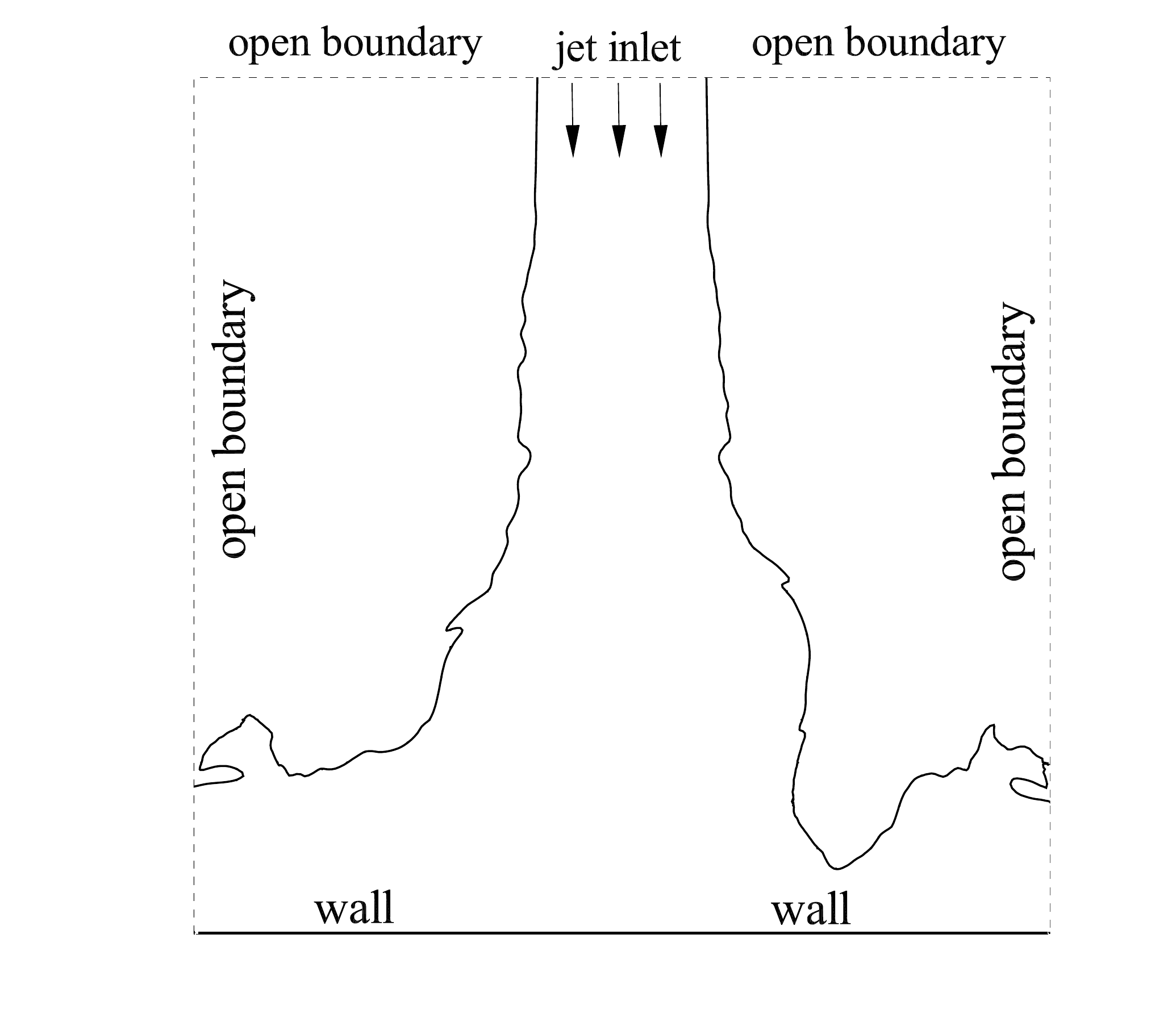}
  }
  \caption{
    Problem configuration of the impinging jet on a wall.
  }
  \label{fig:jet_wall_config}
\end{figure}

In this section we test further the current methods
with another flow problem,
a jet impinging on a solid wall, using two-dimensional simulations.
Due to the open boundaries and the physical instability of the jet,
the open boundary condition is
critical to the successful simulation of this flow.

Specifically,
we study a fluid jet of diameter $d$ impinging on a wall
in two dimensions.
Figure \ref{fig:jet_wall_config} illustrates the configuration
of this problem. Consider a rectangular domain,
$-\frac{5}{2}d\leqslant x\leqslant \frac{5}{2}d$ and
$0\leqslant y\leqslant 5d$,
where $x$ and $y$ axes are along the horizontal and
vertical directions, respectively.
The bottom side of the domain is a solid wall.
The inlet of the jet (with diameter $d$) is located
in the middle of the top side of the domain,
namely, $-R_0\leqslant x\leqslant R_0$ and $y=5d$,
where $R_0$ is the radius of the inlet ($R_0=\frac{d}{2}$).
The jet velocity is assumed to have the following profile
at the inlet,
\begin{equation}
  \left\{
  \begin{split}
    &
    u = 0 \\
    &
    v = -U_0\left[
      \tanh\frac{1-x/R_0}{\sqrt{2}\epsilon/d}\left(H(x,0)-H(x,R_0) \right)
      + \tanh\frac{1+x/R_0}{\sqrt{2}\epsilon/d}\left(H(x,-R_0)-H(x,0) \right)
      \right]
  \end{split}
  \right.
  \label{equ:jet_profile}
\end{equation}
where $U_0$ is the velocity scale ($U_0=1$),
$\epsilon=\frac{1}{40}d$, and
$H(x,a)$ is the unit step function, taking the unit value if
$x\geqslant a$ and vanishing otherwise.
The rest of the domain boundaries, on the top and on the left and
right sides, are all open, where the fluid can freely enter
or leave the domain.
The jet enters the domain through the inlet on the top,
impinges on the bottom wall and splits into two streams, which then
flow sideways out of the domain.
The goal is to simulate and study this process.

% discretization, BCs, simulation parameters

We discretize the domain using a spectral element
mesh of $400$ quadrilateral elements, with
$20$ uniform elements along the $x$ and $y$ directions.
No-slip condition (i.e.~Dirichlet condition with zero velocity)
is imposed on the bottom wall. At the jet inlet we
impose the Dirichlet condition \eqref{equ:dbc},
in which the boundary velocity $\mathbf{w}$ is
given by \eqref{equ:jet_profile}.
On the rest of the domain boundary the open
boundary condition \eqref{equ:obc} is imposed,
and the three boundary conditions (OBC-A, OBC-B and OBC-C) are employed and tested.
Long-time simulations have been performed, and the flow
has reached a statistically stationary state.
So the initial condition is immaterial and will have no effect
on the long-term behavior of the flow.
The problem and the physical variables are normalized
based on the jet diameter $d$ and the velocity scale $U_0$
in the simulations. So the Reynolds number is defined based on
these scales accordingly.
In accordance with the previous simulations of a variant of
this problem~\cite{DongS2015}, we employ an element order $12$ and a
time step size $\Delta t=2.5e-4$ for the current simulations.

% velocity fields

\begin{figure}
  \centerline{
    \includegraphics[width=2in]{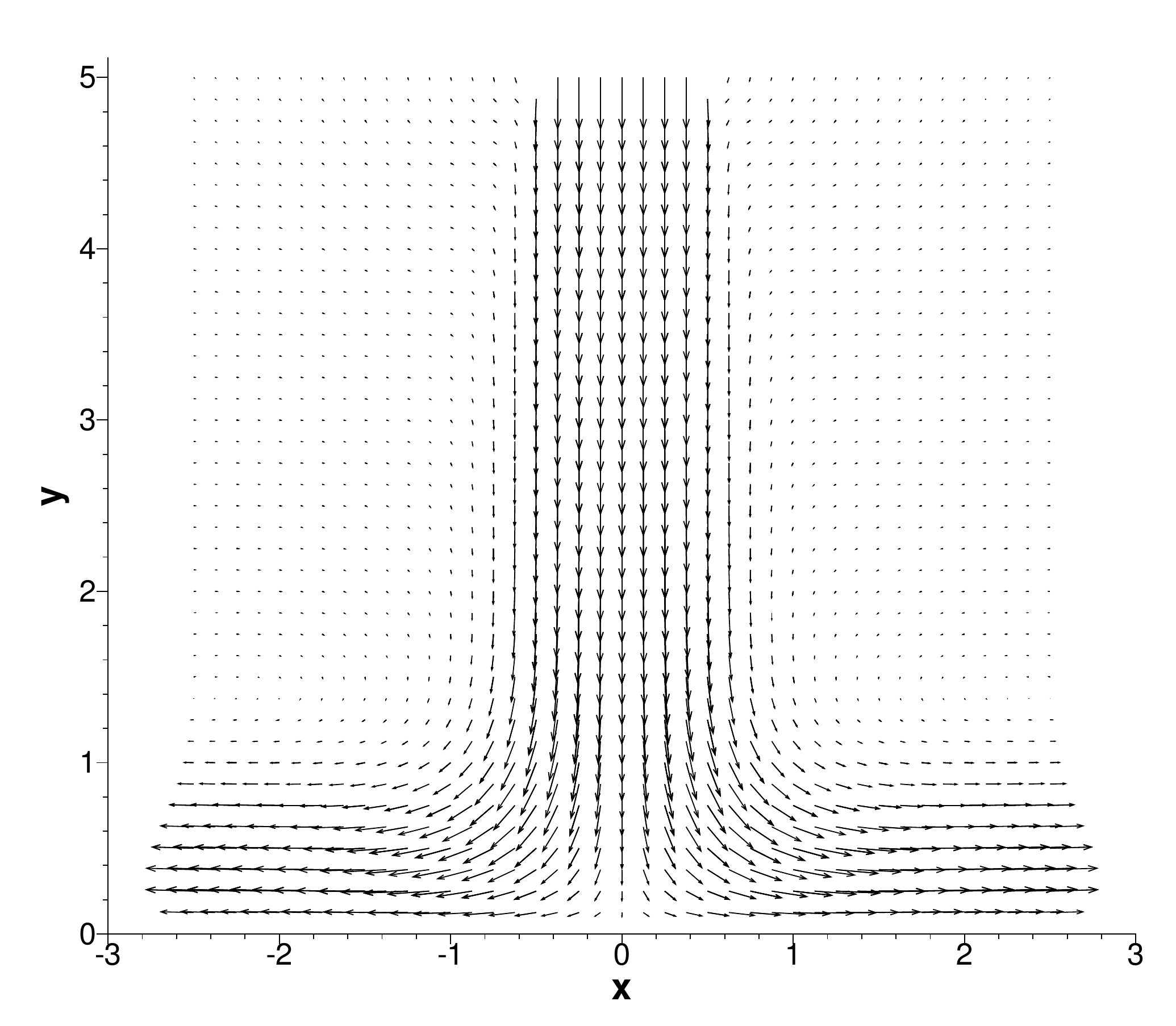}(a)
    \includegraphics[width=2in]{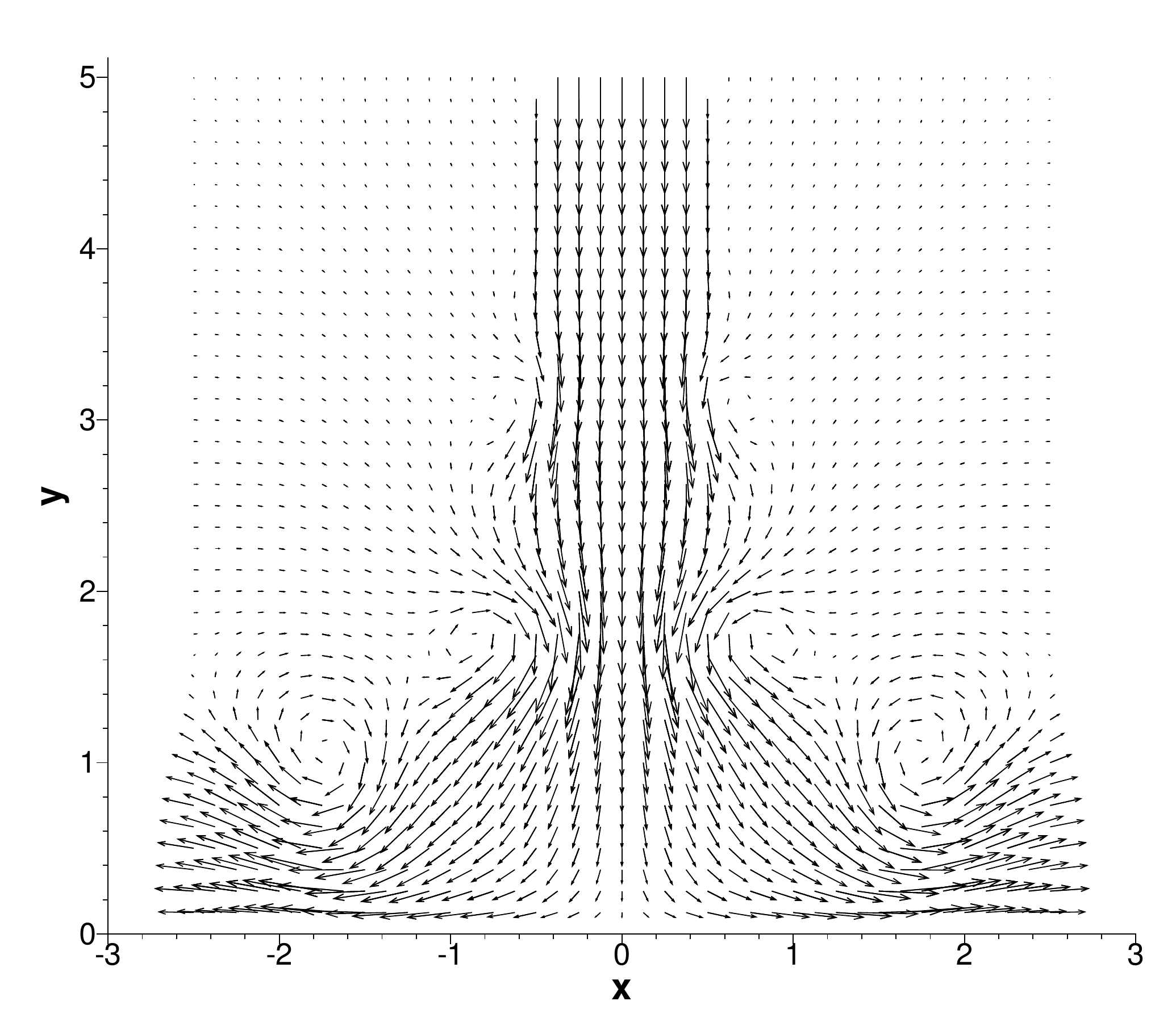}(b)
    \includegraphics[width=2in]{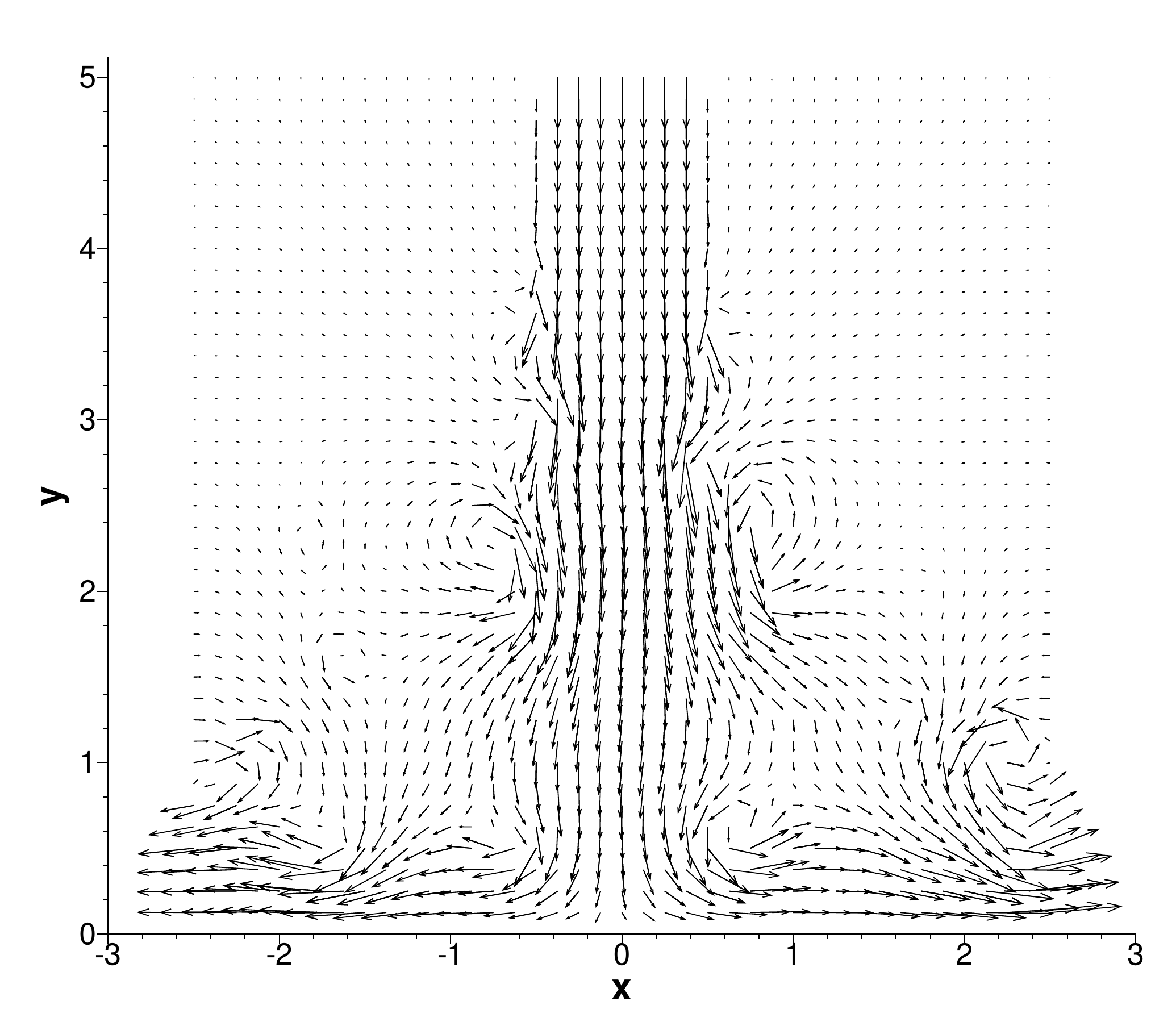}(c)
  }
  \caption{
    Characteristics of the impinging jet: instantaneous velocity
    distribution at Reynolds numbers
    (a) $Re=300$, (b) $Re=2000$, and (c) $Re=10000$.
    Velocity vectors are plotted on every eighth quadrature point
    in each direction within each element.
    Results are computed using OBC-C as the open boundary condition.
  }
  \label{fig:jet_overview}
\end{figure}

% overview of flow features

An overview of the characteristics of this flow is provided
by Figure \ref{fig:jet_overview}.
This figure shows the instantaneous velocity fields
at three Reynolds numbers: $Re=300$, $2000$ and $10000$,
which are computed using OBC-C
as the open boundary condition.
At a sufficiently low Reynolds number (e.g.~$Re=300$)
this flow is at a steady state. After impinging on the wall,
the vertical jet splits into two horizontal streams,
and flow in opposite directions parallel to the wall
until they exit the domain (Figure \ref{fig:jet_overview}(a)).
In regions of the domain outside the jet stream
the velocity appears to be negligibly small.
As the Reynolds number increases the flow becomes unsteady.
The vertical jet stream appears to be stable within some
distance downstream of the inlet, and then the
Kelvin-Helmholtz instability develops and the jet becomes
physically unstable.
Successive pairs of vortices form along the profile of the jet,
and they are convected downstream and eventually out of the domain
along with the jet (Figure \ref{fig:jet_overview}(b)).
For even higher Reynolds numbers,
the region downstream of the inlet with a stable jet profile shrinks,
and the onset of instability moves markedly upstream toward
the inlet. The vortices forming along the jet profile
appear more irregular and numerous, and their interactions
lead to more complicated dynamics (Figure \ref{fig:jet_overview}(c)).

% effects of R and alpha

% (1) plots of velocity field with R=0, -0.2, 0.2 and traction-free OBC
%     at Re=500, to show R effects
% (2) table of Fy, Fy' forces at Re=500 and Re=5000, computed with
%     different R values, optimal R, and traction-free OBC
% (3) table of Fy, Fy' forces at Re=5000, computed with optimal R
%     but with different alpha values, alpha=0.5, 0.25, 0, -0.5
%

\begin{figure}
  \centerline{
    \includegraphics[width=2.in]{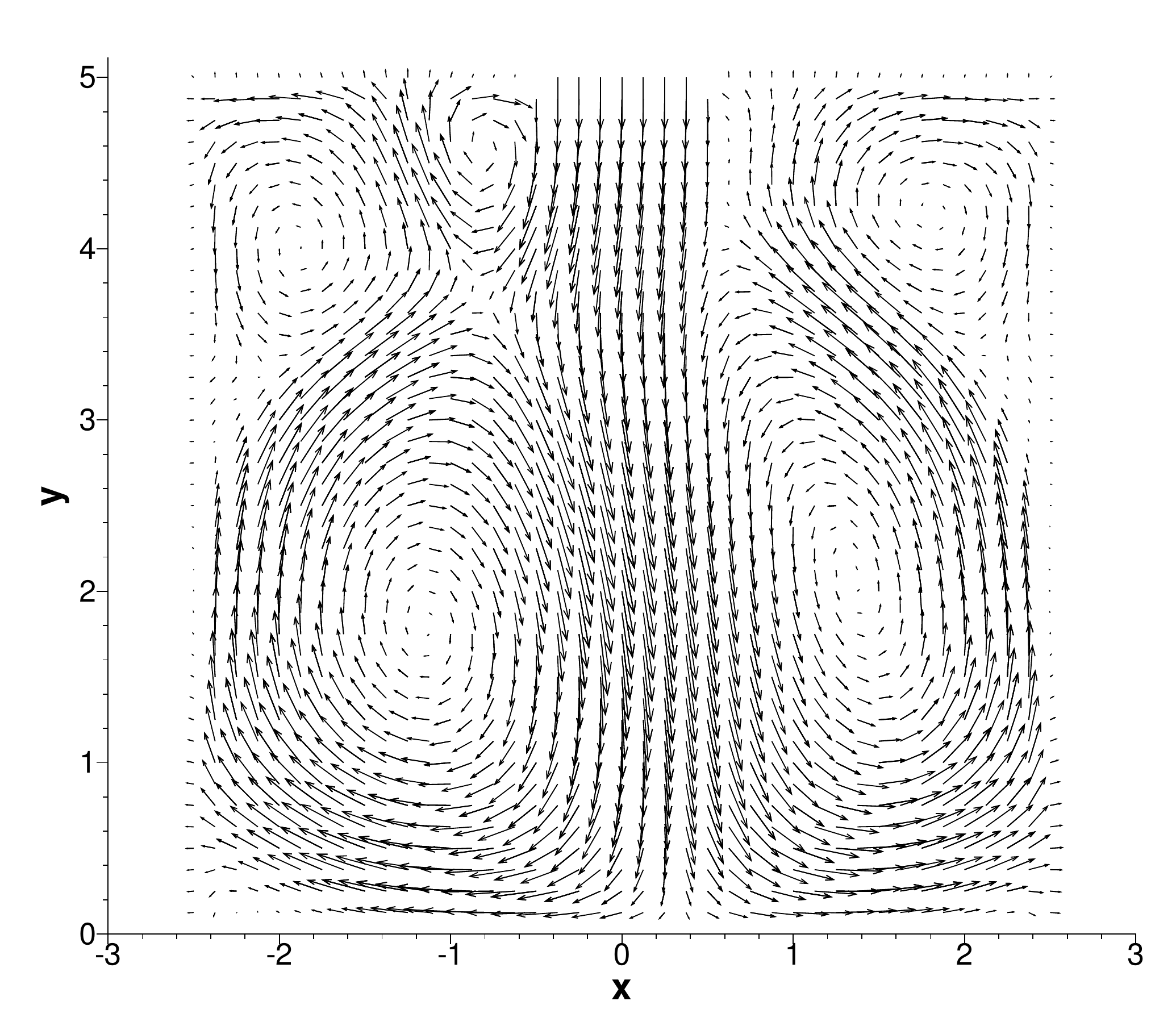}(a)
    \includegraphics[width=2.in]{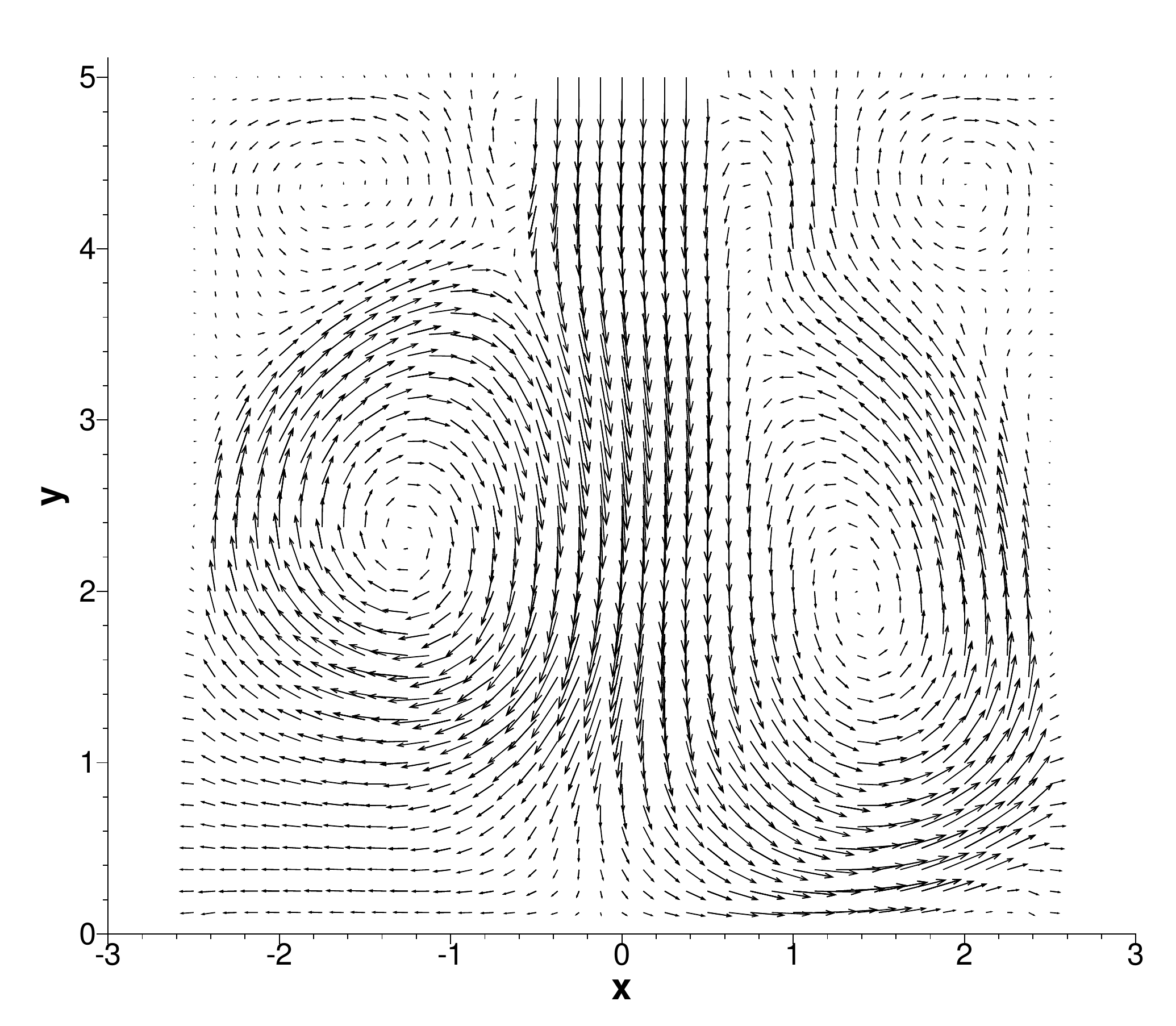}(b)
    \includegraphics[width=2.in]{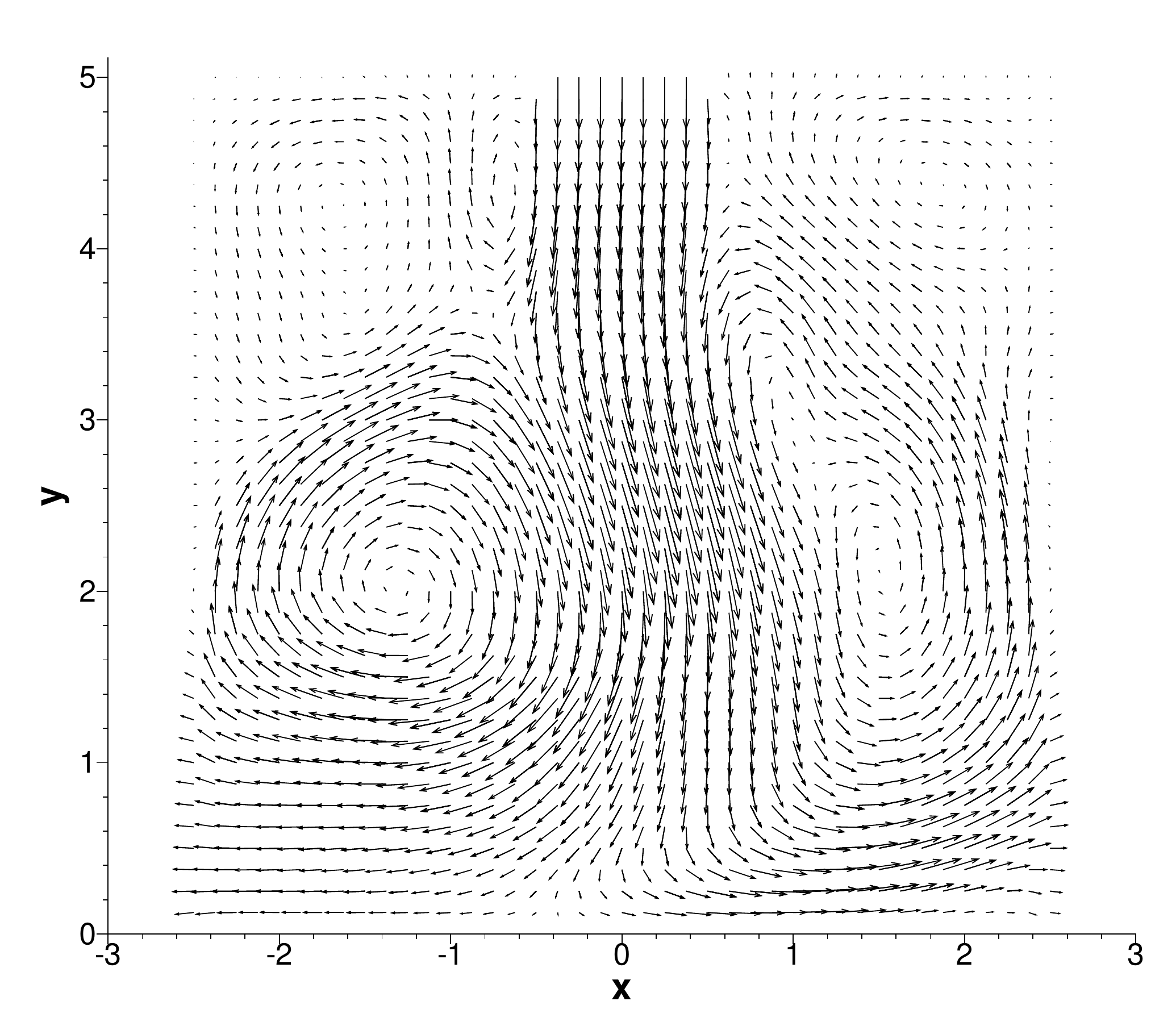}(c)
  }
  \centerline{
    \includegraphics[width=2.in]{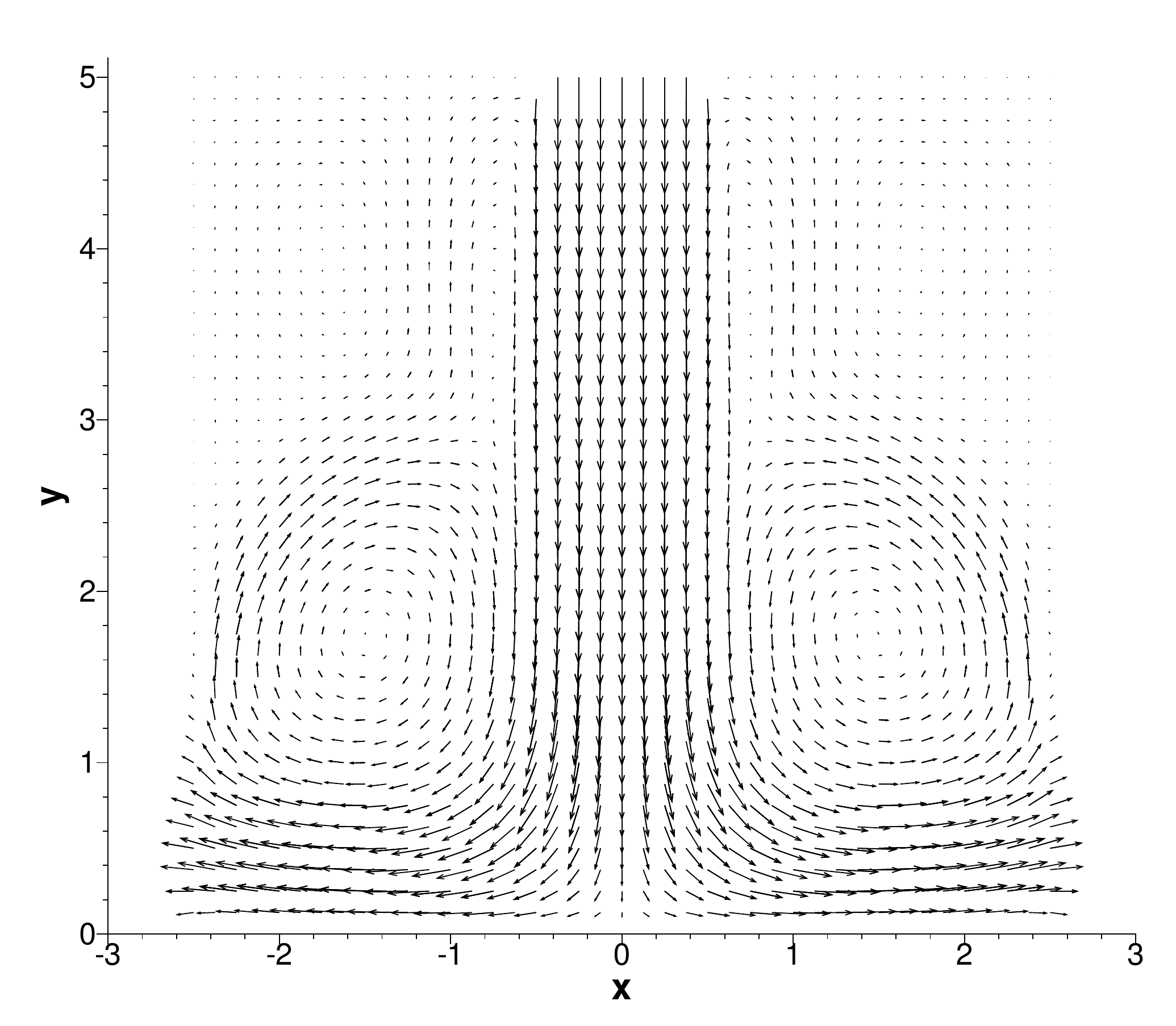}(d)
    \includegraphics[width=2.in]{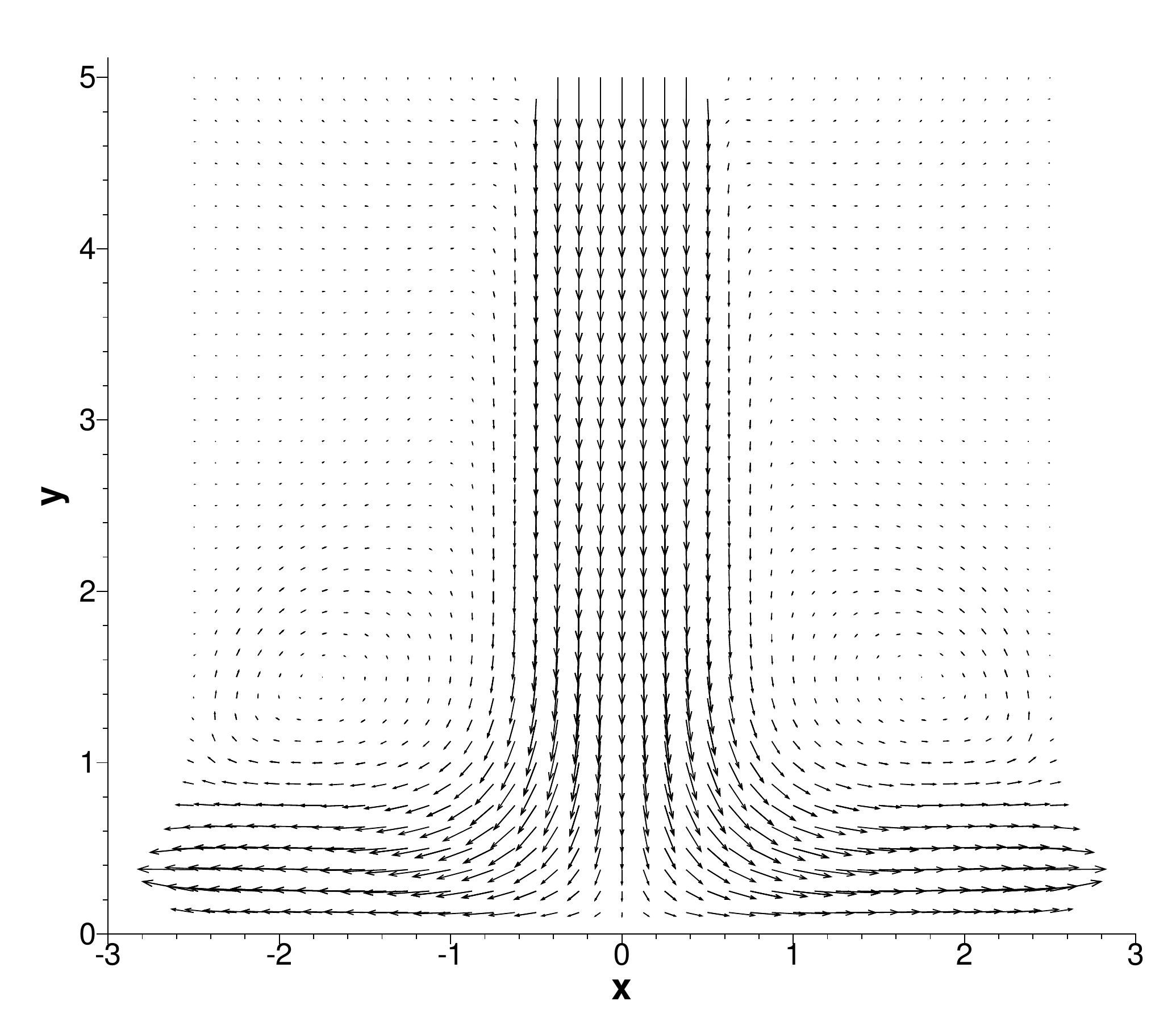}(e)
    \includegraphics[width=2.in]{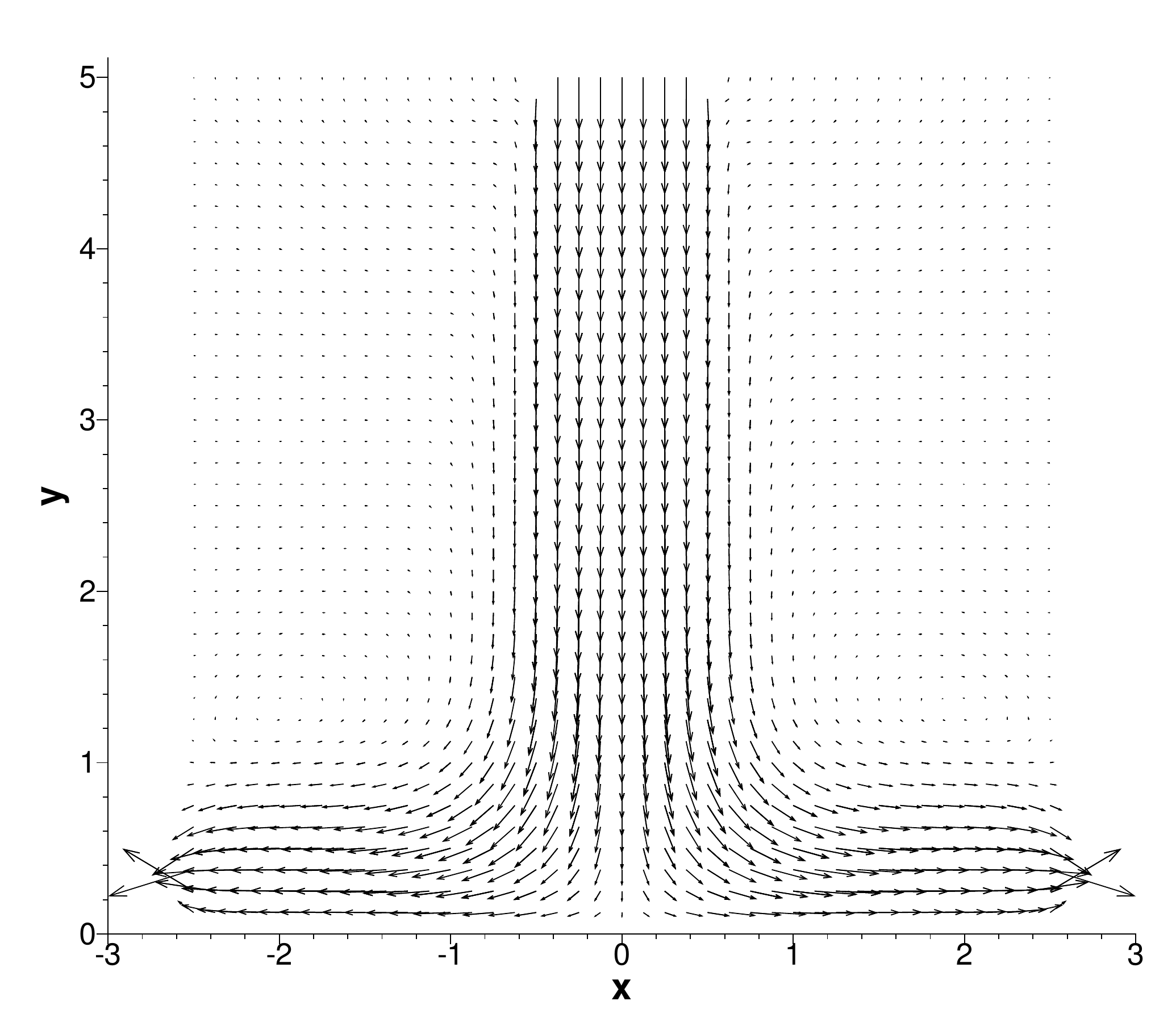}(f)
  }
  \centerline{
    \includegraphics[width=2.in]{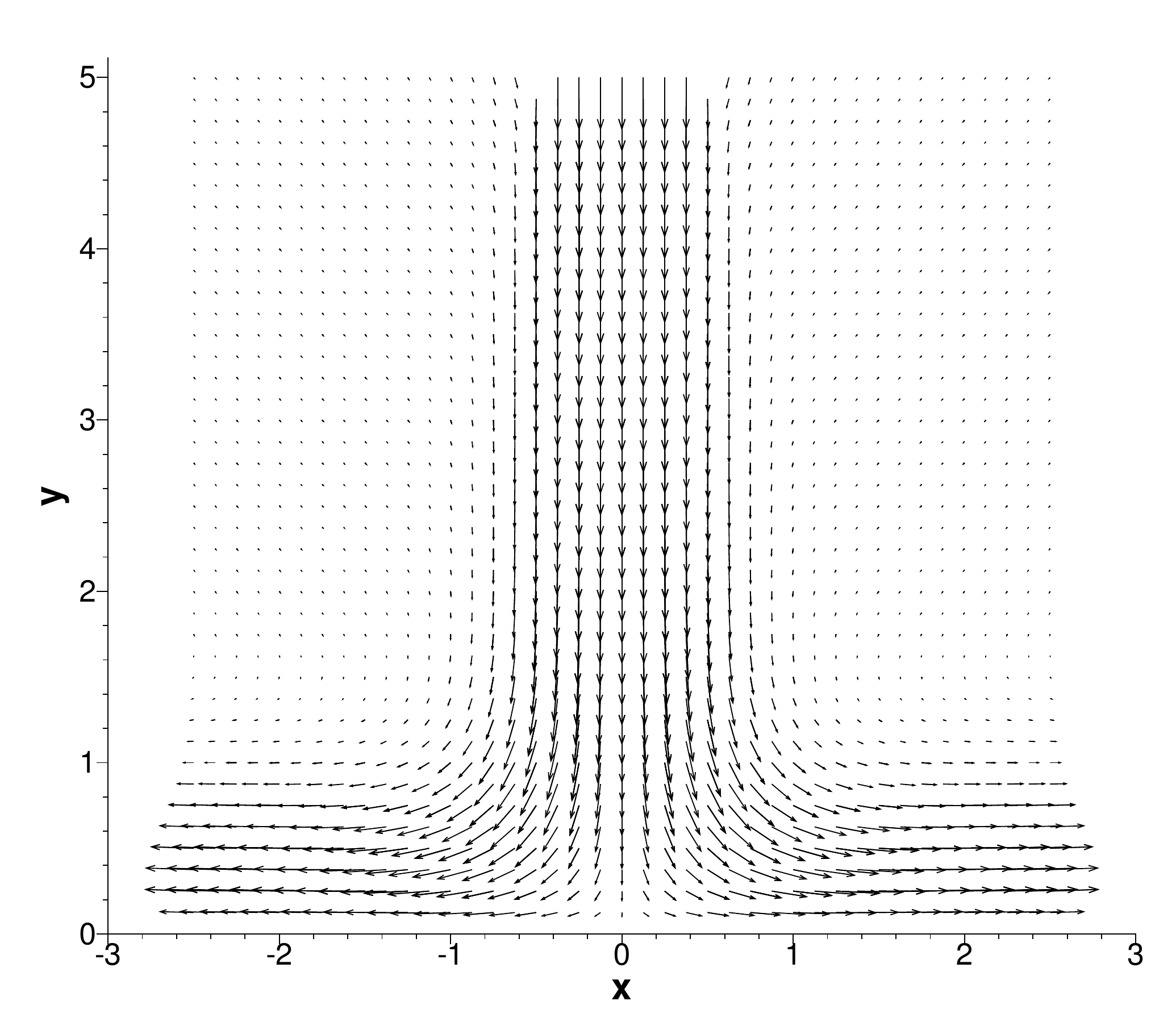}(g)
    \includegraphics[width=2.in]{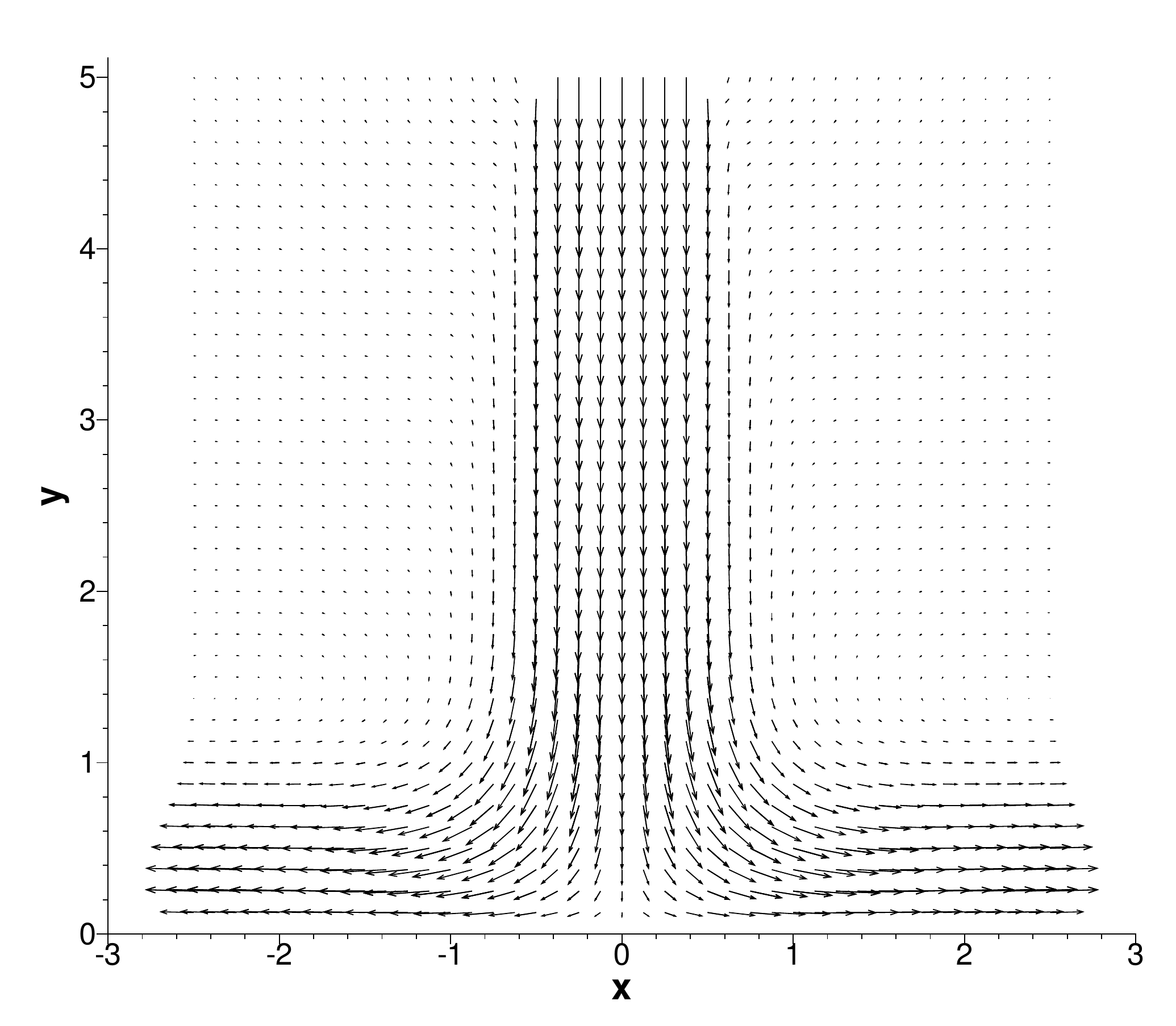}(h)
  }
  \caption{
    Impinging jet ($Re=300$): velocity field distributions
    computed using OBC-A with $\alpha=\frac{1}{2}$ ((a)-(f)),
    the traction-free condition (g), and OBC-B (h).
    Different parameter values for $a_{11}$ (and $a_{22}$, with $a_{22}=a_{11}$) 
    have been tested with OBC-A.
    They are 
    $a_{11}=a_{22}=0.5$ (a), $0.2$ (b), $0.0$ (c),
    $-0.2$ (d), $-0.5$ (e), and $-0.75$ (f).
    Velocity vectors are plotted on every eighth quadrature points in each
    direction within each element.
  }
  \label{fig:jet_re300_obcA}
\end{figure}

% Re=300, OBC-A

% table of forces at Re=300 computed using different methods

\begin{table}
  \centering
  \begin{tabular}{l | r | c l}
    \hline
    method &  parameters & $f_y$ (or mean-$f_y$) & rms-$f_y$ \\ \hline
    OBC-A & $a_{11}=a_{22}=-0.75$  & $-0.912$ & 0 \\
     & $-0.5$  & $-0.986$ & 0 \\
     & $-0.2$  & $-1.189$ & 0 \\
     & $0$  & $-1.384$ & $0.0408$ \\
     & $0.2$  & $-1.653$ & $0.102$ \\
     &  $0.5$  & $-2.343$ & $0.0922$ \\ \hline
    OBC-B &  & $-0.994$ & 0 \\ \hline
    OBC-C &  & $-0.994$ & 0 \\ \hline
%    Dong (2015)~\cite{Dong2015obc} & & $-0.996$ & 0 \\ \hline
    Traction-free OBC & & $-1.026$ & 0 \\ 
    \hline
  \end{tabular}
  \caption{
    Impinging jet ($Re=300$): vertical force on the wall computed
    using OBC-A (with $\alpha=1/2$ and various $a_{11}=a_{22}$ values),
    OBC-B, OBC-C, and the traction-free condition.
  }
  \label{tab:jet_fy_hist_re300}
\end{table}

Let us first focus on a low Reynolds number $Re=300$ and
study the effects of different open boundary conditions
on the simulation results. 
Figure \ref{fig:jet_re300_obcA} is a comparison of
the velocity field distributions at $Re=300$ computed
using OBC-A with $\alpha=\frac{1}{2}$ and a range of
values for $a_{11}$ (and $a_{22}$, with $a_{22}=a_{11}$).
The result obtained using the traction-free open boundary
condition \eqref{equ:obc_tractfree} and OBC-B 
are also included
for comparison. The results in this figure
can be compared with that of Figure \ref{fig:jet_overview}(a),
which is also for $Re=300$ but computed using OBC-C
as the open boundary condition.
We can make the following observations from these results:
\begin{itemize}

\item
  OBC-B and OBC-C result in velocity field distributions similar to
  the traction-free condition.

\item
  The $a_{11}$ (and $a_{22}$) values strongly
  influence the velocity fields computed with OBC-A.
  The velocity distributions obtained using OBC-A with different
  $a_{11}$ (and $a_{22}$) values are qualitatively different.

\item
  The velocity distributions obtained using OBC-A
  with $a_{11}=a_{22}=0.5$, $0.2$, $0$, and $-0.2$
  exhibit a pair (or more) of large vortices filling up 
  the domain, which is unphysical. With the larger
  $a_{11}$ (and $a_{22}$) values, the velocity fields even indicate
  that the flow and the vortices go out of the domain through the upper
  open boundary.

\item
  The flow fields obtained using OBC-A with
  $a_{11}=a_{22}=0.5$, $0.2$ and $0$ are not a steady flow for this Reynolds
  number. The forces on the wall obtained with these methods fluctuate
  over time, albeit in a narrow range.

\item
  The velocity distributions computed using OBC-A with
  $a_{11}=a_{22}=-0.5$ and $-0.75$ exhibit a similarity
  to that obtained with the traction-free condition in the overall
  characteristics. However, in the horizontal jet streams obtained with
  these methods,
  the directions of the velocity vectors
  seem un-natural at the open
  boundary (Figure \ref{fig:jet_re300_obcA}(e)-(f)).
  In addition, although it appears quite weak,
  a pair of large vortices can be discerned from
  the velocity field obtained using OBC-A with $a_{11}=a_{22}=-0.5$
  (Figure \ref{fig:jet_re300_obcA}(e)).

\item
  Using the velocity field resulting from the traction-free condition
  as a reference, the best result for OBC-A seems to correspond to
  a parameter value around $a_{11}=a_{22}=-0.5$ for this problem.
  
\end{itemize}

%The OBC-B and OBC-C with the algorithmic parameter $\alpha$ ranging from
%$\alpha=0.5$ to $\alpha=-0.5$ have been employed
%to simulate this flow at $Re=300$.
%The velocity field distributions obtained with
%these $\alpha$ values are essentially the same, as
%given by Figures \ref{fig:jet_overview}(a) and \ref{fig:jet_re300_obcA}(h) 
%corresponding to$\alpha=0.5$.

% forces

Table \ref{tab:jet_fy_hist_re300} lists the forces ($y$-component)
on the wall obtained using different methods at $Re=300$.
Since the flow computed using OBC-A with $a_{11}=a_{22}=0.5$, $0.2$ and
$0.0$ is unsteady, listed in the table are the mean and
rms forces corresponding to these methods.
We observe that with increasing $a_{11}$ (and $a_{22}$), the force computed using OBC-A
increases substantially in magnitude. The discrepancy in the forces
between OBC-A and the traction-free condition is significant.
Compared with the traction-free condition, the best result
obtained using OBC-A appears to correspond to a value around
$a_{11}=a_{22}=-0.5$.
On the other hand,
the forces obtained using OBC-B and OBC-C are the same,
and they are very close to the that obtained using
the traction-free condition.

% temporal sequence of velocity snapshots

\begin{figure}
  \centerline{
    \includegraphics[width=2in]{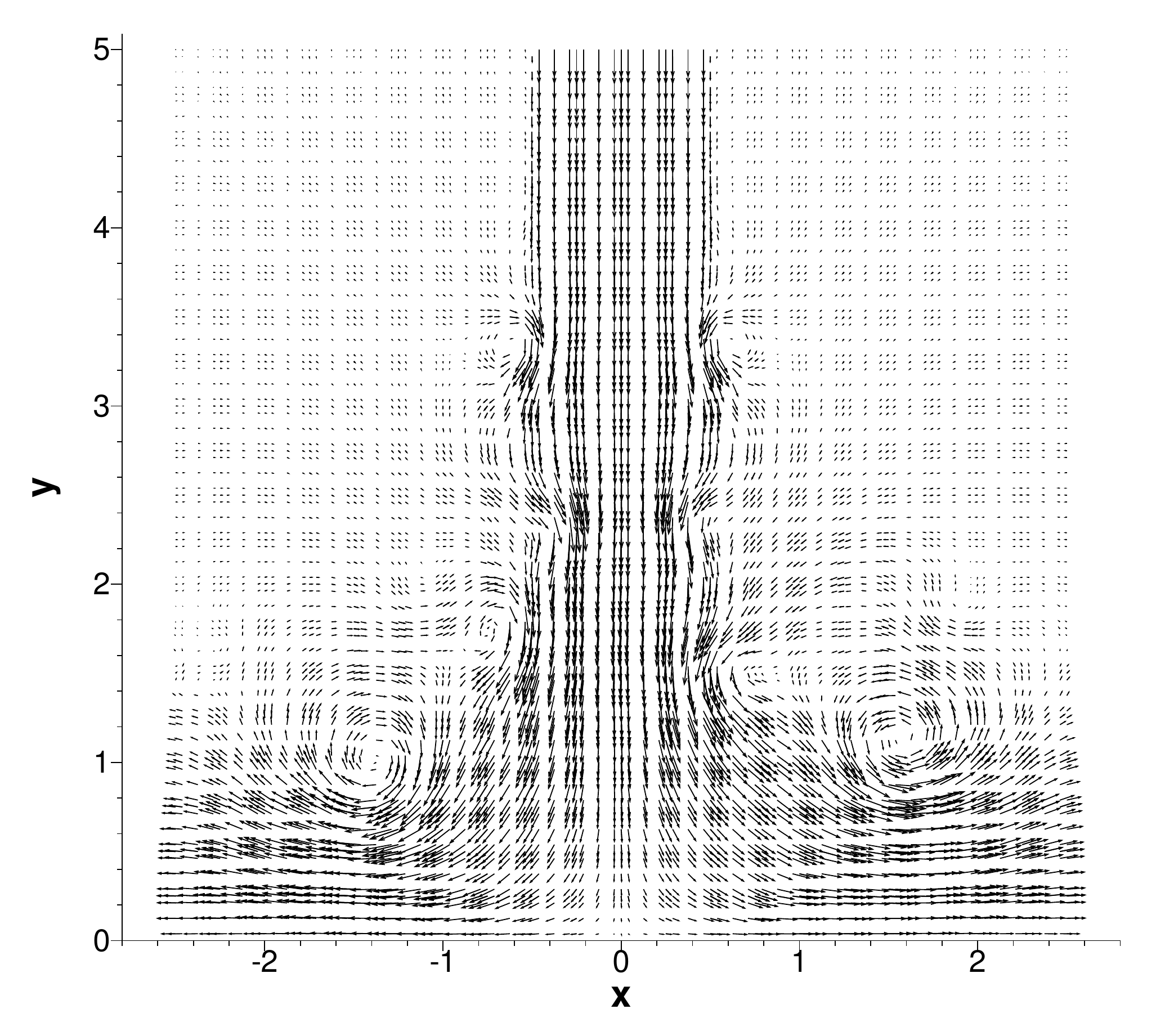}(a)
    \includegraphics[width=2in]{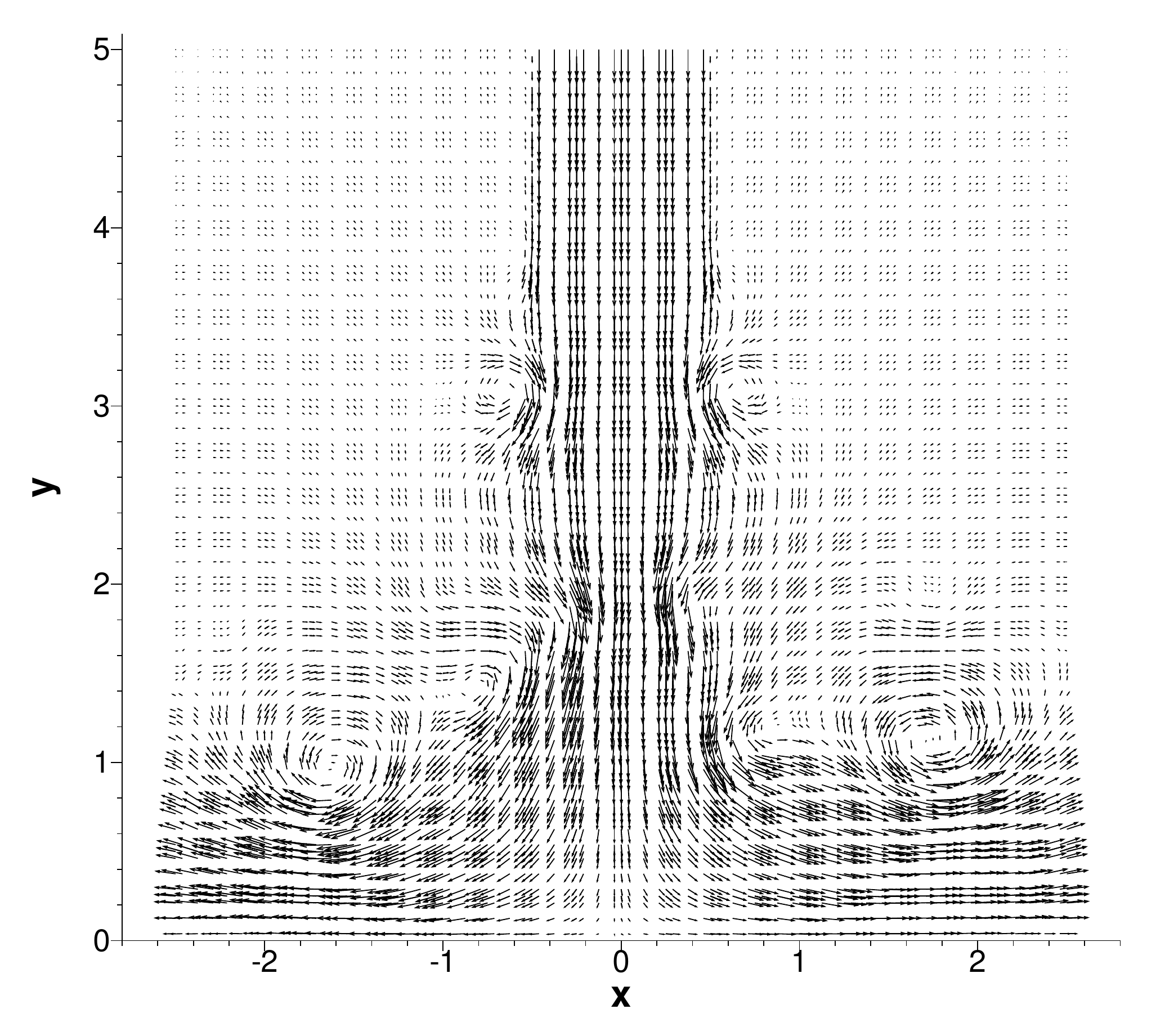}(b)
    \includegraphics[width=2in]{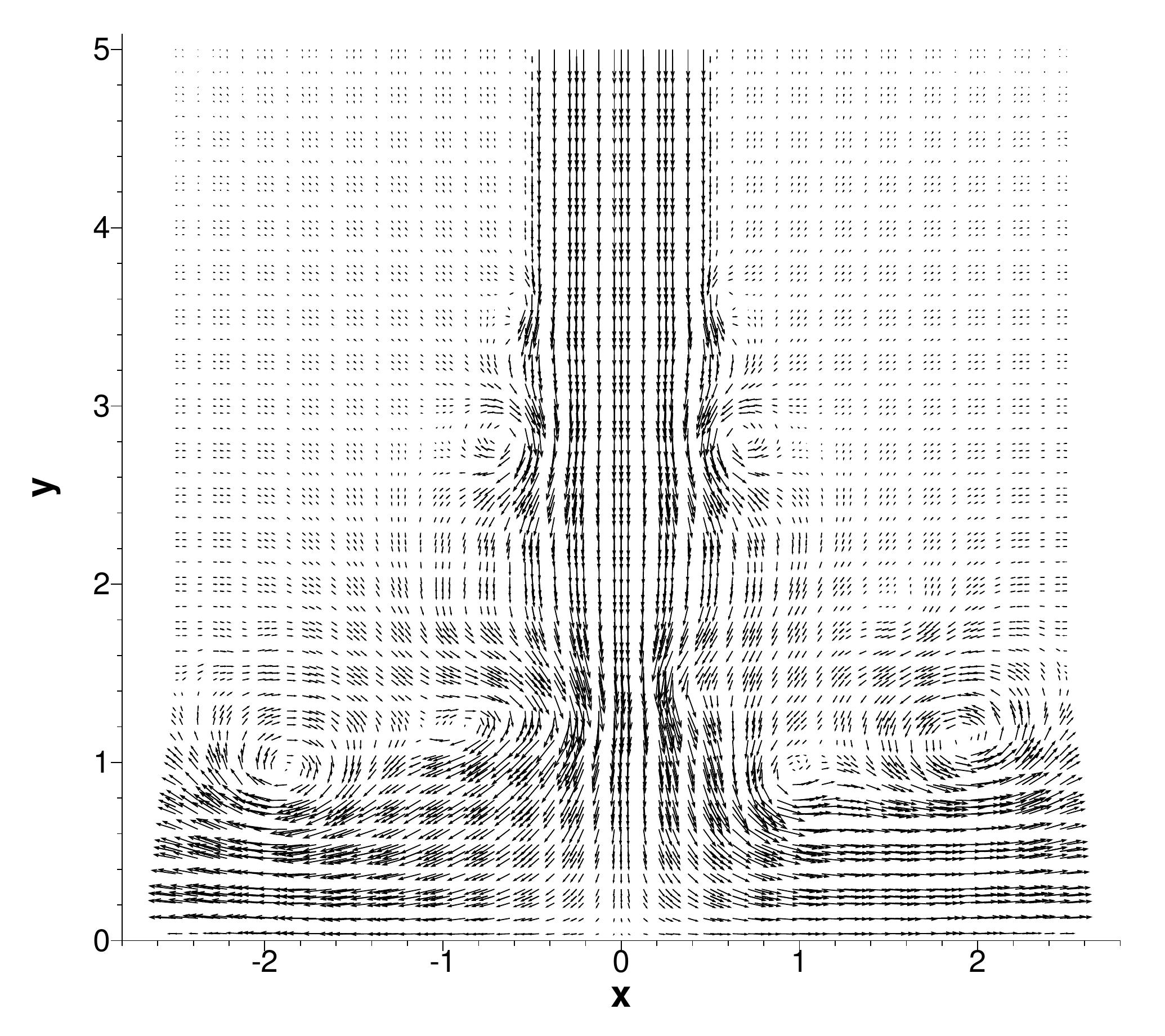}(c)
  }
  \centerline{
    \includegraphics[width=2in]{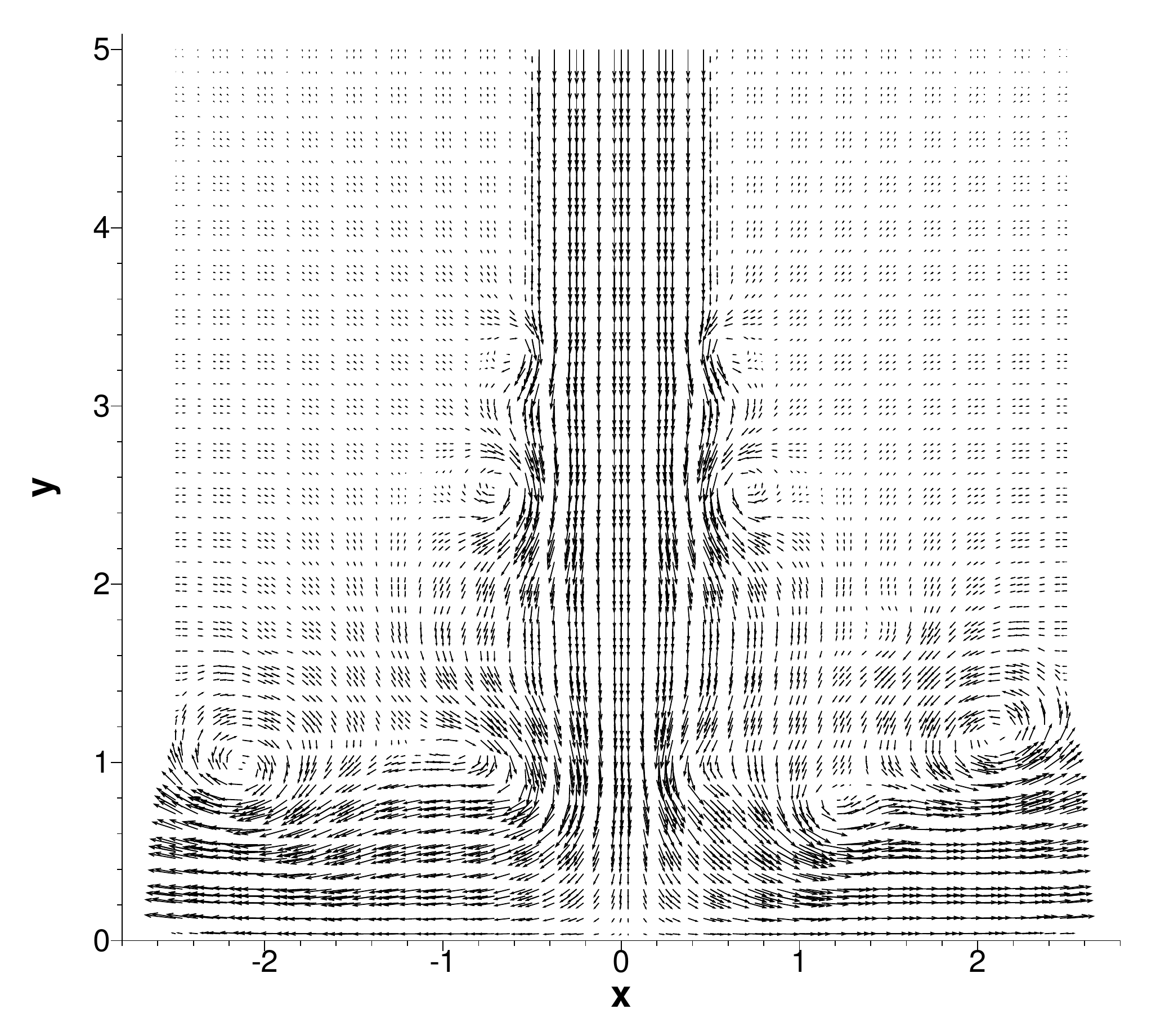}(d)
    \includegraphics[width=2in]{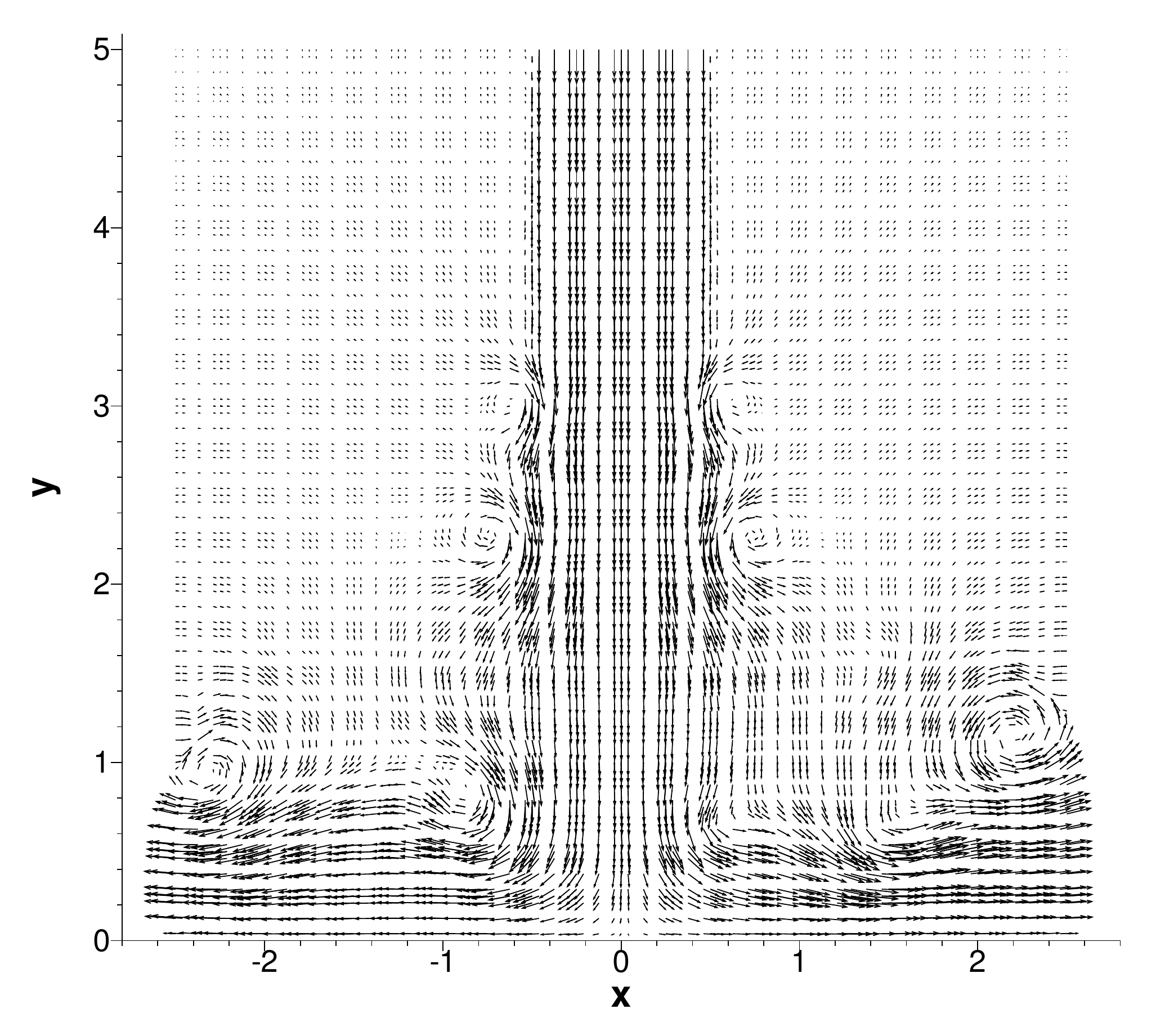}(e)
    \includegraphics[width=2in]{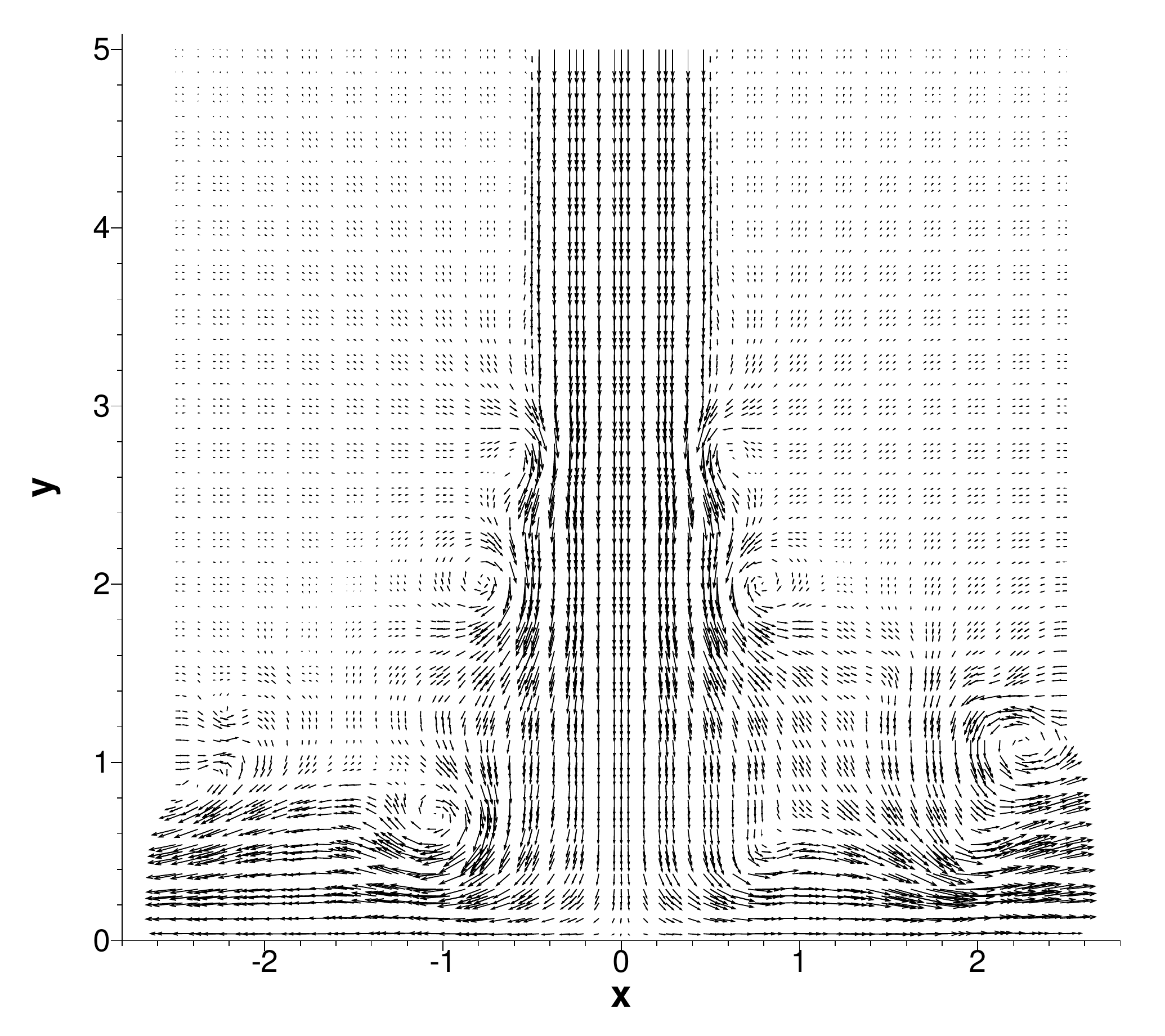}(f)
  }
  \centerline{
    \includegraphics[width=2in]{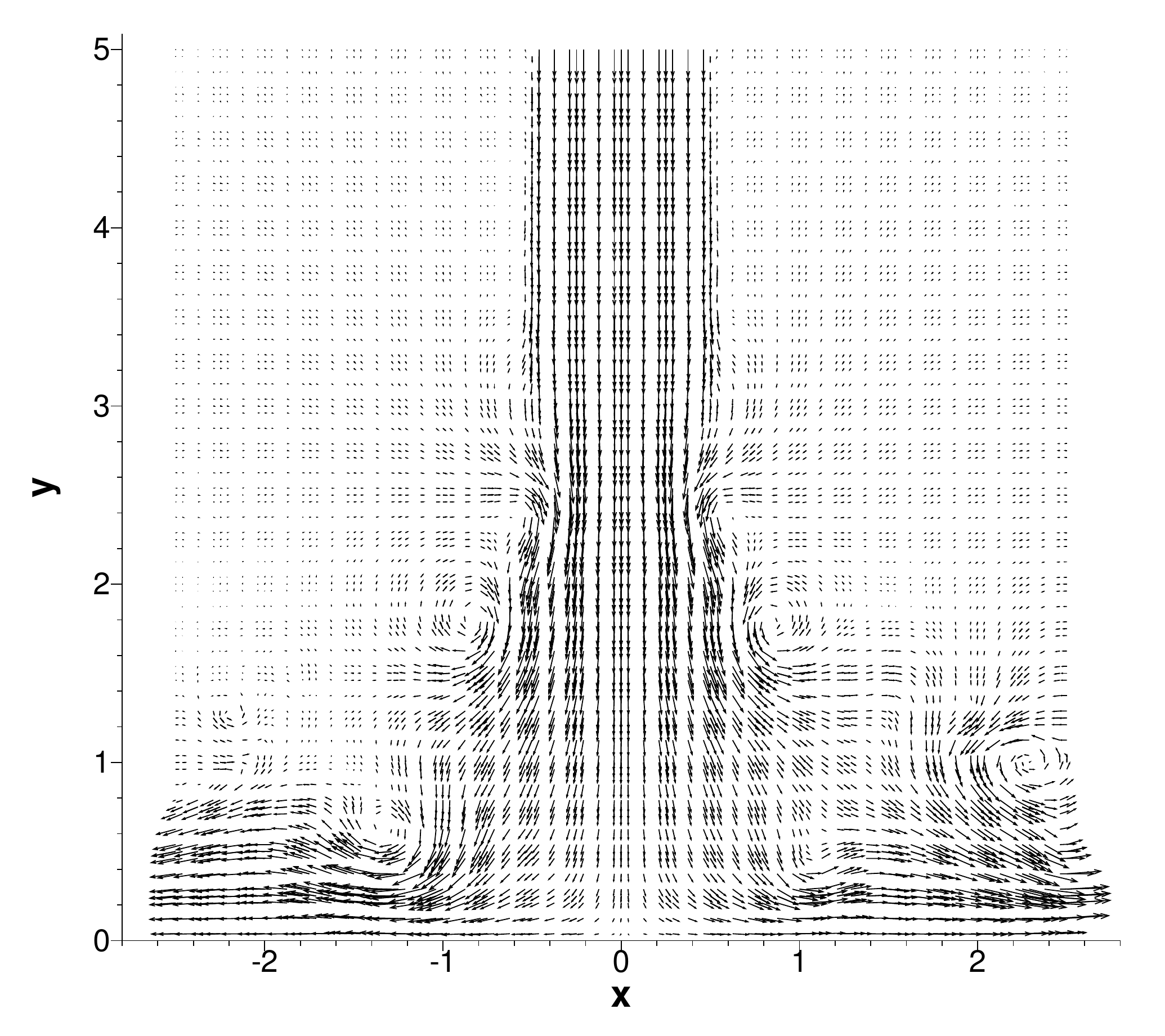}(g)
    \includegraphics[width=2in]{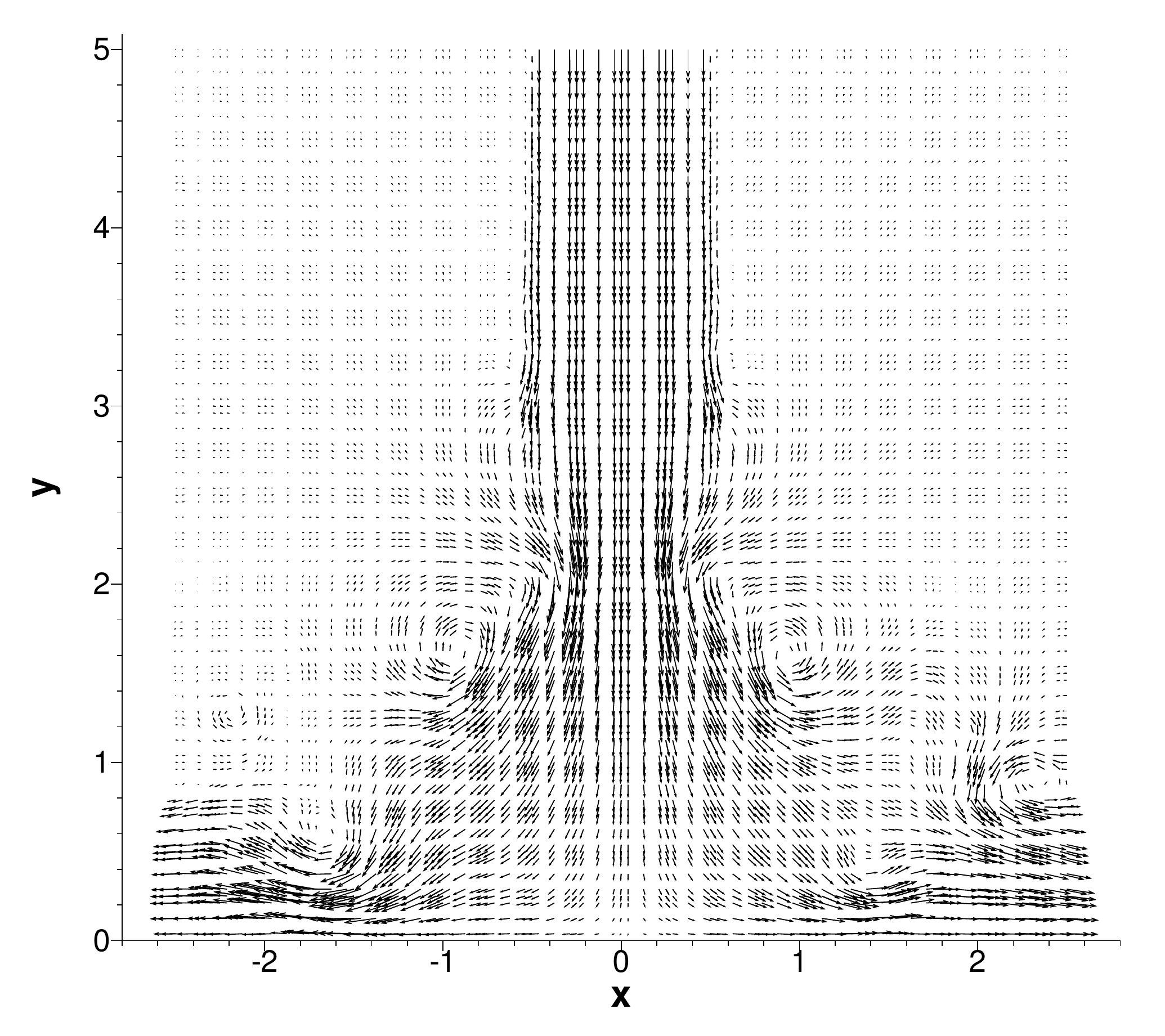}(h)
    \includegraphics[width=2in]{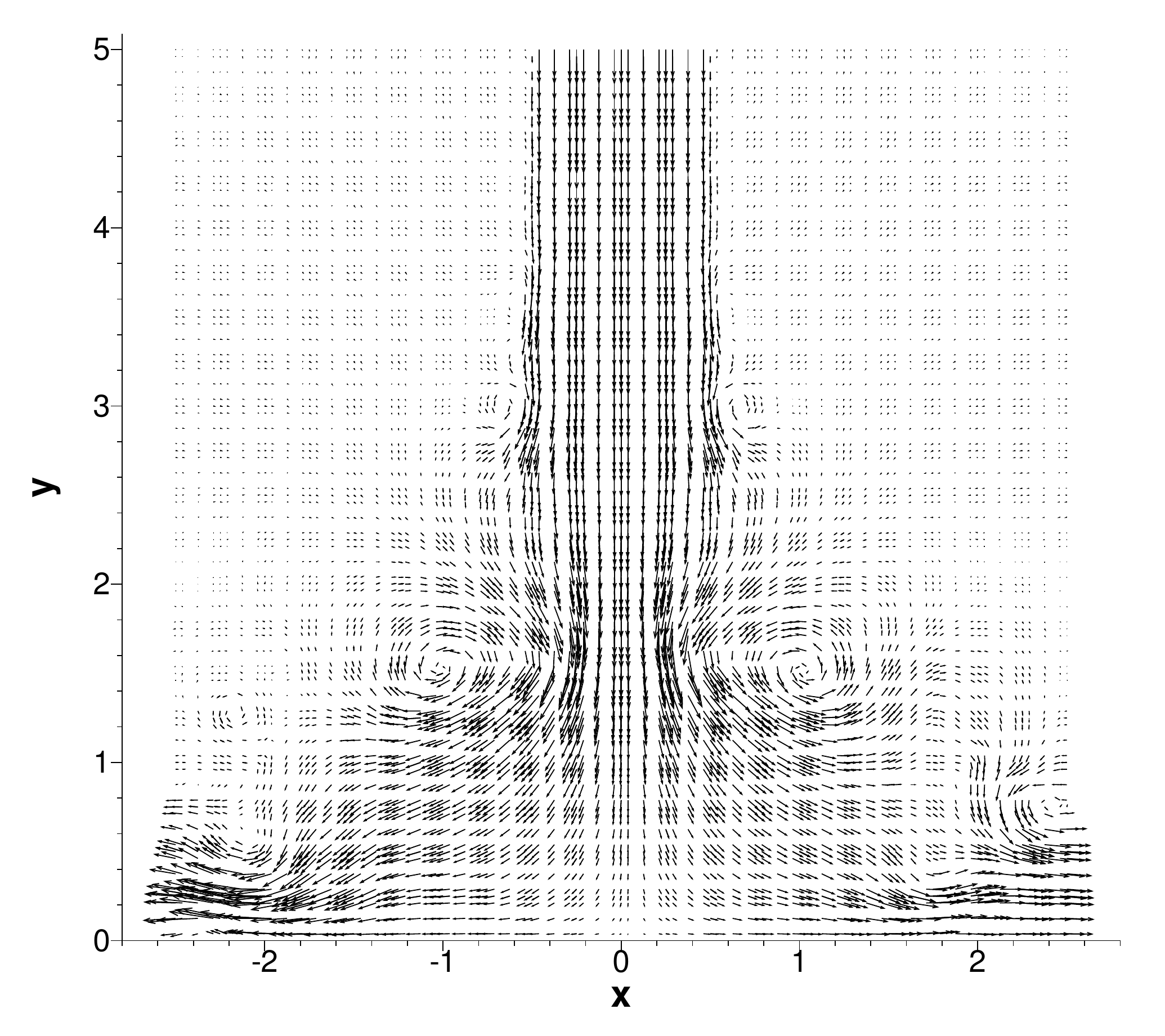}(i)
  }
  \caption{
    Impinging jet ($Re=5000$): temporal sequence of snapshots
    of the instantaneous velocity fields at
    (a) $t=619.05$,
    (b) $t=619.65$,
    (c) $t=620.25$,
    (d) $t=620.85$,
    (e) $t=621.45$,
    (f) $t=622.05$,
    (g) $t=622.65$,
    (h) $t=623.25$,
    (i) $t=623.85$.
    Velocity vectors are plotted on every fourth quadrature point in each
    direction within each element. Results are computed using
    OBC-B as the open boundary condition.
  }
  \label{fig:vel_re5k}
\end{figure}

% higher Re

Let us next look into the impinging jet at higher Reynolds numbers. 
Figure \ref{fig:vel_re5k} shows a temporal sequence of snapshots of
the velocity fields at $Re=5000$ computed using OBC-B
as the open boundary condition.
These results illustrate the vortex-pair formation and the transport
of the train of vortices downstream along the jet profile in
the dynamics of the flow.
They also signify that the method herein can allow the vortices
to pass through the open boundary in a smooth and fairly natural
fashion; see the left boundary in Figures \ref{fig:vel_re5k}(c)-(f).
On the other hand, a certain degree of distortion to the vortices
as they cross the open boundary can also be observed (Figure \ref{fig:vel_re5k}(f)).
The physical instability of the jet and the presence of vortices
on the open boundaries make these simulations very challenging.
The current open boundary conditions 
are very effective for such problems.
It is noted that the traction-free condition
is unstable in these simulations.

% distortion of vortices in more detail
% compare with convective OBC from Dong (2015)

\begin{figure}
\centerline{
  \includegraphics[width=1.8in]{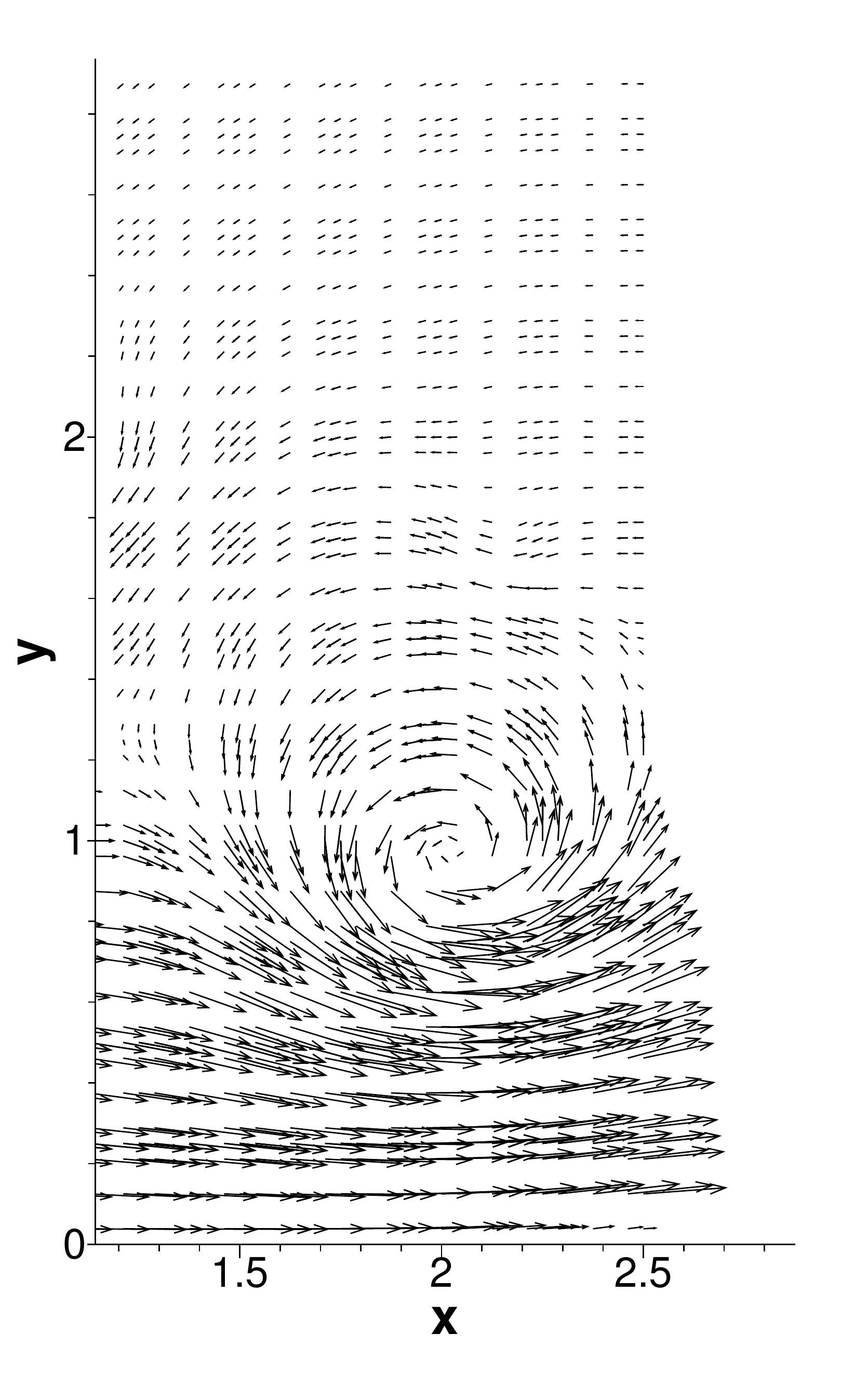}(a)
  \includegraphics[width=1.8in]{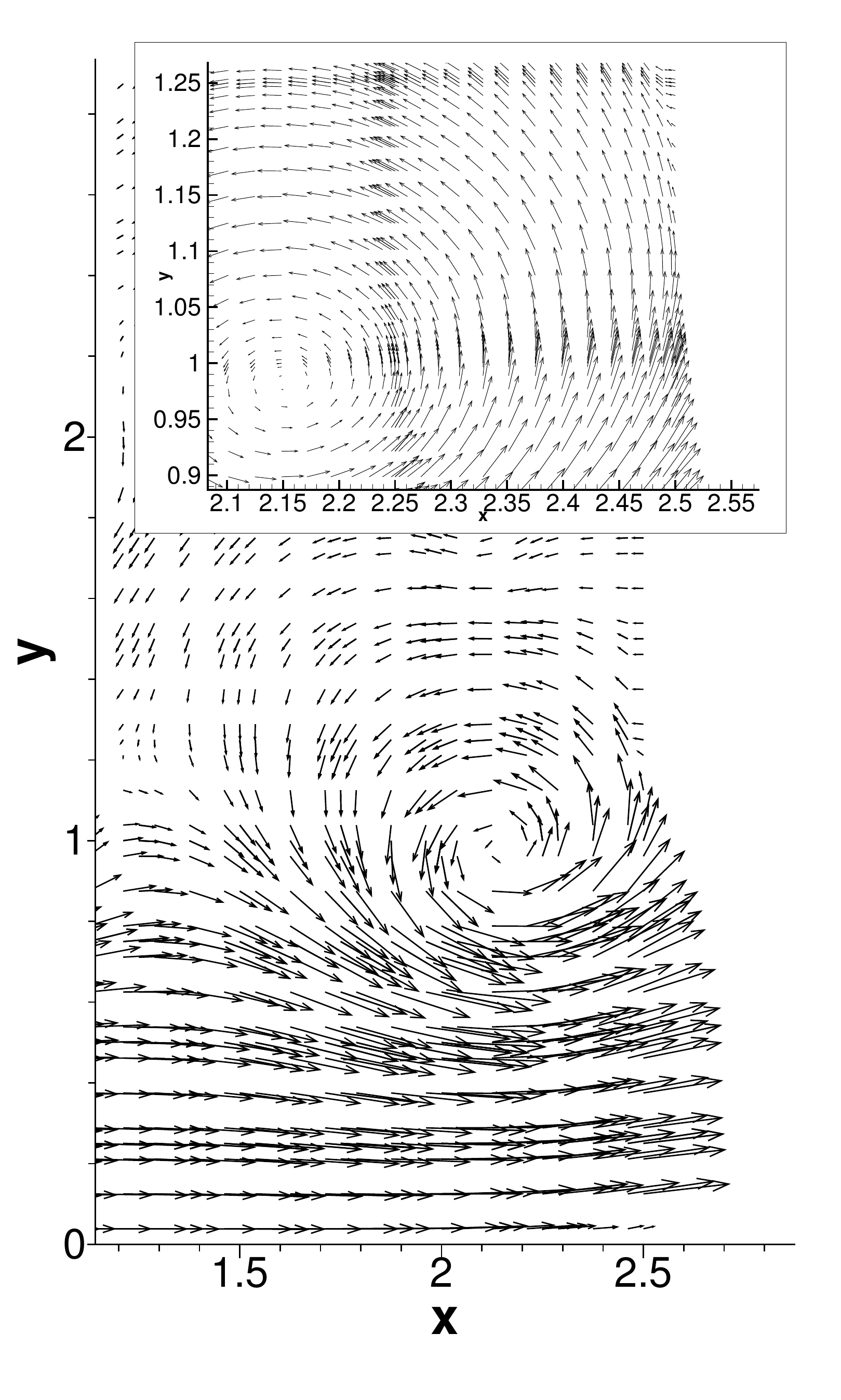}(b)
  \includegraphics[width=1.8in]{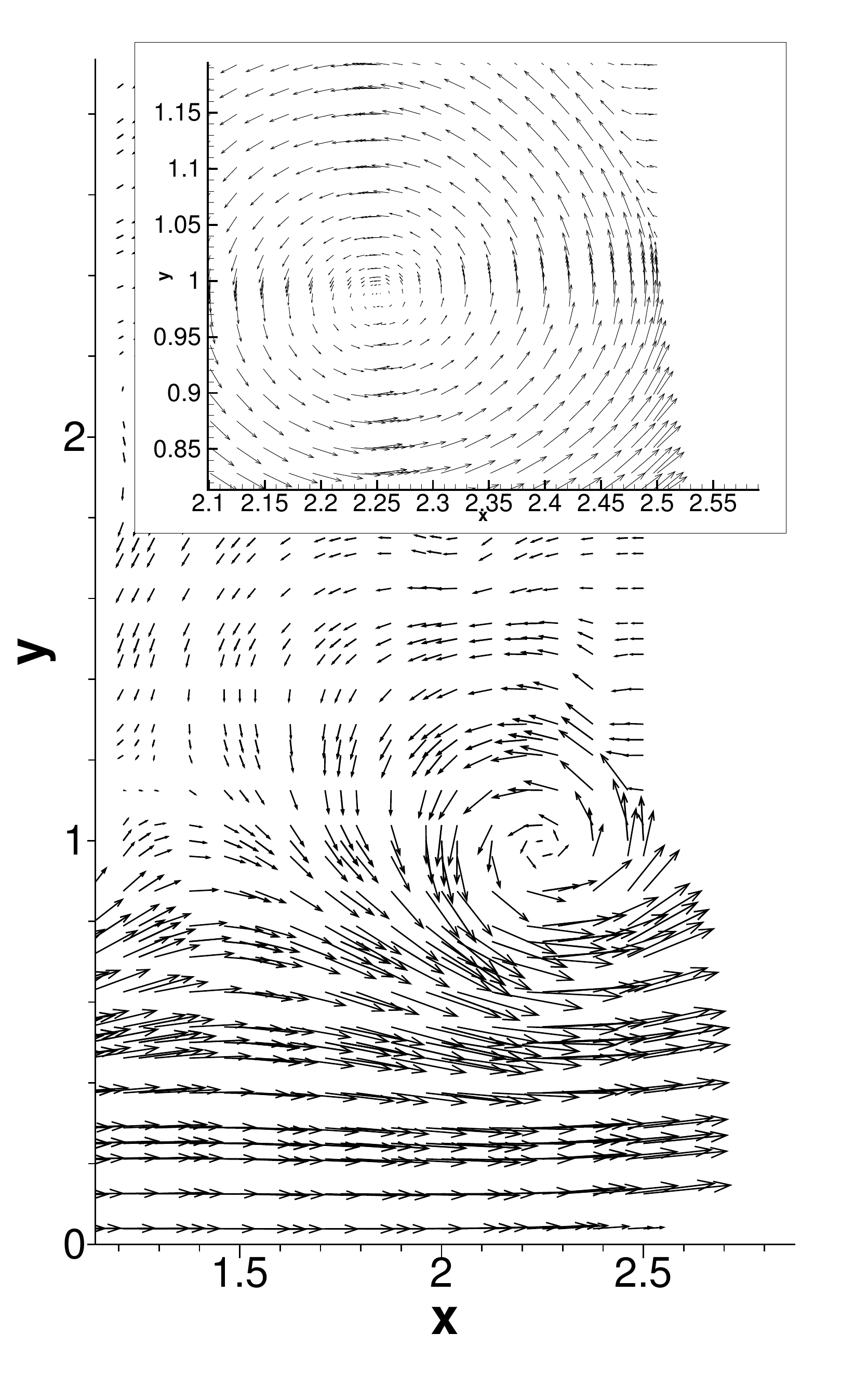}(c)
}
\centerline{
  \includegraphics[width=1.8in]{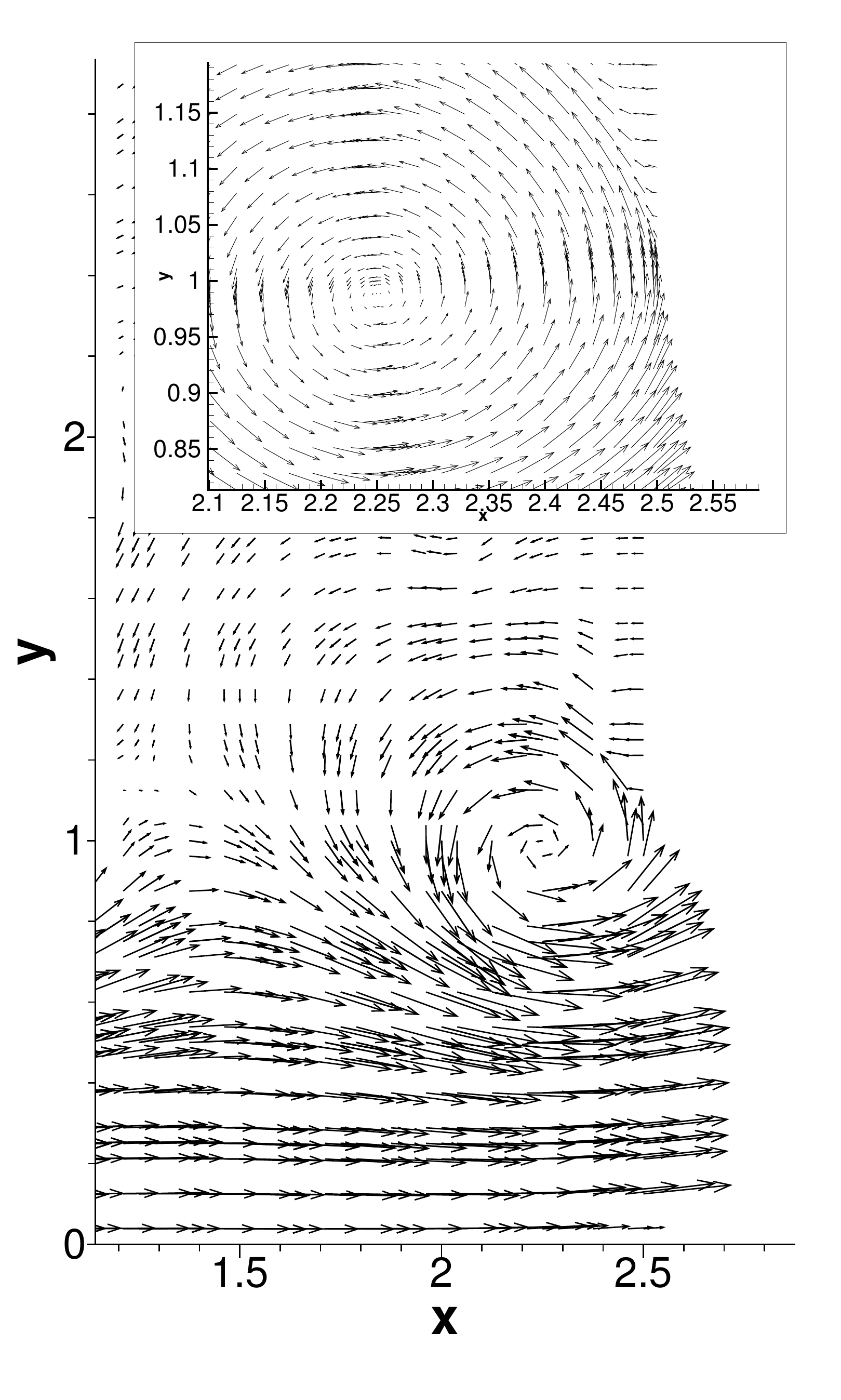}(d)
  \includegraphics[width=1.8in]{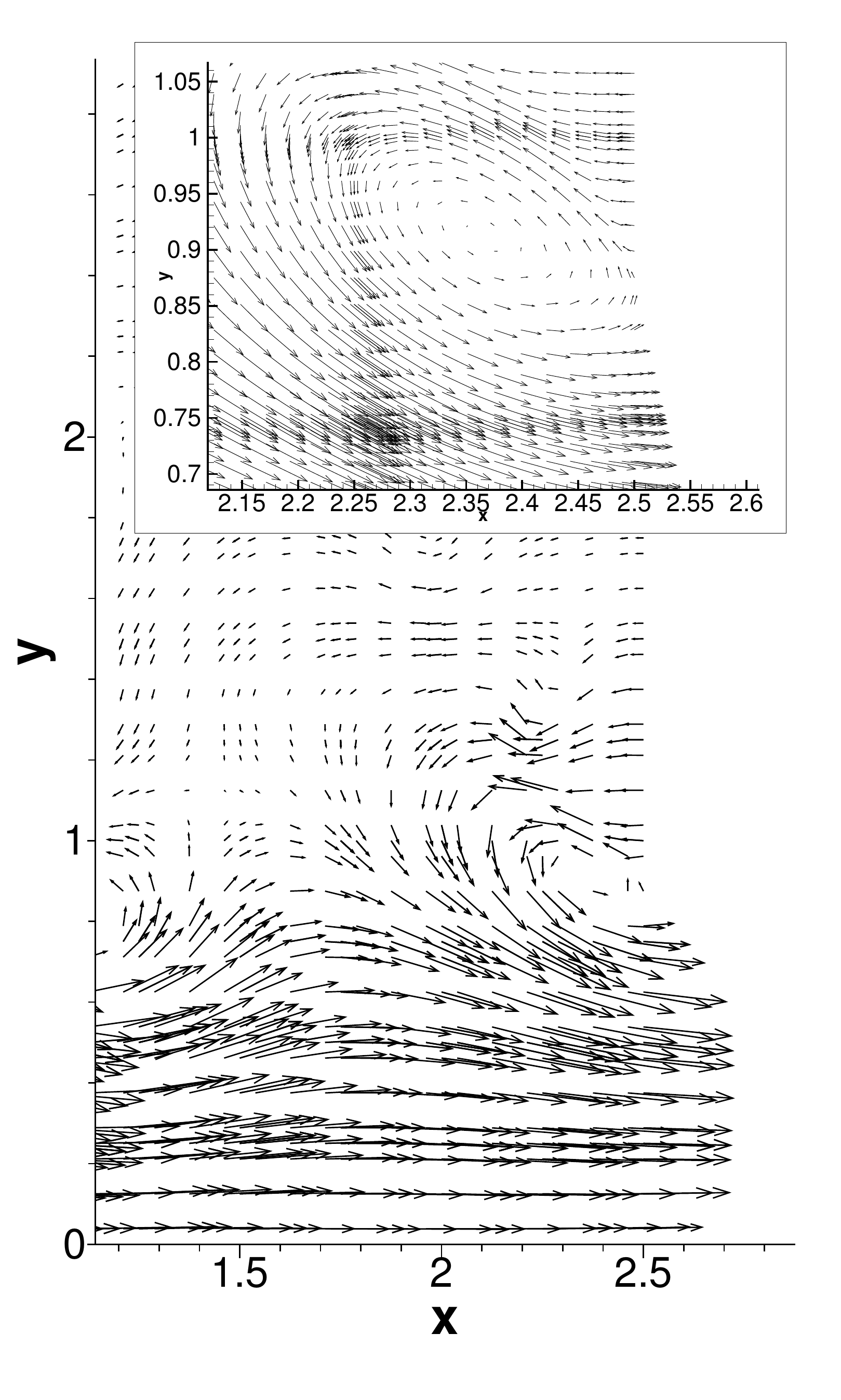}(e)
  \includegraphics[width=1.8in]{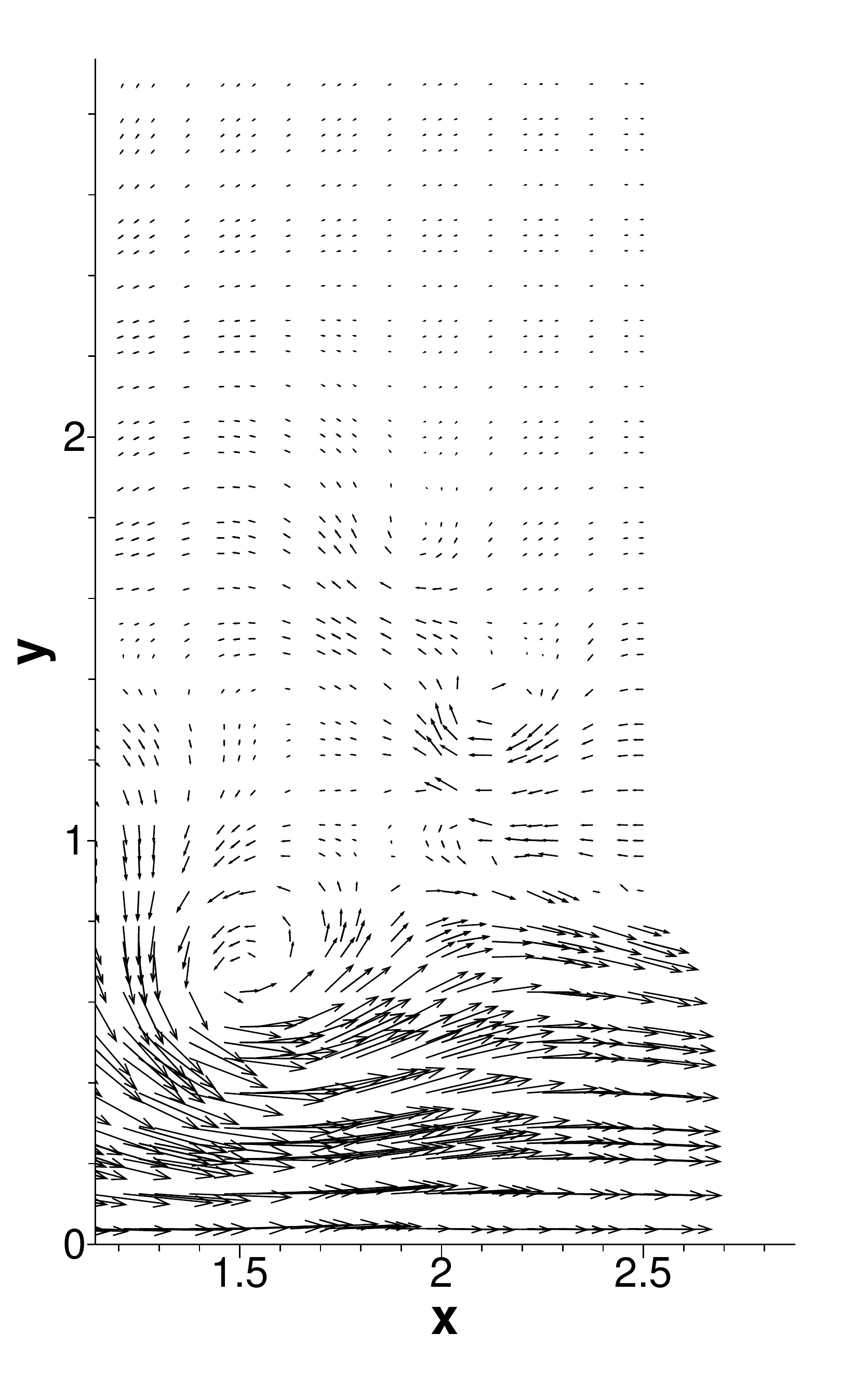}(f)
}
\caption{
  Impinging jet ($Re=5000$): a temporal sequence of snapshots of the velocity fields near
the right open boundary showing the discharge of a vortex from the domain,
computed using OBC-C as the open boundary condition.
(a) $t=593.4$,
(b) $t=593.7$,
(c) $t=594$.
(d) $t=594.3$,
(e) $t=594.6$.
%(f) $t=594.9$,
%(g) $t=595.2$,
(f) $t=595.5$.
Velocity vectors are plotted on every fourth quadrature points in each direction
within each element.
The insets of plots (b)--(e) show a blow-up view of the velocity vectors (shown
on every quadrature point) near the boundary.
}
\label{fig:vortex_distort_obc_C}
\end{figure}

\begin{figure}
\centerline{
  \includegraphics[width=1.8in]{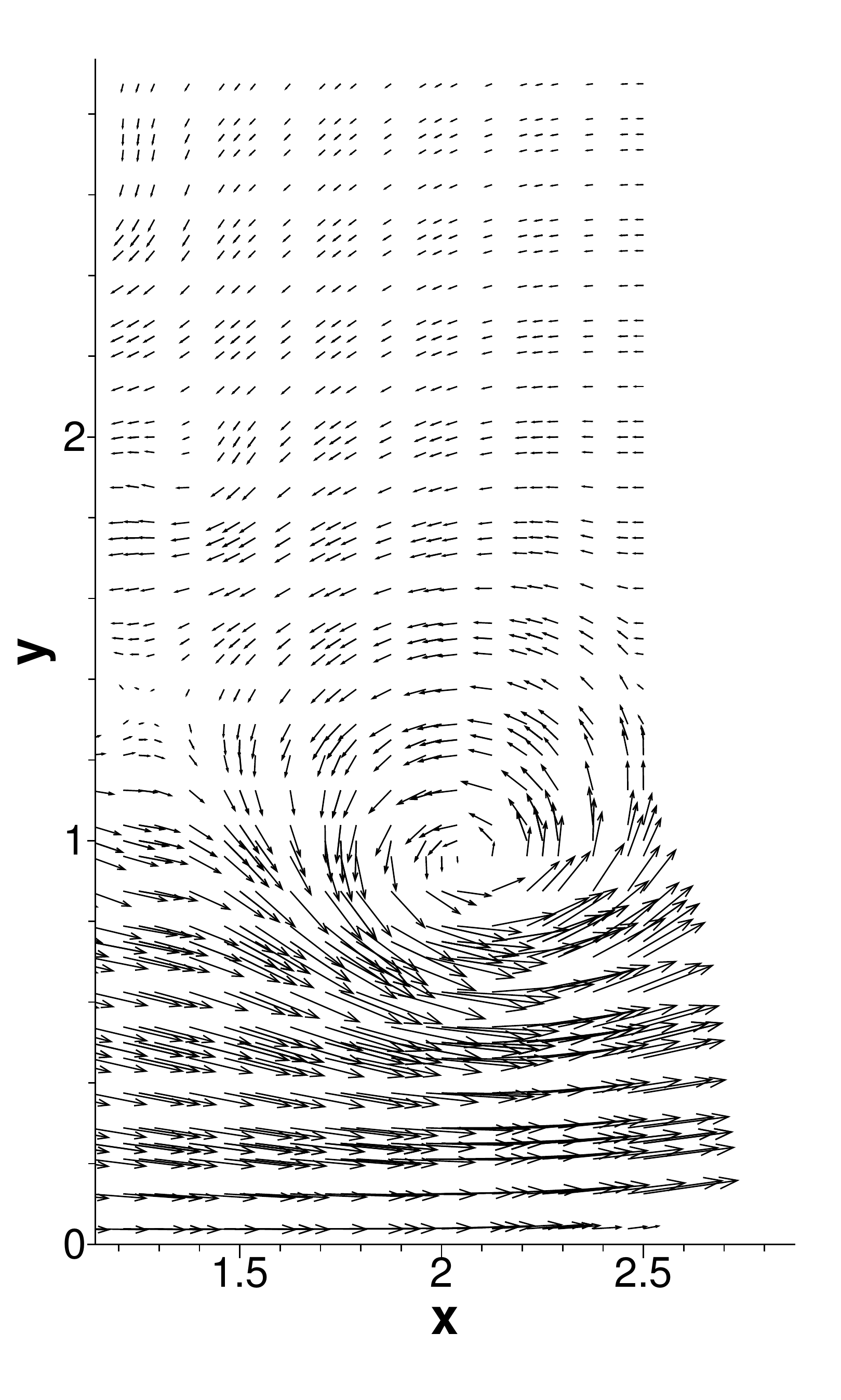}(a)
  \includegraphics[width=1.8in]{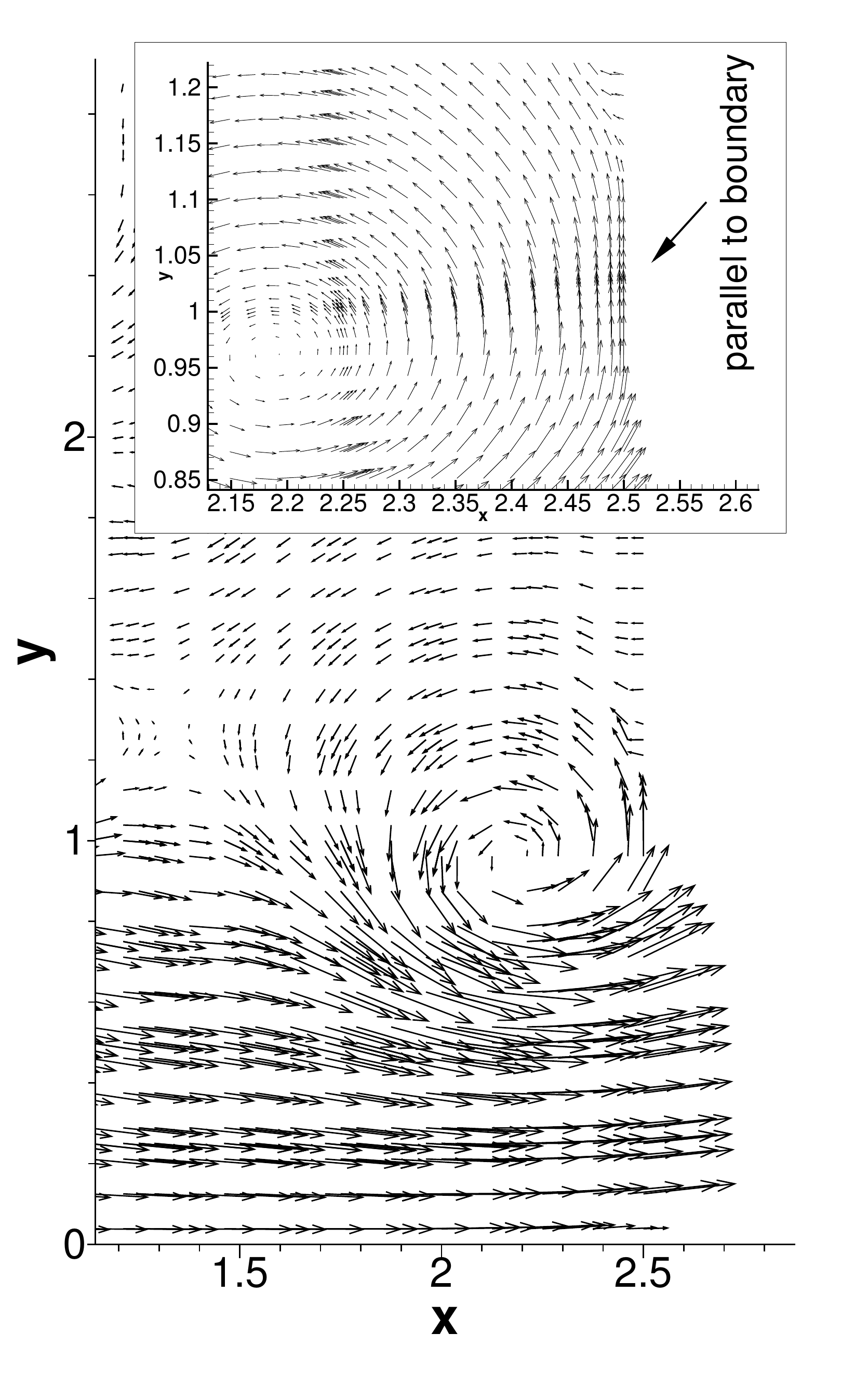}(b)
  \includegraphics[width=1.8in]{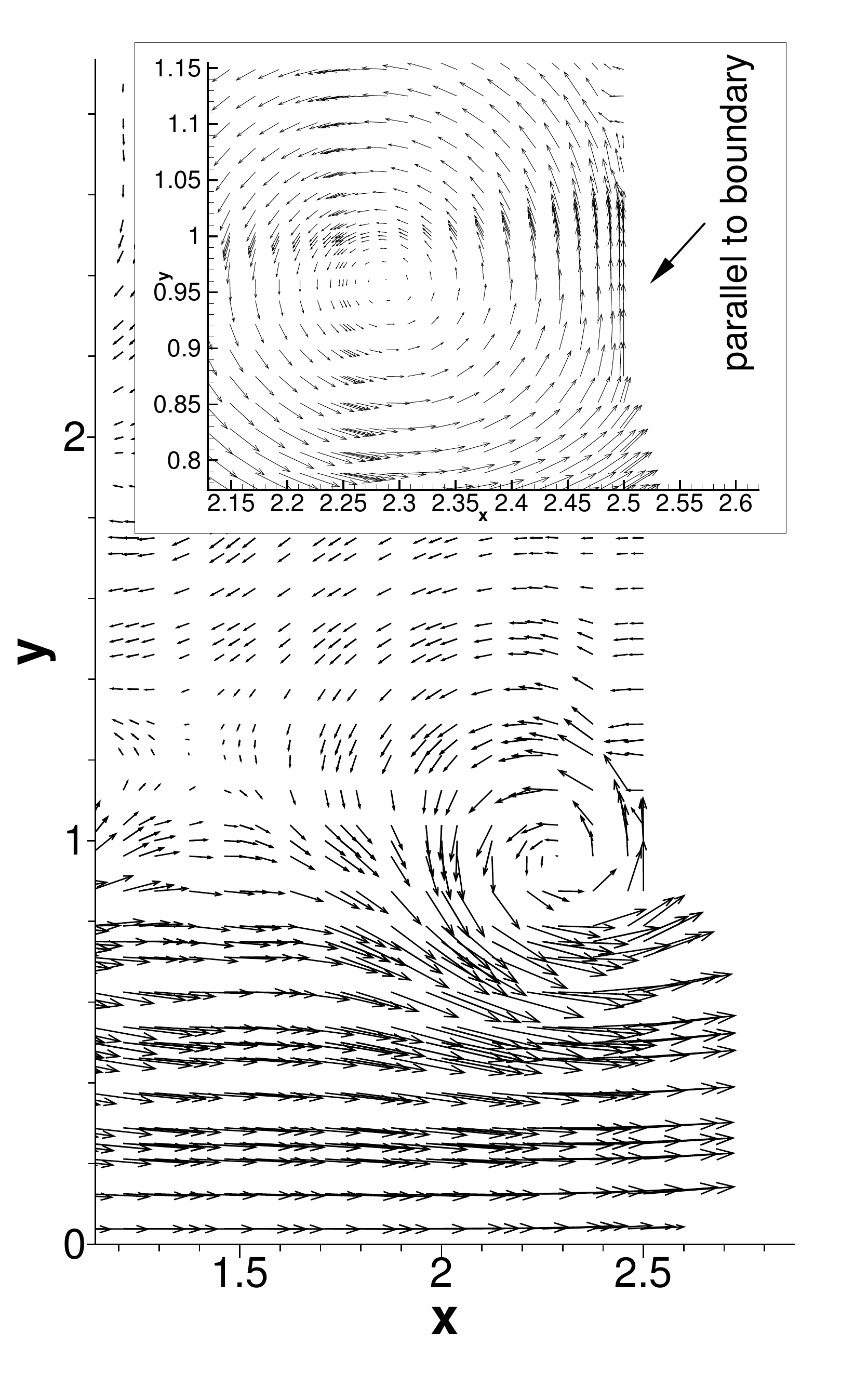}(c)
}
\centerline{
  \includegraphics[width=1.8in]{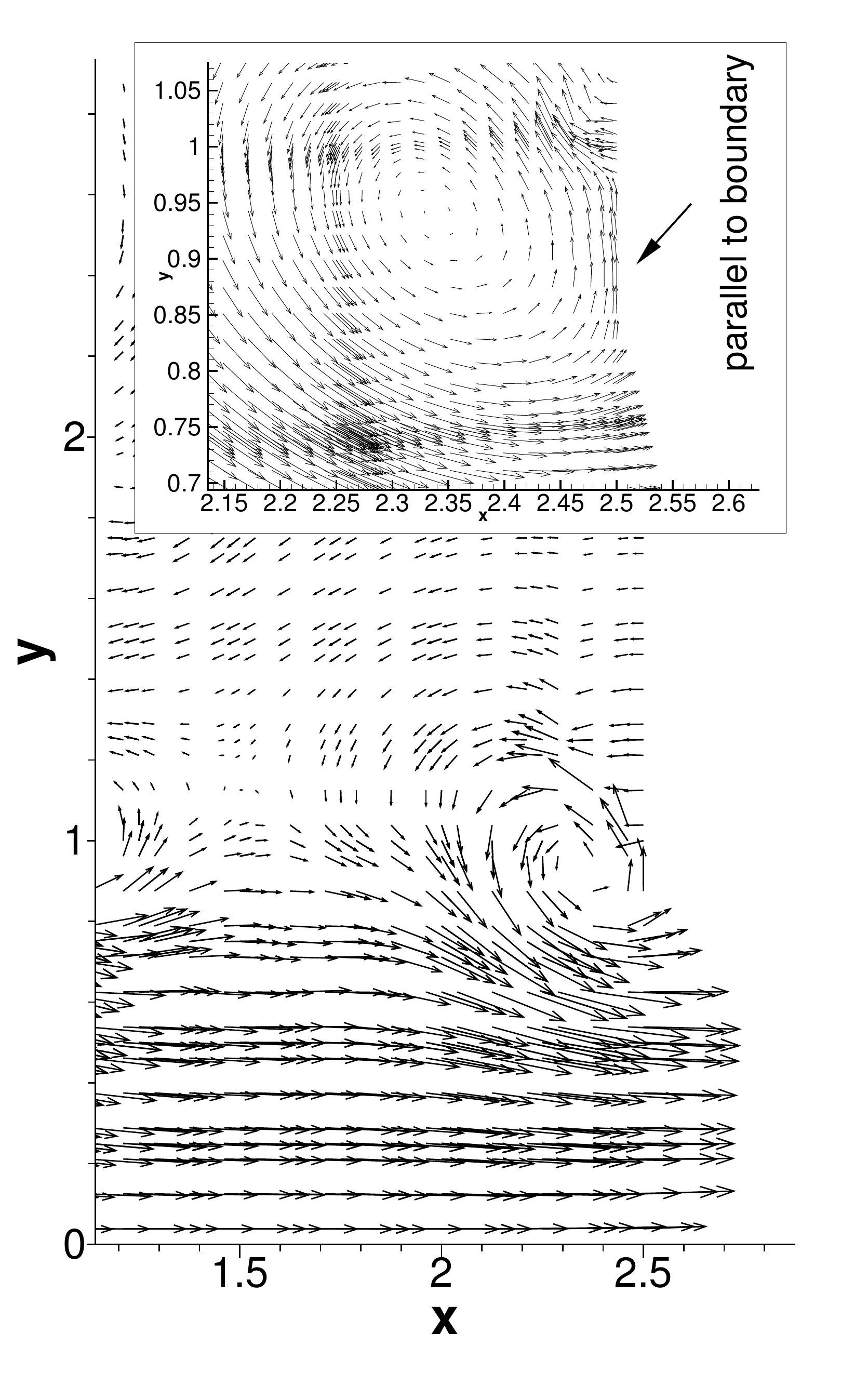}(d)
  \includegraphics[width=1.8in]{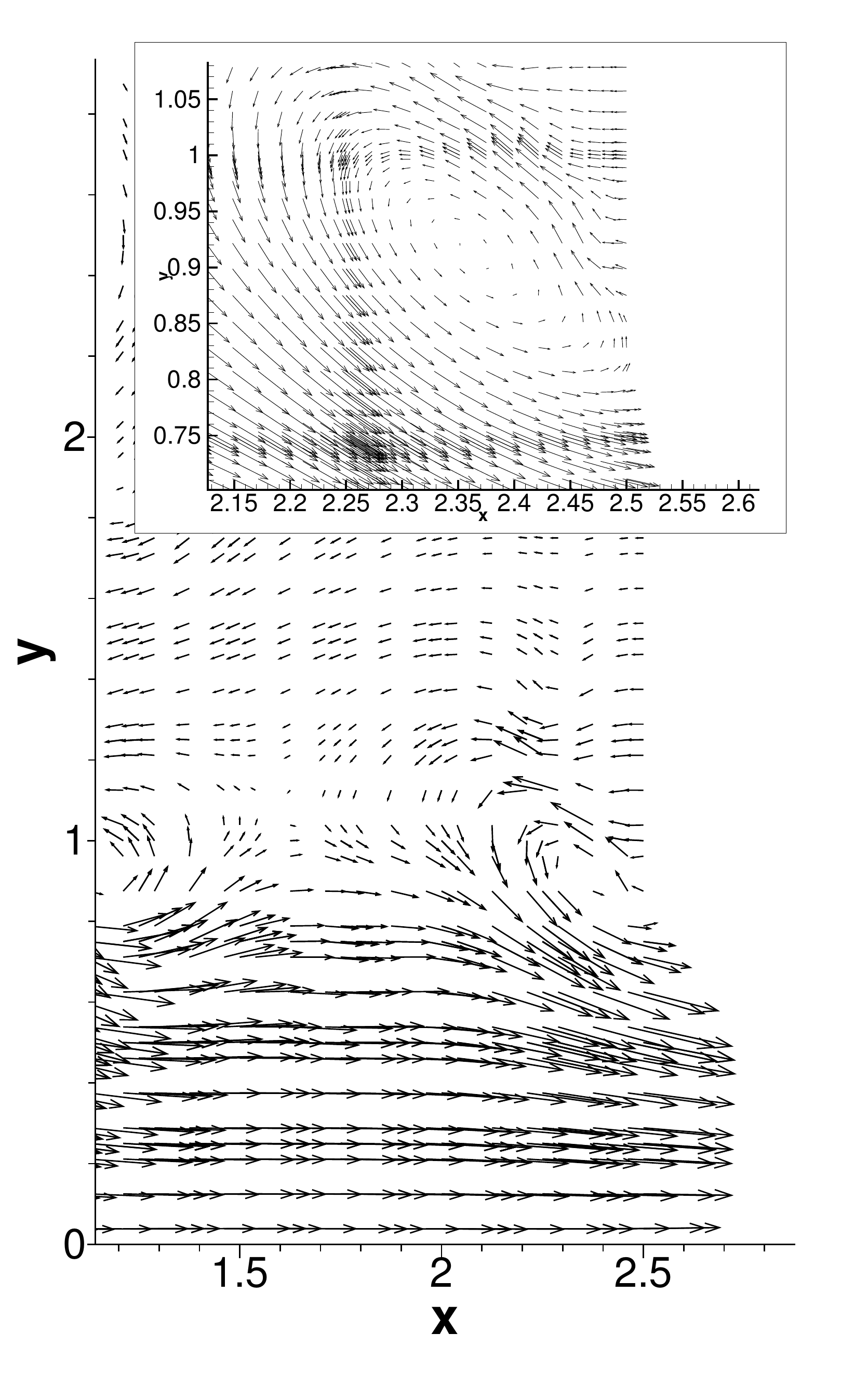}(e)
  \includegraphics[width=1.8in]{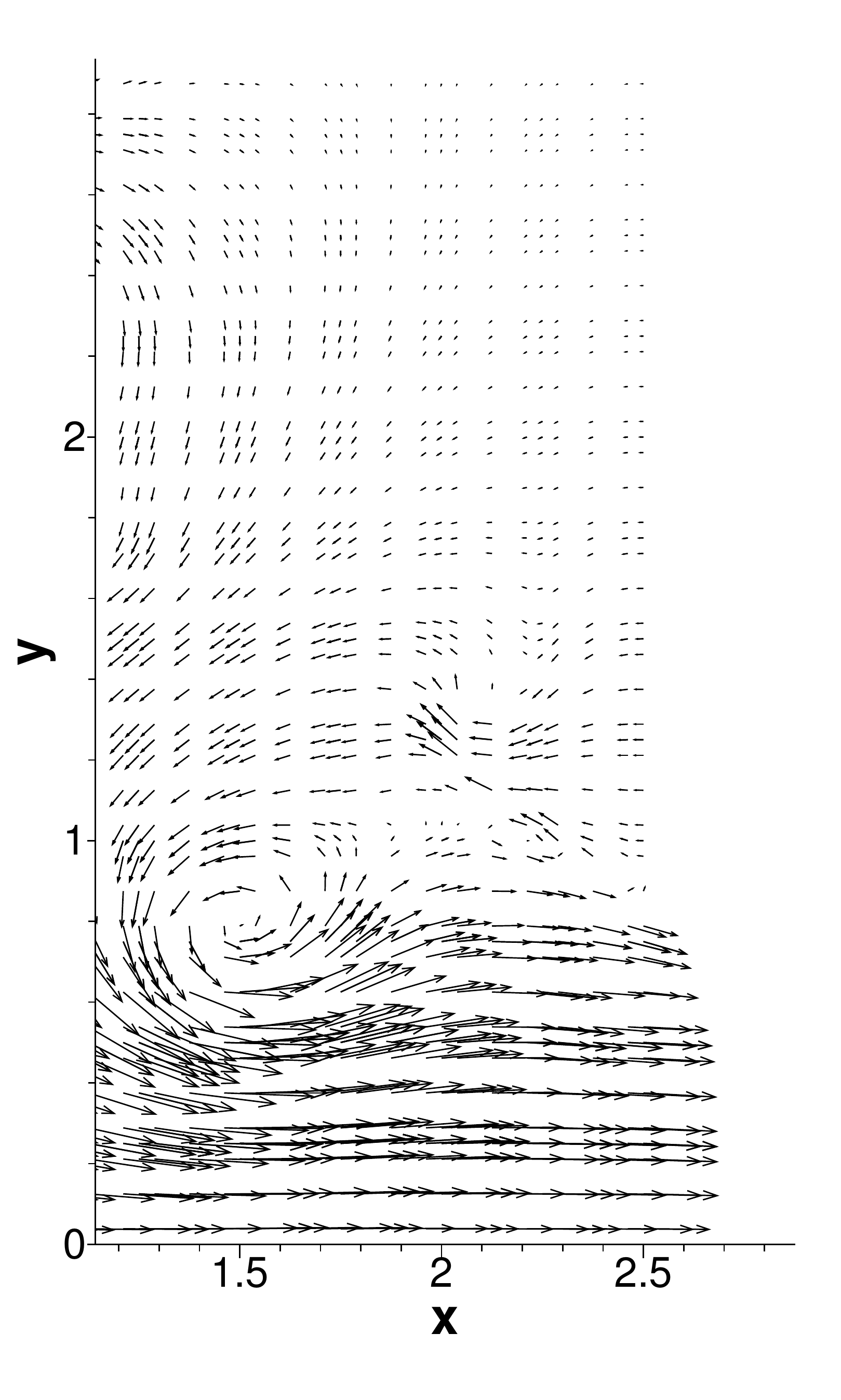}(f)
}
\caption{
  Impinging jet ($Re=5000$): a temporal sequence of snapshots of the velocity fields near
the right open boundary showing the discharge of a vortex from the domain,
computed using the open boundary condition from \cite{Dong2015obc} (without
the inertia term).
(a) $t=1137.75$,
(b) $t=1138.05$,
(c) $t=1138.35$.
(d) $t=1138.65$,
(e) $t=1138.95$.
%(f) $t=1139.25$,
%(g) $t=1139.55$,
(f) $t=1139.85$.
Velocity vectors are plotted on every fourth quadrature points in each direction
within each element.
The insets of plots (b)--(e) show a blow-up view of the velocity vectors (shown
on every quadrature point) near the boundary.
}
\label{fig:vortex_distort_convec_obc}
\end{figure}

Let us next take a closer view of the distortion to
the vortices as they exit the domain through the open boundary.
Figure \ref{fig:vortex_distort_obc_C} illustrates
the typical scenario when a vortex crosses the open boundary, obtained
with the current open boundary
condition OBC-C. This figure shows a temporal sequence of
velocity fields near the right open boundary and
the bottom wall. The insets of Figures \ref{fig:vortex_distort_obc_C}(b)--(e)
are magnified views of a section of the open boundary
near the vortex core.
As the vortex approaches the open boundary,
the velocity patterns show that
the vortex maintains an almost perfect circular shape,
with essentially no or very little
distortion (Figure \ref{fig:vortex_distort_obc_C}(a)-(d)).
Then as the vortex core moves very close to the boundary
a notable deformation to the vortex becomes evident (Figure \ref{fig:vortex_distort_obc_C}(e)).
The vortex deforms into an oval and is elongated in an oblique
direction to the boundary. 
The vortex retains an oval shape until it discharges
completely from the domain.
For comparison, Figure \ref{fig:vortex_distort_convec_obc}
shows a comparable and typical scenario of the vortex exiting
the domain obtained
using the open boundary condition from \cite{Dong2015obc},
but without the inertia (i.e.~time derivative) term therein
so that the boundary condition is also a traction-type condition.
We observe a similar process, with the initial circular
vortex distorted into an oval shape as it moves
out of the domain (Figure \ref{fig:vortex_distort_convec_obc}(e)).
But the velocity patterns of Figures \ref{fig:vortex_distort_obc_C}
and \ref{fig:vortex_distort_convec_obc} also reveal
a notable difference. 
The vortex in Figure \ref{fig:vortex_distort_convec_obc}
experiences another type of distortion, even before
the distortion into an oval becomes evident.
More specifically, we observe that, as the vortex approaches
the open boundary, on the section
of the boundary influenced by the vortex rotation 
and in its vicinity, the velocity vectors tend to point along the
tangential direction to the boundary.
This is evident from the insets of
Figures \ref{fig:vortex_distort_convec_obc}(b)-(d).
This makes the velocity pattern in that region less congruent
or incongruent with those outside the region,
thus causing an apparent distortion to the vortex.
This is especially evident
from Figures \ref{fig:vortex_distort_convec_obc}(c) and (d).
As the vortex further evolves in time,
this distortion seems to disappear and gives way to
the distortion into an oval vortex
(Figures \ref{fig:vortex_distort_convec_obc}(d)-(e)). 
By contrast, from the velocity patterns obtained using
the current open boundary condition we observe that
the vortex retains an essentially perfect shape
(see Figures \ref{fig:vortex_distort_obc_C}(b)-(d))
and does not experience such a distortion
as evidenced from Figure \ref{fig:vortex_distort_convec_obc},
before the oval deformation  kicks in.
These results suggest that the current open boundary conditions
can be more favorable compared with that of \cite{Dong2015obc}
in the sense that they can produce more congruent and more natural
velocity distributions near/at the open boundary
and cause less distortion to the vortices as they
pass through the boundary and exit the domain.

% force fy history

\begin{figure}
  \centering
  \includegraphics[height=2in]{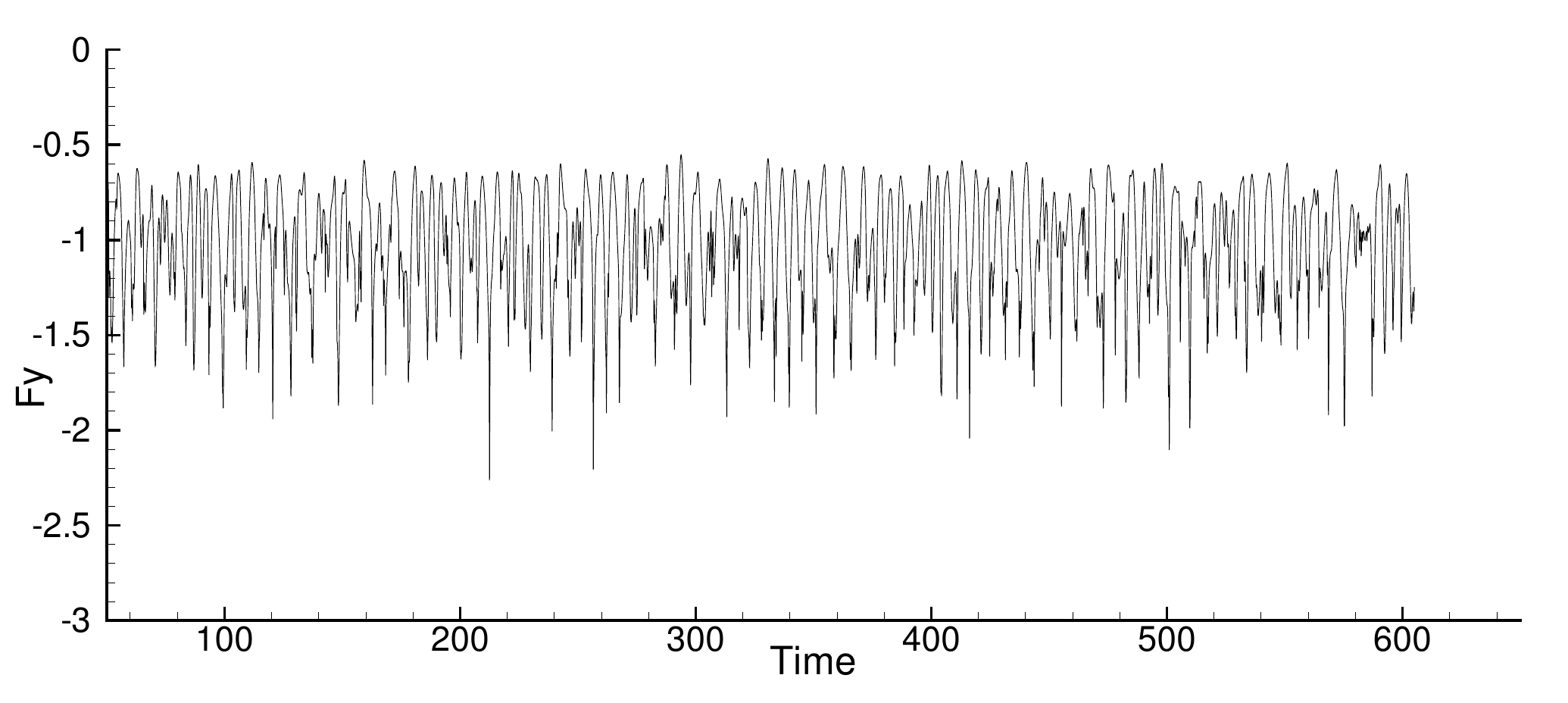}(a)
  \includegraphics[height=2in]{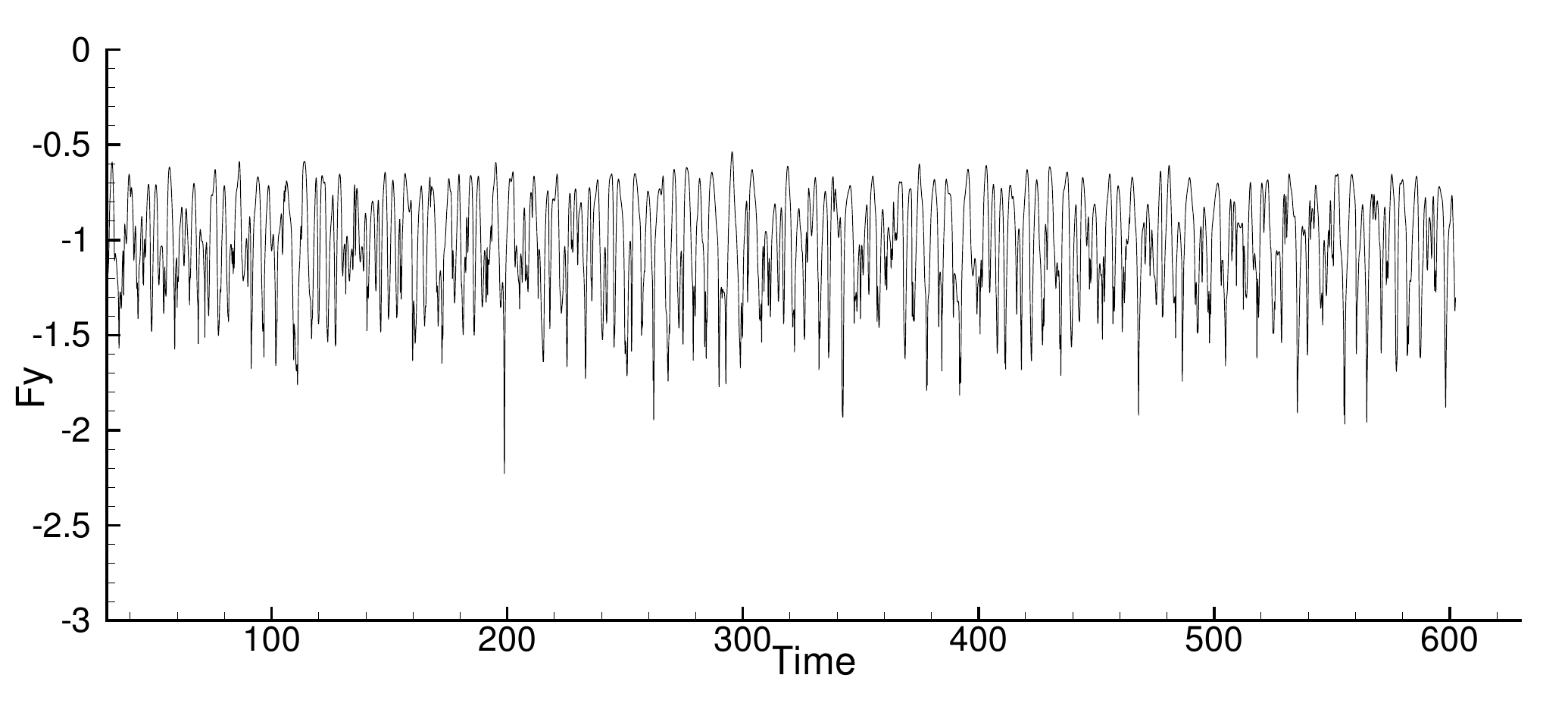}(a)
  \caption{
    Impinging jet: time histories of the vertical force on the bottom wall at
    $Re=10000$ computed using OBC-B (a) and OBC-C (b)
    as the open boundary condition.
  }
  \label{fig:jet_fy_hist}
\end{figure}

Long-time simulations have been performed using the current methods
for the impinging jet flow.
Figure \ref{fig:jet_fy_hist} illustrates the time histories of the vertical
force on the wall at $Re=10000$,
%at three Reynolds numbers,
%$Re=2000$, $5000$ and $10000$, 
which are
obtained using OBC-B and OBC-C 
as the open boundary condition, respectively.
The long history signals demonstrate the long-term
stability of the methods developed in the current
work, and that the flow has reached a statistically stationary
state.

% show mean and rms fy for high Re?

%% Summary, discuss computational cost briefly
\section{Concluding Remarks}
\label{sec:summary}

% what have you done in this paper?
% what is your contribution?
% what are the results?
% what are the implications of the results?
% what are outstanding problems?
% 

In this paper we have developed a set of new energy-stable 
open boundary conditions for simulating 
outflow/open-boundary problems of incompressible flows.
These boundary conditions can effectively overcome the backflow instability,
and give rise  to stable and accurate simulation results
when strong vortices or backflows occur at the open/outflow
boundary. 

% how are they developed?
% importance of two steps

The development of these boundary conditions involves two
steps. First, we devise a general form of the boundary conditions
that ensure the energy stability on the open boundary
by reformulating the boundary contribution into a quadratic form
involving a symmetric matrix. 
%based on a procedure 
%proposed by~\cite{Nordstrom2017}. 
Second, we 
require that the scale of the boundary dissipation,
 upon imposing the boundary conditions from
the previous step, should match a physical scale and thus
attain the final boundary conditions.

Both of these two steps are crucial. The first step ensures energy stability,
and the second step ensures physical correctness and accuracy.
Those boundary conditions resulting from the first step only are
referred to as ``OBC-A'' in the current paper, and the boundary
conditions ``OBC-B'' and ``OBC-C'' studied in the current paper
involve both steps
in the development. 
Extensive numerical experiments have been conducted to test
the accuracy and performance of these boundary conditions.
These tests show that, as expected,
all three conditions (OBC-A, OBC-B and OBC-C) produce stable simulations,
 even when strong vortices or backflows occur at
the open/outflow boundary. 
However, it is observed that
OBC-A in general gives rise to
poor or even unphysical simulation results, unless 
the algorithmic parameters take some ``optimal'' value for
the flow problem under study. It is further observed 
that the ``optimal'' parameter values for OBC-A 
are unfortunately flow-problem dependent. 
For example, for the cylinder
flow the ``optimal'' values for $a_{11}$ ($a_{22}$)
are around $a_{11}=a_{22}=-0.2$, while for the impinging jet
problem they are around $a_{11}=a_{22}=-0.5$ with OBC-A.
It is further noted that OBC-A gives rise to poor simulation results
with $a_{11}=a_{22}=-0.5$ for the cylinder flow
and with $a_{11}=a_{22}=-0.2$ for the impinging jet problem.
These observations suggest that OBC-A, i.e.~the energy-stable boundary conditions
resulting from the first step only, may only be of limited use in practice.
While OBC-A leads to stable computations, the simulation results in general
can deviate from the physical results considerably.
In contrast, the open boundary conditions  OBC-B and OBC-C
lead to favorable results in terms of both stability and accuracy.
Numerical experiments show that they  produce stable and
accurate simulation results, in terms of comparisons
with both the experimental measurements and the results produced by
other methods.

% what are the effects of alpha on OBC-B?
% do these methods work better? what are the advantages?
% how do they compare with previous such methods?

The open boundary conditions devised based on the quadratic form
and the eigen problem of the symmetric matrix involved therein
can be formulated into a traction-type condition; see equation \eqref{equ:obc}.
It is noted that the boundary conditions  herein
in general give rise to a non-zero
traction on the entire open boundary, in both the backflow
regions (if any) and the normal outflow regions.
This is in contrast to the energy-stable boundary conditions
in previous works in the literature (see e.g.~\cite{DongS2015}),
which results in a non-zero traction in
the backflow regions only and a zero traction in the normal 
outflow regions of the open boundary.

% is there any advantage with the methods?
% what is the advantage?
% how do OBC-B and OBC-C compare?

While their formulations are different, numerical experiments
indicate that OBC-B and OBC-C
tend to produce very similar or the same simulation results.
The various numerical experiments appear to give a sense that
the fluctuations in the physical quantities (e.g.~forces)
produced by OBC-B can at times be somewhat larger
than that by OBC-C, resulting in e.g.~higher values
in the largest amplitude in the time-history signals.
In such a sense, OBC-C may be somewhat more favorable when compared
with OBC-B. But the difference in their  simulation results, if any, is minor.

While the current open boundary conditions can allow the vortices
to cross the open/outflow boundary and exit the domain
in a smooth and fairly natural fashion, even at quite high Reynolds numbers,
a certain level of
distortion to the vortices is also evident during the process.
For example, an otherwise circular vortex can deform into
an oval while exiting the domain through the open boundary. 
This type of distortion is also observed with previous methods
(see e.g.~\cite{Dong2015obc}). However, as shown by the results in
Section \ref{sec:jet}, the current open boundary condition
seems more favorable compared with that of \cite{Dong2015obc}
(without the inertia term) in terms of the distortions, 
because the current condition
leads to velocity distributions more congruent on the open boundary
and in its vicinity.
In contrast, the boundary condition of \cite{Dong2015obc}
can lead to less congruent or incongruent velocity distributions
on sections of the open boundary where the vortices cross,
causing additional distortions to the vortices.

% convection-like OBCs

It should be pointed out that, since the current open boundary conditions
are formulated in a traction form, 
it is not difficult to extend 
these conditions to arrive at 
a set of corresponding ``convective-like'' energy-stable
open boundary conditions, by e.g.~incorporating an appropriate inertia
term in a similar way to \cite{Dong2015obc}. 
This may be desirable in term of the control over
the velocity field on the open/outflow boundaries.

% what are applications?
% what are the implications?

Backflow instability is one of the primary issues
encountered when scaling up the Reynolds number 
in the simulations of a large class of incompressible
flows, such as wakes, jets, shear layers,
and cardiovascular and respiratory flows. 
The method developed in the current work
provides a new effective technique to algorithmically 
eliminate the backflow instability.
Algorithmic elimination of the backflow instability
can be critical to and will be instrumental in 
flow simulations at high (and moderate) Reynolds numbers.
For example, it enables one to employ a significantly 
smaller computational domain (permitted by accuracy consideration),
thus leading to a much higher spatial resolution than otherwise, even with
the same mesh size. 
The current work contributes a useful and effective tool 
toward the numerical simulations of such challenging problems.

% what else to discuss here?

\section*{Appendix A.~Proof of Theorem \ref{thm:thm_1}}

Since $\mathbf{A}$ and $\mathbf{G}$ are symmetric and real matrices, all
their eigenvalues are real.
None of the eigenvalues of $\mathbf{A}$ is zero
because $\text{det}(\mathbf{A})\neq 0$.
Suppose $\lambda$ ($\lambda\neq 0$) is an eigenvalue of $\mathbf{A}$. Then
\begin{equation}
  \begin{split}
  0 &= \text{det}\begin{bmatrix}-\lambda\mathbf{I} & -\mathbf{I} \\
    -\mathbf{I} & \mathbf{G}-\lambda\mathbf{I}\end{bmatrix}
  = \text{det}(-\lambda\mathbf{I})
  \text{det}\left[(\mathbf{G}-\lambda\mathbf{I})-
    (-\mathbf{I})(-\lambda\mathbf{I})^{-1}(-\mathbf{I})  \right] \\
  &= (-\lambda)^m\text{det}\left[\mathbf{G}-
    \left(\lambda-\frac{1}{\lambda} \right)\mathbf{I}  \right] \\
  &\Longrightarrow \quad
  \text{det}\left[\mathbf{G}-
    \left(\lambda-\frac{1}{\lambda} \right)\mathbf{I}  \right] = 0,
  \end{split}
  \label{equ:app_A_1}
\end{equation}
where we have used the Schur complement.
Therefore $\left(\lambda-\frac{1}{\lambda} \right)$ is an eigenvalue
of $\mathbf{G}$.
Suppose $\xi$ is an eigenvalue of $\mathbf{G}$,
and $\lambda-\frac{1}{\lambda}=\xi$.
Then
$
\lambda = \frac{\xi}{2} \pm \sqrt{\left(\frac{\xi}{2} \right)^2+1}
\neq 0.
$
So the steps in equation \eqref{equ:app_A_1} can be reversed.
We conclude that $\lambda$ is an eigenvalue of the matrix $\mathbf{A}$.

Suppose
$
\begin{bmatrix}\mathbf{Z} \\ -\lambda\mathbf{Z}  \end{bmatrix}
$
is an eigenvector of $\mathbf{A}$ corresponding to the eigenvalue $\lambda$.
Then $\mathbf{Z}\neq 0$ and
\begin{equation}
  \begin{split}
0 &= \begin{bmatrix}-\lambda\mathbf{I} & -\mathbf{I} \\
  -\mathbf{I} & \mathbf{G}-\lambda\mathbf{I}\end{bmatrix}
\begin{bmatrix}\mathbf{Z} \\ -\lambda\mathbf{Z}  \end{bmatrix}
= \begin{bmatrix} 0 \\
  -\mathbf{Z}-\lambda(\mathbf{G}-\lambda\mathbf{I})\mathbf{Z}  \end{bmatrix}
= \begin{bmatrix} 0 \\
  -\lambda\left(\mathbf{G}-\left(\lambda-\frac{1}{\lambda}\right)\mathbf{I}\right)\mathbf{Z}  \end{bmatrix} \\
&\Longrightarrow \quad
\left(\mathbf{G}-\left(\lambda-\frac{1}{\lambda}\right)\mathbf{I}\right)\mathbf{Z} =0.
  \end{split}
  \label{equ:app_A_2}
\end{equation}
We conclude that $\mathbf{Z}$ is an eigenvector of $\mathbf{G}$
corresponding to the eigenvalue $(\lambda-\frac{1}{\lambda})$.
Now suppose $(\lambda-\frac{1}{\lambda})$ is an eigenvalue
of $\mathbf{G}$ and $\mathbf{Z}$ is the corresponding eigenvector.
Then $\lambda\neq 0$, and
the steps in \eqref{equ:app_A_2} can be reversed.
So we conclude that
$
\begin{bmatrix}\mathbf{Z} \\ -\lambda\mathbf{Z}  \end{bmatrix}
$
is an eigenvector of $\mathbf{A}$ corresponding
to the eigenvalue $\lambda$.

\section*{Appendix B.~Numerical Algorithm}

In this appendix we provide a summary of our algorithm 
for numerically solving the equations \eqref{equ:nse}--\eqref{equ:continuity},
together with the boundary conditions \eqref{equ:dbc} and \eqref{equ:obc}.
This algorithm is based on the scheme originally developed
in  \cite{Dong2015obc} (presented in Section 2.4 of \cite{Dong2015obc}).

We modify the outflow/open boundary condition \eqref{equ:obc} slightly
by adding a source term as follows,
\begin{equation}
  -p\mathbf{n} + \nu\mathbf{n}\cdot\nabla\mathbf{u}
  - \mathbf{E}(\mathbf{u},\partial\Omega_o) = \mathbf{f}_b(\mathbf{x},t),
  \quad \text{on} \ \partial\Omega_o
  \label{equ:obc_mod}
\end{equation}
where $\mathbf{f}_b$ is a prescribed source term on $\partial\Omega_o$
for the purpose of numerical testing only, and
it will be set to $\mathbf{f}_b=0$ in actual simulations.
$\mathbf{E}(\mathbf{u},\partial\Omega_o)$ is given by
either \eqref{equ:def_obc_E}.
%or \eqref{equ:def_obc_E2}.

The following algorithm is for equations \eqref{equ:nse}--\eqref{equ:continuity},
together with the boundary conditions \eqref{equ:dbc} on $\partial\Omega_d$
and \eqref{equ:obc_mod} on $\partial\Omega_o$.
Let $n\geqslant 0$ denote the time step index,
and $(\cdot)^n$ denote the variable $(\cdot)$
at time step $n$. Let $J$ ($J=1$ or $2$) denote the temporal
order of accuracy of the scheme.
Given $\mathbf{u}^n$, we compute
$(p^{n+1},\mathbf{u}^{n+1})$ successively in a de-coupled manner in two steps: \\
\underline{For $p^{n+1}$:}
\begin{subequations}
\begin{equation}
\frac{\gamma_0\tilde{\mathbf{u}}^{n+1}-\hat{\mathbf{u}}}{\Delta t}
+ \mathbf{u}^{*,n+1}\cdot\nabla\mathbf{u}^{*,n+1}
+ \nabla p^{n+1}
+ \nu \nabla\times\nabla\times\mathbf{u}^{*,n+1}
= \mathbf{f}^{n+1}
\label{equ:pressure_1}
\end{equation}
\begin{equation}
\nabla\cdot\tilde{\mathbf{u}}^{n+1} = 0
\label{equ:pressure_2}
\end{equation}
\begin{equation}
\mathbf{n}\cdot\tilde{\mathbf{u}}^{n+1}
 = \mathbf{n} \cdot \mathbf{w}^{n+1},
\quad \text{on} \ \partial\Omega_d
\label{equ:pressure_3}
\end{equation}
\begin{equation}
p^{n+1} =
\nu\mathbf{n}\cdot\nabla\mathbf{u}^{*,n+1}\cdot\mathbf{n}
- \mathbf{n}\cdot\mathbf{E}(\mathbf{u}^{*,n+1},\partial\Omega_o)
-\mathbf{f}_b^{n+1}\cdot\mathbf{n},
\quad \text{on} \ \partial\Omega_o.
\label{equ:pressure_4}
\end{equation}
\end{subequations}
\underline{For $\mathbf{u}^{n+1}$:}
\begin{subequations}
\begin{equation}
\frac{\gamma_0\mathbf{u}^{n+1}-\gamma_0\tilde{\mathbf{u}}^{n+1}}{\Delta t}
- \nu\nabla^2\mathbf{u}^{n+1}
= \nu \nabla\times\nabla\times\mathbf{u}^{*,n+1}
\label{equ:velocity_1}
\end{equation}
\begin{equation}
\mathbf{u}^{n+1} = \mathbf{w}^{n+1},
\quad \text{on} \ \partial\Omega_d
\label{equ:velocity_2}
\end{equation}
\begin{equation}
  \mathbf{n}\cdot\nabla\mathbf{u}^{n+1} = \frac{1}{\nu}\left[
p^{n+1}\mathbf{n}
+ \mathbf{E}(\mathbf{u}^{*,n+1},\partial\Omega_o)
- \nu\left(\nabla\cdot\mathbf{u}^{*,n+1}  \right)\mathbf{n}
+ \mathbf{f}_b^{n+1}
\right], 
\quad \text{on} \ \partial\Omega_o.
\label{equ:velocity_3}
\end{equation}
\end{subequations}
In the above equations, $\Delta t$ is the time step size,
$\mathbf{n}$ is the outward-pointing unit vector normal
to the boundary, and $\tilde{\mathbf{u}}^{n+1}$ is an
auxiliary variable approximating $\mathbf{u}^{n+1}$.
$\mathbf{u}^{*,n+1}$ is a $J$-th order explicit approximation
of $\mathbf{u}^{n+1}$ given by
\begin{equation}
\mathbf{u}^{*,n+1} = \left\{
\begin{array}{ll}
\mathbf{u}^n, & J=1, \\
2\mathbf{u}^n - \mathbf{u}^{n-1}, & J=2.
\end{array}
\right.
\end{equation}
$\hat{\mathbf{u}}$ and the constant $\gamma_0$ are such that
the expressions
$
\frac{1}{\Delta t}(\gamma_0\mathbf{u}^{n+1}-\hat{\mathbf{u}})
$
and
$
\frac{1}{\Delta t}(\gamma_0\tilde{\mathbf{u}}^{n+1}-\hat{\mathbf{u}})
$
approximate
$
\left.
\frac{\partial\mathbf{u}}{\partial t}
\right|^{n+1}
$
with a $J$-th order backward differentiation
formula, and  they
are given by
\begin{equation}
\hat{\mathbf{u}} = \left\{
\begin{array}{ll}
\mathbf{u}^n, & J=1, \\
2\mathbf{u}^n-\frac{1}{2}\mathbf{u}^{n-1}, & J=2,
\end{array}
\right.
\qquad\quad
\gamma_0 = \left\{
\begin{array}{ll}
1, & J=1, \\
\frac{3}{2}, & J=2.
\end{array}
\right.
\end{equation}

By taking the $L^2$ inner products between a test function
and equation \eqref{equ:pressure_1}, one can obtain
the weak form about $p^{n+1}$. By taking the $L^2$ between
a test function and the equation obtained by summing up
equations \eqref{equ:pressure_1} and \eqref{equ:velocity_1},
one can get the weak form about $\mathbf{u}^{n+1}$.
Let
$
H_{p0}^1(\Omega) = \left\{ \
  v\in H^1(\Omega) \ : \
  v|_{\partial\Omega_o} = 0
\ \right\},
$
and
$
H_{u0}^1(\Omega) = \left\{ \
  v \in H^1(\Omega) \ : \
  v|_{\partial\Omega_d} = 0
\ \right\}.
$
Let $q\in H_{p0}^1(\Omega)$ and $\varphi\in H_{u0}^1(\Omega)$
denote the test functions for the pressure and velocity.
Then the weak form for $p^{n+1}$ is
\begin{equation}
\begin{split}
\int_{\Omega} \nabla p^{n+1}\cdot \nabla q
= & \int_{\Omega} \mathbf{G}^{n+1}\cdot\nabla q
- \nu\int_{\partial\Omega_d\cup\partial\Omega_o}
     \mathbf{n}\times \bm{\omega}^{*,n+1}\cdot\nabla q 
- \frac{\gamma_0}{\Delta t}\int_{\partial\Omega_d}
     \mathbf{n}\cdot \mathbf{w}^{n+1} q, 
\ \forall q\in H_{p0}^1(\Omega),
\end{split}
\label{equ:p_weakform}
\end{equation}
where $\bm{\omega}^{*,n+1} = \nabla\times\mathbf{u}^{*,n+1}$, and
\begin{equation}
  \mathbf{G}^{n+1} = \mathbf{f}^{n+1} + \frac{\hat{\mathbf{u}}}{\Delta t}
- \mathbf{u}^{*,n+1}\cdot\nabla\mathbf{u}^{*,n+1}.
\label{equ:G_expr}
\end{equation}
The weak form for $\mathbf{u}^{n+1}$ is
\begin{equation}
\begin{split}
\frac{\gamma_0}{\nu\Delta t} \int_{\Omega}\mathbf{u}^{n+1}\varphi
&+ \int_{\Omega}\nabla\varphi\cdot\nabla\mathbf{u}^{n+1}
= \frac{1}{\nu}\int_{\Omega}\left(
  \mathbf{G}^{n+1}-\nabla p^{n+1}
\right)\varphi \\
&
+ \frac{1}{\nu} \int_{\partial\Omega_o}
   \left[
    p^{n+1}\mathbf{n} + \mathbf{E}(\mathbf{u}^{*,n+1},\partial\Omega_o)
    + \mathbf{f}_b^{n+1}
    - \nu \left(\nabla\cdot\mathbf{u}^{*,n+1}  \right)\mathbf{n}
  \right] \varphi,
\quad \forall \varphi \in H_{u0}^1(\Omega).
\end{split}
\label{equ:u_weakform}
\end{equation}
The weak forms \eqref{equ:p_weakform} and \eqref{equ:u_weakform}
can be discretized in space using a high-order spectral
element method~\cite{Dong2015obc}
or a combined spectral-element and Fourier-spectral
method~\cite{DongK2005,Dong2007}.

Given $\mathbf{u}^n$,
the following operations are involved in the final algorithm
within a time step for computing $p^{n+1}$ and $\mathbf{u}^{n+1}$:
\begin{itemize}

\item
  Solve equation \eqref{equ:p_weakform}, together with the Dirichlet
  condition \eqref{equ:pressure_4} on $\partial\Omega_o$,
  for $p^{n+1}$.

\item
  Solve equation \eqref{equ:u_weakform}, together with the Dirichlet
  condition \eqref{equ:velocity_2} on $\partial\Omega_d$,
  for $\mathbf{u}^{n+1}$.

\end{itemize}

\section*{Acknowledgement}
This work was partially supported by 
 NSF (DMS-1318820, DMS-1522537).

%\newpage
%
\bibliographystyle{plain}
\bibliography{obc,mypub,nse,sem,cyl}
%contact_line,interface,mypub,basis,nse,time_integration,sem}
%pfem,time_integration,mypub,sem,taylor_couette,precon}

\end{document}